\documentclass[12pt,oneside]{book}
\usepackage[top=1.3in,left=1.5in,bottom=1in,right=1in,headsep=0.5in]{geometry}

\usepackage{setspace}
\usepackage{fancyhdr}

\newcommand\startchapter[1]{\chapter{#1}\thispagestyle{myheadings}}
\newcommand\startappendix[1]{\chapter{#1}\thispagestyle{myheadings}}
\newcommand\startfirstchapter[1]{\chapter{#1}}

\newcommand\TOCadd[1]{\newpage\phantomsection\addcontentsline{toc}{chapter}{#1}}

\usepackage{tocloft}

\newcommand{\HRule}{\rule{\linewidth}{0.5mm}}

\newcommand\de{\partial}
\newcommand{\lr}[1]{\left( #1 \right)}
\newcommand{\pder}[2]{\frac{\partial #1}{\partial #2}}

\newcommand{\oser}[1]{{\cal O}\lr{#1}}

\def\Im{\text{Im}}
\def\Re{\text{Re}}

\def\feq{f^{\rm eq.}}
\def\dperp{D^{\perp}}
\def\LC{{\cal L}}
\newcommand\ce{\varepsilon}
\def\AdS5{AdS$_{\rm 5}$}
\def\Lied{\mathsterling}

\def\TE{{\cal E}}
\def\TP{{\cal P}}
\def\TQ{{\cal Q}}
\def\TT{{\cal T}}
\def\JN{{\cal N}}
\def\JJ{{\cal J}}

\def\tr{\text{tr}\,}

\def\mmu{\upmu}
\def\muphi{\mu_{\Phi}}
\def\muphio{\mu_{\Phi,0}}
\def\rhophi{\rho_{\Phi}}
\def\rhophio{\rho_{\Phi,0}}
\def\delperp{\Delta_{\perp}}
\def\Jform{{\bf J}}

\def\mbb{\mathbb}

\def\mc{\mathcal}
\def\mf{\mathfrak}

\def\nnl{\nonumber\\}

\usepackage{dcolumn}
\usepackage{longtable}
\usepackage{multicol}
\usepackage{tipa}

\usepackage{amsmath}
\usepackage{amsthm}
\usepackage{amssymb}
\usepackage{braket}
\usepackage{bm}
\usepackage{upgreek}
\usepackage{bbm}
\usepackage{bbold}

\usepackage{xspace}
\usepackage{textcase}

\usepackage{wasysym}
\usepackage{graphics}
\usepackage{graphicx}   
\usepackage{subcaption}
\usepackage{xcolor}

\usepackage{hyperref}
\hypersetup{
    colorlinks,
    citecolor=black,
    filecolor=black,
    linkcolor=black,
    urlcolor=black
}

\usepackage{blindtext}

\begin{document}

\newcommand\thesistitle{Causal Theories of Relativistic Hydrodynamics}
\newcommand\nameanddegrees{%
Raphael E. Hoult\\
B.Sc., University of Winnipeg, 2018\\
M.Sc., University of Victoria, 2020}
\newcommand\panel{%
\HRule\\\panelist{Dr. P. Kovtun}{Supervisor}{Department of Physics and Astronomy, University of Victoria}
\HRule\\\panelist{Dr. A. Ritz}{Departmental Member}{Department of Physics and Astronomy, University of Victoria}
\HRule\\\panelist{Dr. B. Khouider}{External Member}{Department of Mathematics, University of Victoria}}
\newcommand\tpbreak{\\[\baselineskip]}

\newpage
\thispagestyle{empty}

\pagestyle{myheadings}
\pagenumbering{roman}
\fancypagestyle{plain}{%
\fancyhf{}
\fancyhead[R]{\thepage}
\renewcommand{\headrulewidth}{0pt}
\renewcommand{\footrulewidth}{0pt}
}

\pagebreak
{
\centering
{\LARGE \textbf{\thesistitle}}
\tpbreak
by
\tpbreak
\nameanddegrees
\tpbreak
A Dissertation Submitted in Partial Fulfillment of the \\
Requirements for the Degree of
\tpbreak
DOCTOR OF PHILOSOPHY
\tpbreak
in the Department of Physics and Astronomy\\
\vfill
\begin{figure}[h!]
    \centering
    \includegraphics[width=0.4\linewidth]{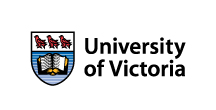}
    \label{fig:uvic-logo}
\end{figure}
\begin{tabular}{cl}
& \copyright\ Raphael E. Hoult, 2025\\
& \phantom{\copyright} University of Victoria
\end{tabular}
\tpbreak
All rights reserved. This dissertation may not be reproduced in whole or in part, by \\
photocopying or other means, without the permission of the author.\hfill
\newline\newline
\hfill We acknowledge and respect the L\textschwa $\stackrel{\textrm{,}}{\textrm{k}}$\textsuperscript{w}\textschwa \textipa{\ng}\textschwa n (Songhees and X\textsuperscript{w}seps\textschwa m/Esquimalt)\\
Peoples on whose territory the university stands, and the L\textschwa $\stackrel{\textrm{,}}{\textrm{k}}$\textsuperscript{w}\textschwa \textipa{\ng}\textschwa n and\hfill\\
\hfill \b WS\'ANE\'C Peoples whose historical relationships with the land continue to this day.\hfill
}
\pagebreak

\newpage
\TOCadd{Supervisory Committee}

{
\centering
{\large \textbf{\thesistitle}}
\tpbreak
by
\tpbreak
\nameanddegrees
\tpbreak
}

\newcommand\panelist[3]{\noindent #1, #2\\\noindent(#3)\tpbreak}
\vfill
\noindent {\Large \textbf{Supervisory Committee}}
\tpbreak
\panel
\vfill
\pagebreak

\newpage
\TOCadd{Abstract}

\noindent \textbf{Supervisory Committee}
\tpbreak
\panel

\begin{center}
\textbf{ABSTRACT}
\end{center}

The first-order textbook formulations of relativistic viscous hydrodynamics are unstable and acausal. These shortcomings may be rectified by using effective theories which maintain stability and causality. In this dissertation, which is intended to also serve as an introduction to the field, causal theories of relativistic hydrodynamics are developed and explored. Conditions are obtained for a linearized analysis to predict the non-linear causality of a theory, and constraints are found on short-wavelength dispersion relations as a consequence of ensuring stability in all reference frames. First-order causal theories of hydrodynamics are extracted from kinetic theory and holography descriptions. Finally, causal theories describing charged plasmas (one-form magnetohydrodynamics), and describing superfluids are developed.

\TOCadd{Table of Contents}\tableofcontents
\newpage

\hspace{0pt}
\vfill
\begin{center}
    {\Large \textbf{Note for the arXiv version:}}\\
    \vspace{1em}
    \fbox{\begin{minipage}{25em} Due to this document's status as a dissertation, a great deal of emphasis has been placed on my own work. Nevertheless, the document has also been written with a prospective student in mind, and I hope that those interested in learning more about causal theories of relativistic hydrodynamics will find it a useful addition to the literature.
    \end{minipage}}
\end{center}
\vfill
\hspace{0pt}
\newpage
\TOCadd{Contributions}
{\Huge \textbf{Contributions}}\\\\
\vspace{1em}\\
My results which contributed to the content of this dissertation are contained in four papers~\cite{Hoult:2021gnb,Hoult:2023clg,Hoult:2024cyx,Hoult:2024qph}. A more specific breakdown follows:
\begin{itemize}
    \item The results of~\cite{Hoult:2021gnb} are related in Chapter~\ref{chapter:micro}.
    \item The results of~\cite{Hoult:2023clg} are related in Chapter~\ref{chapter:math}.
    \item The results of~\cite{Hoult:2024cyx} are related in Chapter~\ref{chapter:extensions}.
    \item The results of~\cite{Hoult:2024qph} are related in Chapters~\ref{chapter:math} and~\ref{chapter:extensions}.
\end{itemize}
Additionally, work from my Masters' degree, which was published in~\cite{Hoult:2020eho}, contributed in part to Chapter~\ref{chapter:background}.
\TOCadd{List of Tables}\listoftables
\setcounter{lofdepth}{2}
\TOCadd{List of Figures}\listoffigures
\newpage
\TOCadd{Glossary of Notation}
{\Huge \textbf{Glossary of Notation}}\\
\begin{multicols*}{2}
[
The following is a glossary of terms and notation frequently used throughout the dissertation, included here for convenience and reference.
]
\begin{itemize}
    \item $\omega$ is the frequency, $k^j$ is the ($d$-dimensional) wavevector.
    \item $\oser{\de^n}$ refers to a term that comes with a parameter $\ce^n$ given the scaling $\de_\mu \to \ce \de_\mu$.
    \item $\nabla_\mu$ is the general relativistic covariant derivative.
    \item $\braket{\cdot}$ refers to an expectation value taken in some (usually thermal) ensemble, unless stated otherwise.
    \item The BDNK framework is a class of causal first-order theories of hydrodynamics. It works via a choice of hydrodynamic frame.
    \item A hydrodynamic frame is a choice for how to define the out-of-equilibrium effective variables of the hydrodynamic theory.
    \item The MIS framework is a class of causal theories of hydrodynamics. It works by introducing additional relaxational equations.
    \item The hydrostatic generating functional is a tool for obtaining hydrostatic constitutive relations.
    \item A perfect fluid is described by a fluid theory with only $\oser{\de^0}$ terms in the constitutive relations.
    \item A system is covariantly stable if it is both stable and causal.
    \item A physical transport coefficient is a coefficient in the constitutive relations which can be obtained from a microscopic theory via a Kubo formula.
    \item A transport parameter is any parameter appearing in the constitutive relations at first order, especially in a general fluid frame.
\end{itemize}
\end{multicols*}
\newpage
\TOCadd{Acknowledgements}

\begin{center}
ACKNOWLEDGEMENTS
\end{center}

\noindent I would like to thank:
\begin{description}
\item[My family,]
	for supporting me through this whole process, both the dissertation and the years leading up to it. Their belief and love has meant the world to me.
\item[My supervisor, Prof. Pavel Kovtun,]
	for his support and encouragement, and for his careful guidance and challenging questions. I have been incredibly enriched as a scientist and as a person from my time working with him.
\item[My committee members,]
    for their support and valuable input on this dissertation, as well as challenging questions over the years.
\item[The Natural Science and Engineering Research Council of Canada (NSERC),]
	for supporting my studies with a Canada Graduate Scholarship -- Doctoral (CGS-D) during the course of my studies.
\end{description}

\begin{figure}[b]
    \flushright
    \includegraphics[width=0.5\linewidth]{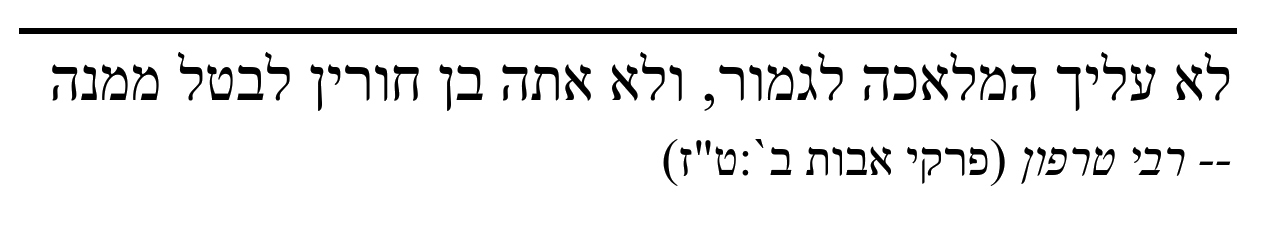}\\
    {\tiny It is not incumbent upon you to complete the work, but neither are you free to abstain. \textit{R. Tarfon, Pirkei Avot 2:15}}
\end{figure}

\newpage
\TOCadd{Dedication}

\begin{center}
DEDICATION
\end{center}

\begin{center}
To Kasey, for being my bedrock and my shelter in troubled waters. Without your support and love, this dissertation would not exist.
\end{center}

\newpage
\pagestyle{myheadings}
\pagenumbering{arabic}
\fancypagestyle{plain}{%
\fancyhf{}
\fancyhead[R]{\ifnum\thepage=1\relax\else\thepage\fi}
\renewcommand{\headrulewidth}{0pt}
\renewcommand{\footrulewidth}{0pt}
}

\newpage

	\startfirstchapter{Introduction}
\label{chapter:introduction}

Since ancient times, humanity has been fascinated by water. Correspondingly, the study of fluids is one of the oldest areas of physics, with contributions dating back to Archimedes, Newton, Bernoulli, and Euler. This culminated in the works of Navier (1822), Cauchy (1823), Poisson (1829), Saint-Venant (1837) and then Stokes (1845)~\cite{Bistafa:NS2024,DARRIGOL2002}, who wrote down (all more or less independently) the viscous theory of fluid mechanics. These equations, today dubbed the Navier-Stokes equations, form the basis for most quotidian fluid-dynamic simulations and models. The relativistic extension of the Navier-Stokes equations were written down by Eckart~\cite{Eckart-original} and by Landau and Lifshitz~\cite{LL6}.

Despite this long history, hydrodynamics remains an active area of research on a number of different fronts. In this dissertation, our interest resides primarily in relativistic hydrodynamics. Heavy ion collisions lead to the formation of the quark-gluon plasma, a microscopic phase which is amenable in some regimes to a hydrodynamic description~\cite{Romatschke:2017ejr,Rezzolla-Zanotti}. Binary neutron star mergers may be described in part using relativistic hydrodynamics~\cite{Faber:2012rw}. Black hole accretion disks are simulated using hydrodynamics coupled to magnetic fields (magnetohydrodynamics)~\cite{Abramowicz:2011xu}. Hydrodynamics has been found to describe the perturbations of black branes in asymptotically Anti-de Sitter spacetimes~\cite{Policastro:2001yc,Policastro:2002se,Kovtun:2003wp,Kovtun:2004de,Kovtun:2005ev}, leading to a holographic description of the hydrodynamic behaviour of strongly-coupled theories via the fluid/gravity correspondence~\cite{Bhattacharyya:2008jc,Banerjee:2008th,Erdmenger:2008rm}. Field theoretic techniques are being applied to understand statistical fluctuations which are not captured by the classical equations of motion for one-point functions, using the closed time path (CTP) and Schwinger-Keldysh formalisms~\cite{Liu:2018kfw}. It is an exciting time to be a theorist working in relativistic hydrodynamics.

This begs the question: what exactly \textit{is} hydrodynamics? The answer may be simply stated as follows:
\begin{center}
    \textit{Hydrodynamics is the long-wavelength, low-frequency effective description of physical systems which have conserved currents.}
\end{center}
In practice, this amounts to the statement that over very long times and long distances compared to some microscopic scale, systems which have conservation laws may be effectively described by the transport of effective degrees of freedom which obey those conservation laws. In a zero-temperature field theory, the most relevant degrees of freedom are gapless degrees of freedom which are easily excited. In a thermal field theory, on the other hand, such degrees of freedom are not the appropriate description -- after a fluctuation, they rapidly decohere~\cite{Liu:2018kfw}. With a finite decay time, fluctuations which are not subject to conservation laws acquire an effective mass, leading to gapped modes. The only relevant degrees of freedom are a (generally much smaller) collection of effective variables which are long-lived compared to the decohering modes, due to the requirement that they relax via transport. These are the subjects of the conservation laws.

Hydrodynamics was referred to as a ``long-wavelength, low-frequency effective description", as one usually considers variables which vary slowly in time and space.  This ``hydrodynamic regime" is sometimes characterized by a number called the Knudsen number, which is defined by
\begin{equation}
    {\rm Kn} = \frac{\ell_{\rm micro}}{L_{\rm macro}}
\end{equation}
where $\ell_{\rm micro}$ is the microscopic scale (usually taken to be the mean free path for those theories that admit such a quantity), and $L_{\rm macro}$ is the macroscopic scale of the setup under consideration. The hydrodynamic regime is strictly given by Kn $\ll1$. In terms of a plane-wave description $\exp(-i \omega t + i k_j x^j)$, with the speed of light $c=1$, this limit is that of $\omega/\Gamma \ll1$, $|k|/\Gamma \ll 1$ for some scale $\Gamma$, hence the name. Questions about the exact regime of validity of the hydrodynamic approximation remain open; in particular, the discovery that the hydrodynamic description seems to still work in some cases when ${\rm Kn} \lesssim 1$ has led to discussion of the ``unreasonable success of hydrodynamics"~\cite{Romatschke:2017ejr}.

Relativistic hydrodynamics is formulated in terms of conservation laws for one-point functions of microscopic operators. In most relativistic hydrodynamic theories, the main object under consideration is the one-point function of the stress-energy tensor operator, $\braket{T^{\mu\nu}}$. This one point function obeys the conservation equation
\begin{equation}\label{intro:Tmunu-eq}
    \nabla_\mu \braket{T^{\mu\nu}} = 0\,,
\end{equation}
where $\nabla_\mu$ is the covariant derivative. In theories with a global $U(1)$ symmetry such as those we consider in this dissertation, this equation is supplemented by the conservation of the charge current one-point function:
\begin{equation}\label{intro:Jmu-eq}
    \nabla_\mu \braket{J^\mu} =0\,.
\end{equation}
In what follows, we will usually neglect the brackets; the one-point function will be assumed unless otherwise stated. It is clear from simple counting arguments that the equations~\eqref{intro:Tmunu-eq},~\eqref{intro:Jmu-eq} are not ``closed" in the sense that the number of variables does not match the number of equations. There exist two philosophically distinct means of closing the equations. The first is to write the one-point functions $\braket{T^{\mu\nu}}$, $\braket{J^\mu}$ in terms of a smaller set of effective auxiliary fields. The second method is to write down additional, non-conservation equations for the components of the stress-energy tensor and charge current one-point functions. At various points in this dissertation, we will make use of both methods.

The expression of a conserved one-point function in terms of a smaller set of effective variables is called a constitutive relation. We additionally note that the equilibrium forms of $\braket{T^{\mu\nu}}$ and $\braket{J^\mu}$ may be calculated from a partition function, and are known. The equilibrium state for a theory with a global $U(1)$ symmetry may be parametrized by a temperature $T$, and $U(1)$ chemical potential $\mu$, and a fluid velocity $u^\mu$. These parameters are then assumed to sensibly describe the theory out of equilibrium as well.

More precisely, given the slow variation of the hydrodynamic variables, we assume that the one-point functions may be expressed in terms of a derivative expansion in $T$, $\mu$, and $u^\mu$, such that
\begin{subequations}\label{intro:deriv_expansion}
    \begin{align}
        T^{\mu\nu} &= T^{\mu\nu}_{(0)} + T^{\mu\nu}_{(1)} + T^{\mu\nu}_{(2)} + ...\,,\\
        J^\mu &= J^{\mu}_{(0)} + J^{\mu}_{(1)} + J^{\mu}_{(2)} + ...\,,
    \end{align}
\end{subequations}
where $T^{\mu\nu}_{(n)}$ denotes the $n$-th order contribution to the derivative expansion, etc. Truncating the expansion at zeroth order yields (after taking the divergence) the relativistic Euler equations. Truncating at first order gives (after taking the divergence) the relativistic Navier-Stokes equations, while second order yields the relativistic version of the Burnett equations, and third order yields the super-Burnett equations. The first-order constitutive relations are given in the formulation of Landau and Lifshitz~\cite{LL6} by
\begin{subequations}\label{intro:ns}
\begin{align}
    \braket{T^{\mu\nu}} &= \epsilon u^\mu u^\nu + (p - \zeta \nabla_\mu u^\mu) \Delta^{\mu\nu} \nonumber\\
    &- \eta \lr{\Delta^{\mu\alpha} \Delta^{\nu\beta} + \Delta^{\mu\beta} \Delta^{\nu\alpha} - \frac{2}{d} \Delta^{\mu\nu} \Delta^{\nu\beta}}\nabla_\alpha u_\beta\,,\\
    \braket{J^\mu} &= n u^\mu - \sigma T \Delta^{\mu\nu} \nabla_\nu \lr{\frac{\mu}{T}}\,,
\end{align}
\end{subequations}
where $\Delta^{\mu\nu} = u^\mu u^\nu + g^{\mu\nu}$ is the projector orthogonal to the fluid velocity $u^\mu$, $\epsilon$ is the energy density, $n$ is the charge density, $p$ is the isotropic pressure, $\zeta$ is the bulk viscosity, $\eta$ is the shear viscosity, and $\sigma$ is the charge conductivity. All of the scalar functions above are considered functions of the temperature $T$ and the chemical potential $\mu$. 

The relativistic Navier-Stokes equations, which one obtains by inserting~\eqref{intro:ns} into the conservation equations, have a dirty secret -- they exhibit parabolic behaviour, which means that they are acausal. Furthermore, and perhaps more troublingly, the relativistic Navier-Stokes equations are linearly unstable when considering perturbations in any Lorentz frame aside from the fluid rest frame. This problem has been known for many years~\cite{Hiscock:1985zz,Hiscock:1987zz}, and numerous methods of rectifying the acausality and its accompanying problems have been proposed in the literature. In this dissertation, we discuss the various means of introducing causal theories of relativistic hydrodynamics. One method that has been in use for a long time (and is subsequently the most common remedy) is the ``M\"uller-Israel-Stewart" (MIS) theory of hydrodynamics~\cite{Muller:1967zza,Israel:1976tn,Israel-Stewart}, which renders the system of equations causal by supplementing the conservation equations with relaxation-type equations. Another more recent development is the ``Bemfica-Disconzi-Noronha-Kovtun" (BDNK) theory of hydrodynamics~\cite{Bemfica:2017wps,Kovtun:2019hdm,Bemfica:2019knx,Hoult:2020eho,Bemfica:2020zjp}, which renders the system of equations causal by adding time-derivative terms to~\eqref{intro:ns}. Both types of theories will be discussed in the course of this dissertation.

The structure of the dissertation is as follows. In Chapter~\ref{chapter:background}, we introduce in more detail the background of hydrodynamics, and the instability and acausality of the relativistic Navier-Stokes equations. We also formally introduce the MIS and BDNK theories of hydrodynamics. This chapter is based in part on~\cite{Hoult:2020eho}. In Chapter~\ref{chapter:math}, we discuss some of the more mathematical results regarding causality and stability, in particular focusing on dispersion relations. This chapter is based in part on~\cite{Hoult:2023clg} and~\cite{Hoult:2024qph}. In chapter~\ref{chapter:micro}, we discuss how causal theories of hydrodynamics can be obtained by considerations of microscopic theories, in particular kinetic theory and holography. This chapter is based in part on~\cite{Hoult:2021gnb}. In chapter~\ref{chapter:extensions}, we discuss some extensions of the causal theories of hydrodynamics discussed in Chapter~\ref{chapter:background} to new systems; specifically, we focus on the one-form theory of relativistic magnetohydrodynamics, and viscous theory of relativistic superfluids. This chapter is based in part on~\cite{Hoult:2024cyx} and~\cite{Hoult:2024qph}. In Chapter~\ref{chapter:concl}, we conclude the dissertation, summarize the results contained herein, and discuss some potential future lines of inquiry. Finally, in Appendix~\ref{app:species_MIS}, we add more detail to some of the MIS theories introduced in Chapter~\ref{chapter:background}, in Appendix~\ref{app:linear response}, we review some of the basic ideas behind linear response theory (LRT), and in Appendix~\ref{app:RH-criteria}, we explicitly lay out criteria for stability for various-order controlling polynomials.

\paragraph{Notation:} In this dissertation, we will use the mostly positive definition of the metric: the flat space metric is given by $g_{\mu\nu} = {\rm diag}(-1,+1,...,+1)$. We also will set $c=k_B =\hbar=1$. In general, greek letters will denote spacetime indices, while lower-case latin letters will denote spatial indices. In the context of the gauge/gravity duality in Chapter~\ref{chapter:micro}, capital latin letters will refer to indices in the AdS bulk. Einstein summation notation is employed. 
The coordinate four-vector is given by e.g. $x^\mu = (t,x^j)$. The momentum four-vector is given by $p^\mu = (E,p^i)$, while the fluid velocity is defined by $u^\mu = \gamma(1,v^i)$, where $\gamma = (1-v^2)^{-1/2}$ is the Lorentz factor, and $v^i$ is the physical velocity. A vector squared is defined by e.g. $p^2 = p_\mu p^\mu$.
We will use $d$ for the number of (boundary) spatial dimensions. Finally, the symmetrization of two indices will be denoted with the shorthand $A^{(\mu}B^{\nu)} = \frac{1}{2} \lr{A^\mu B^\nu + A^\nu B^\mu}$ and the anti-symmetrization by $A^{[\mu} B^{\nu]} = \frac{1}{2} \lr{A^\mu B^\nu - A^\nu B^\mu}$.
	\startchapter{MIS and BDNK}
\label{chapter:background}

\newlength{\savedunitlength}
\setlength{\unitlength}{2em}

In this chapter, in order to better understand the philosophical differences between the M\"uller-Israel-Stewart (MIS) theory of causal relativistic hydrodynamics and the Bemfica-Disconzi-Noronha-Kovtun (BDNK) formulation, we will begin by investigating a far simpler case -- namely, diffusion. Once we have thoroughly worked that example, we shall proceed to review both the MIS and BDNK theories of relativistic hydrodynamics.

\section{Diffusion}
\label{ch2:sec_diffusion}
In the following we restrict ourselves to Minkowski spacetimes. Let us consider a theory which has a global $U(1)$ symmetry; a simple example is complex $\varphi^4$ theory. This global $U(1)$ symmetry has a charge current associated with it; in the quantum theory, this is a charge current operator, $J^\mu$. The one-point function of this operator is conserved, $\de_\mu \braket{J^\mu} = 0\,.$ The conservation equation is insufficient to uniquely determine the components of $\braket{J^\mu}$; in $d+1$ spacetime dimensions, there are $d+1$ components to $\braket{J^\mu}$, but only one equation. In order to find a unique solution for $\braket{J^\mu}$, one must introduce additional assumptions into the description (one must ``close" the system of equations).

From here on, we neglect the brackets about $J^\mu$, and assume we are describing the one-point function. There exist two philosophically distinct means by which we will close the system of equations. The first is to demand that $J^\mu$ is in fact a function of only one variable, so as to match the number of equations. The other option is to introduce more equations, so as to match the number of variables. 

We begin by taking the first route. Let us first consider an equilibrium state described by an external observer with four-velocity $u^\mu$ such that $u^2 = -1$. Then we may write the equilibrium charge current as $J^\mu_{\rm eq.} = \rho u^\mu$, where $\rho$ is the $U(1)$ charge density $\rho \equiv - J^\mu_{\rm eq.}  u_\mu$. We note that the $U(1)$ charge density may be coupled to a classical source called the chemical potential $\mu$, and $\rho = \rho(\mu)$. Therefore, $J^\mu_{\rm eq.}  = J^\mu_{\rm eq.}(\mu)$\,. We will also assume that the temperature of the system, $T$, is held constant.

We will now depart from equilibrium. Let us promote $\mu$ (and therefore also $\rho$) to be a function of spacetime; $u^\mu$, being external, remains constant, as does the temperature of the system. We now postulate that, near equilibrium, the charge current $J^\mu$ is a local functional\footnote{Functional, as opposed to function, because $\mu$ is itself a function of spacetime.} of $\mu(x^\mu)$ and the external parameter $u^\mu$, i.e. $J^\mu = J^\mu[\mu(x^\mu), u^\mu]$. This expression, which describes one-point functions of operators in terms of variables which parametrize the equilibrium state, is called a ``constitutive relation". An important caveat, one to which we will later return: the variable $\mu$ was well-defined in equilibrium. Out-of-equilibrium, the field $\mu(x)$ upon which $J^\mu$ depends is not well-defined. Instead, $\mu(x)$ is simply a variable brought in to parametrize the theory. The only restriction imposed on $\mu(x)$ is that, upon returning to equilibrium, $\mu(x) \to \mu$. We will often suppress the spacetime dependence of $\mu(x)$; from here-on, equilibrium parameters will be denoted by a subscript $0$, e.g. $\mu_0$.

As $J^\mu$ is a local functional of $\mu(x)$, it may be expressed in terms of $\mu(x)$ and spacetime derivatives of $\mu(x)$. We then impose another assumption, namely that the field $\mu(x)$ in a slowly varying function of spacetime. With this assumption, we can organize $J^\mu[\mu(x)]$ into a derivative expansion in $\mu(x)$. One can then write
\begin{equation}
    \label{ch2:diffusion:deriv_expansion_v2}
    J^\mu = J^\mu_{(0)} + J^{\mu}_{(1)} + J^\mu_{(2)} + \oser{\de^3}\,,
\end{equation}
where the subscript denotes the order of the contribution in the derivative expansion. Given the scaling $\partial_\mu \to \varepsilon \,\partial_\mu$, where $\varepsilon$ is a counting parameter, a term is said to be of order $\oser{\de^n}$ if it is proportional to $\varepsilon^n$. For example, a term\footnote{Note that, throughout the dissertation, $\mu$ will be used both as an index and as a variable. When in a subscript or superscript, it serves as an index; otherwise, it serves as a variable.} such as $u^\mu u^\lambda \de_\lambda \mu$ is first order in derivatives ($\oser{\de}$) and would contribute to $J^\mu_{(1)}$. We fix the zeroth-order ($\oser{1}$) term of this expansion, $J^\mu_{(0)}$, by demanding consistency of the solution with the equilibrium charge current $J^\mu_{\rm eq.}$. We therefore write that $J^\mu_{(0)} = \rho(\mu(x)) u^\mu\,,$ where the charge density\footnote{If the higher-derivative terms in $J^\mu$ are such that $J^\mu_{(0)}u_\mu = J^\mu u_\mu$, then $\rho$ is the $U(1)$ charge density. If not, then it is only equal to the charge density up to some order in the derivative expansion. Nevertheless, we will usually retain the name.} is given by $\rho = -J^\mu_{(0)} u_\mu$. With the constitutive relation~\eqref{ch2:diffusion:deriv_expansion_v2}, the conservation equation $\de_\mu J^\mu = 0$ yields (assuming that $\chi \equiv \de \rho/ \de \mu \neq 0$) that 
$
u^\lambda \de_\lambda \mu = \oser{\de^2}\,.
$
We will call this equation for $\mu$ the ``zeroth-order equation of motion", as it amount to the statement $\de_\mu J^\mu_{(0)} = \oser{\de^2}$.

We have no \textit{a priori} idea what the first-order contribution $J^\mu_{(1)}$ ought to be. As such, we will take an effective theory approach and write down all possible terms of first-order in derivatives which transform covariantly under Lorentz transformations, parity, and $U(1)$ charge conjugation symmetry. There is one such scalar $(u^\lambda \de_\lambda \mu)$ and one transverse vector $(\Delta^{\mu\nu}\de_\nu \mu)$. The most general constitutive relation is then given by:
\begin{equation}
\label{ch2:diffusion:first_order_conrel_v2}
    J^\mu_{(1)} = \lr{a u^\lambda \de_\lambda \mu} u^\mu - \sigma \Delta^{\mu\nu} \de_\nu \mu\,,
\end{equation}
where $\Delta^{\mu\nu} = u^\mu u^\nu + g^{\mu\nu}$ is the projector orthogonal to $u^\mu$, and $a$, $\sigma$ are (as yet) arbitrary $\oser{1}$ functions of $\mu$. However, we found that $u^\lambda \de_\lambda \mu = \oser{\de^2}$ by the zeroth order equation of motion, meaning that the first term of equation~\eqref{ch2:diffusion:first_order_conrel_v2} is actually higher-order in derivatives. Such a term may be neglected, leaving
\begin{equation}
\label{ch2:diffusion:full}
    J^\mu = \rho(\mu) u^\mu - \sigma(\mu) \Delta^{\mu\nu} \de_\nu \mu + J^\mu_{(2)} + \oser{\de^3}\,.
\end{equation}
We will not go to second order. Truncating the derivative expansion at first order and substituting it into the conservation equation yields
\begin{equation}
\label{ch2:diffusion:covar_diff_eq_v2}
    u^\lambda \de_\lambda \mu - \frac{1}{\chi}\Delta^{\alpha\beta}\de_\alpha\lr{\sigma  \de_\beta \mu} = 0\,.
\end{equation}
Let us now consider the specialized case where $D \equiv \sigma/\chi$ is a constant. Evaluating in the rest frame of the external observer $u^\mu = \delta^\mu_0$ then yields
\begin{equation}
\label{ch2:diffusion:rest_frame_diff_eq_v2}
    \de_t \mu - D \nabla^2 \mu = 0\,,
\end{equation}
where $\nabla^2$ is the Laplacian. The constant $D$ is called the diffusion constant. If $D > 0$, then equation~\eqref{ch2:diffusion:rest_frame_diff_eq_v2} is the celebrated diffusion equation. If $D<0$, this is the anti-diffusion equation, which is unstable to perturbations away from homogeneous equilibrium. This is an undesirable property, as we would like stable equilibria; we therefore enforce that $D \geq 0$. Moreover, we would like stable equilibria for all observers, a property we will dub ``covariant stability". This seemingly reasonable demand proves to be problematic; in fact, even the regular diffusion equation is unstable upon boosting due to its parabolic nature.

To show this, we introduce machinery which will be of constant use going forward -- a linearized plane-wave analysis. Let us perturb $\mu(x)$ away from a homogeneous equilibrium solution $\mu(x) = \mu_0 + \delta \mu(x)$. Further, let us take the perturbations to be in the form of plane waves, such that $\delta \mu(x) = \widetilde{\delta \mu}(\omega,k) e^{-i \omega t + i k_j x^j}$, where $\omega$ is the (angular) frequency and $k_j$ is the wave-vector. Inserting this plane-wave solution into the diffusion equation~\eqref{ch2:diffusion:rest_frame_diff_eq_v2} and linearizing in $\delta \mu$ (which is redundant in the case of equation~\eqref{ch2:diffusion:rest_frame_diff_eq_v2}, but would not be for e.g.~\eqref{ch2:diffusion:covar_diff_eq_v2}) yields
\begin{equation}
    \lr{-i \omega + D k^2 } \widetilde{\delta \mu} e^{-i \omega t + i k x} = 0\,.
\end{equation}
We see that there can only be a non-trivial perturbation if $\omega$ and $k_j$ are not independent. Such a relation $\omega = \omega(k)$ is known as a ``dispersion relation"; for the diffusion equation, $\omega = - i D k^2$, where $k \equiv \sqrt{k_j k^j}$. We will frequently refer to solutions $\omega = \omega(k)$ as ``modes". This form, $\omega \propto k^2$, is ubiquitous in dissipative systems. The perturbation is then proportional to $e^{- D k^2 t}$, and decays with time so long as $D > 0$. More generally, we can see that for $\omega = \omega' + i \omega''$ with real $\omega', \omega''$, the plane wave perturbations take the form $\exp\lr{-i \omega' t + i k_j x^j} \exp\lr{\omega'' t}$. We therefore have a more general stability condition, 
\begin{equation}
    \Im(\omega) \leq 0\,.
\end{equation}

Let us now repeat the procedure for equation~\eqref{ch2:diffusion:covar_diff_eq_v2}, setting $u^\mu = \gamma(1,v^j)$, where $\gamma$ is the Lorentz factor $\gamma = (1-v^2)^{-1/2}$. Similarly perturbing $\mu(x)$ and linearizing the equations in $\delta \mu$ yields the dispersion relations
\begin{equation}
    \omega_{\pm}(k) = \frac{\gamma \lr{i + 2 D \gamma k_i v^i} \pm \sqrt{4 i D \gamma k_i v^i - \gamma^2 + 4 D^2 \lr{\lr{1-\gamma^2} k^2 + \gamma^2 \lr{k_i v^i}^2}}}{2 D \lr{\gamma^2 - 1}}
\end{equation}
There are now two possible solutions $\omega_{\pm}$. To get a feel for their behaviour, let us expand both solutions in small $|v|$ for finite $k$. We then find that, to leading order in $|v|$, the two solutions become
\begin{equation}
    \omega_{\pm} = \begin{cases}
        - i D k^2 + \oser{v}& \textbf{Stable}\,,\\
        \frac{i}{D v^2} + \oser{1/v} & \textbf{Unstable}\,.
    \end{cases}
\end{equation}
We see that for an observer in motion, the diffusion equation exhibits an unstable mode which was not present for the observer at rest. Moreover, for a slower observer, the blow-up appears to be worse. Such behaviour is undesirable; if a physical instability were to exist in the system, its existence should surely not depend on the velocity of an external observer. We would like to find a way to fix this, while still keeping the physical small-$k$ behaviour $\omega = - i D k^2 + ...$, the $...$ denoting higher-order terms in $k$. We will discuss two means of doing so.

\subsection{Maxwell-Cattaneo theory}
\begin{figure}[t]
    \centering
    \begin{subfigure}[t]{0.49\linewidth}
    \includegraphics[width=\linewidth]{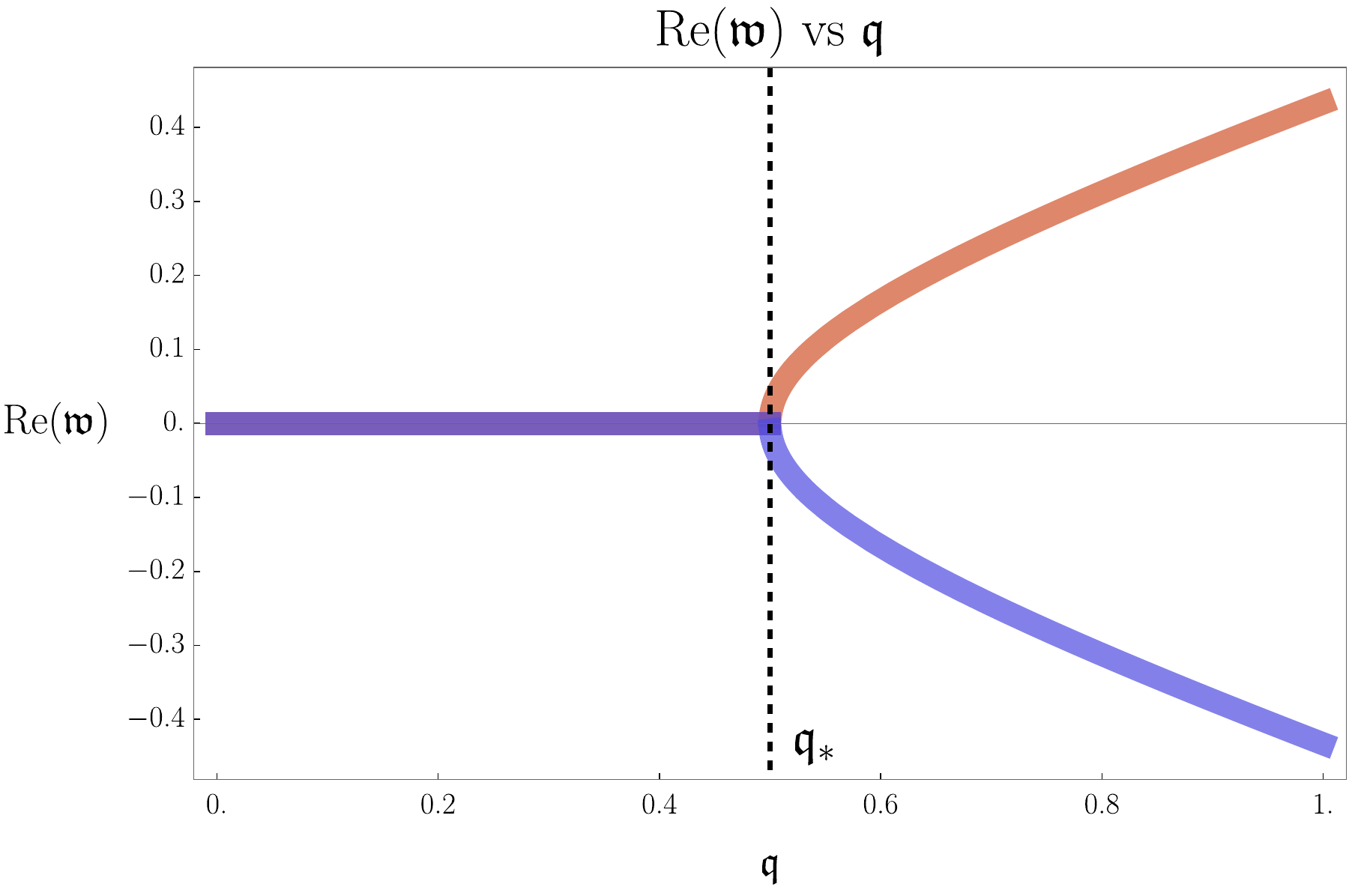}
    \end{subfigure}
    \begin{subfigure}[t]{0.49\linewidth}
    \includegraphics[width=\linewidth]{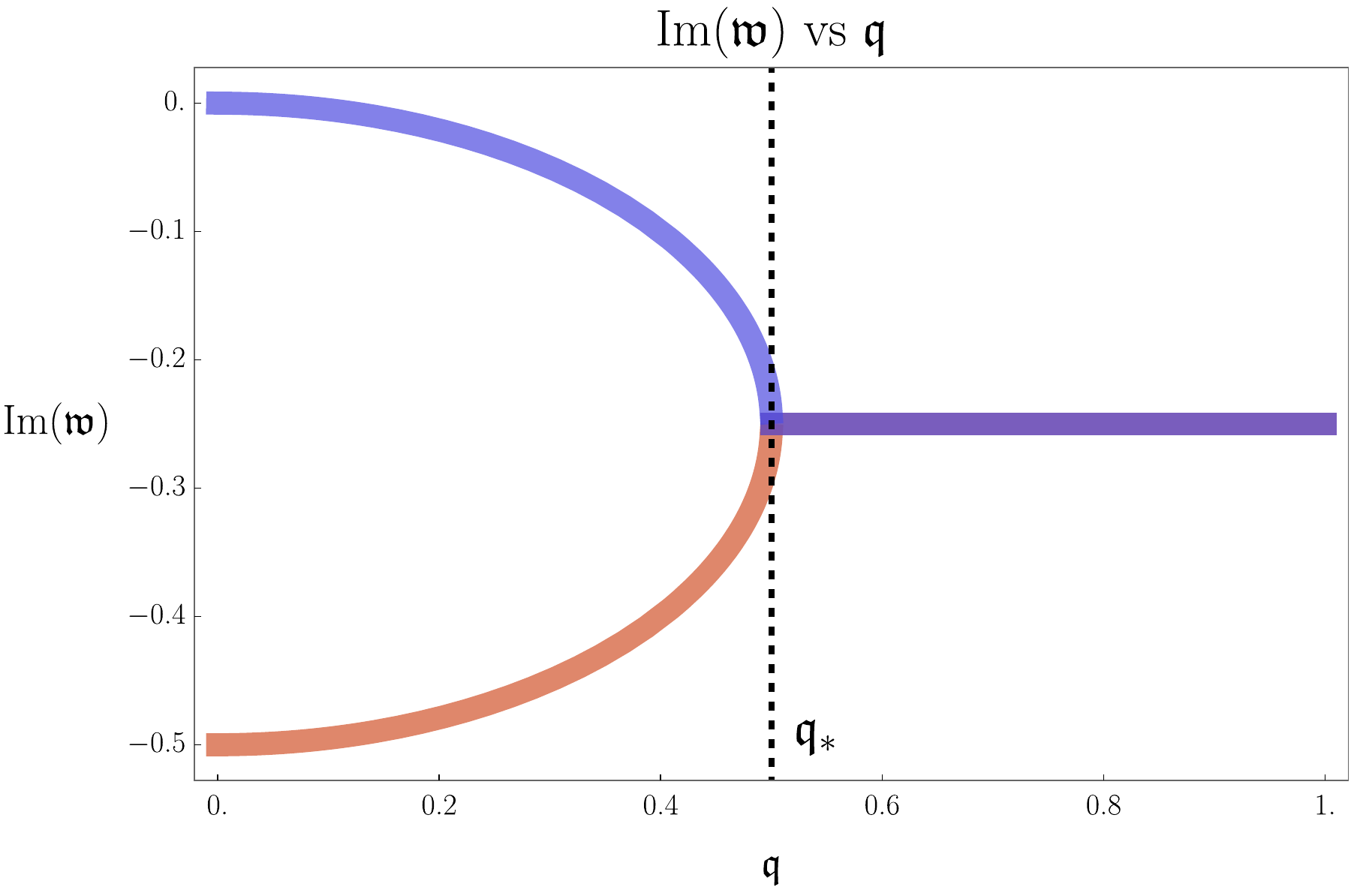}
    \end{subfigure}
    \caption{A plot of the $\omega_{\pm}$ modes of the Maxwell-Cattaneo model around an equilibrium state with $v^i = 0$. In the plot, we have introduced the dimensionless variables $\mathfrak{w} = \omega/T_0$, $\mathfrak{q} = k/T_0$, where $T_0>0$ is the (constant) temperature of the system, brought in here solely to serve as a dimensionful scale. The above is plotted for $\tau = 2/T_0$, $D = 1/(2 T_0)$. The orange curve represents $\omega_+$, while the blue curve is $\omega_-$. The constant modes $\omega_{1,2}$ are not pictured.}
    \label{fig:MC_modes}
\end{figure}

\begin{figure}[t]
    \centering
    \begin{subfigure}[t]{0.49\linewidth}
    \includegraphics[width=\linewidth]{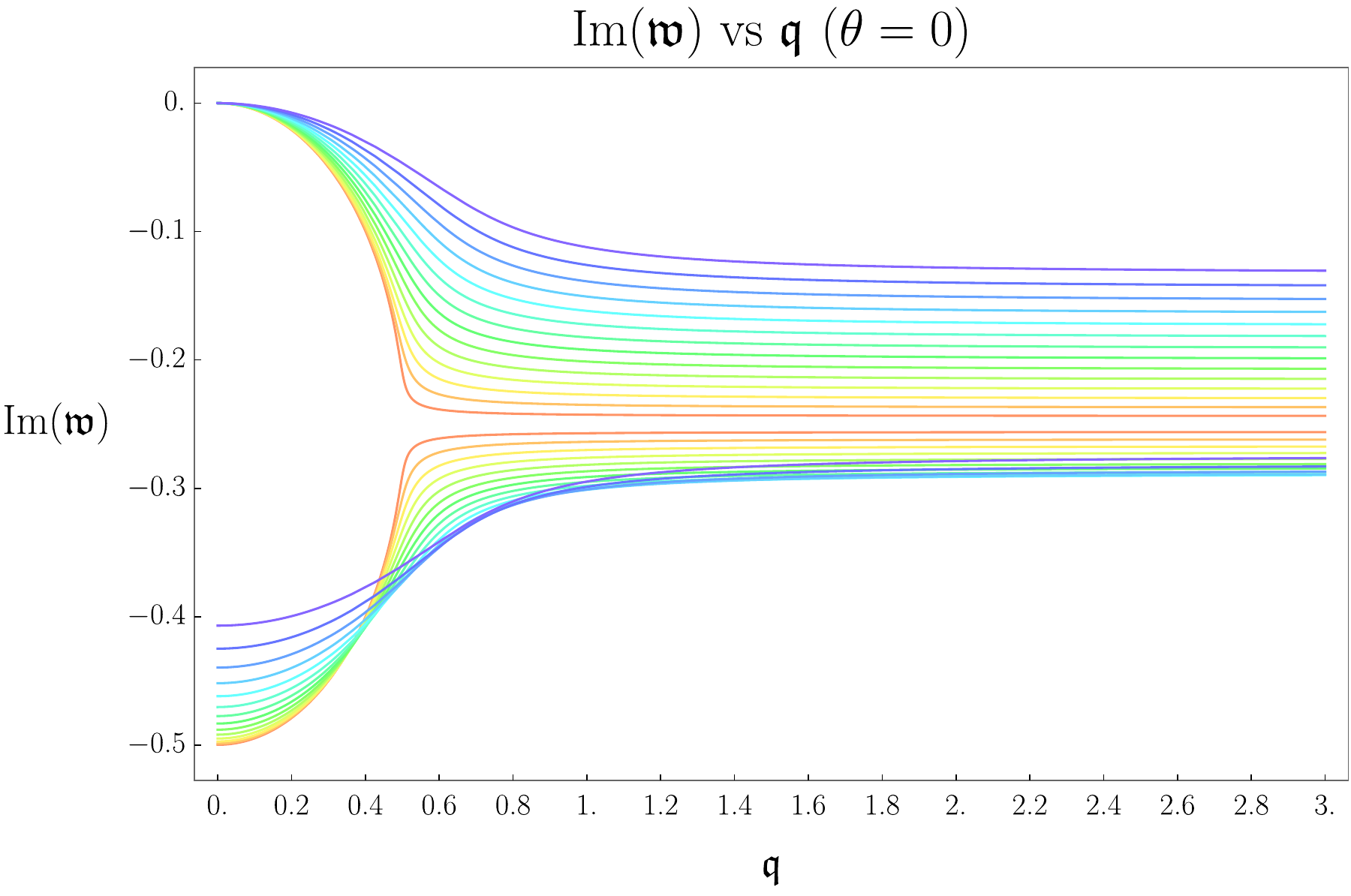}
    \end{subfigure}
    \begin{subfigure}[t]{0.49\linewidth}
    \includegraphics[width=\linewidth]{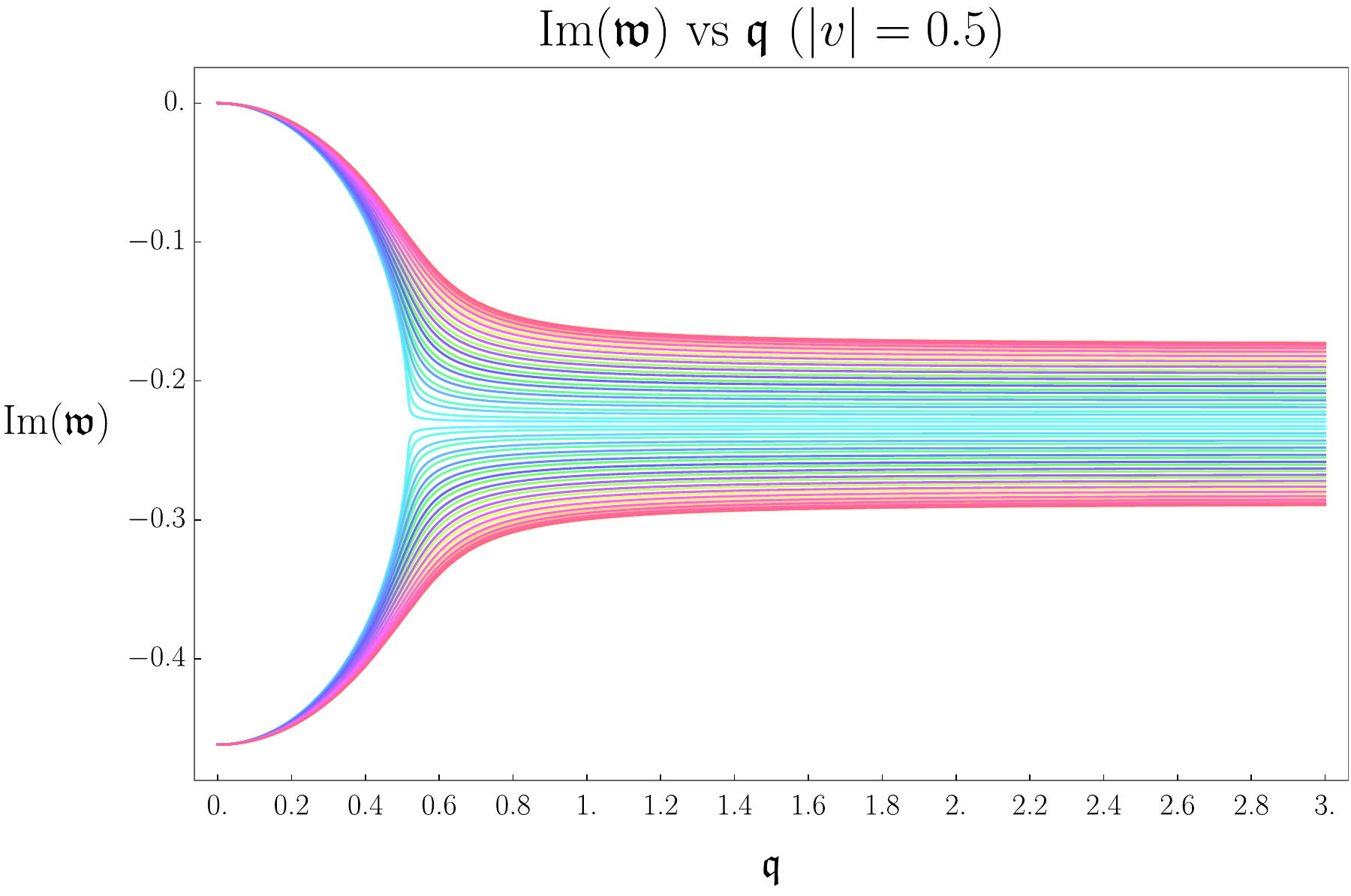}
    \end{subfigure}
    \caption{Two plots of the imaginary parts of the $\omega_{\pm}$ modes of the Maxwell-Cattaneo model for $v^i \neq 0$. In the first plot, the angle $\theta = \arccos(v{\cdot}k)$ is held fixed at zero, and $|v|$ is varied from $0.1$ to $0.7$. In the second, $|v|$ is fixed at $0.5$, and $\theta$ is varied from $0.1$ to $\pi-0.1$. In both cases, $\omega$ and $k$ have been replaced with $\mathfrak{w} = \omega/T_0$ and $\mathfrak{q} = |k|/T_0$, where $T_0$ is the background temperature brought in solely to act as a dimensionful scale.}
    \label{fig:MC_modes_boosted}
\end{figure}

Instead of implementing the derivative expansion~\eqref{ch2:diffusion:deriv_expansion_v2}, let us define the charge density $\rho = -J^\mu u_\mu$ and then write the charge current as
\begin{equation}
    J^\mu = \rho u^\mu + n^\mu\,,
\end{equation}
where $n^\mu$ is the transverse charge current which satisfies $n^\mu u_\mu = 0$. Note that, from the perspective of the power counting used in~\eqref{ch2:diffusion:deriv_expansion_v2}, the quantity $n^\mu$ contains all orders. Now, we will supply a set of $d$ equations so as to close the system of equations (i.e. we ensure there are an equal number of variables and equations that evolve those variables). For now, we simply propose the additional equations; in later sections, we will do more to motivate them. We take the additional equations to be
\begin{equation}
\label{ch2:diffusion:MC_additional_eq}
    \tau  \Delta^\lambda_{\,\,\,\,\nu}\,\dot{n}^\nu + n^\lambda  = -  \sigma(\mu) \Delta^{\lambda\nu} \de_\nu \mu\,,
\end{equation}
where we use the notation $\dot{n}^\nu = u^\mu \de_\mu n^\nu$. The combined system of equations $\de_\mu J^\mu=0$ and equations~\eqref{ch2:diffusion:MC_additional_eq} is then given in matrix form by
\begin{equation}
\label{ch2:diffusion:MC_matrix_v2}
    \begin{split}
        \begin{pmatrix}
            \chi u^\lambda & \delta^\lambda_{\,\,\,\,\beta}\\
             \sigma \Delta^{\alpha\lambda}&\tau \Delta^{\alpha}_{\,\,\,\,\beta} u^\lambda           
        \end{pmatrix}\de_\lambda \begin{bmatrix}
            \mu\\
            n^\beta
        \end{bmatrix} + \begin{bmatrix}
            0\\
            n^\alpha
        \end{bmatrix} = \begin{bmatrix}
            0\\
            0^\alpha
        \end{bmatrix}\,.
    \end{split}
\end{equation}
We refer to the column of variables $U^B = (\mu, n^\beta)^T$ as the ``state vector". We can more compactly write this equation as ${\cal M}_{AB}^\lambda[U] \de_\lambda  U^B + {\cal N}_A[U] = 0\,,$ where $(A,B)$ run from $1$ to $d+1$. We see that in the limit $\tau \to 0$, we recover the diffusion equation. Now, let us consider a perturbation about a homogeneous solution to the equations~\eqref{ch2:diffusion:MC_matrix_v2}. The chemical potential may take on any constant value $\mu_0$, but the only constant solution for $n^\mu$ is $n^\mu_0 = 0$, as may be seen from the algebraic term in the equations. We therefore introduce the plane-wave perturbations $U^B = U_0^B + \widetilde{\delta U}^B e^{-i \omega t + i k_j x^j} =  (\mu_0, 0)^T + (\widetilde{\delta \mu}, \widetilde{\delta n}^\mu)^T e^{-i \omega t + i k_j x^j}$. For $u^\mu = \delta^\mu_0$ there only exist non-trivial perturbations $\widetilde{\delta U}^B$ so long as (in $d=3$)
\begin{equation}
    \det\begin{vmatrix}
         - i \chi\omega & i k_j\\
        i \sigma k^i & \lr{1 - i \omega \tau} \delta^i_j
    \end{vmatrix} = \lr{\omega + \frac{i}{\tau}}^2 \lr{\chi \tau \omega^2 + i \chi \omega - \sigma k^2} = 0\,.
\end{equation}
Unlike in the regular diffusion equation, where we had one mode ($\omega = - i D k^2$), we now have \textit{four} modes. Two of them are identical, while the other two differ:
\begin{subequations}
\begin{align}
\omega_{1,2} &= - \frac{i}{\tau}\label{ch2:diffusion:MC_gap_1_v2}\,,\\
    \omega_{\pm} &=  -\frac{i}{2 \tau}\lr{1 \pm \sqrt{1 - 4 D \tau k^2}}\label{ch2:diffusion:MC_gap_2_v2}\,.
\end{align}
\end{subequations}
These modes are stable for $\tau >0$, $D \geq 0$. Note that, unlike the diffusion equation, three of the modes are gapped ($\lim_{k \to 0} \omega(k) \neq 0$), a hallmark of relaxational physics (for $\Im \,\omega \leq 0$). We call these modes ``non-hydrodynamic modes". On the other hand, one mode is gapless, a hallmark of transport phenomena: we will refer to these modes as ``hydrodynamic modes". Note that if we expand the modes~\eqref{ch2:diffusion:MC_gap_2_v2} in small $k$, they are of the form
\begin{equation}\label{ch2:diffusion:small_k_v2}
    \omega_{+} = - \frac{i}{\tau} + i D k^2 + \oser{k^4}, \quad \omega_- = - i D k^2 + \oser{k^4}\,.
\end{equation}
This na\"ively looks concerning at first glance, as $\omega_+$ appears to be moving towards the upper-half complex plane with increasing $k$. The situation is saved by a ``pole collision"\footnote{In our current presentation; there are no poles anywhere. However, one can show (see Appendix~\ref{app:linear response}) that dispersion relations appear as poles of retarded two-point functions.}, where the modes $\omega_{\pm}$ collide and exhibit non-analytic behaviour. This occurs\footnote{In this example, the collision occurs at a real value of $k$. In general, collisions can happen at complex values of $k$, see e.g.~\cite{Grozdanov:2019uhi}.} at a critical value of $k_* = (4 D \tau)^{-1/2}$, which is the branch point of the square root in~\eqref{ch2:diffusion:MC_gap_2_v2}. For $k>k_*$, the imaginary part freezes out and the real part of the solution begins to grow. For an illustration refer to figure~\ref{fig:MC_modes}.
Let us now repeat the analysis for $u^\mu = \gamma(1,v^j)$ as before. The modes $\omega_{1,2}$ remain stable, given by $\omega_{1,2} = - i \frac{\sqrt{1-v^2}}{\tau} + v^j k_j$. The modes $\omega_{\pm}$ become quite complex to write down; instead, they have been plotted in Figure~\ref{fig:MC_modes_boosted} for varying values of the boost velocity $v^j$ and the angle $\theta = \arccos\lr{v{\cdot}k}$ between $v^j$ and $k^j$. As one can see, the modes remain stable for all values plotted.
The Maxwell-Cattaneo model is therefore stable in all reference frames, or ``covariantly stable". Let us now take the other approach.

\subsection{Telegrapher's equation}
\begin{figure}[t]
    \centering
    \includegraphics[width=0.7\linewidth]{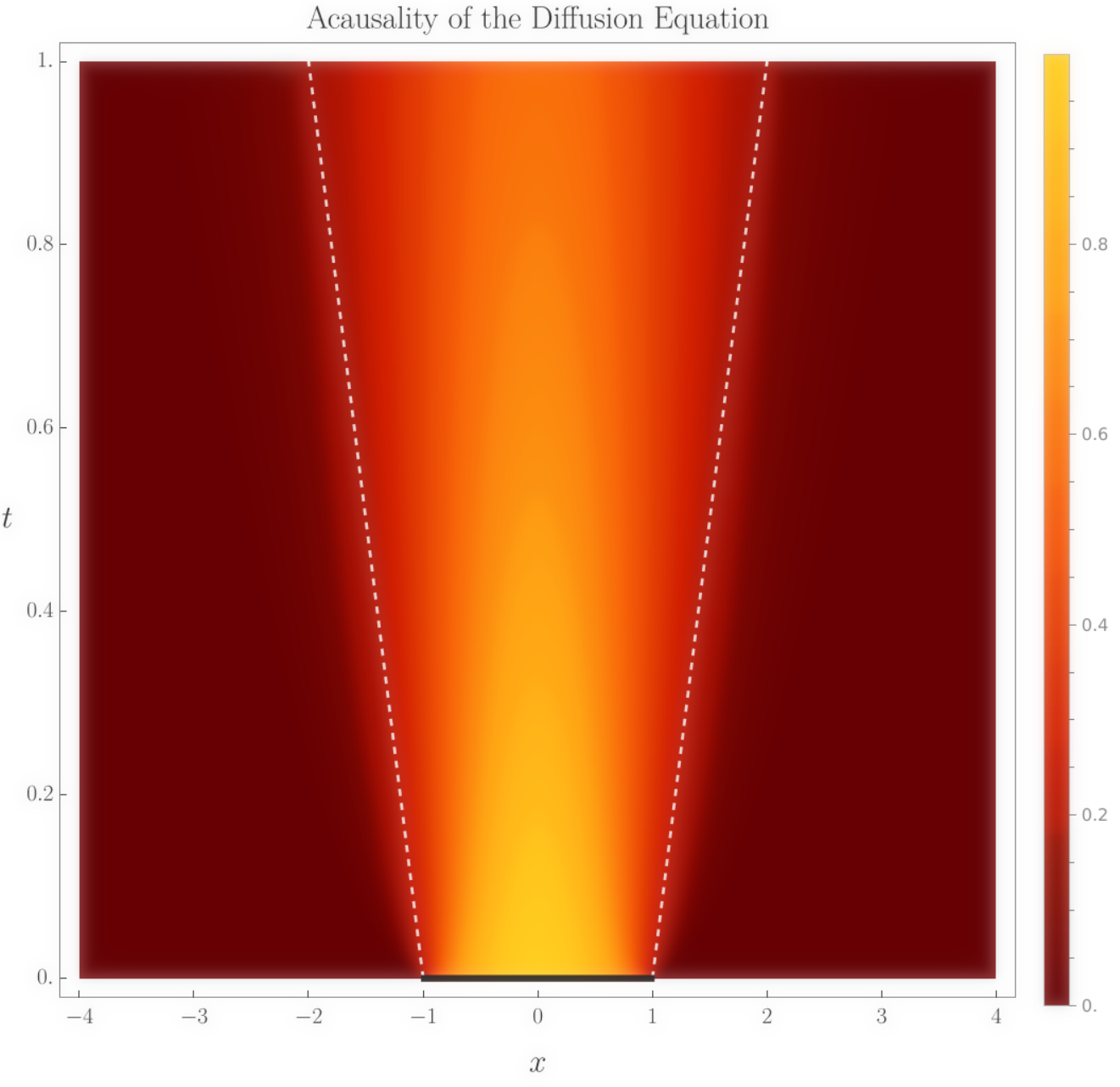}
    \caption{Density plot of the solution to the $1+1$D diffusion equation $\de_t \rho(t,x) - D \de_x^2 \rho(t,x) = 0$ given an initial condition with compact support $\rho(0,x) = \sqrt{1-x^2}$. The region of initial compact support is shown in grey. The dotted lines denote the lightcone extending out from the edges of compact support for the initial conditions at $x=\pm 1$. As one can clearly see, the solution extends outside the lightcone; in fact, the region of support extends out to spatial infinity for all $t>0$. See also~\cite{Gavassino:2024EAS,Gavassino:2025hwz} for similar plots.}
    \label{fig:Diffusion-plot}
\end{figure}

Returning back to the procedure we followed to obtain the diffusion equation, we note that we removed a term of the form $u^\lambda \de_\lambda \mu$ via an application of the zeroth-order equation of motion. More precisely, the equation enforced that the term was of order $\oser{\de^2}$ in derivatives, and could be neglected. In field theory parlance, this was the effect of taking the theory ``on-shell" with respect to the zeroth-order equations of motion. Let us now instead consider the theory ``off-shell" with respect to the zeroth-order equation of motion; in other words, we retain a term which is, strictly speaking, higher-order in derivatives. This term will act, in a sense, as a UV\footnote{UV in the sense that higher derivatives correspond roughly to shorter length scales and higher energies in the derivative expansion.} regulator to ensure that the equations are well-behaved (i.e. covariantly stable). Retaining the term gives back
\begin{equation}
\label{ch2:diffusion:telegrapher_diffusion_v2}
    J^\mu = \lr{\rho(\mu)  + a u^\lambda \de_\lambda \mu} u^\mu  - \sigma \Delta^{\mu\lambda} \de_\lambda \mu + \oser{\de^2}\,.
\end{equation}
Inserting the constitutive relation~\eqref{ch2:diffusion:telegrapher_diffusion_v2} into the conservation equation yields
\begin{equation}
    \chi u^\mu \de_\mu \mu + a u^\mu u^\lambda \de_\mu \de_\lambda \mu - \sigma \Delta^{\mu\lambda}\de_\mu \de_\lambda \mu + \pder{a}{\mu} \lr{u^\lambda \de_\lambda \mu}^2 - \pder{\sigma}{\mu} \Delta^{\mu\lambda} \lr{\de_\mu \mu}\lr{\de_\lambda \mu} = 0\,.
\end{equation}
Taking $a$ and $\sigma$ to be constants (or, alternatively, linearizing in $\mu$) yields the telegrapher's equation:
\begin{equation}
\label{ch2:diffusion:telegrapher-eq}
    u^\mu \de_\mu \mu + \tau u^\mu u^\lambda \de_\mu \de_\lambda \mu - D \Delta^{\mu\nu}\de_\mu \de_\lambda \mu = 0\,,
\end{equation}
where $\tau \equiv a/\chi$, and $D=\sigma/\chi$ as before. Considering once again plane-wave perturbations of $\mu$, we find that there only exist non-trivial perturbations if the dispersion relations
\begin{equation}\label{ch2:diffusion:telegrapher_disp_v2}
    \omega_{\pm} = - \frac{i}{2 \tau}\lr{1 \pm \sqrt{1-4 D \tau k^2}}
\end{equation}
hold. This is, of course, the exact same set of dispersion relations as $\omega_{\pm}$ in the Maxwell-Cattaneo theory; therefore, we already know that it is stable. We will instead attempt to answer a different question, namely whether the telegrapher's equation is causal. In order to answer this question, we must address what it actually means for a system of equations to be causal. We will postpone the full question to the next section, but for now will satisfy ourselves by imposing two conditions on the dispersion relations:
\begin{equation}\label{ch2:diffusion:causality_mom_v2}
    -1 \leq v_f \equiv \lim_{k \to \infty} \frac{\Re(\omega)}{|k|} \leq 1, \qquad \lim_{k \to \infty} \frac{\Im(\omega)}{|k|} = 0\,,
\end{equation}
where $v_f$ is the phase velocity. We will later show that these conditions correspond to the demand that wavefronts for (linearized) perturbations lie inside the lightcone. One can straightforwardly show that in the large-$k$ limit, $\omega_{\pm} \sim \pm \sqrt{\frac{D}{\tau}} k + ...$, where the $...$ denote subleading terms in large $k$. Therefore, the conditions~\eqref{ch2:diffusion:causality_mom_v2} are satisfied so long as $\tau \geq D \geq 0$. Comparison to the diffusion equation evaluated in the rest frame of the external observer shows that, for equations~\eqref{ch2:diffusion:covar_diff_eq_v2},
\begin{equation}
    v_f = 0, \quad \lim_{k \to \infty}\frac{\Im(\omega)}{k} \to - \infty\,.
\end{equation}
We see that the diffusion equation is acausal. This is the reason the diffusion equation becomes unstable upon performing a Lorentz boost; spacelike separated events do not have an invariant notion of temporal ordering. Diffusion takes a perturbation (which propagates acausally) from an excited state to a diffuse state; yet, a boost can reverse the ordering, taking a diffuse state to an excited state. We therefore see an instability. For an visual illustration of the acausality, refer to Figure~\ref{fig:Diffusion-plot}.

Let us now briefly consider a motivation for the term $u^\lambda \de_\lambda \mu$ that appears in the telegrapher's equation. Suppose this term were not present in~\eqref{ch2:diffusion:telegrapher-eq}; it could then be generated by an application of the transformation $\mu \to \mu' = \mu + a u^\lambda \de_\lambda \mu$, where we discard terms of second order in derivatives. Such a transformation is easily motivated; the variable $\mu$ is only well-defined in equilibrium. Out of equilibrium, there is no unique definition of $\mu$, and one can define a new $\mu'$ which differs from $\mu$ by terms which vanish in equilibrium (so that $\mu'\to \mu_0$). It is not yet clear that $u^\lambda \de_\lambda \mu$ does vanish in equilibrium, but it will turn out to be the case. We will make use of this re-definition freedom frequently when we discuss relativistic hydrodynamics.
\section{Relativistic Hydrodynamics}
\label{ch2:sec_uncharged}
\subsection{Thermodynamics}
Before discussing hydrodynamics, which is the study of dynamics out of equilibrium, we should attempt to understand equilibrium first. To begin with, we can consider the first law of thermodynamics:\footnote{In this subsection, we temporarily drop the $0$ subscript on equilibrium values for notational brevity.}
\begin{equation}\label{ch2:hydro:first-law}
    dU = - p \,dV + T \,dS + \mu \,dQ
\end{equation}
where $U$ is the internal energy, $p$ is the isotropic pressure, $V$ is the volume, $T$ is the temperature, $S$ is the entropy, $\mu$ is the $U(1)$ chemical potential, and $Q$ is the $U(1)$ charge. We will be primarily interested in intensive quantities; therefore, we will instead consider the energy density $\epsilon = U/V$, the entropy density $s = S/V$, and the charge density\footnote{Note that in the previous section, this was denoted by $\rho$.} $n = Q/V$. It is then easy to show that
\begin{equation}
     d\epsilon  =  T  ds + \mu  dn + \lr{- p + s T + \mu n - \epsilon} \frac{dV}{V}
\end{equation}
However, one can show\footnote{Specifically, by demanding that the internal energy transform homogeneously under scaling. If $U(\lambda V, \lambda S,\lambda Q) = \lambda U(V,S,Q)$, then $U = - p V + T S + \mu Q$ by Euler's homogeneous function theorem.} from the first law~\eqref{ch2:hydro:first-law} that $\epsilon = -p + s T + n \mu$, and so we arrive at the intrinsic version of the first law, $d\epsilon = T ds + \mu dn$. This can be written instead as a Gibbs-Duhem relation
\begin{equation}\label{ch2:hydro:Gibbs-Duhem}
    dp = s\, dT + n\, d\mu
\end{equation}
which is the form we will be using for the majority of the remainder of the dissertation. One can view $T$ and $\mu$ as sources, with $s$ and $n$ being the response of the pressure to these external sources. This allows us to make the connection $s = \lr{\pder{p}{T}}_\mu,$ $n = \lr{\pder{p}{\mu}}_T$, and $\lr{\pder{s}{\mu}}_T = \lr{\pder{n}{T}}_\mu$, where the subscript on the derivative indicates that quantity is being held constant. Finally, we also include a unit timelike vector $u^\mu$ which describes the bulk motion of the homogeneous equilibrium state, which we will call the fluid velocity. This makes the collection of independent degrees of freedom $\{T, \mu, u^\mu\}$. Since $u^\mu$ is normalized, this means that the equilibrium state is characterized by $d+2$ independent parameters.

As indicated in Chapter~\ref{chapter:introduction}, given a theory with diffeomorphism invariance and a $U(1)$ symmetry, there are conservation equations for the stress-energy tensor $T^{\mu\nu}$ and the charge current $J^\mu$:
\begin{equation}\label{ch2:hydro:conservation-eqs}
    \nabla_\mu T^{\mu\nu} = 0, \quad \nabla_\mu J^{\mu} = 0\,,
\end{equation}
where $\nabla_\mu$ is the covariant derivative. The energy-momentum tensor $T^{\mu\nu}$ is the symmetrized version, not the ``canonical" energy-momentum tensor that arises from considering the Noether current for spacetime translations. The ``canonical" energy-momentum tensor can be symmetrized\footnote{For theories in curved spacetimes with torsion (i.e. spacetimes with connections such that $\Gamma^\mu_{\rho\sigma} - \Gamma^{\mu}_{\sigma\rho} \neq 0$), this procedure can't be done, and one also ends up with a ``spin current". Defining the spin current and stress-energy tensor in the torsional case and then taking the torsion-free limit gives a unique definition of both quantities~\cite{Gallegos:2022jow}. Investigations of hydrodynamics involving a spin current, i.e. ``spin hydrodynamics", are an active area of research~\cite{Gallegos:2022jow,Becattini:2023ouz}; however, these theories are outside the scope of this dissertation.}~\cite{Belinfante:1940,Rosenfeld} (for a textbook description, see e.g.~\cite{DiFrancesco:1997nk}) via the use of a Belinfante-Rosenfeld tensor to arrive at $T^{\mu\nu}$; in what follows, we will always only consider the symmetrized version.

Just as in the case of the diffusion equation, the conservation equations~\eqref{ch2:hydro:conservation-eqs} are not ``closed"; the stress-energy tensor $T^{\mu\nu}$ is a symmetric rank-two contravariant tensor which therefore has $(d+1)(d+2)/2$ independent components, while the charge current $J^\mu$ is a $(d+1)$-dimensional contravariant vector. Together, $T^{\mu\nu}$ and $J^\mu$ have $(d+1)(d+4)/2$ independent components; in $d=3$, this gives fourteen components. The equations~\eqref{ch2:hydro:conservation-eqs}, on the other hand, number only $(d+2)$. Generically, $(d+1)(d+4)/2 > (d+2)$ for $d \geq 1$, and the conservation equations~\eqref{ch2:hydro:conservation-eqs} are not closed.

The form of the equilibrium stress-energy tensor and charge current in an isotropic, homogeneous medium has been known for a long time~\cite{LL6,LL2}, and is given by
\begin{equation}\label{ch2:hydro:eq-const-first}
    T^{\mu\nu}_{\rm eq.} = \epsilon u^\mu u^\nu + p \Delta^{\mu\nu}, \quad J^\mu_{\rm eq.} = n u^\mu\,,
\end{equation}
where $\Delta^{\mu\nu} = g^{\mu\nu} + u^\mu u^\nu$ is again the projector orthogonal to $u^\mu$, now for a general metric. For a more general equilibrium state, we can find the constitutive relations by the use of a generating functional.

\subsubsection{Hydrostatic Generating Functional}
``Hydrostatics" refers to equilibrium contributions to flow (or the absence thereof). These contributions to the constitutive relations need not be algebraic; for example, in the study of magnetohydrodynamics, certain magnetovortical terms containing derivatives remain non-zero in the constitutive relations even after equilibration~\cite{Hernandez:2017mch}. These hydrostatic contributions can be determined using techniques from field theory. In particular, one of the techniques in field theory for computing correlation functions $\braket{{\cal O}_1 {\cal O}_2 ... {\cal O}_n}$ is the use of a generating functional. In this section, we will focus primarily on an operational understanding of the generating functional, rather than focusing on its origins, or questions of \textit{why} it works. The interested reader may instead refer to~\cite{Jensen:2012jh,Banerjee:2012iz}.

Let us consider an operator ${\cal O}$ which is sourced by a classical source $S_{{\cal O}}$. We can define an equilibrium generating functional $W[S_{{\cal O}}]$, such that that
\begin{align*}
    \braket{{\cal O}} &= \frac{\delta W[S_{{\cal O}}]}{\delta S_{{\cal O}}}\biggl|_{S_{{\cal O}} = 0}\,,    
\end{align*}
where $S_{{\cal O}} = 0$ denotes turning the sources off, and $\braket{{\cdot}}$ denotes the one-point function of the operator ${\cal O}$. We are interested in obtaining the one-point\footnote{In this dissertation, we will only be using this generating functional approach to obtain one-point functions; however, there is a huge world of application for field theoretic approaches to hydrodynamics. In particular, a framework for applying field-theoretic techniques to hydrodynamics called the Schwinger-Keldysh EFT (SK-EFT) is of great interest. The SK-EFT may be used to consider statistical fluctuations which are not gaussian in nature, among other things. This is an active area of current investigation -- the interested reader may learn more by reading the review of Liu and Glorioso~\cite{Liu:2018kfw}.} functions for $T^{\mu\nu}$ and $J^\mu$. The source for the stress-energy tensor is the metric, while the source for the charge current $J^\mu$ is the $U(1)$ gauge field. We will therefore introduce a background metric perturbation $g_{\mu\nu}$ and a background gauge field $A_\mu$. The generating functional can be written
\begin{equation}
    W[g_{\mu\nu}, A_\mu] = \int d^{d+1}x \sqrt{-g} \,{\cal F}(g,A)
\end{equation}
where $g$ is the metric determinant, and ${\cal F}$ is the free energy density. The metric above is taken to be Lorentzian, rather than Euclidian. The variation of the generating functional is given by
\begin{equation}\label{ch2:hydro:genfunc_T_J}
    \delta W = \int d^{d+1}x \sqrt{-g} \biggl[\frac{1}{2}T^{\mu\nu} \delta g_{\mu\nu} + J^\mu\delta A_\mu \biggr]
\end{equation}
where
\begin{equation}
    T^{\mu\nu} = \frac{2}{\sqrt{-g}} \frac{\delta \lr{\sqrt{-g} {\cal F}}}{\delta g_{\mu\nu}}, \qquad J^\mu = \frac{\delta {\cal F}}{\delta A_\mu}\,.
\end{equation}

Let us take a moment to consider what it means to be in equilibrium. Equilibrium states must be time-translation invariant, which implies that there exists a time-like Killing vector of the background metric $K^\mu$ which generates the time-translation symmetry. This Killing vector by definition obeys the Killing equation $\Lied_{K}g_{\mu\nu} = 0$, which implies that
\begin{equation}
\label{ch2:hydro:Killing_Eq}
    \nabla_\mu K_\nu + \nabla_\nu K_\mu = 0\,.
\end{equation}
Let us now define $\beta^\mu = \beta_0 K^\mu$, where $\beta_0>0$ is a dimensionful constant with units of inverse energy. In homogeneous equilibrium, we may then construct the temperature and the fluid velocity from $\beta_\mu$~\cite{Jensen:2012jh}:
\begin{equation}\label{ch2:hydro:def_T}
    T \equiv \frac{1}{\sqrt{- \beta^\mu g_{\mu\nu} \beta^\nu}}, \quad u^\mu \equiv \frac{\beta^\mu}{\sqrt{- \beta^\mu g_{\mu\nu} \beta^\nu}}\,.
\end{equation}
When $K^\mu = \delta^\mu_0$, the equation for $T$ is referred to as Tolman's law~\cite{Tolman:1930zza,Tolman}, $T = T_0/\sqrt{-g_{00}}$, where $T_0 = 1/\beta_0$. Combining these back together, we find
\begin{equation}
    \beta^\mu = \frac{u^\mu}{T}
\end{equation}
We may similarly define the chemical potential $\mu$ from the gauge field by introducing~\cite{Jensen:2013kka} a gauge parameter $\Lambda$,
\begin{equation}\label{ch2:hydro:def_mu}
    \alpha \equiv \frac{\mu}{T} = \beta^\mu A_\mu + \Lambda\,.
\end{equation}
Under a background gauge transformation $A_\mu \to A_\mu + \de_\mu \lambda$, the parameter $\Lambda$ transforms as $\Lambda \to \Lambda -\beta^\mu \de_\mu \lambda$. Therefore, equation~\eqref{ch2:hydro:def_mu} is gauge invariant by the definition of $\Lambda$. We may now determine what the equilibrium constitutive relations for the stress-energy tensor and charge current are. Let us assume we can organize the constitutive relations into a derivative expansion; then to leading order in the derivative expansion, the generating functional is given by
\begin{equation}
    W[g, A] = \int d^{d+1}x \sqrt{-g} \lr{ p(T,\mu) + ...}
\end{equation}
where the $...$ denotes the possible higher-derivative contributions. Noting that
\begin{equation}
\begin{split}
        \frac{\delta \sqrt{-g}}{\delta g_{\mu\nu}} &= \frac{1}{2} \sqrt{-g} \,g^{\mu\nu}\, , \quad \frac{\delta T}{\delta g_{\mu\nu}} = \frac{1}{2} T u^\mu u^\nu\,, \quad \frac{\delta \mu}{\delta g_{\mu\nu}} = \frac{1}{2} \mu u^\mu u^\nu\,, \quad \frac{\delta \mu}{\delta A_\mu} = u^\mu\,,
\end{split}
\end{equation}
where the final three equalities follow from the definitions~\eqref{ch2:hydro:def_T},~\eqref{ch2:hydro:def_mu} of $T$ and $\mu$ respectively, we can vary $W$ to get
\begin{equation}
    T^{\mu\nu}_{\rm eq.} = \epsilon u^\mu u^\nu + p \Delta^{\mu\nu}, \qquad J^\mu_{\rm eq.} = n u^\mu\,,
\end{equation}
as anticipated in equations~\eqref{ch2:hydro:eq-const-first}, where $\epsilon \equiv - p + s T + n \mu$, $s \equiv \de p/ \de T\vert_\mu$, and $n \equiv \de p/\de \mu\vert_T$. We will extensively make use of the generating functional approach going forward for understanding hydrostatics. Of particular interest will be determining what sorts of terms vanish when the fluid approaches equilibrium, and which do not. Writing out $\beta^\mu = u^\mu/T$ and multiplying through by $T$, the Killing equation~\eqref{ch2:hydro:Killing_Eq} (which must hold in equilibrium) becomes
\begin{equation}\label{ch2:hydro:Lie-g}
    \nabla_\mu u_\nu + \nabla_\nu u_\mu - \lr{\frac{u_\nu \nabla_\mu T + u_\mu \nabla_\nu T}{T}} = 0
\end{equation}
We can now project equation~\eqref{ch2:hydro:Lie-g} with $u^\mu u^\nu$, $\Delta^{\mu\nu}$, $\frac{1}{2} \lr{\Delta^{\alpha \mu} u^\nu + \Delta^{\alpha\nu} u^\mu}$, and $\Delta^{\alpha\beta\mu\nu} \equiv \frac{1}{2} \lr{\Delta^{\mu\alpha}\Delta^{\nu\beta} + \Delta^{\mu\beta} \Delta^{\nu\alpha} - \frac{2}{d} \Delta^{\mu\nu} \Delta^{\alpha\beta}}$, where $\Delta^{\alpha\beta\mu\nu}$ is the transverse traceless projector. Applying each of these to the Killing equation, we find (recalling that $u^\nu \nabla_\mu u_\nu = 0$) that
    \begin{align}
\frac{u^\mu \nabla_\mu T}{T} &= 0\,, \quad \nabla_\mu u^\mu = 0\,,\quad  \Delta^{\alpha \mu} \lr{u^\nu \nabla_\nu u_\mu + \frac{\nabla_\mu T}{T}} = 0\,, \quad   \sigma^{\alpha\beta} = 0\,,
    \end{align}
where we have defined the shear tensor $\sigma^{\alpha \beta} \equiv 2 \Delta^{\alpha\beta\mu\nu}\nabla_\mu u_\nu$. Similarly, we can consider the demand that the Lie derivative of the gauge field with respect to $\beta^\mu$ vanish in equilibrium. However, $\Lied_{\beta^\mu} A_\mu = 0$ is not a gauge-invariant constraint, and so we instead write
\begin{equation}\label{ch2:hydro:Lie-A}
    \Lied_{\beta^\mu} A_\mu + \nabla_\mu \Lambda = 0\,.
\end{equation}
Let us project equation~\eqref{ch2:hydro:Lie-A} with $u^\mu$ and $\Delta^{\alpha\mu}$. We then arrive at
\begin{subequations}
    \begin{align}
u^\mu \nabla_\mu \lr{\frac{\mu}{T}} = 0\,,\quad \Delta^{\alpha\mu} \lr{-\frac{E_\mu}{T} + \nabla_\mu \lr{\frac{\mu}{T}}} = 0\,,
    \end{align}
\end{subequations}
where $E_\mu = F_{\mu\nu} u^\nu$ is the electric field associated with the background gauge field. We therefore have the following collection of one-derivative terms which vanish in equilibrium for a $U(1)$ charged fluid:
\begin{equation}
    \begin{gathered}\label{ch2:hydro:the-basis}
        s_1 \equiv \frac{u^\mu \nabla_\mu T}{T}, \quad s_2 \equiv \nabla_\mu u^\mu, \quad s_3 \equiv u^\mu \nabla_\mu \lr{\frac{\mu}{T}}\,,\\
        V_1^\mu \equiv \Delta^{\mu\nu} \lr{u^\lambda \nabla_\lambda u_\nu + \frac{\nabla_\nu T}{T}}, \quad V_2^\mu \equiv \Delta^{\mu\nu} \lr{-\frac{E_\nu}{T} + \nabla_\nu \lr{\frac{\mu}{T}}}\,,\\
        \sigma^{\mu\nu}  = 2 \Delta^{\mu\nu\alpha\beta} \nabla_\alpha u_\beta\,.
    \end{gathered}
\end{equation}
In particular, note that the acceleration $a^\mu \equiv u^\lambda \nabla_\lambda u^\mu$ and the transverse derivative of temperature $(\Delta^{\mu\nu}\nabla_\nu T)/T$ do not go to zero independently in equilibrium. Similarly, $E^\mu/T$ and $\Delta^{\mu\nu}\nabla_\nu \lr{\mu/T}$ are not independently zero in equilibrium. 

There is one more equilibrium detail we would like to introduce, which is the static susceptibility. The static susceptibility describes the change of densities in the theory (in the rest frame energy density $\epsilon =T^{00}$, charge density $n = J^0$, and the momentum density $\pi^j = T^{0j}$ which we usually do not explicitly write down) due to a change in the classical sources associated with those densities. One can show~\cite{Kovtun:2012rj} that the sources for perturbations $\varphi_A = (\delta \epsilon,\delta n,\delta \pi^j)$ are given by $\lambda_B = (\delta T/T,\delta v^j, T \delta \alpha)$. We therefore can write the matrix relation
\begin{equation}
    \varphi_A = \chi_{AB} \lambda_B\,.
\end{equation}
The matrix $\chi_{AB}$ is the static susceptibility. In the case of diffusion, where temperature and velocity were not perturbed, the only relevant density was the charge density, and $T \delta \alpha = \delta \mu$ meant that (using the notation of the previous section), $\delta \rho = \lr{\pder{\rho}{\mu}} \delta \mu$, which was why we defined $\chi = \pder{\rho}{\mu}$. 

\subsection{``Canonical" formulation of relativistic hydrodynamics}
\label{ch4:hydro:sec_Landau_Eckart}
Let us now depart from equilibrium. We will choose to parametrize the stress-energy tensor and charge current in terms of the ``hydrodynamic variables" $T(x)$, $u^\mu(x)$, and $\mu(x)$. Out of equilibrium, the hydrodynamic variables do not have a unique definition, and may be redefined by terms that vanish in equilibrium (e.g. $T \to T' = T + \frac{u^\mu \de_\mu T}{T}$). We once again assume that the hydrodynamic variables are slowly varying functions of spacetime, such that the constitutive relations may be organized into a derivative expansion\footnote{One immediate question that may spring to mind is whether the derivative expansion is convergent. This is an active area of research, with remaining open questions. If it is convergent, what is its radius of convergence~\cite{Grozdanov:2019kge,Grozdanov:2019uhi}? If it is divergent, what is the optimal truncation point~\cite{Heller:2021oxl,Heller:2021yjh}?}, as in equation~\eqref{intro:deriv_expansion}:\!
\vspace{-1.5em}\begin{subequations}\label{ch2:hydro:deriv_expansion}
    \begin{align}
        T^{\mu\nu} &= T^{\mu\nu}_{(0)} + T^{\mu\nu}_{(1)} + \oser{\de^2}\,,\\
        J^\mu &= J^{\mu}_{(0)} + J^\mu_{(1)} + \oser{\de^2}\,,
    \end{align}
\end{subequations}
where the subscript indicates the order in the derivative expansion. We can write down the form of the stress-energy tensor and charge current based on symmetry. Given a unit timelike vector $u^\mu$, one can decompose a generic symmetric 2-tensor and generic vector according to
\begin{subequations}\label{ch2:hydro:gen-decomposition}
    \begin{align}
        T^{\mu\nu} &= \TE u^\mu u^\nu + \TP \Delta^{\mu\nu} + 2 \TQ^{(\mu} u^{\nu)} + \TT^{\mu\nu}\,,\\
        J^\mu &= \JN u^\mu + \JJ^\mu\,,
    \end{align}
\end{subequations}
where $\TQ^{\mu}$, $\JJ^\mu$ are transverse vector to $u^\mu$, and $\TT^{\mu\nu}$ is the transverse traceless contribution to the stress-energy tensor. The curly variables are defined by
\begin{equation}
    \begin{gathered}
        \TE = T^{\mu\nu}u_\mu u_\nu, \quad \TP = \frac{1}{d} T^{\mu\nu} \Delta_{\mu\nu}, \quad \JN = - J^\mu u_\mu\,,\\
        \TQ^\alpha = \Delta^{\alpha}_{\,\,\,\,\mu} u_\nu T^{\mu\nu},\quad \JJ^\alpha = \Delta^{\alpha}_{\,\,\,\,\mu} J^\mu, \quad  \TT^{\mu\nu} = \Delta^{\mu\nu\alpha\beta} T_{\alpha\beta}\,.
    \end{gathered}
\end{equation}
Combining the decomposition~\eqref{ch2:hydro:gen-decomposition} with the derivative expansion~\eqref{ch2:hydro:deriv_expansion} leads to the statement
\begin{subequations}\label{ch2:hydro:deriv-decomp}
    \begin{alignat}{4}
        &\,\,\,\,\,\TE &&= \TE_{(0)} + \TE_{(1)} + \oser{\de^2}\,,\quad &&\,\,\,\,\,\TP &&= \TP_{(0)} + \TP_{(1)} + \oser{\de^2}\,,\\
        &\,\,\,\JN &&= \JN_{(0)} + \JN_{(1)} + \oser{\de^2}\,,\quad &&\,\,\TQ^\mu &&= \TQ^\mu_{(0)} + \TQ^\mu_{(1)} + \oser{\de^2}\,,\\
        &\JJ^\mu &&= \JJ^\mu_{(0)} + \JJ^\mu_{(1)} + \oser{\de^2}\,,\quad &&\TT^{\mu\nu} &&= \TT^{\mu\nu}_{(0)} + \TT^{\mu\nu}_{(1)} + \oser{\de^2}\,.
    \end{alignat}
\end{subequations}
We will now proceed to consider the derivative expansion at each order.

\subsubsection{Zeroth Order (or, Ideal Order)}
At zeroth order, the stress-energy tensor and the charge current take on their equilibrium forms, albeit with the hydrodynamic variables $T(x^\mu)$, $\mu(x^\mu)$, and $u^\mu(x^\mu)$:
\begin{subequations}\label{ch2:hydro:ideal-conrel}
    \begin{align}
        T^{\mu\nu} = \epsilon u^\mu u^\nu + p\, \Delta^{\mu\nu}\,,\quad J^\mu = n u^\mu\,.
    \end{align}
\end{subequations}
Matching the constitutive relations~\eqref{ch2:hydro:ideal-conrel} with the form~\eqref{ch2:hydro:gen-decomposition}, we can say
\begin{equation}
\begin{gathered}
    \TE_{(0)} = \epsilon, \quad \TP_{(0)} = p, \quad \JN_{(0)} = n\,,\\
    \TQ^\mu_{(0)} = \JJ^{\mu}_{(0)} = \TT^{\mu\nu}_{(0)} = 0\,.
\end{gathered}
\end{equation}
Inserting the constitutive relations~\eqref{ch2:hydro:ideal-conrel} into the conservation equations yields (projecting the stress-energy conservation equation with $u^\mu$ and $\Delta^{\mu\nu}$)
\begin{subequations}\label{ch2:hydro:euler_eqs}
    \begin{align}
        u^\mu \nabla_\mu\epsilon + \lr{\epsilon + p} \nabla_\mu u^\mu &= 0,\\
        \Delta^{\alpha}_{\,\,\,\,\nu} \lr{ \lr{\epsilon + p} u^\mu \nabla_\mu u^\nu + \Delta^{\nu\mu}\nabla_\mu p} &= 0\,,\\
        \nabla_\mu \lr{n \,u^\mu} &= 0\,.
    \end{align}
\end{subequations}
where we have used the fact that $u_\nu u^\mu \nabla_\mu u^\nu = 0$. A system which obeys the relativistic Euler equations~\eqref{ch2:hydro:euler_eqs} is called a perfect fluid. A perfect fluid is characterized by a lack of viscosity; this is not to be confused with the ideal gas, which has infinite viscosity\footnote{Broadly speaking, viscosity goes as the mean free path. A perfect fluid has a mean free path of zero, corresponding to zero Knudsen number. An ideal gas has an infinite mean free path.}. Let us now investigate the perfect fluid.

Consider perturbations about a homogeneous equilibrium state characterized by thermodynamic parameters $(T_0,\mu_0, u^\mu_0)$:
\begin{equation}\label{ch2:hydro:perturbs_euler}
\begin{split}
    T(x^\mu) = T_0 + \delta T(x^\mu), &\quad \mu(x^\mu) = \mu_0 + \delta \mu(x^\mu),\quad u^\mu(x^\mu) = u^\mu_0 +\delta u^\mu(x^\mu)\,.
    \end{split}
\end{equation}
We demand that $u^\mu$ remain normalized to linear order, and so $u_\mu^0 \delta u^\mu = 0$. Let us now insert the perturbations~\eqref{ch2:hydro:perturbs_euler} into the relativistic Euler equations~\eqref{ch2:hydro:euler_eqs}, restrict to flat space, and linearize in the perturbations. Then, we find (defining $\epsilon_0 = \epsilon(T_0,\mu_0)$, $p_0 = p(T_0, \mu_0)$, $n_0 = n(T_0,\mu_0)$, $s_0 = s(T_0,\mu_0)$ and $\Delta^{\mu\nu}_0 = \eta^{\mu\nu} + u_0^\mu u_0^\nu$)
    \begin{subequations}
        \begin{align}
             T_0\lr{\pder{\epsilon_0}{T_0}}_{\alpha_0}\frac{u_0^\mu \de_\mu \delta T}{T_0} + \lr{\pder{\epsilon_0}{\mu_0} }_{T_0} u_0^\mu T_0 \de_\mu \delta \alpha + \lr{\epsilon_0 + p_0} \de_\mu \delta u^\mu &= 0\,,\\
        \Delta^\alpha_{0,\nu} \lr{ \lr{\epsilon_0 + p_0} u_0^\mu \de_\mu \delta u^\nu + \Delta^{\nu\mu}_0 \lr{\lr{\epsilon_0 + p_0} \frac{\de_\mu \delta T}{T_0} + n_0 T_0 \de_\mu \delta \alpha}} &= 0\,,\\
        n_0 \de_\mu \delta u^\mu + T_0\lr{\pder{n_0}{T_0}}_{\alpha_0} \frac{u^\mu_0 \de_\mu \delta T}{T_0} + \lr{\pder{n_0}{\mu_0}}_{T_0} T_0 u^\mu_0 \de_\mu \delta \alpha &=0\,.
        \end{align}
    \end{subequations}
where we have written the perturbations in terms of $T_0 \delta \alpha= \delta \mu - \alpha_0 \delta T $, and we have used the fact that $\epsilon_0 + p_0 = s_0 T_0 + n_0 \mu_0$. We also note that $T_0\lr{\pder{x}{T_0}}_{\alpha_0} = \lr{\pder{x}{T_0}}_{\mu_0}T_0 + \lr{\pder{x}{\mu_0}}_{T_0} \mu_0$ for thermodynamic variable $x$. Re-writing this equation in matrix form, we find
\begin{equation}\label{ch2:hydro:Euler_eq_matrix}
    X_{AB} u^\mu_0 \de_\mu U^B + B_{AB}^\nu \Delta^{\mu}_{0,\nu} \de_\mu  U^B = 0_{A}
\end{equation}
where $U^B = (\frac{\delta T}{T_0},\delta u^\beta,T_0 \delta \alpha)^T$, and
\begin{subequations}
    \begin{align}
        X_{AB} &= \begin{pmatrix}
            T_0\lr{\pder{\epsilon_0}{T_0}}_{\alpha_0}  &  0_\beta & \lr{\pder{\epsilon_0}{\mu_0}}_{T_0} \\
            0^\alpha& \Delta^{\alpha}_{0,\beta} \lr{\epsilon_0 + p_0} & 0^\alpha \\
            T_0\lr{\pder{n_0}{T_0}}_{\alpha_0}   & 0_\beta & \lr{\pder{n_0}{\mu_0}}_{T_0}
        \end{pmatrix}\,,\label{ch2:hydro:static_susceptibility}\\ 
        B_{AB}^\nu &= \begin{pmatrix}
            0^\nu & \lr{\epsilon_0 + p_0} \Delta^\nu_{0,\beta} & 0^\nu \\
            \Delta^{\alpha\nu}_0 (\epsilon_0 + p_0) & 0^{\alpha\nu}_\beta& \Delta^{\alpha\nu}_0 n_0 \\
            0^\nu & n_0 \Delta^\nu_{0,\beta} & 0^\nu 
        \end{pmatrix}\,,
    \end{align}
\end{subequations}
where we have made use of the identity $\de_\mu \delta u^\mu = \Delta^{\alpha}_{0,\beta} \de_\alpha \delta u^\beta$. The vector of variables $U^B$ is sometimes called the ``state vector". Due to the choice of variables in $U^B$, the matrix $X_{AB}$ is equal to the susceptibility matrix $\chi_{AB}$ for the fluid. One can show (see Appendix~\ref{app:linear response}) that the susceptibility matrix has some desirable properties:
\begin{itemize}
    \item The susceptibility matrix $\chi_{AB}$ is symmetric, i.e. $\chi_{AB} = \chi_{BA}$. Therefore, $\lr{\pder{\epsilon_0}{\mu_0}}_{T_0} = T_0\lr{\pder{n_0}{T_0}}_{\alpha_0} =  T_0 \lr{\pder{n_0}{T_0}}_{\mu_0} + \mu_0 \lr{\pder{n_0}{\mu_0}}_{T_0}$.
    \item Secondly, one can show that the diagonal entries must be all non-negative, i.e. $\chi_{AA} \geq 0$ for all $A$. Therefore, $p_0+\epsilon_0 \geq 0$, $\lr{\pder{n_0}{\mu_0}}_{T_0} \geq 0$, and $T_0\lr{\pder{\epsilon_0}{T_0}}_{\alpha_0} \geq 0$.
    \item From these two, one arrives at the demand that the susceptibility matrix be positive semi-definite.
\end{itemize}
Returning to the relativistic Euler equations~\eqref{ch2:hydro:Euler_eq_matrix}, let us specialize to the case of plane-wave perturbations, i.e. $U^B(x^\mu) = U^B (K_\mu) e^{i K_\mu x^\mu}$, where\footnote{When appearing in the linearized analysis, $K^\mu$ is the momentum four-vector, not to be confused with the Killing vector $K_\mu$ of the previous subsection; we hope context is enough to disambiguate the two.} $K_\mu = (-\omega, k_j)$. We then obtain the system of equations
\begin{equation}
    \lr{X_{AB} \lr{i u_0^\mu K_\mu} + B_{AB}^\nu \Delta_{0,\nu}^\mu \lr{i K_\mu}}  U^B(\omega, k_i) e^{i K_\mu x^\mu} \equiv M_{AB}  U^B(\omega,k_i) e^{i K_\mu x^\mu} = 0_A
\end{equation}
These equations only have non-trivial solutions for the perturbations if $F(\omega,k_i) \equiv \det\lr{M} = 0$. The zeroes of the function $F(\omega, k_i)$ determine the dispersion relations of the fluid, i.e. they determine $\omega_n = \omega_n(k_i)$, where $n$ labels the frequency mode. We will refer to the function $F(\omega, k_i)$ as the spectral curve, as it is an algebraic curve in $\omega$ and $k_i$. For an equilibrium state at rest (i.e. $u_0^\mu = \delta^\mu_0$),
\begin{equation}\label{ch2:hydro:euler-matrix-fourier}
    M_{AB} = \begin{pmatrix}
        - i \omega T_0\lr{\pder{\epsilon_0}{T_0}}_{\alpha_0} & i \lr{\epsilon_0 + p_0}k_j & -i\omega \lr{\pder{\epsilon_0}{\mu_0} }_{T_0}\\
        i k^i \lr{\epsilon_0 + p_0} & -i \omega \delta^i_j \lr{\epsilon_0 + p_0} & i k^i n_0\\
        -i \omega T_0\lr{\pder{n_0}{T_0}}_{\alpha_0} & i n_0 k_j & - i \omega \lr{\pder{n_0}{\mu_0}}_{T_0}
    \end{pmatrix}\,.
\end{equation}
Isotropy of the equilibrium state allows us to fix (without loss of generality) $k_j$ to point in the $x$-direction. The determinant of matrix~\eqref{ch2:hydro:euler-matrix-fourier} will then factorize, yielding $(d-1)$ copies of a mode $\omega = 0$. Looking ahead, we will refer to these modes as the ``shear modes", and denote them by $\omega_{\eta}$. They describe the propagation (or lack there-of) of momentum perturbations transverse to the direction of the wave-vector $k_i$. The remaining factor of the spectral curve is then given by
\begin{equation}\label{ch2:hydro:block_matrix_Euler}
   \det \begin{vmatrix}
        - i \omega T_0\lr{\pder{\epsilon_0}{T_0}}_{\alpha_0} & i \lr{\epsilon_0 + p_0}k_x & -i\omega \lr{\pder{\epsilon_0}{\mu_0}}_{T_0}\\
        i k_x \lr{\epsilon_0 + p_0} & -i \omega  \lr{\epsilon_0 + p_0} & i k_x n_0\\
        -i \omega T_0\lr{\pder{n_0}{T_0}}_{\alpha_0} & i n_0 k_x & - i \omega \lr{\pder{n_0}{\mu_0}}_{T_0}
    \end{vmatrix}
\end{equation}
Taking the determinant of this matrix and setting it to zero yields
\begin{equation}\label{ch2:hydro:euler_longitudinal}
    \omega \lr{ \omega^2 - \frac{B}{A} k_x^2} = 0
\end{equation}
where
\begin{subequations}
    \begin{align}
        A &=  (\epsilon_0 + p_0)  \lr{\lr{\pder{n_0}{\mu_0}}_{T_0} \lr{\pder{\epsilon_0}{T_0}}_{\alpha_0} T_0- \lr{\pder{\epsilon_0}{\mu_0}}_{T_0}^2}\,,\\
        B &=  \lr{\pder{\epsilon_0}{T_0}}_{\alpha_0} T_0 n_0^2 + \lr{p_0 + \epsilon_0}^2 \lr{ \lr{\pder{n_0}{\mu_0}}_{T_0}-\frac{2 n_0}{p_0 + \epsilon_0} \lr{\pder{\epsilon_0}{\mu_0}}_{T_0}}\,,
    \end{align}
\end{subequations}
where in the above we have made use of the fact $T_0\lr{\pder{n_0}{T_0}}_{\alpha_0} = \lr{\pder{\epsilon_0}{\mu_0}}_{T_0}$. Inspecting~\eqref{ch2:hydro:euler_longitudinal}, we see that another factor of $\omega$ factors out of the block matrix~\eqref{ch2:hydro:block_matrix_Euler}. Looking ahead, we will call this mode the $U(1)$ charge diffusion mode, or just the charge mode for short, and will denote it by $\omega_\sigma$. Finally, there are two modes of the form
$    \omega = \pm v_s |k|$
where $v_s = \sqrt{B/A}$. These modes, which we will denote by $\omega_{\pm}$, describes the propagation of a mixture of longitudinal (with respect to $k_i$) momentum perturbations and temperature perturbations. These modes are sound modes, and the coefficient $v_s$ the is speed of sound. 

At first glance, the expression for $v_s^2 = B/A$ looks somewhat complicated. However, this is only due to the choice of hydrodynamic variables. If we had used $\epsilon$ and $n$ instead of $T$ and $\mu$, the expressions would have taken on a different form (though the physics remains the same):
\begin{equation}\label{ch2:hydro:speed_of_sound}
    v_s^2 = \lr{\pder{p_0}{\epsilon_0}}_{n_0} + \frac{n_0}{\epsilon_0 + p_0} \lr{\pder{p_0}{n_0}}_{\epsilon_0}\,.
\end{equation}
In a theory without a $U(1)$ charge, this expression simplifies further, to
$    v_s^2 = \pder{p_0}{\epsilon_0}\,.$
With the dispersion relations now established, let us consider the stability and causality of the relativistic Euler equations. First of all, they are stable so long as $\lr{\pder{p_0}{\epsilon_0}}_{n_0} + \frac{n_0}{\epsilon_0 + p_0} \lr{\pder{p_0}{n_0}}_{\epsilon_0} \geq 0$, as all of the modes are real. Next, we demand causality. For simplicity, we simply consider the two conditions~\eqref{ch2:diffusion:causality_mom_v2} -- we will take a more nuanced approach at first order in the derivative expansion. The shear mode $\omega_\eta$ and the charge mode $\omega_\sigma$ are identically zero, and so satisfy both conditions. For the sound mode, we find
\begin{equation}
   \lim_{|k| \to \infty} \frac{|\Re(\omega)|}{|k|} = |v_s|, \quad \lim_{|k| \to \infty} \frac{\Im(\omega)}{|k|} = 0\,.
\end{equation}
We therefore see that the (linearized) relativistic Euler equations are causal so long as the speed of sound is less than the speed of light. 

Finally, let us consider the fate of the second law of thermodynamics for an out-of-equilibrium system. The second law of thermodynamics states that the total entropy may never decrease with time: therefore, for a time $t$ and a later time $t'$,
\begin{equation}
    S(t') - S(t) \geq 0\,.
\end{equation}
$S(t')$ and $S(t)$ may be written as the fluxes going through spacelike hypersurfaces $\Sigma_{t'}$ and $\Sigma_{t}$ respectively. Then
\begin{equation}\label{ch2:hydro:entropy-difference}
    \int d\Sigma_{\mu}^{t'} S^\mu - \int d\Sigma^t_\mu S^\mu \geq 0\,.
\end{equation}
Let us denote the volume contained between $\Sigma_\mu^{t'}$ and $\Sigma_\mu^t$ as $V_{\Delta t}$. Assuming standard boundary conditions, we can use Gauss' law to re-write the boundary integrals~\eqref{ch2:hydro:entropy-difference} as a volume integral~\cite{Rezzolla-Zanotti}
\begin{equation}
    \int_{V_{\Delta t}} d^{d+1}x \nabla_\mu S^\mu \geq 0\,.
\end{equation}
However, since $t'$ and $t$ were arbitrary, we must have that
\begin{equation}\label{ch2:hydro:second-law}
    \nabla_\mu S^\mu  \geq 0\,.
\end{equation}
This is the local version of the second law of thermodynamics. The current $S^\mu$ is called the entropy current, and it is one of the most important objects in relativistic hydrodynamics. Based on the equilibrium statement that $T_0 s_0 = p_0 + \epsilon_0 - \mu_0 n_0$, we can define a canonical form for the entropy current:
\begin{equation}\label{ch2:hydro:canonical-entropy}
   S^\mu = \frac{p}{T} u^\mu - T^{\mu\nu}\beta_\nu - \alpha J^\mu
\end{equation}
For the ideal order constitutive relations, this takes on the far simpler form
$    S^\mu = s u^\mu$.
The relativistic Euler equations~\eqref{ch2:hydro:euler_eqs} then enforce that
\begin{equation}
    \nabla_\mu S^\mu = \nabla_\mu \lr{s u^\mu} = 0
\end{equation}
We arrive at the interesting statement that for a perfect fluid, there is no entropy production except via discontinuous processes such as shocks.\footnote{This last point brings up an interesting property of perfect fluids, which was originally brought to my attention by D. Wagner. Equilibrium is the global entropy maximum, and a given out-of-equilibrium configuration must, necessarily, have a smaller overall entropy. However, in a perfect fluid entropy cannot be generated via dissipation, and so the fluid will eventually arrive at a homogeneous entropy density profile with a total entropy less than that of global equilibrium. Therefore, in order to return to the global entropy maximum, a perfect fluid must equilibrate via genuinely entropy-producing processes, i.e. shocks. This process is, of course, unphysical: all real fluids produce entropy via dissipative processes, which also serve to smooth out shock profiles. That said, the study of relativistic shocks in both perfect and dissipative fluids is an active area of research due to their interesting mathematical and physical properties~\cite{Taub:1948zz,LL6,IsraelAnileChoquetBruhat:1989,Pandya:2021ief,Pandya:2022pif,Pandya:2022sff,Freistuhler:2021lla}.} With the details of the ideal-order case established, we now turn to the more complicated consideration of the first-order case. In what follows, the difference between the hydrodynamic variables $T(x)$, $\mu(x)$, $u^\mu(x)$ and their thermodynamic counterparts will play a central role.

\subsubsection{First-order}
We will now endeavour to determine the first-order corrections to the decompositions~\eqref{ch2:hydro:deriv-decomp}. As a matter of first principles, we may divide the corrections into two categories -- those that arise from the generating functional (``hydrostatic contributions") and those that vanish in equilibrium (``non-hydrostatic contributions"):
\begin{align*}
    T^{\mu\nu} &= T^{\mu\nu}_{(0)} + T^{\mu\nu}_{(1), \rm n.h.s.} + T^{\mu\nu}_{(1),\rm  h.s.} + \oser{\de^2}\,,\\
    J^\mu &= J^\mu_{(0)} + J^\mu_{(1), \rm n.h.s.} + J^\mu_{(1),\rm  h.s.} + \oser{\de^2}\,.
\end{align*}
Let us begin by considering the hydrostatic contributions. A first-order hydrostatic contribution would arise due to a non-vanishing first-order contribution to the grand canonical free energy, i.e.
\begin{equation}
    {\cal F} =  p + {\cal F}_{(1)} + \oser{\de^2}
\end{equation}
We do not \textit{a priori} know what these contributions should be, and so we should write down any scalars at first order in derivatives which do not vanish in equilibrium. In the theory under consideration in this section (i.e. the theory of a $U(1)$ charged fluid which respects time-reversal, parity, and charge-conjugation symmetry), there turn out not to be any contributions at first order. However, in MHD, there exist non-trivial contributions e.g. $\Omega_\mu B^\mu$, where $B^\mu$ is the magnetic field, and $\Omega_\mu = \epsilon_{\mu\nu\rho\sigma}u^\nu \de^\rho u^\sigma$ is the vorticity vector. This magnetovortical term is generically present in the constitutive relations of first-order MHD unless charge-conjugation symmetry is imposed to remove it~\cite{Hernandez:2017mch}.

Next, let us consider non-hydrostatic terms in the constitutive relations at first order.\footnote{Technically, non-hydrostatic contributions can come in two types -- dissipative contributions (terms that describe effects that generate entropy) and non-dissipative contribution. In what follows, we will be somewhat imprecise, and use the terms non-hydrostatic and dissipative interchangeably; we will not be discussing non-hydrostatic non-dissipative terms in this dissertation.} The form of the non-hydrostatic contribution is constrained by the requirements that it be built out of first derivatives of the hydrodynamic variables, and that it vanish in equilibrium. We may therefore construct the most general expression using the standard approach of effective field theory: write down everything consistent with the symmetries of the theory under consideration, and then try to constrain the parameters accompanying each contribution.

We have already determined which one-derivative quantities vanish in equilibrium -- the ``building blocks"~\eqref{ch2:hydro:the-basis}. The most general one-derivative dissipative correction would then be of the form
\begin{subequations}\label{ch2:hydro:gen-con-eq}
    \begin{alignat}{4}
       & \,\,\,\,\,\,\TE &&= \epsilon + \sum_{n=1}^3\ce_n s_n +\oser{\de^2}\,,\quad &&\,\,\,\,\,\TP &&= p + \sum_{n=1}^3\pi_n s_n + \oser{\de^2}\,,\\
        &\,\,\,\JN &&= n + \sum_{n=1}^4\nu_n s_n+ \oser{\de^2}\,,\quad &&\,\,\TQ^\mu &&= \sum_{n=1}^2\theta_n V_n^\mu+\oser{\de^2}\,,\\
        &\JJ^\mu &&= \sum_{n=1}^2\gamma_n V_n^\mu +\oser{\de^2}\,,\quad &&\TT^{\mu\nu} &&= - \eta \sigma^{\mu\nu}\,.
    \end{alignat}
\end{subequations}
where the sign on $\eta$ has been chosen for future convenience. There are 14 \textit{a priori} independent parameters in the theory. This is a somewhat unpleasant situation to be in; it would be good if we could find some way to eliminate some of these parameters. We can do this by considering two facts about the hydrodynamic variables $\{T, \mu, u^\mu\}$.

\paragraph{Frame redefinitions.} The first fact is that the hydrodynamic variables have no unique definition out of equilibrium. A set of variables $U^B = \{T,\mu,u^\mu\}$ and a different set of variables $U'^B = \{T',\mu',u'^\mu\}$ are both equally valid so long as $\delta U^B = \{T'-T,\mu'-\mu,u'^\mu-u^\mu\} = \{\delta T,\delta \mu, \delta u^\mu\}$ vanishes in equilibrium\footnote{In curved space, there remains ambiguity even in equilibrium, see e.g.~\cite{Kovtun:2022vas}}. Any particular choice of variable is called a ``fluid frame" or a ``hydro frame" -- not to be confused with a Lorentz frame. A change from one set of variables to another set of variables is called a ``frame transformation" or a ``frame redefinition". While the names are unfortunately somewhat confusing, they are standard. 

Let us take $\delta U^B$ to be first order in derivatives. Applying the frame change $U^B \to U'^B$ to~\eqref{ch2:hydro:deriv-decomp} after inserting $T^{\mu\nu}_{(0)}$ and $J^\mu_{(0)}$, and demanding the invariance of the one point functions under frame transformations\footnote{The one-point function of a microscopic operator is independent of which effective variables we use.} leads to the first-order contributions undergoing (up to $\oser{\de^2}$) shifts:
\begin{subequations}
   \begin{alignat}{4}
        &\TE_{(1)} &&\to \TE'_{(1)} - \biggl[ \lr{\pder{\epsilon}{T}}_{\mu} \delta T + \lr{\pder{\epsilon}{\mu}}_{T} \delta \mu\biggr]\,,\quad &&\JJ_{(1)}^\mu &&\to \JJ'^\mu_{(1)} - n \delta u^\mu\,, \\
        &\JN_{(1)} &&\to \JN'_{(1)} - \biggl[ \lr{\pder{n}{T}}_{\mu} \delta T + \lr{\pder{n}{\mu}}_{T} \delta \mu \biggr]\,,\quad &&\TQ_{(1)}^\mu &&\to \TQ'^\mu_{(1)} - (\epsilon + p) \delta u^\mu\,,\\
        &\TP_{(1)} &&\to \TP'_{(1)} - \biggl[ \lr{\pder{p}{T}}_{\mu} \delta T + \lr{\pder{p}{\mu}}_{T} \delta \mu \biggr]\,,\quad &&\TT^{\mu\nu}_{(1)} &&\to \TT'^{\mu\nu}_{(1)}\,.
   \end{alignat}
\end{subequations}
We see that all of the dissipative contributions are shifted under the change of variables except for the tensor contribution $\TT^{\mu\nu}_{(1)}$. There exist certain linear combinations of the dissipative contributions that are invariant under reparametrizations. One way to write these down these so-called ``frame invariants" is the following:
\begin{subequations}\label{ch2:hydro:frame-invar-full}
    \begin{align}
        {\cal F} &= \TP_{(1)} - \lr{\pder{p}{\epsilon}}_{n} \TE_{(1)} - \lr{\pder{p}{n}}_{\epsilon} \JN_{(1)}\,,\\
        {\cal L}^\mu &= \JJ^\mu_{(1)} - \frac{n}{\epsilon + p} \TQ^{\mu}_{(1)} \,.
    \end{align}
\end{subequations}
The frame invariants~\eqref{ch2:hydro:frame-invar-full} may be decomposed with respect to the building blocks~\eqref{ch2:hydro:the-basis}, such that
    \begin{align}
        {\cal F} = f_1 s_1 + f_2 s_2 + f_3 s_3\,,\qquad {\cal L}^\mu = \ell_1 V_1^\mu + \ell_2 V_2^\mu\,.
    \end{align}
Therefore, by~\eqref{ch2:hydro:gen-con-eq}, we may identify
\begin{subequations}\label{ch2:hydro:frame-invar-basis}
    \begin{align}
        f_i &= \pi_i - \lr{\pder{p}{\epsilon}}_n \ce_i - \lr{\pder{p}{n}}_\epsilon \nu_i \quad (i \in\{1..3\})\,,\\
        \ell_i &= \gamma_i - \frac{n}{\epsilon + p} \theta_i \quad (i \in \{1..2\})\,.
    \end{align}
\end{subequations}
Since the one-point function should be invariant under changes of frame, the underlying physics ought to be invariant under a change of frame. Let us therefore use the redefinition freedom inherent in the equations to write them in a convenient form. Let us pick a frame such that $T^{00} = \epsilon$ for a fluid at rest, $J^0 = n$, and the fluid velocity $u^\mu$ is aligned with the flow of energy, so that there is no transverse heat flow. Then $\ce_i = \nu_i = \theta_i = 0$, and the number of free parameters in the theory has been reduced from fourteen to six:
\begin{subequations}
\label{ch2:hydro:partial-ns}
    \begin{align}
        T^{\mu\nu} &= \epsilon u^\mu u^\nu + \lr{p + f_1 s_1 + f_2 s_2 + f_3 s_3} \Delta^{\mu\nu} - \eta \sigma^{\mu\nu}\,,\\
        J^\mu &= n u^\mu + \ell_1 V_1^\mu + \ell_2 V_2^\mu\,.
    \end{align}
\end{subequations}
This choice of hydrodynamic frame is usually called the ``Landau frame", or ``Landau-Lifshitz frame", so-called because of its use in the famous textbook series by Landau and Lifshitz in volume 6, ``Fluid Mechanics"~\cite{LL6}. In writing down the frame invariants~\eqref{ch2:hydro:frame-invar-full}, we tailored them for use with the Landau frame.\footnote{Another frame which is frequently used in the literature is the ``Eckart frame", which was first used by Carl Eckart in~\cite{Eckart-original}, and is also used in the textbook ``Gravitation and Cosmology" by Weinberg~\cite{weinberg:1972}. This frame choice is the same in the scalar section ($\ce_i = \nu_i = 0$), but instead aligns the fluid velocity with the flow of $U(1)$ charge, so that there is no transverse flow ($\gamma_i = 0$). The constitutive relations in the Eckart frame are given by
    \begin{align*}
        T^{\mu\nu} &= \epsilon u^\mu u^\nu + \lr{p + f_1 s_1 + f_2 s_2 + f_3 s_3}\Delta^{\mu\nu} - \frac{2(\epsilon+p)}{n}\lr{\ell_1 V_1^{(\mu} + \ell_2 V_2^{(\mu} } u^{\nu)} - \eta \sigma^{\mu\nu}\,,\\
        J^\mu &= n u^\mu\,.
    \end{align*}} There still remain six parameters $(f_1,f_2,f_3,\ell_1,\ell_2,\eta)$ in equations~\eqref{ch2:hydro:partial-ns}; we can do better.

\paragraph{Equations of motion.}The second point we can take advantage of is that the relativistic Euler equations~\eqref{ch2:hydro:euler_eqs} may be used to write down relations between the respective scalar and vector ``building blocks". There are two scalar equations $(u_\nu \nabla_\mu T^{\mu\nu} = 0$ and $\nabla_\mu J^\mu=0$) and one vector equation ($\Delta^\alpha_\nu \nabla_\mu T^{\mu\nu} = 0$). The relativistic Euler equations~\eqref{ch2:hydro:euler_eqs} may therefore be used to eliminate two scalars, and one vector:
\begin{subequations}
    \begin{align}
        \lr{\pder{\epsilon}{T}}_{\alpha} T s_1 + \lr{\epsilon+p} s_2 + \lr{\pder{\epsilon}{\mu}}_{T} T s_3 &= \oser{\de^2}\,,\\
        \lr{\pder{n}{T}}_\alpha T s_1 + n s_2 + \lr{\pder{n}{\mu}}_{T} T s_3 &= \oser{\de^2}\,,\\
        \lr{\epsilon+p} V_1^\mu + n T V_2^\mu &= \oser{\de^2}\,.
    \end{align}
\end{subequations}

We choose to eliminate $s_1$, $s_3$, and $V_1^\mu$, i.e. all of the variables that contain time derivatives of the fields in the rest frame. In field-theory parlance, this puts the theory ``on-shell" with respect to the ideal-order equations. After applying the ideal-order equations of motion, the Landau-frame constitutive relations become
\begin{equation}\label{ch2:hydro:ns}
        T^{\mu\nu} = \epsilon u^\mu u^\nu + \lr{p - \zeta s_2} \Delta^{\mu\nu} - \eta \sigma^{\mu\nu}\,,\quad J^\mu = n u^\mu - \sigma V_2^\mu\,.
\end{equation}
where we have defined new parameters $\zeta$ and $\sigma$. These are given by
\begin{subequations}
    \begin{align}
        \zeta &\equiv - f_2 + \lr{\lr{\pder{n}{\mu}}_{T} (\epsilon + p) - \lr{\pder{\epsilon}{\mu}}_{T} n} \frac{f_1}{T \partial(\epsilon,n)/\partial(T,\mu)} \nonumber\\
        &+ \lr{n \lr{\pder{\epsilon}{T}}_{\alpha} - \lr{\epsilon
         + p} \lr{\pder{n}{T}}_{\alpha}}\frac{f_3}{T\partial(\epsilon,n)/\partial(T,\mu)} \,,\\
        \sigma &\equiv - \ell_2 + \frac{n T}{\epsilon + p} \ell_1\,,
    \end{align}
\end{subequations}
where $\partial(\epsilon,n)/\partial(T,\mu)$ is the Jacobian of $(\epsilon,n)$ and $(T,\mu)$. The negative signs, similarly to the negative sign in front of $\eta$, have been chosen for convenience. We will give the coefficients appearing in equations~\eqref{ch2:hydro:ns} names: $\zeta$ is the bulk viscosity, $\eta$ is the shear viscosity, and $\sigma$ is the charge conductivity. The constitutive relations~\eqref{ch2:hydro:ns} are the relativistic Navier-Stokes equations. The fluid described by these equations is dissipative, meaning that for arbitrary non-equilibrium configurations, there should be an increase of entropy with time as the fluid returns to equilibrium. That the entropy always increases may be enforced via the entropy current~\eqref{ch2:hydro:canonical-entropy} and the local second law. For a generic first-order perturbation from equilibrium, the divergence of the entropy current may be written
\begin{equation}
    \nabla_\mu S^\mu = - T^{\mu\nu}_{(1)} \nabla_\mu \beta_\nu - J^\mu_{(1)} \nabla_\mu \alpha + \oser{\de^2}\,,
\end{equation}
where we have used the fact $\nabla_\mu \lr{s u^\mu} = \oser{\de^2}$, and the fact that $T^{\mu\nu}$ and $J^\mu$ satisfy the conservation equations. Inserting the constitutive relations for the relativistic Navier-Stokes equations~\eqref{ch2:hydro:ns} yields (imposing the ideal-order equations on $\beta_\mu$ and $\alpha$)
\begin{equation}
    \nabla_\mu S^\mu = \frac{\zeta}{T} s_2^2 +\frac{\sigma}{T} V_{2,\mu} V_2^\mu + \frac{\eta}{2T} \sigma_{\mu\nu} \sigma^{\mu\nu} + \oser{\de^2}\,.
\end{equation}
Demanding the non-negativity of the entropy current for all possible field configurations immediately leads to the demand that
$    \zeta \geq 0, \, \sigma \geq 0,\, \eta \geq 0\,.$
These coefficients $\zeta,\,\sigma,\,\eta$ are examples of what we call ``physical transport coefficients". We differentiate the physical transport coefficients from other possible parameters in the theory (such as those in equations~\eqref{ch2:hydro:gen-con-eq}) by the fact that the physical transport coefficients can be obtained from (and are therefore fixed by) the underlying microscopic theory. This may be achieved via ``Kubo formulae" -- the interested reader may again refer to Appendix~\ref{app:linear response} for more details. Any parameter in the theory which cannot be obtained from a Kubo formula is not considered a physical transport coefficient, and we simply call it a ``transport parameter".

\subsubsection{Conformal fluids} 
If the underlying theory enjoys a conformal symmetry, then the theory is invariant under Weyl transformations, i.e. rescalings of the metric of the form $g_{\mu\nu} \to e^{-2 \phi} g_{\mu\nu}$. A field $\varphi$ which transforms according to $\varphi \to e^{\boldsymbol{w} \phi}\varphi$ is said to transform with Weyl weight $\boldsymbol{w}$. It is a well known property of conformal theories that the stress-energy tensor is traceless, $T^\mu_{\,\,\,\,\mu} = 0$. For the relativistic Euler equations, this leads to the celebrated conformal condition $\epsilon = d\,p$, and so
    \begin{align}\label{ch2:hydro:conformal-euler}
        T^{\mu\nu} = d \,p u^\mu u^\nu + p \Delta^{\mu\nu} + \oser{\de}\,,\quad J^\mu = n u^\mu\,.
    \end{align}
By the definitions~\eqref{ch2:hydro:def_T},~\eqref{ch2:hydro:def_mu} one can show that $T$, $\mu$, and $u^\mu$ transform with $\boldsymbol{w}=1$, while $u_\mu$ transforms with $\boldsymbol{w}=-1$. By demanding the invariance of the generating functional under Weyl transformations, we can also determine from~\eqref{ch2:hydro:genfunc_T_J} that $T^{\mu\nu}$ transforms with $\boldsymbol{w}=d+3$, while $J^\mu$ transforms with $\boldsymbol{w}=d+1$. In order for the constitutive relation~\eqref{ch2:hydro:conformal-euler} to transform correctly under the Weyl transformation, the equation of state is heavily constrained, and must be of the form
\begin{equation}
    p(T,\mu) = T^{d+1} f\lr{\frac{\mu}{T}}\,.
\end{equation}
In particular, this can be shown to lead to the result that $(\partial p/\partial n)_\epsilon = 0$, $(\partial p/\partial \epsilon)_n = 1/d$. We therefore arrive at the statement that for a conformal fluid, the speed of sound~\eqref{ch2:hydro:speed_of_sound} is given by
$    v_s = \pm 1/\sqrt{d}\,.$
Turning now to the first-order theory, the tracelessness condition imposes that $\TE_{(1)} = d \,\TP_{(1)}$, which imposes from~\eqref{ch2:hydro:gen-con-eq} that $\ce_i = d \,\pi_i$ for $i \in \{1..3\}$.

We can also note that the building blocks $s_3$, $V_1^\mu$, $V_2^\mu$, and $\sigma^{\mu\nu}$ all transform homogeneously under Weyl transformations. However, $s_1$ and $s_2$ do not -- only the linear combination $\tilde{s}_1 = s_1 - \frac{1}{d} s_2$ transforms homogeneously. We must then also have the relation $\ce_1 = d \ce_2$, $\pi_1 = d \pi_2$, and $\nu_1 = d \nu_2$. Combining this with the previous relation between $\ce_i$ and $\pi_i$ there are nine \textit{a priori} independent parameters in the constitutive relations for a charged conformal fluid:
\begin{subequations}
    \begin{align}
        T^{\mu\nu} &= d\lr{p + d \pi_2 \tilde{s}_1 + \pi_3 s_3} u^\mu u^\nu + \lr{p + d \pi_2 \tilde{s}_1 + \pi_3 s_3}\Delta^{\mu\nu} \nonumber\\
        &+ 2 \lr{\theta_1 V_1^{(\mu} + \theta_2 V_2^{(\mu}}u^{\nu)} - \eta \sigma^{\mu\nu}\,,\\
        J^\mu &=  \lr{n + d \nu_2 \tilde{s}_1 + \nu_3 s_3} u^\mu + \gamma_1 V_1^\mu + \gamma_2 V_2^\mu\,.
    \end{align}
\end{subequations}
The frame invariants $\ell_i$ are unchanged when conformality is imposed; however, by the definition~\eqref{ch2:hydro:frame-invar-basis}, we can see that $(\partial p/\partial n)_{\epsilon} = 0$ and $\ce_i = d \pi_i$ implies $f_i = 0$ for all $i$. This has the immediate consequence that in the relativistic Navier-Stokes equations~\eqref{ch2:hydro:ns}, the bulk viscosity is identically zero, $\zeta = 0$. This makes physical sense, as the bulk viscosity is associated with changes in scale, and conformal symmetry requires scale-invariance.

\subsection{Stability and causality at first order}
\subsubsection{Characteristics and causality}
We will now more formally discuss causality. The notions of causality discussed here are local, rather than statements about any global structure of spacetime. To do so, we will need to understand some results from the theory of partial differential equations. This section mostly follows the theory as outlined in~\cite{Courant-Hilbert}; other references include~\cite{Leray:1958,Lax:2006}. A generic second-order quasilinear partial differential equation (PDE) in some variable $\Psi$ can be written in the following way:
\begin{equation}\label{ch2:hydro:gen-quasilinear-pde}
    L[U] = {\cal A}^{\mu\nu}[\Psi, \de \Psi] \de_\mu \de_\nu \Psi + {\cal B}^{\mu}[\Psi, \de \Psi] \de_\mu \Psi + {\cal C}[\Psi] = 0 \,.
\end{equation}
Note that the coefficient ${\cal A}$ depends on $\Psi$ and the first derivative of $\Psi$, but not the second derivative. We assume here that ${\cal A}$, ${\cal B}$, and ${\cal C}$ are analytic if $\Psi$ is analytic. Let us now consider supplying initial conditions for the PDE~\eqref{ch2:hydro:gen-quasilinear-pde} on some spacelike initial hypersurface $\Sigma_0$. Let us further foliate the spacetime\footnote{Of course, there is nothing special about time from a mathematical point of view; we could just as well supply a timelike ``initial hypersurface" and look for solutions given conditions on that surface. The same procedure applies.} with a whole family of such hypersurfaces, parametrized by a scalar parameter $\phi(x^\mu)$ -- the initial hypersurface is the solution to the condition $\phi = 0$. We assume that $\phi$ is an analytic function of $x^\mu$. The normal to the hypersurface is given by $\xi_\mu \equiv \nabla_\mu \phi$. The parameter $\phi$ could be time, or it could be something else. We also consider $d$ parameters which characterize the $d$-directions internal to the hypersurface: we will call these $\eta^i(x^\mu)$, and will also assume that they are analytic functions of $x^\mu$. Finally, we assume that $\phi(x^\mu)$, $\eta^i(x^\mu)$ are such that the relationship can be inverted, and we can write $x^\mu = x^\mu(\phi,\eta^i)$ so that $\Psi = \Psi(x^\mu(\phi,\eta^i))$.

The initial conditions are then given on the initial hypersurface by
\begin{equation}
    \Psi_0(\eta^i) = \Psi\vert_{\phi = 0}, \quad \dot{\Psi}_0(\eta^i) = \pder{\Psi}{\phi} \vert_{\phi = 0}\,,
\end{equation}
where we employ the notation $\dot{\Psi} \equiv \pder{\Psi}{\phi}$, and $\Psi_0$, $\dot{\Psi}_0$ are some prescribed functions of $\eta^i$. We now apply the chain rule to the PDE~\eqref{ch2:hydro:gen-quasilinear-pde} to rewrite the equation in coordinates $(\phi,\eta^i)$. Evaluated on the hypersurface $\Sigma_0$, equation~\eqref{ch2:hydro:gen-quasilinear-pde} is then written as
\begin{equation}\label{ch2:hydro:origin_characteristic_eq}
    \lr{{\cal A}^{\mu\nu}\xi_\mu \xi_\nu} \frac{\de^2\Psi}{\de\phi^2}\biggl|_{\phi = 0} = {\cal I}[\Psi_0, \dot{\Psi}_0]\,,
\end{equation}
where ${\cal I}$ contains all terms in equation~\eqref{ch2:hydro:gen-quasilinear-pde} which may be determined entirely from $\Psi_0(\eta^i)$ and $\dot{\Psi}_0(\eta^i)$. The quantity $\xi_\mu = \nabla_\mu \phi$ is, again, the normal to the hypersurface $\Sigma_0$. We would now like to answer a very simple question: can we use the PDE~\eqref{ch2:hydro:gen-quasilinear-pde} to construct a local solution some finite $\delta \phi$ off of the initial hypersurface? In other words, can we use the PDE to find $\de^2\Psi/\de \phi^2\vert_{\phi = 0}$, $\de^3 \Psi/\de \phi^3\vert_{\phi = 0}$, and so on up to infinite order? 

It is clear from equation~\eqref{ch2:hydro:origin_characteristic_eq} that we can determine $\de^2 \Psi/\de \phi^2 \vert_{\phi = 0}$ from the initial conditions so long as ${\cal A}^{\mu\nu} \xi_\mu \xi_\nu \neq 0$. If this condition is satisfied, higher derivatives may be found by successive applications of $\pder{}{\phi}$ to equation~\eqref{ch2:hydro:origin_characteristic_eq}, and a local solution may be constructed: this is the Cauchy-Kowalevski\footnote{Sophie Kowalevski, for whom the theorem is named, is considered the first woman to have received a modern doctorate in mathematics.} theorem~\cite{Courant-Hilbert}. Since $\xi_\mu$ denotes the normal to the hypersurface, whether the condition ${\cal A}^\mu \xi_\mu \xi_\nu \neq 0$ holds or not depends entirely on the hypersurface $\Sigma_0$ under consideration. In particular, we can identify one-parameter \textit{families} of hypersurfaces for which ${\cal A}^{\mu\nu} \xi_\mu \xi_\nu$ vanishes. We refer to these surfaces as ``characteristic surfaces", or sometimes just ``characteristics" for short; we will also sometimes imprecisely refer to $\xi_\mu$ as the characteristics. As the equation that determines the characteristics, ${\cal A}^{\mu\nu}\xi_\mu \xi_\nu = 0$ is called the characteristic equation, and ${\cal A}^{\mu\nu} \xi_\mu \xi_\nu$ is called the characteristic form. For a point $p$, the envelope of the characteristics passing through $p$ will form a conoid for each family of solutions to the characteristic equations. This conoid is called the Monge conoid~\cite{Courant-Hilbert}. 

As a simple example, for the wave equation in flat space ($\eta^{\mu\nu} \de_\mu \de_\nu \Psi = 0$), the characteristics are given by the set of hypersurfaces for which $\eta^{\mu\nu} \xi_\mu \xi_\nu  = \xi_\mu \xi^\mu = 0\,.$ The Monge conoid is then the light cone.

Characteristic surfaces are critical for understanding the dynamics of classical field theories. One can show (see e.g.~\cite{Courant-Hilbert}, Ch.~6 Sec.~2) that characteristic surfaces are the only hypersurfaces across which the solution $\Psi$ may experience non-analyticities. In a causal theory, any solution which is defined on a region of compact support must still have compact support at a later time. A solution often\footnote{A counterexample is the so-called bump function.} experiences a non-analyticity when passing outside of a region of compact support, and so the edges of the region of support are often marked by characteristics. The Monge conoids for each family of characteristics therefore act as ``spherical" \textit{wavefronts} for a perturbation localized to a point $p$, and in a causal theory they must lie inside of the lightcone at every point. The technical expression is that characteristics define the ``domain of dependence" and ``domain of influence" for a point $p$. The Monge conoids determine the regions of spacetime which a disturbance at $p$ affects (and is affected by)  -- hence the demand that the Monge conoids lie inside the lightcone.

In what follows, we will usually discuss the characteristics, rather than directly referring to the Monge conoids. Let us now consider a system of partial differential equations:
\begin{equation}\label{ch2:hydro:quasilinear-gen-system}
    L[U] = {\cal A}^{\mu\nu}_{AB}[U, \de U] \de_\mu \de_\nu U^B + {\cal B}^\mu_{AB}[U, \de U] \de_\mu U^B + {\cal C}_A[U] = 0\,,
\end{equation}
where $U^B$ is once again the state vector. The capital latin labels $A$, $B$ run from $1$ to $n$, where $n$ is the number of components in $U^B$. In the above, the greek indices $\mu$, $\nu$ label the different matrices ${\cal A}_{AB}$, ${\cal B}_{AB}$. The same arguments\footnote{In obtaining equation~\eqref{ch2:hydro:characteristic equation}, there is a subtlety -- we have assumed that all of the elements of $U^B$ obey differential equations of the same order, i.e. second order. This may not be true -- in that case, one must consider mixed-order systems of equations, and the situation complicates. We will not consider such systems of equations in this dissertation; the more mathematically inclined reader may instead refer to~\cite{Disconzi:2023rtt}.} lead to an only slightly modified equation: the characteristics are obtained from the characteristic form ${\cal A}^{\mu\nu}_{AB} \xi_\mu \xi_\nu$ via the characteristic equation
\begin{equation}\label{ch2:hydro:characteristic equation}
    Q(\xi_0, \xi_j) = \det\lr{{\cal A}^{\mu\nu}_{AB}\xi_\mu \xi_\nu} = 0\,.
\end{equation}
We are interested in causal theories. In particular, we demand that the system of PDEs~\eqref{ch2:hydro:quasilinear-gen-system} be (at least weakly) hyperbolic, meaning that the characteristics are real and lead to finite (i.e. less than infinite) propagation speeds. Causality may be quantified by the following three demands~\cite{Hoult:2024qph} on the solutions $\xi_0 = \xi_0(\xi_j)$ to the characteristic equation~\eqref{ch2:hydro:characteristic equation} for real $\xi_j$:
\begin{subequations}\label{ch2:hydro:real-space-causality}
    \begin{align}
        \frac{|\Re(\xi_0(\xi_j))|}{|\xi_j|} \leq 1\quad \forall \,\xi_j\,,&\qquad \Im(\xi_0(\xi_j)) = 0\quad \forall \,\xi_j\,,\\
        {\cal O}_{|\xi_j|}\lr{Q(\xi_0 = a |\xi_j|, \xi_j = s_j |\xi_j|)} &= {\cal O}_{\xi_0} \lr{Q(\xi_0,\xi_j)} \quad (\xi_0, |\xi_j| \neq 0)\,,
    \end{align}
\end{subequations}
where ${\cal O}_X(Q)$ refers to the order of the polynomial $Q$ in $X$, $a$ is an arbitrary non-zero constant, and $s_j$ is a unit vector. 

The first condition is the demand that the characteristic equation~\eqref{ch2:hydro:characteristic equation} give rise to characteristic surfaces that lie inside the lightcone. The second condition enforces reality of the characteristics (and therefore that the system of equations is not elliptic). Finally, the third condition\footnote{As far as I am aware, the third condition was first written down in this form in~\cite{Hoult:2023clg}. While it may have previously been known in the mathematics literature, it was not present in the literature for hydrodynamics.} demands that all solutions to the characteristic equation can be written in the form $\xi_0 = \xi_0(\xi_j)$. In particular, the third condition eliminates $\xi_0 = 0, \xi_j \neq 0$ as a solution, which corresponds to a parabolic system of equations with an infinite propagation speed. Note that this condition is a constraint on the characteristic equation itself, not on its solutions.

Let us consider two brief illustrations of these constraints. We previously found that the characteristics for the wave equation are the light cone. The characteristic equation is $-\xi_0^2 + \xi_j \xi^j = 0$, and the solutions are given by $\xi_0 = \pm |\xi_j|$. Therefore,
\begin{equation}
    \frac{|\Re(\xi_0)|}{|\xi_j|} = 1, \quad \Im(\xi_0) = 0, \quad 2 = {\cal O}_{|\xi|}\lr{(-a^2 + 1)|\xi|^2} = {\cal O}_{\xi_0}\lr{-\xi_0^2 + |\xi|^2} = 2\,.
\end{equation}
All three conditions are satisfied, and so the wave equation is causal (as expected). We can also consider the diffusion equation in the observer rest frame,
$    \de_t \rho - D \de_j \de^j \rho = 0\,.$
The characteristic equation is given by $\xi_j \xi^j = 0$. This violates the third condition of~\eqref{ch2:hydro:real-space-causality}:
\begin{equation}
    2 = {\cal O}_{|\xi|}( |\xi_j|^2) \neq {\cal O}_{\xi_0} (|\xi_j|^2) = 0
\end{equation}
The diffusion equation is therefore acausal as expected. To illustrate this, we note that any $\xi_\mu$ with $\xi_j = 0$ and $\xi_0\neq0$ satisfies the characteristic equation, leading immediately to
\begin{equation}
    \frac{|\Re(\xi_0)|}{|\xi_j|} = \infty
\end{equation}
We therefore find that the diffusion equation has infinite propagation speed. This defines the diffusion equation as being parabolic. Since infinite speed is (slightly) faster than the speed of light, we find once again that the diffusion equation is acausal. 

\paragraph{Linear theories.} Let us now specialize to the case of a homogeneous linear system of partial differential equations. This may arise either naturally for the full system of equations (such as in the wave equation), or it may be due to the linearization of a non-linear (or quasilinear) system of equations. Regardless, let us consider the specific case where ${\cal A}^{\mu\nu}_{AB}$ and ${\cal B}_{AB}^\mu$ are constant, and ${\cal C}_A[U] = {\cal C}_{AB} U^B$. Then the characteristics are the same for every point in spacetime, and form a family of constant hyperplanes.

Let us assume that condition three of~\eqref{ch2:hydro:real-space-causality} is satisfied by the characteristic equation. Then $|\xi_j| = 0$, $\xi_0 \neq 0$ is not a valid solution to the characteristic equation. We subsequently take $\xi_j$ real and $|\xi_j| > 0$. Defining $V_\mu \equiv \xi_\mu/|\xi_j|$, the characteristic equation can be re-written as
\begin{equation}\label{ch2:hydro:char-eq-V-1}
    \det\lr{{\cal A}^{\mu\nu} V_\mu V_\nu} = 0\,.
\end{equation}
The first two causality constraints may also be re-written as $|\Re(V_0)| \leq 1$, $\Im(V_0) = 0$. Now, let us consider a plane-wave solution $U^B = \tilde{U}^B(K_\mu) \exp(i K_\mu x^\mu)$, where $K_\mu = (-\omega, k_j)$ as before. Inserting the plane-wave solution into the system of PDEs~\eqref{ch2:hydro:quasilinear-gen-system}, we find
\begin{equation}\label{ch2:hydro:lin_general_PDE}
    \lr{{\cal A}^{\mu\nu}_{AB} K_\mu K_\nu  + {\cal B}^\mu_{AB} K_\mu + {\cal C}_{AB}} \tilde{U}^B \exp\lr{i K_\mu x^\mu} = 0
\end{equation}
The dispersion relations are given by the roots of the spectral curve, such that
\begin{equation}
    F(\omega,k_j) = \lr{{\cal A}^{\mu\nu}_{AB} K_\mu K_\nu  + {\cal B}^\mu_{AB} K_\mu + {\cal C}_{AB}} = 0\,.
\end{equation}
In a causal theory, the dispersion relations $\omega = \omega(k)$ must obey:
\begin{subequations}\label{ch2:hydro:full_momentum_causcon}
    \begin{align}
        \lim_{|k| \to \infty} \frac{|\Re(\omega)|}{|k|} \leq 1, \quad  \lim_{|k| \to \infty} \frac{\Im(\omega)}{|k|} = 0&\,,\\
        {\cal O}_{|k|}\lr{F(\omega = a |k|,\, k_j = s_j |k|)} = {\cal O}_\omega \lr{F(\omega,k_j)}&\,.
    \end{align}
\end{subequations}
Note the addition of the third condition compared to the previous causality conditions on dispersion relations. As a simple example of the third condition's necessity, let us consider\footnote{This equation was brought to my attention by L. Gavassino.} the linearized Benjamin-Bona-Mahoney equation in $(1+1)$-dimensions, a PDE intended for studying surface gravity waves (not to be confused with gravitational waves)~\cite{BBM:1972}. The linearized equation is given by
\begin{equation}
    \de_t \phi + \de_x \phi-\de_x^2 \de_t \phi = 0 
\end{equation}
Looking at plane wave solutions $\phi = \tilde{\phi} \exp\lr{i K_\mu x^\mu}$, we find
\begin{equation}\label{ch2:hydro:BBM-SC}
    F(\omega,k) = \lr{- i \omega + i k - i k^2 \omega}  = 0
\end{equation}
which is solved by
\begin{equation}\label{ch2:hydro:BBM}
    \omega = \frac{k}{1+k^2}
\end{equation}
The dispersion relation~\eqref{ch2:hydro:BBM} appears to be totally well-behaved at first glance. In the limit of large $k$, we find
\begin{equation}
   \lim_{|k| \to \infty} \frac{|\Re(\omega)|}{|k|} =0, \quad \lim_{|k| \to \infty} \frac{|\Im(\omega)|}{|k|} = 0\,.
\end{equation}
In fact, $\Im(\omega) = 0$ for all real values of $k$, and so the solution is stable as well. However, if we consider the third condition, we find that
\begin{equation}
    3 = {\cal O}_{|k|}(F(a |k|, s_j |k|)) \neq {\cal O}_\omega (F(\omega,k)) = 1\,.
\end{equation}
The Benjamin-Bona-Mahoney equation is therefore acausal -- and indeed, we can see this directly by boosting $\omega \to \gamma(\omega' - v k')$, $k \to \gamma(k' - v \omega')$ in equation~\eqref{ch2:hydro:BBM-SC}, leading to the modified equation
\begin{equation}
    - i \gamma (\omega' - v k') + i \gamma (k' - v \omega') - i \gamma^3 (k' - v \omega')^2(\omega' - v k') = 0
\end{equation}
This equation now has three solutions $\omega' = \omega'(k')$ -- and two of them go as $\omega' = \frac{1}{v} k' + ...$ in the large-$k'$ limit, where the $...$ indicate subleading contributions in $k'$. Since $v$ is the boost parameter, these new solutions are clearly acausal even as accounted for by the first two conditions. The third condition of~\eqref{ch2:hydro:full_momentum_causcon} is necessary for preventing these acausal modes from appearing in the system upon boosting.

Returning to the linear PDE~\eqref{ch2:hydro:lin_general_PDE}, we consider now the same equation divided through by $k^2$. We will be interested in the large-$k$ limit, and so we also define the vector $V'_\mu = K_\mu/k = \{- \frac{\omega}{k}, \frac{k_j}{k}\}$. The linear PDE may then be written
\begin{equation}
    \lr{{\cal A}^{\mu\nu}_{AB} V'_\mu V'_\nu + \frac{1}{k}{\cal B}^\mu_{AB} V_\mu  + \frac{1}{k^2}{\cal C} } \tilde{U}^B \exp\lr{i K_\mu x^\mu} = 0
\end{equation}
Let us now assume that the theory satisfies the causality constraints~\eqref{ch2:hydro:full_momentum_causcon}. The first two conditions may be straightforwardly re-written as $\lim_{|k| \to \infty} |\Re\lr{V'_0}| \leq 1$, $\lim_{|k| \to \infty} \Im(V'_0) = 0$; it is then clear that as $k$ grows, $V'_\mu$ remains finite, and $V'_\mu/k \to 0$. Therefore, in the large-$k$ limit, the dispersion relations are given by
\begin{equation}
    \det\lr{{\cal A}^{\mu\nu}_{AB} V'_\mu V'_\nu} = 0
\end{equation}
However, this is simply the characteristic equation~\eqref{ch2:hydro:char-eq-V-1} with $V_\mu \to V'_\mu$! We therefore have found that the large-$k$ limit of the dispersion relations of the linear theory computes the characteristics of the linear theory -- and it is in this sense that we can claim that they are truly causality constraints.

A natural follow-up question might be the following: when the linear theory arises from linearizing a quasilinear system of PDEs, can this relationship between large-$|k|$ dispersion relations and characteristics for the linearized theory tell us anything about the characteristics of the original quasilinear theory? The answer is ``sometimes". We will discuss this in more detail in Chapter~\ref{chapter:math}.

Let us now note a result~\cite{Bemfica:2020zjp,Gavassino:2021owo}: if a theory is causal, and it is stable in one reference frame, it is causal in all reference frames. Conversely, if a theory is acausal and has $\Im(\omega) < 0$, it will be unstable in at least one reference frame. This means that the analysis of the diffusion equation where we boosted and investigated the stability of the theory was overkill -- we could, in fact, have simply established that the theory was acausal, and left it at that. A theory that is both causal and stable in one reference frame is therefore covariantly stable. We will discuss this property in more detail in Chapter~\ref{ch3:sec_covarstable}.

\subsubsection{Causality of the Navier-Stokes equations}
We will now investigate the causality of the relativistic Navier-Stokes equations in the Landau frame (all of the following results will also hold true for the Eckart frame as well). To begin with, let us restrict to $d=3$ for definiteness, and consider linearized perturbations about equilibrium in the rest frame of the form
\[
T = T_0 + \delta T \,e^{-i \omega t + i k x}, \quad \mu = \mu_0 + \delta \mu \,e^{-i \omega t + i k x}, \quad u^\mu = u_0^\mu + \delta u^\mu e^{-i \omega t + i k x}\,,
\]
where again $u_0^\mu \delta u_\mu = \oser{\de^2}$. We have also taken advantage of the isotropy of the equilibrium state to align $k_j$ in the $x$-direction -- we use the shorthand $k= |k_j|$ as before. Inserting these into the relativistic Navier-Stokes equation, we find
\begin{equation}\label{ch2:hydro:Landau-spectral}
    \begin{split}
        F(\omega,k) &= \lr{ \eta k^2 - i \omega (\epsilon_0 + p_0)}^2 \biggl[a \omega^3 + b k^2 \omega^2 + (c_4 k^4 + c_2 k^2) \omega + d k^4\biggr] = 0
    \end{split}
\end{equation}
where $a$, $b$, $c_4$, $c_2$, and $d$ depend on the transport parameters $\zeta, \eta, \sigma$ and thermodynamic parameters ($T_0, \mu_0, \epsilon_0, p_0, n_0$), but not on $k$ or $\omega$. We see that, as in the relativistic Euler equations, two copies of the shear mode factorize, giving dispersion relations
\begin{equation}\label{ch2:hydro:shear-mode-landau}
     \omega = - i \frac{\eta}{(\epsilon_0+p_0)} k^2\,.
\end{equation}
We see now why the mode was called the shear mode -- it describes the diffusion of shear perturbations. However, looking at equation~\eqref{ch2:hydro:shear-mode-landau}, there are already alarm bells ringing. We know from investigating the diffusion equation that the dispersion relation $\omega = - i D k^2$ is acausal if taken to large-$k$ -- and equation~\eqref{ch2:hydro:shear-mode-landau} is of the same form, with $D \to \eta/(\epsilon_0 + p_0)$. 

Even without solving for the charge mode and sound modes, we can see that the longitudinal modes will also have issues, because they violate the third condition of~\eqref{ch2:hydro:full_momentum_causcon}. Specifically,
\begin{equation}
    4 = {\cal O}_{k} \lr{F(\omega = a k, k_j = s_j k)} \neq {\cal O}_\omega \lr{F(\omega,k_j)} = 3
\end{equation}
Solving for the charge mode and sound mode in the limit of small $k$, we find
\begin{equation}
    \omega = \pm v_s \, k - i \frac{\Gamma_s}{\epsilon_0 + p_0} k^2 + \oser{k^3}, \quad \omega = - i \Gamma_{\sigma} k^2 + \oser{k^4}
\end{equation}
where
\begin{subequations}
    \begin{align}
        \Gamma_s &=\frac{\zeta + \frac{4}{3} \eta}{2} + \lr{\pder{p_0}{n_0}}_{\epsilon_0}^2 \frac{\sigma}{2 T_0 v_s^2}\,,\\
        \Gamma_\sigma &= \frac{(p_0+\epsilon_0) \sigma}{T_0^2 v_s^2 \lr{\partial(\epsilon_0,n_0)/\partial(T_0,\mu_0)}}\,.
    \end{align}
\end{subequations}
In the limit of large $k$, the three modes in the longitudinal part go as
\begin{equation}
        \omega = - i \lr{\frac{\zeta + (4/3) \eta}{p_0+\epsilon_0}}k^2 + ...\,, \quad \omega = -i \,A \,\sigma k^2+...\,,
        \quad \omega = - i B + ...\,,
\end{equation}
where $A$ and $B$ are thermodynamic functions, and the $...$ indicate subleading behaviour in large $k$. We can see that two of the modes go as $k^2$, which immediately violates the causality conditions~\eqref{ch2:hydro:full_momentum_causcon}. We therefore arrive at the troubling conclusion that the relativistic Navier-Stokes equations are acausal. Worse, because they are acausal, there must be a frame in which they are unstable. In fact, thanks to violating the third condition of~\eqref{ch2:hydro:full_momentum_causcon}, it turns out that the relativistic Navier-Stokes equations are unstable in \textit{every} reference frame apart from the fluid rest frame. As in the diffusion equation, the blow up time is proportional to $v$, the boost velocity -- and so, the blow-up gets worse the closer to rest one gets.

Whether or not this acausality and instability is a problem depends on what questions one is trying to answer. If the question is only one of response, e.g. determining what the various viscosities $\zeta$, $\eta$, $\sigma$ are from the microscopic theory, or looking at stochastic fluctuations, then this is perhaps not such a big issue. However, if one wishes to actually \textit{solve} the full quasilinear hydrodynamic equations, as in a hydrodynamic simulation, this is a massive issue -- because one cannot numerically distinguish the true physical hydrodynamic behaviour from the (unphysical) blowup due to the acausality of the modes. 

This problem has been known for a long time~\cite{Hiscock:1985zz,Hiscock:1987zz}, and solutions were soon established. In the remainder of this chapter, we will discuss the two main means of rendering the relativistic Navier-Stokes equations stable and causal. The first is the M\"uller-Israel-Stewart (MIS) theory of hydrodynamics~\cite{Muller:1967zza,Israel:1976tn,Israel-Stewart}, which is spiritually the same as the Maxwell-Cattaneo equation for diffusion. The second is the more recent Bemfica-Disconzi-Noronha-Kovtun (BDNK) formulation of causal hydrodynamics~\cite{Bemfica:2017wps,Kovtun:2019hdm,Bemfica:2019knx,Hoult:2020eho,Bemfica:2020zjp}, which is spiritually the same as the telegrapher's equation. Both have their strengths and their weaknesses, as we shall see.

\subsection{M\"uller-Israel-Stewart (MIS) theory}
\label{ch2:hydro:sec_MIS}
The stress-energy current and charge current may be written as
\begin{subequations}
    \begin{align}
        T^{\mu\nu} &= \lr{\epsilon + \delta \TE}u^{\mu} u^\nu + \lr{p + \delta P}\Delta^{\mu\nu} + 2 \delta \TQ^{(\mu}u^{\nu} + \delta \TT^{\mu\nu}\,,\\
        J^\mu &= \lr{n + \delta \JN}u^\mu + \delta \JJ^\mu\,,
    \end{align}
\end{subequations}
where by the delta of a curly variable, we mean all derivative corrections from first order up to infinite order. Now, using the field-redefinition freedom built into the theory, let us put the constitutive relations into the Landau frame. We may then parametrize the theory as
\begin{subequations}\label{ch2:hydro:MIS-gen-constitutive}
    \begin{align}
        T^{\mu\nu} &= \epsilon u^\mu u^\nu + \lr{p + \Pi} \Delta^{\mu\nu} + \pi^{\mu\nu}\,,\\
        J^\mu &= n u^\mu + n^\mu\,.
    \end{align}
\end{subequations}
where we will (imprecisely) refer to $\Pi$ as the bulk viscosity, $\pi^{\mu\nu}$ as the shear viscosity tensor, and $n^\mu$ as the charge current. As with $\JJ^\mu$ and $\TT^{\mu\nu}$, $n^\mu$ and $\pi^{\mu\nu}$ are both transverse to $u^\mu$, and $\pi^{\mu}_{\,\,\,\,\mu} = 0$. Now, as in the Maxwell-Cattaneo model, we promote the corrections $\Pi$, $n^\mu$, and $\pi^{\mu\nu}$ to degrees of freedom on par with $T(x^\mu)$, $\mu(x^\mu)$, and $u^\mu(x^\mu)$. There are now $(d+2) + 1 + d + (d^2+d-2)/2 = (d+1)(d+4)/2$ degrees of freedom, which comprises the degrees of freedom in the hydrodynamic variables, $\Pi$, $n^\mu$, and $\pi^{\mu\nu}$ respectively. In order to close the equations (i.e. introduce enough equations to solve for the variables of interest), we must introduce $d(d+3)/2$ new equations. These equations are sometimes referred to as ``phenomenological equations", as they are not associated with any symmetry of the theory.

We briefly present here three means of obtaining MIS-type equations. A fourth, the DNMR theory, will be discussed in Chapter~\ref{chapter:micro} when we approach kinetic theory. For more detail on the three approaches presented here, the interested reader may refer to Appendix~\ref{app:species_MIS}.

\subsubsection{Canonical Approach to MIS}
The canonical approach uses the entropy current to fix the phenomenological equations~\cite{Israel:1976tn,Israel-Stewart,Hiscock:1983zz}. We present here a summary modified from~\cite{Noronha:2021syv}. Let us modify the entropy current with a piece that is quadratic in $\Pi, n^\mu, \pi^{\mu\nu}$, which we denote by $Q^\mu$. Then, taking the constitutive relations~\eqref{ch2:hydro:MIS-gen-constitutive} into account, the entropy current is of the form
\begin{equation}
    S^\mu = s u^\mu - \alpha n^\mu - Q^\mu\,,
\end{equation}
where $\alpha=\mu/T$. We do not, \textit{a priori}, know the form of $Q^\mu$, and so we write down a general quadratic form:
\begin{equation}\label{ch2:hydro:second-order-Q}
\begin{split}
    Q^\mu &= \frac{1}{2T}\biggl[ \beta_\pi \pi^{\alpha\beta} \pi_{\alpha\beta} + \beta_{\Pi} \Pi^2 + \beta_n n_\alpha n^\alpha \biggr]u^\mu - \frac{1}{T} \biggl[ \alpha_{\Pi n} \Pi  n^\mu + \alpha_{\pi n} \pi^{\mu\alpha} n_\alpha\biggr]\,.
\end{split}
\end{equation}
We can also use the equations of motion to find an expression for $\nabla_\mu \lr{s u^\mu}$:
\begin{equation}\label{ch2:hydro:MIS-1-sumu}
    \nabla_\mu \lr{s u^\mu} = \alpha \nabla_\mu n^\mu + \beta_\nu \nabla_\mu \pi^{\mu\nu} - \Pi \Delta^{\mu\nu}\nabla_\mu \beta_\nu\,,
\end{equation}
Combining these together, we can write the divergence of the entropy current as
\begin{equation}
    \nabla_\mu S^\mu = - \frac{\Pi}{T} A- \frac{n^\mu}{T} \Delta_{\mu\nu}B^\nu - \frac{\pi^{\mu\nu}}{T} \Delta_{\mu\nu\alpha\beta}C^{\alpha\beta}\,,
\end{equation}
where $A$, $B_\mu$, and $C_{\mu\nu}$ are lengthy expressions in terms of $\Pi, n^\mu, \pi^{\mu\nu}, T, \mu, u^\mu$, and their first derivatives. We would like to enforce the positivity of entropy production; we may do this by relating
\begin{equation}\label{ch2:hydro:canonical-1}
    \Pi = - \zeta A, \quad n^\mu = - \sigma \Delta^{\mu\nu}B_\nu, \quad \pi^{\mu\nu} = - \eta \Delta^{\mu\nu\alpha\beta}C_{\alpha\beta}\,.
\end{equation}
The equations~\eqref{ch2:hydro:canonical-1} are the phenomenological equations in the canonical approach. They may be cast in a more familiar form by pulling terms out of $A$, $B^\mu$, and $C^{\mu\nu}$:
\begin{subequations}\label{ch2:hydro:canonical-2}
    \begin{align}
        \tau_{\Pi} u^\lambda \nabla_\lambda \Pi + \Pi &= - \zeta \nabla_\mu u^\mu + ...\,,\\
        \tau_n u^\lambda \Delta^{\mu\nu}\nabla_\lambda n_\nu + n^\mu &= - \sigma T \Delta^{\mu\nu} \nabla_\nu \alpha + ...\,,\\
        \tau_\pi \Delta^{\mu\nu\alpha\beta}u^\lambda \nabla_\lambda \pi_{\alpha\beta} + \pi^{\mu\nu} &= - \eta \sigma^{\mu\nu} + ...\,,
    \end{align}
\end{subequations}
where the $...$ contain the remaining terms in $A, B^\mu, C^{\mu\nu}$. In the above we defined $\tau_{\Pi} = \zeta \beta_{\Pi}$, $\tau_n = \beta_n \sigma  $, and $\tau_\pi = 2 \eta \beta_\pi$. The equations~\eqref{ch2:hydro:canonical-2} look very similar to the equation we wrote down in equation~\eqref{ch2:diffusion:MC_additional_eq} for the Maxwell-Cattaneo model. After the identification~\eqref{ch2:hydro:canonical-1}, the entropy current becomes
\begin{equation}
    T\nabla_\mu S^\mu = \frac{\Pi^2}{\zeta} + \frac{n^\mu n_\mu}{\sigma} + \frac{\pi^{\mu\nu}\pi_{\mu\nu}}{2\eta}\,,
\end{equation}
which is positive so long as $\zeta, \sigma, \eta > 0$. The full equations~\eqref{ch2:hydro:canonical-2}, which are related in Appendix~\ref{app:species_MIS}, can be shown to be causal linearly~\cite{Hiscock:1983zz} for sufficiently large relaxation times $\tau_{\Pi,n,\Pi}$. In the case of an uncharged fluid, they have also been shown to be causal non-linearly~\cite{Bemfica:2020xym} subject to more complicated causality constraints.

\subsubsection{Resummed BRSSS Approach}
The resummed BRSSS~\cite{Baier:2007ix} approach is not directly a method for deriving MIS-type equations. Rather, it began as a method of obtaining second-order hydrodynamics. We consider here a conformal fluid with no conserved $U(1)$ charge. Comparing to~\eqref{ch2:hydro:MIS-gen-constitutive}, this means there is no charge current (and therefore no $n^\mu$), and $\Pi = 0$ identically. At first order, the constitutive relations are given by
\begin{equation}
    T^{\mu\nu} = \epsilon u^\mu u^\nu + p \Delta^{\mu\nu} - \eta \sigma^{\mu\nu}\,,
\end{equation}
where $p=\epsilon/d$. At second order, there are five transverse traceless corrections one could write, which we denote by ${\cal O}^{\mu\nu}_n$. Then we can write
\begin{equation}
    \pi^{\mu\nu} = - \eta \sigma^{\mu\nu} + \tau_\pi \eta {\cal O}_1^{\mu\nu} + \kappa {\cal O}_2^{\mu\nu} + \lambda_1 {\cal O}_3^{\mu\nu} + \lambda_2 {\cal O}_4^{\mu\nu} + \lambda_3 {\cal O}_5^{\mu\nu} + \oser{\de^3}\,.
\end{equation}
Now, let us note that to first order, $\sigma^{\mu\nu} = -\pi^{\mu\nu}/\eta$. We can then replace $\sigma^{\mu\nu}$ with $-\pi^{\mu\nu}/\eta$ anywhere it appears in the ${\cal O}_n^{\mu\nu}$ to get equations for $\pi^{\mu\nu}$. Schematically, after this resummation the equations are of the form
\begin{equation}
    \tau_{\pi} \Delta^{\mu\nu\alpha\beta} u^\lambda \nabla_\lambda \pi_{\alpha\beta} + \pi^{\mu\nu} = - \eta \sigma^{\mu\nu} + \,...\,,
\end{equation}
where the $...$ indicate terms which are non-linear in $\pi^{\mu\nu}$ and derivatives of the fluid velocity. We see that, once again, we have obtained MIS-like equations. These equations will be (at least linearly) causal for large enough values of $\tau_\pi$~\cite{Baier:2007ix}.

\subsubsection{Generating Functional Approach}
The final MIS-type theory we consider in this chapter is a more recent development due to~\cite{Jain:2023obu}, which extracts MIS-like equations using a combination of the entropy current and the generating functional. To begin with, we propose auxiliary fields $\kappa_{\mu\nu}$ and $v_\mu$, where $\kappa_{\mu\nu}$ is symmetric. We then postulate that the pressure is a function of scalars formed from $\kappa_{\mu\nu}$ and $v_\mu$:
\begin{equation}
    W[g,A] = \int d^{d+1}x\sqrt{-g} \lr{p\lr{T,\mu,v^2, (\tr\kappa)^2, \kappa^2-\frac{1}{d}(\tr \kappa)^2} + ...}\,,
\end{equation}
where the $...$ denote potential higher-order terms in derivatives of the fields $(T,\mu,v_\mu, \kappa_{\mu\nu})$. Writing $T^{\mu\nu} = \epsilon u^\mu u^\nu + p\Delta^{\mu\nu} + \Pi^{\mu\nu}$, where (in Landau frame) $\Pi^{\mu\nu}= \Pi \Delta^{\mu\nu} + \pi^{\mu\nu}$, we then define $\Pi^{\mu\nu}$ and $n^\mu$ in terms of $v^\mu$ and $\kappa$:
\begin{subequations}
\label{ch2:hydro:JK_Pimn}
    \begin{align}
        \Pi^{\mu\nu} &= \alpha_\kappa^S \tr\kappa \Delta^{\mu\nu} + \alpha_\kappa^T \kappa^{\braket{\mu\nu}} + \Pi^{\mu\nu}_{\rm h.s.}\,,\\
        n^\mu &= \alpha_v v^\mu\,,
    \end{align}
\end{subequations}
where we introduce the notation $\kappa^{\braket{\mu\nu}} = \Delta^{\mu\nu\alpha\beta}\kappa_{\alpha\beta}$. The $\Pi^{\mu\nu}_{\rm h.s.}$ is the hydrostatic contribution to $\Pi^{\mu\nu}$ arising due to the dependence of $p$ on the fields $\kappa$ and $v^\mu$, while $\alpha_\kappa^{S,T}$, $\alpha_v$ are arbitrary parameters.

The entropy current is given by
\begin{equation}
    S^\mu = s u^\mu - \alpha n^\mu\,,
\end{equation}
where $\alpha = \mu/T$. Taking the divergence, using the equations of motion to express $\nabla_\mu \lr{s u^\mu}$ in terms of $\Pi_{\mu\nu}$ and $n^\mu$, and then finally inserting the definitions of $\Pi^{\mu\nu}$ and $n^\mu$ in terms of $\kappa^{\mu\nu}$ and $v^\mu$, we find
\begin{equation}
    \nabla_\mu S^\mu = - \frac{\alpha_\kappa^S}{2T} \tr\kappa \Delta^{\mu\nu} A_{\mu\nu} - \frac{\alpha^T_\kappa}{2T} \kappa_{\braket{\mu\nu}}\Delta^{\mu\nu\alpha\beta} B_{\mu\nu} - \frac{\alpha_v}{T} v^\mu C_\mu\,,
\end{equation}
where $A_{\mu\nu}$, $B_{\mu\nu}$, and $C_\mu$ are functions of $\kappa_{\mu\nu}$, $v_\mu$, $\alpha$, and $\beta^\mu=u^\mu/T$. We can enforce positivity of entropy production by identifying
\begin{equation}
\label{ch2:hydro:kappa_v_eq}
   \alpha_{\kappa}^S \tr \kappa \Delta^{\mu\nu} = - \zeta \Delta^{\mu\nu} A_{\mu\nu}, \quad \alpha_{\kappa}^T  \kappa^{\braket{\mu\nu}} = - \eta \Delta^{\mu\nu\alpha\beta}B_{\alpha\beta}, \quad \alpha_v v^\mu = - \sigma \Delta^{\mu\nu}C_\nu\,,
\end{equation}
where the parameters $(\zeta, \eta, \sigma)$ are the usual bulk viscosity, shear viscosity, and charge conductivity. This is almost good enough, but we note that we want equations for $\Pi^{\mu\nu}$ and $n^\mu$, not $\kappa^{\mu\nu}$ and $v^\mu$. We would ideally like to invert the relations~\eqref{ch2:hydro:JK_Pimn} to express $\kappa_{\mu\nu}$ and $v_\mu$ in the equations~\eqref{ch2:hydro:kappa_v_eq} in terms of $\Pi_{\mu\nu}$ and $n_\mu$; however, the $\Pi^{\mu\nu}_{\rm h.s.}$ contribution to $\Pi^{\mu\nu}$ makes inverting~\eqref{ch2:hydro:JK_Pimn} non-trivial. We can at least find a solution to linear order in $\Pi_{\mu\nu}$ and $n_\mu$, yielding
\begin{subequations}
\label{ch2:hydro:Lied_MIS}
    \begin{align}
        \tau_{\Pi}^S T \Lied_{\beta} \lr{\frac{1}{d} \tr \Pi^{\mu\nu}} + \frac{1}{d} \tr \Pi^{\mu\nu} &= - \zeta \nabla_\mu u^\mu + ...\,,\\
        \tau_{\Pi}^T T \Delta^{\mu\nu\alpha\beta} \Lied_{\beta} \Pi_{\braket{\alpha\beta}} + \Pi^{\braket{\mu\nu}} &= - \eta \sigma^{\mu\nu} + ...\,,\\
        \tau_n T \Delta^{\mu\nu} \Lied_\beta n_\nu + n^\mu &= - \sigma T \Delta^{\mu\nu} \nabla_\nu \lr{\frac{\mu}{T}} + ...\,,
    \end{align}
\end{subequations}
where the $...$ once again denote a complicated combination of terms involving non-linear combinations of the fields (both hydrodynamics and $\Pi^{\mu\nu}$, $n^\mu$) and their derivatives. Note that, unlike in the other approaches, the time-derivatives of the viscous corrections naturally package themselves as Lie derivatives. 

Due to the relative youth of this formulation of the MIS theory, a causality analysis has not yet been performed -- however, given the MIS-like structure of the equations~\eqref{ch2:hydro:Lied_MIS}, they will presumably be at least linearly causal for large enough values of $\tau_\Pi^S$, $\tau_\Pi^T$, and $\tau_n$.

\subsection{Bemfica-Disconzi-Noronha-Kovtun (BDNK) theory}
\label{ch2:hydro:sec_BDNK}
We now come to the second, philosophically distinct method of rendering the relativistic Navier-Stokes equations~\eqref{ch2:hydro:ns} stable and causal. This method, by Bemfica, Disconzi, Noronha, and Kovtun \cite{Bemfica:2017wps}\cite{Kovtun:2019hdm}\cite{Bemfica:2019knx}, is morally the same as the idea behind the telegrapher's equation for the diffusion equation: one retains terms that are formally higher-order on-shell, which act as UV regulators for the theory. This subsection is based primarily on my paper~\cite{Hoult:2020eho}, which found causality constraints for the charged $U(1)$ fluid; see also~\cite{Hoult2020:thesis}\cite{Bemfica:2020zjp}.

The constitutive relations for the BDNK theory of a charged $U(1)$ fluid are given by equations~\eqref{ch2:hydro:gen-con-eq}, i.e.
\begin{subequations}\label{ch2:hydro:conrel-BDNK}
    \begin{align}
        T^{\mu\nu}&= \lr{\epsilon + \sum_{n=1}^3 \ce_n s_n}u^\mu u^\nu + \lr{p + \sum_{n=1}^3 \pi_n s_n}\Delta^{\mu\nu} + 2 \sum_{n=1}^2 \theta_n V_n^{(\mu} u^{\nu)} - \eta \sigma^{\mu\nu}\,,\\
        J^\mu &= \lr{n + \sum_{n=1}^3 \nu_n s_n}u^\mu + \sum_{n=1}^2 \gamma_n V_n^\mu\,.
    \end{align}
\end{subequations}
Before, we used the relativistic Euler equations to remove $s_1$, $s_3$, and $V_1^\mu$. Now, let us instead keep all possible terms -- subject to one constraint. The physical transport coefficients $\zeta$, $\sigma$, and $\eta$ are not free for us to choose. Instead, they have values fixed by the underlying microscopic theory, values which may be obtained via the use of Kubo formulae. Therefore, we must limit ourselves to values for the transport parameters that are consistent with these microscopic values:
\begin{subequations}
    \begin{align}
        \zeta =& - f_2 + \lr{\frac{\lr{\pder{n}{\mu}}_T\lr{\epsilon+p} - \lr{\pder{\epsilon}{\mu}}_T n}{T \partial(\epsilon,n)/\partial(T,\mu)}} f_1+ \lr{\frac{n \lr{\pder{\epsilon}{T}}_{\alpha} - \lr{\epsilon + p} \lr{\pder{n}{T} }_{\alpha}}{T^2 \partial(\epsilon,n)/\partial(T,\mu)}}f_3\,,\\
               \sigma =& - \ell_2 + \frac{n T}{\epsilon + p} \ell_1\,,
    \end{align}
\end{subequations}
and, of course, $\eta = \eta$, where $f_{1,2,3}$, $\ell_{1,2}$ are as defined in equations~\eqref{ch2:hydro:frame-invar-basis}. With the constitutive relations in hand, we will now follow two paths. First, we will look at dispersion relations in a linearized analysis, especially in the limit of large $k$. Then, we will consider causality conditions on the full quasi-linear system of equations from the point of view of the theory of partial differential equations. As the full theory is not linear, there is no guarantee the two notions of causality agree.\footnote{In fact, in the MIS theory of hydrodynamics, they do not; there exist field configurations which the linearized theory would suggest are causal, but are in fact acausal.}  We will show in Chapter~\ref{chapter:math} that, for most fluid frames, such a disparity does not exist in the BDNK theory of hydrodynamics.

\subsubsection{Linearized Analysis}
Let us begin by considering linearized plane-wave perturbations about the rest frame for $d=3$, i.e.
\begin{equation}
    T = T_0 + \delta T(\omega,k) e^{-i \omega t + i k x}, \quad \mu = \mu_0 + \delta \mu(\omega,k) e^{-i \omega t + i k x}, \quad u^\mu = \delta^\mu_0 + \delta u^\mu(\omega,k) e^{-i \omega t + i k x}\,.
\end{equation}
In the above, we have made use of the $SO(3)$ invariance of the equilibrium state to align $k_j$ with the $x$-direction. Inserting these into the equations of motion (obtained from substituting the constitutive relations~\eqref{ch2:hydro:conrel-BDNK} into the conservation equations), we find
\begin{equation}\label{ch2:hydro:bdnk-matrix}
    \begin{pmatrix}
        a & b & 0 & 0 & c\\
        d & e & 0 & 0 & f\\
        0 & 0 & k^2 \eta - i \omega (p_0+\epsilon_0) - \theta_1 \omega^2 & 0 & 0\\
        0 & 0 & 0 & k^2 \eta - i \omega (p_0+\epsilon_0) - \theta_1 \omega^2 & 0\\
        g & h & 0 & 0 &I
    \end{pmatrix}\begin{bmatrix}
        \delta T\\
        \delta u^x\\
        \delta u^y\\
        \delta u^z\\
        \delta \mu
    \end{bmatrix}= \begin{bmatrix}
        0\\
        0\\
        0\\
        0\\
        0
    \end{bmatrix}
\end{equation}
where
the parameters $a...I$ are polynomials in $\omega$ and $k$ which depend on the transport parameters and the thermodynamics. As before, the spectral curve $F(\omega,k)$ is given by the determinant of the matrix in equation~\eqref{ch2:hydro:bdnk-matrix}:
\begin{equation}
    F(\omega,k) = \lr{k^2 \eta - i \omega (p_0+\epsilon_0) - \theta_1 \omega^2}^2 \det\begin{vmatrix}
        a & b & c\\
        d & e & f\\
        g & h & I
    \end{vmatrix} = 0
\end{equation}
As before, two copies of the shear mode factor out. The shear mode is given explicitly by
\begin{equation}
    \omega =-i \lr{\frac{p_0+\epsilon_0}{2 \theta_1}} \lr{1 \pm \sqrt{1 - \frac{4 \eta \theta_1}{\lr{p_0+\epsilon_0}^2}k^2}}\,.
\end{equation}
We see that this is identical to equation~\eqref{ch2:diffusion:telegrapher_disp_v2} upon the identification $\tau = \theta_1/(p_0+\epsilon_0)$ and $D = \eta/(p_0+\epsilon_0)$. By direct comparison with~\eqref{ch2:diffusion:small_k_v2}, we can see that the stability of the $k \to 0$ gap leads to the demand that $\theta_1 > 0$. Note that this already excludes the Landau frame -- in the Landau frame, $\theta_1 = 0$! We also know, qualitatively, what the behaviour of this mode is by referring back to Figure~\ref{fig:MC_modes}. When $k^2 = \lr{p_0 + \epsilon_0}^2/(4 \eta \theta_1)$, the hydrodynamic mode and non-hydrodynamic mode collide. After the collision, $\Im(\omega)$ ``freezes out" and becomes independent of $k$, while $\Re(\omega)$ gains $k$ dependence that was previously absent. This ``pole collision" behaviour is generic in causal systems, and marks the transition from diffusive behaviour at small $k$ to transmissive behaviour at large $k$.

The remaining determinant in the spectral curve is due to the perturbation of temperature, charge density, and the longitudinal (with respect to $k_j$) component of the fluid velocity. There are three hydrodynamic modes (the two sound modes and the charge mode), and three non-hydrodynamic modes. In the $k\to0$ limit, the non-hydrodynamic modes are given by $\omega = - i \lr{\epsilon_0 + p_0}/\theta_1$, and the roots of
\begin{equation}\label{ch2:hydro:long-gap}
    \det\biggl[ -i \begin{pmatrix}
        \lr{\pder{\epsilon_0}{T_0}}_{\mu_0} & \lr{\pder{\epsilon_0}{\mu_0}}_{T_0}\\
        \lr{\pder{n_0}{T_0}}_{\mu_0} & \lr{\pder{n_0}{\mu_0}}_{T_0}
    \end{pmatrix} - \frac{\omega}{T_0} \begin{pmatrix}
        \ce_1 - \frac{\mu_0}{T_0} \ce_3 & \ce_3\\
        \nu_1 - \frac{\mu_0}{T_0} \nu_3 & \nu_3
    \end{pmatrix} \biggr] = 0\,.
\end{equation}
These ``gaps" must have negative imaginary part to be stable. One of them is already fixed by $\theta_1 > 0$. For the other, we must choose a hydrodynamic frame (i.e. values for the transport parameters) such that the gaps are in the lower half of the complex plane. There exists a criterion to enforce that roots of polynomials lie in a particular half-plane, which is called the Routh-Hurwitz criterion. One can show via an application of the Routh-Hurwitz criterion that necessary and sufficient conditions for the gap~\eqref{ch2:hydro:long-gap} to be stable are~\cite{Hoult:2020eho}
\begin{subequations}
    \begin{align}
        \ce_1 \nu_3 - \ce_3 \nu_1 &\geq 0\,,\\
        \lr{\pder{n_0}{\mu_0}}_{T_0} \ce_1 - \ce_3 \lr{\pder{n_0}{T_0}}_{\alpha_0} - \lr{\pder{\epsilon_0}{\mu_0}}_{T_0} \nu_1 + \lr{\pder{\epsilon_0}{T_0}}_{\alpha_0}\nu_3 &> 0\,.
    \end{align}
\end{subequations}
Ensuring the stability of the equilibrium state for arbitrary $k$ is quite difficult, because of the heavy dependence of the spectral curve on the equation of state. For any particular theory, it can be done numerically. Turning now to the large-$k$ dispersion relations and causality, the causality constraints on the shear mode are given by
\begin{equation}
    0 \leq \frac{\eta}{\theta_1} \leq 1.
\end{equation}
For the remaining modes, the leading order behaviour of the large-$k$ limit is controlled by the roots of
\begin{equation}
\label{ch2:hydro:large-k-controlling-eq}
    \det\begin{pmatrix}
        -\frac{k^2}{T_0} \lr{\theta_1 - \frac{\mu_0}{T_0} \theta_2} - \frac{\omega^2}{T_0} \lr{\ce_1 - \frac{\mu_0}{T_0} \ce_3} & k \omega \lr{\ce_2 + \theta_1} & - \frac{k^2}{T_0} \theta_2 - \frac{\omega^2}{T_0} \ce_3\\
        \frac{k \omega}{T_0} \lr{\theta_1 - \frac{\mu_0}{T_0} \theta_2 + \pi_1 - \frac{\mu_0}{T_0} \pi_3} & k^2 \lr{\frac{4 \eta}{3} - \pi_2} - \omega^2 \theta_1 & \frac{\omega k}{T_0} \lr{\theta_2 + \pi_3}\\
        - \frac{k^2}{T_0} \lr{\gamma_1 - \frac{\mu_0}{T_0} \gamma_2} - \frac{\omega^2}{T_0} \lr{\nu_1 - \frac{\mu_0}{T_0} \nu_3} & \omega k \lr{\gamma_1 + \nu_2} & - \frac{k^2}{T_0} \gamma_2 - \frac{\omega^2}{T_0} \nu_3
    \end{pmatrix} = 0
\end{equation}
which were obtained by scaling $\omega \to \lambda \omega$, $k \to \lambda k$, and taking the limit of large $\lambda$. This is justified by the fact that we demand causality, meaning that $\omega$ must grow (at most) linearly in $k$. In a general frame, the controlling equation~\eqref{ch2:hydro:large-k-controlling-eq} is an order-six polynomial in $\omega$, which therefore does not have a closed-form solution. Nevertheless, we can constrain its roots to lie within a unit disk in the complex plane via a criterion called the Schur-Cohn criterion. That said, it is somewhat simpler to instead restrict from the most general frame to a smaller subset of frames. Two possible choices are the ``decoupled frame" of~\cite{Hoult:2020eho}, or the $\nu_i = \gamma_i = 0$ frame of~\cite{Bemfica:2020zjp}. In the decoupled frame, one sets $\ce_3 = \pi_3 = \theta_2=0$: then the determinant factorizes, and
\begin{equation}\label{ch2:hydro:controlling-soundcharge-mode-largek}
    \det\begin{pmatrix}
        -\frac{k^2}{T_0} \lr{\theta_1} - \frac{\omega^2}{T_0} \lr{\ce_1 } & k \omega \lr{\ce_2 + \theta_1}\\
         \frac{k \omega}{T_0} \lr{\theta_1 + \pi_1} & k^2 \lr{\frac{4 \eta}{3} - \pi_2} - \omega^2 \theta_1
    \end{pmatrix} \lr{- \frac{k^2}{T_0} \gamma_2 - \frac{\omega^2}{T_0} \nu_3} = 0
\end{equation}
The final bracket, the large-$k$ limit of the charge mode and its corresponding non-hydrodynamic mode, gives
\begin{equation}
    \omega = \pm \sqrt{- \frac{\gamma_2}{\nu_3}} k +\,...
\end{equation}
where the $...$ terms are subleading in large-$k$. This is causal so long as
$    0 \leq - (\gamma_2/\nu_3) \leq 1\,.$
The remaining modes are the large-$k$ limit of the sound modes and their corresponding non-hydrodynamic modes. They form a quadratic equation in $\omega^2/k^2$ (a ``biquadratic"):
\begin{equation}\label{ch2:hydro:controlling-sound-mode-largek}
     \ce_1 \theta_1 \lr{\frac{\omega}{k}}^4 - \lr{  \theta_1 \pi_1 +  \ce_2 \lr{\theta_1 + \pi_1} + \ce_1 \lr{\frac{4 \eta}{3} - \pi_2}}  \lr{\frac{\omega}{k}}^2 - \theta_1 \lr{\frac{4 \eta}{3} - \pi_2} = 0
\end{equation}
In general for a biquadratic of the form $a x^4 + b x^2 + c = 0$, sufficient conditions for the roots to be causal by imposing reality and the Schur-Cohn criterion are
\begin{equation}
    b^2 - 4 a c >0, \quad b < 0, \quad 0 < c < a, \quad a+b+c > 0
\end{equation}
Inserting $a = \ce_1 \theta_1$, $b = - \lr{  \theta_1 \pi_1 +  \ce_2 \lr{\theta_1 + \pi_1} + \ce_1 \lr{\frac{4 \eta}{3} - \pi_2}} $, and $c=\theta_1 \lr{\pi_2 - \frac{4 \eta}{3}}$ yields a complicated, non-linear set of constraints on the transport parameters. These constraints are not empty -- there exist frames which can simultaneously satisfy all of the conditions~\cite{Hoult:2020eho}\cite{Bemfica:2020zjp}. We have therefore found that the relativistic Navier-Stokes equations may be rendered (linearly) causal by the BDNK procedure of introducing formally higher-order terms to regulate the acausality in the theory. These higher-order terms enforce the causality and stability of the theory via the introduction of non-hydrodynamic modes. These non-hydrodynamic modes collide with the hydrodynamic modes at some critical $k_*$, transitioning from diffusive behaviour to subluminal propagation

The last thing we will do in this chapter is repeat the same analysis, but for the full quasilinear system of PDEs. We will see that the controlling equations for the modes are (almost) the same between the linearized and quasilinear systems of equations.

\subsubsection{Characteristics}
Let us return to the full equations of motion, restricting ourselves to the $\ce_3 = \pi_3 = \theta_2 = 0$ frame for convenience and again setting $d=3$. Going forward, it will be more convenient to work with the unconstrained vector $\beta^\mu = u^\mu/T$ and with $\alpha = \mu/T$. In terms of these variables, the principal part ${\cal A}_{AB}^{\mu\nu}\de_\mu \de_\nu U^B$ of the system of equations is given by~\cite{Hoult:2020eho}
\begin{subequations}\label{ch2:hydro:non-lin-principal part}
    \begin{align}
        \nabla_\mu T^{\mu\nu} = T \biggl[&\ce_1 u^\mu u^\nu u^\alpha u_\beta + \ce_2 u^\mu u^\nu \Delta^{\alpha}_{\,\,\,\,\beta} + \pi_1 \Delta^{\mu\nu}u^\alpha u_\beta + \pi_2 \Delta^{\mu\nu} \Delta^\alpha_{\,\,\,\,\beta}\nonumber\\
        &+2\theta_1 \lr{ u^\alpha u^{(\nu} \Delta^{\mu)}_{\,\,\,\,\beta} +  u_\beta u^{(\nu}\Delta^{\alpha)\mu}} - \eta \Delta^{\mu\nu\alpha}_{\,\,\,\,\,\,\,\,\,\beta} \biggr]\de_\mu \de_\alpha \beta^\beta\,,\\
        \nabla_\mu J^\mu &= T \biggl[\nu_1 u^\mu u^\alpha u_\beta + \nu_2 u^\mu \Delta^\alpha_{\,\,\,\,\beta} + \gamma_1 u^\alpha \Delta^{\mu}_{\,\,\,\,\beta} + \gamma_2 \Delta^{\mu\alpha}u_\beta\biggr]\de_\mu \de_\alpha \beta^\beta\nonumber\\
        &+ \biggl[ \nu_3 u^\mu u^\alpha + \gamma_3 \Delta^{\mu\alpha}\biggr] \de_\mu \de_\alpha \alpha\,,
    \end{align}
\end{subequations}
or, in matrix form,
\begin{equation}
    {\cal A}^{\mu\alpha}_{AB} \de_\mu \de_\alpha U^B = \begin{pmatrix}
        {\cal M}^{\nu}_{\,\,\,\,\beta} & 0^\nu\\
        {\cal N}_{\beta} & {\cal R}
    \end{pmatrix}^{\mu\alpha} \de_\mu \de_\alpha \begin{pmatrix}
        \beta^\beta\\
        \alpha
    \end{pmatrix}\,,
\end{equation}
where ${\cal M}$, ${\cal N}$, and ${\cal R}$ are implicitly defined by reading off the relevant terms from~\eqref{ch2:hydro:non-lin-principal part}. The characteristic form $Q = {\cal A}^{\mu\nu}\xi_\mu \xi_\nu$ is then obtained by simply taking $\de_\mu \to \xi_\mu$. The characteristic form is given (introducing the shorthand $a = \lr{u{\cdot}\xi}$, $b^\mu = \Delta^{\mu\nu}\xi_\nu$, and $b^2= \xi_\mu \xi_\nu \Delta^{\mu\nu}$) by
\begin{equation}\label{ch2:hydro:BDNK-char-form}
    Q = \begin{pmatrix}
        T\left(\begin{aligned}
        \ce_1 a^2 u^\nu u_\beta + \ce_2 a u^\nu b_\beta + \pi_1 a b^\nu u_\beta + \pi_2 b^\nu b_\beta \\
        + \theta_1 \lr{a u^\nu b_\beta + a^2 \Delta^\nu_{\,\,\,\,\beta} + u_\beta u^\nu b^2 + a u_\beta b^\nu} - \eta \Delta^{\mu\nu\alpha}_{\,\,\,\,\,\beta}\xi_\mu \xi_\alpha
        \end{aligned}\right),& 0^\nu\\
        T\lr{\nu_1 a u_\beta + \nu_2 a b_\beta + \gamma_1 a b_\beta + \gamma_2 b^2 u_\beta}, & \nu_3 a^2 + \gamma_3 b^2
    \end{pmatrix}
\end{equation}
It is now clear that, in this frame, the determinant factors, i.e.
\begin{equation}
    \det(Q) = \det({\cal M}) \lr{\nu_3 a^2 + \gamma_3 b^2} = 0\,.
\end{equation}
The second factor gives
\begin{equation}
    \lr{u{\cdot}\xi} = \pm \sqrt{-\frac{\gamma_3}{\nu_3}}  \sqrt{\xi_\mu \Delta^{\mu\nu}\xi_\nu}\,,
\end{equation}
and we can immediately recognize the charge mode. With regard to $\det\lr{{\cal M}}$, we may note the identity~\cite{Hoult:2020eho}
\begin{equation}
    \begin{split}
        &\det\lr{ A u^\nu u_\beta + B \Delta^\nu_{\,\,\,\,\beta} + C u^\nu \xi_\beta + D \xi^\nu u_\beta + E \xi^\alpha \xi_\beta }\\
        & B^{d-1} [ - AB + B(C+D) a - B E a^2 + (CD-AE) b^2] = 0\,,
    \end{split}
\end{equation}
where $A,B,C,D,E$ can depend on $a$ and $b^2$. Using the definition of $\Delta^{\mu\nu\alpha\beta}$, we find that $B = \theta_1 a^2 - \eta b^2 = 0$, and so we find
\begin{equation}
    \lr{u{\cdot}\xi} = \sqrt{\frac{\eta}{\theta_1}} \sqrt{\xi_\mu \Delta^{\mu\nu}\xi_\nu}\,.
\end{equation}
Since $d=3$, we find two copies of this mode -- and we can once again immediately recognize the shear mode. Finally, evaluating the final factor, we find
\begin{equation}
    \begin{split}
         \ce_1 \theta_1 a^4 - \biggl[ \ce_1 \lr{\frac{4}{3} \eta - \pi_2} + \ce_2(\theta_2 + \pi_1) + \theta_1 \pi_1\biggr]a^2 b^2 - \theta_2 \lr{\frac{4}{3} \eta - \pi_2}b^4 = 0\,.
    \end{split}
\end{equation}
Now, $\det\lr{Q}$ is a Lorentz scalar, and we may (locally) evaluate it in whatever reference frame we like. Therefore, let us consider the system locally in the fluid rest frame, i.e. align $u^\mu = \delta^\mu_0$. This, of course, cannot be done globally -- we will delve deeper into when this does and does not work in Chapter~\ref{chapter:math}. Taking $u^\mu = \delta^\mu_0$ for the moment, we immediately arrive at the fact $a = \xi_0$, $b^2 = |\xi_j|^2$, and
\begin{subequations}\label{ch2:hydro:non-linear-dispersion}
    \begin{alignat}{3}
         &\xi_0^2 - \lr{- \frac{\gamma_3}{\nu_3}} |\xi_j|^2 = 0\,,&&\\
         & \xi_0^2 - \lr{\frac{\eta}{\theta_1}} |\xi_j|^2 = 0\,,&&\\
         &\ce_1 \theta_1 \xi_0^4 - \biggl[ \ce_1 \lr{\frac{4}{3} \eta - \pi_2} + \ce_2(\theta_2 + \pi_1) + \theta_1 \pi_1\biggr]\xi_0^2 |\xi_j|^2  &&- \theta_2 \lr{\frac{4}{3} \eta - \pi_2}|\xi_j|^4 = 0\,.&&
    \end{alignat}
\end{subequations}
We can see that if we use the $SO(3)$ symmetry of the equations to align $\xi_j$ in the $x$-direction, and then identify $\xi_0 \to - \omega$, and $|\xi_j| \to k$, we arrive at \textit{almost} the same controlling equations~\eqref{ch2:hydro:controlling-soundcharge-mode-largek}\eqref{ch2:hydro:controlling-sound-mode-largek} as in the linearized case. We say almost because the transport parameters in the linearized case are constants depending on the background solutions $T_0$, $\mu_0$, while the transport parameters in equations~\eqref{ch2:hydro:non-linear-dispersion} are functions of spacetime via the hydrodynamic variables $T$, $\mu$. However, they can be \textit{made} precisely the same if we promote $T_0 \to T(x^\mu)$, and $\mu_0 \to \mu(x^\mu)$ in the constraints we identified in the linearized case.

We therefore find the remarkable result that, if we show that the linearized theory is causal for all values of $T_0$ and $\mu_0$, then the non-linear theory will be causal as well. This was not guaranteed to hold -- and indeed, there exist certain pathological frames where it does not. The underlying assumption of this section was that all of the equations were second order in derivatives. However, if we set $\ce_3 = \pi_3 = \nu_3 = \theta_2 = \gamma_2 = 0$ (a frame that was first noticed in~\cite{Abboud:2023hos}), then the equation for $\mu$ is only first-order in derivatives. This breaks our underlying assumption of no mixed-order theories, and indeed the equivalence between linearized and non-linear causality does not hold in that case. However, the equivalence \textit{does} hold in a general fluid frame with all transport parameters non-zero and independent. In the next chapter, we will go into more depth as to why that is the case.

\setlength{\unitlength}{\savedunitlength}

	\startchapter{Mathematical Details}
\label{chapter:math}
In this chapter, we will introduce some of the more mathematical results related to causal relativistic hydrodynamics. This chapter is by no means comprehensive -- a great deal of work has been done in recent years pushing the limits of causality in hydrodynamics from a mathematical point of view. Regarding my particular contributions, this chapter is heavily based on my papers~\cite{Hoult:2023clg} and~\cite{Hoult:2024qph}. Section~\ref{ch3:sec_RHandSC} formally introduces in detail the Routh-Hurwitz and Schur-Cohn criteria which were applied in the previous section. In Section~\ref{ch3:sec_equivalence}, we introduce the mathematical details behind the equivalence between linearized and non-linear notions of causality. Finally, in Section~\ref{ch3:sec_covarstable}, we discuss some of the work that has been done regarding covariantly stable systems, including constraints on dispersion relations required to ensure covariant stability.

\section{Routh-Hurwitz and Schur-Cohn criteria}
\label{ch3:sec_RHandSC}
\subsection{Routh-Hurwitz criterion}
The Routh-Hurwitz (RH) criterion is a criterion which may be applied to polynomials with real coefficients, which enforces that the roots of those polynomials lie in the left-hand complex plane. This criterion has its origins in the branch of engineering known as control theory; originally developed by Routh~\cite{routh1877treatise} in 1877, and refined by Hurwitz~\cite{Hurwitz1895} in 1895, the criterion can be straightforwardly applied to any polynomial with real coefficients. See also~\cite{korn2013mathematical} for more detail. In the case of complex coefficients, the story becomes significantly more complicated, and is an ongoing area of research (see e.g.~\cite{Hastir_2023} for recent work). Thankfully, in the theories under consideration in this dissertation, the symmetries mostly only allow $k^j$ to enter the spectral curve in the form $k^2$, meaning the imaginary factor that accompanies it is also squared. In the previously discussed BDNK theory of $U(1)$ charged hydrodynamics, this was due to isotropy. In the theory of magnetohydrodynamics that we will discuss in Chapter~\ref{chapter:extensions}, the contraction of the wavevector with the magnetic flux direction, $k_j h^j$, is a pseudoscalar and can only appear in the spectral curve squared due to the demands of parity. In Chapter~\ref{chapter:extensions}, the contraction of the wavevector with the transverse superfluid velocity $k_j \zeta^j$ is a scalar and can enter the spectral curve linearly, along with an accompanying imaginary factor. We will not attempt to make use of the Routh-Hurwitz criterion in this last case.

On to the actual criterion. This approach calculates the first column of the so-called ``Routh array"~\cite{Ogata:2009}; it is related to the ``alternative formulation" in section 1.6-6 of~\cite{korn2013mathematical}. Let us consider a polynomial in a complex variable $z$ with real coefficients, all of which are different from zero:
\begin{equation}\label{ch3:RHSC:poly}
    P(z) = a_n z^n + a_{n-1} z^{n-1} + ... + a_1 z + a_0
\end{equation}
Let us now split the polynomial~\eqref{ch3:RHSC:poly} into two sub-polynomials: one containing the odd-power terms, and one containing the even-power terms.
\begin{subequations}
    \begin{align}
        P_0(z) &= a_n z^n + a_{n-2} z^{n-2} + ...\,,\\
        P_1(z) &= a_{n-1}z^{n-1} + a_{n-3} z^{n-3} + ...\,.
    \end{align}
\end{subequations}
Let us now define a new polynomial, $P_2(z)$, which is defined to be the remainder of $P_0(z)$ divided by $P_1(z)$:
\begin{equation}
    P_2(z) = \text{Rem}(P_0(z), P_1(z))\,.
\end{equation}
From this polynomial, we may again define a new polynomial $P_3(z)$ given by
$    P_3(z) = \text{Rem}(P_1(z),P_2(z))\,.$
This process is repeated until the resulting polynomial is zeroth order (but non-zero), or we get zero, at which point we stop. If we arrive at zero before getting a polynomial of zeroth order, the stability criterion is violated. Now, let us define by ${\cal P}_k$ the coefficient of the highest-order term of the polynomial $P_k$. The set of all such highest-order terms is given by:
\begin{equation}
   {\cal G} =  \{{\cal P}_0, {\cal P}_1, ..., {\cal P}_k\}
\end{equation}
where $k$ here is the final element of the procedure. The set ${\cal G}$ is the first column of the Routh array. The Routh-Hurwitz criterion then amounts to demanding that all of the entries of ${\cal G}$ have the same sign. For completeness, the RH criterion for polynomials of order two, three, and four has been computed, and may be found in Appendix~\ref{app:RH-criteria}. 

\subsection{Schur-Cohn criterion}
The Schur-Cohn (SC) criterion is a criterion which may be applied to polynomials such that the roots of those polynomials lie within the open unit disk in the complex plane~\cite{Gargantini-Schur}. The SC criterion is, in essence, a dressed-up version of the RH criterion. Let us again consider the generic order-$n$ polynomial with real coefficients~\eqref{ch3:RHSC:poly}. The RH criterion enforces that roots lie in the left-hand complex plane. Therefore, we wish to consider a polynomial which has had the unit disk mapped to the left-hand complex plane. Therefore, let us consider the M\"obius-transformed polynomial
\begin{equation}
    P'(z) = (z-1)^n P\lr{\frac{z+1}{z-1}}\,.
\end{equation}
We then apply the RH criterion to the transformed polynomial $P'(z)$. We should reiterate that this mapping is for the open unit disk; that is to say, demanding the RH stability of $P'(z)$ imposes that the roots of $P(z)$ satisfy $|z|<1$, not $|z| \leq 1$. In the context of causality, this constraint does not allow propagation at exactly the speed of light. If one wishes to impose the constraint that the phase velocity equal the speed of light, some other technique must be used. In particular, we must demand that $P(1)\neq 0$ for the M\"obius transformation to be well-defined.

Generally, ``causality constraints" on polynomials with real coefficients require more than just the SC criterion. They also demand reality of the roots, such that $\Re(z) \in (-1,1)$, and $\Im(z) = 0$. This will generally involve a condition on the discriminant of the polynomial; for $n \geq 4$, the condition is more than just positivity, as positivity of the discriminant only implies that $s$ mod $4=0$ by Brill's theorem (see Lemma 2.2 of~\cite{Washington1982}), where $s$ is the number of non-real roots.

\begin{figure}[t!]
    \centering
    \includegraphics[width=0.49\textwidth]{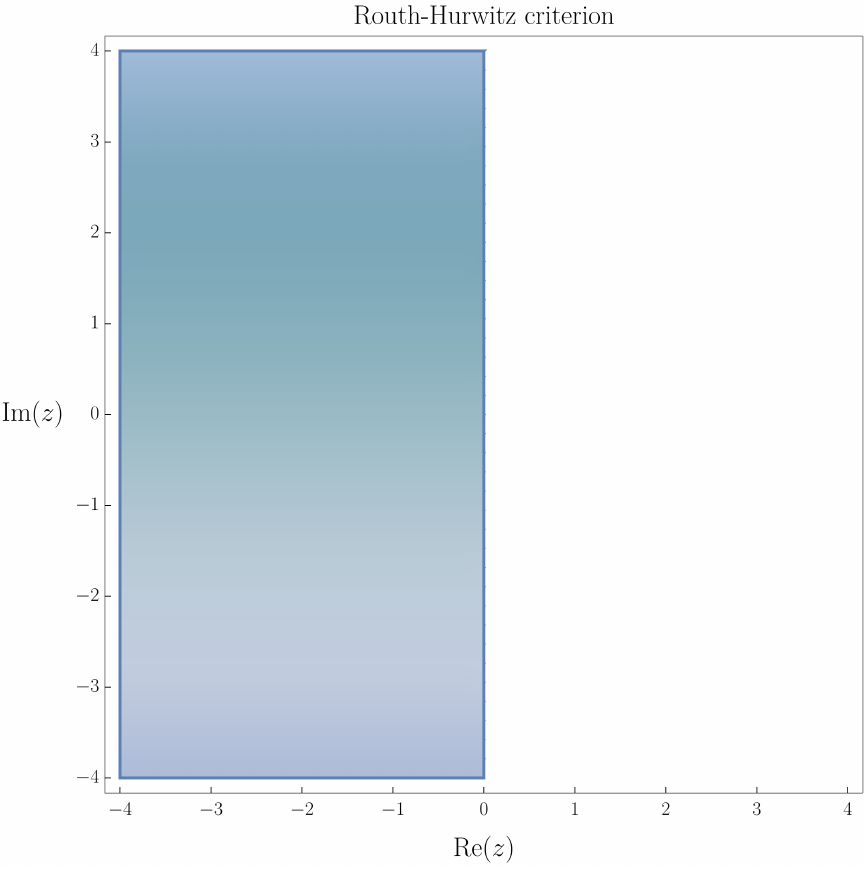}
    \includegraphics[width=0.49\textwidth]{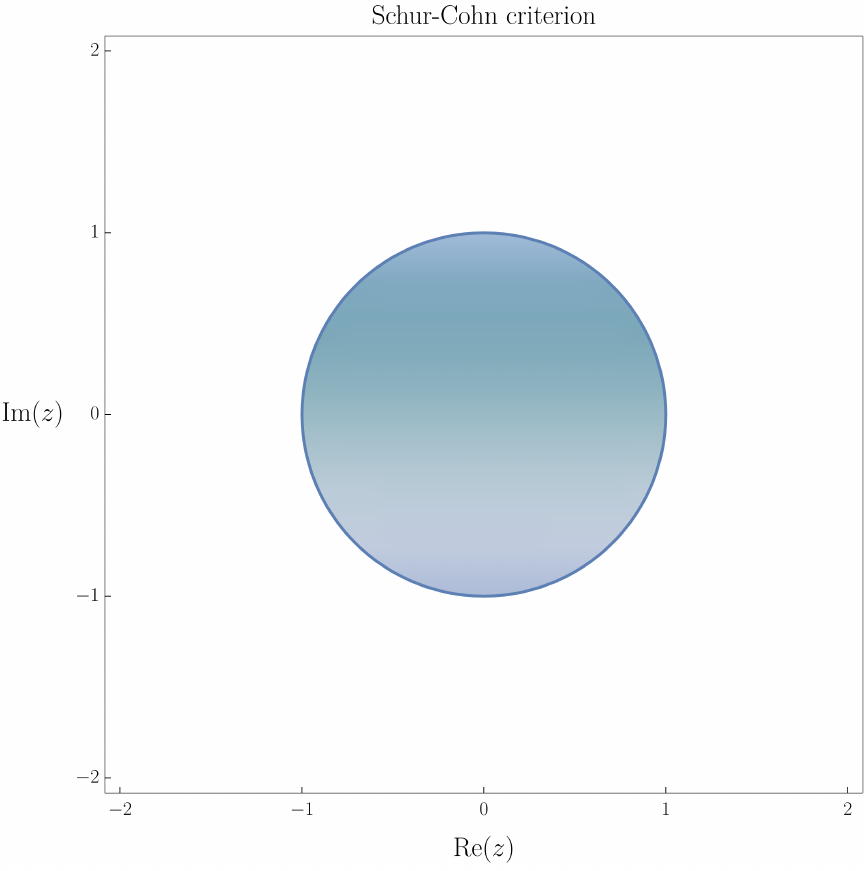}
    \caption{An illustration of the constraints imposed by the Routh-Hurwitz criterion and the Schur-Cohn criterion respectively. The RH-criterion ensures that roots lie in the left-hand complex plane. The Schur-Cohn criterion ensures that roots lie within the open unit disc. In terms of stability and causality, the RH criterion are used to ensure that $\omega$ lies in the lower half-complex plane, and the SC criterion are employed the impose that $\lim_{k \to \infty}\omega/k$ lies in the open unit disk in the complex plane.}
    \label{fig:RHSC}
\end{figure}
\section{Equivalence of linearized and non-linear theories}
\label{ch3:sec_equivalence}
In the previous chapter, we showed that for the BDNK theory of a $U(1)$ charged fluid, non-linear causality conditions were equivalent to linearized causality conditions. In this section, we obtain sufficient conditions for this equivalence to hold for a generic second-order system of partial differential equations. Before evaluating the general case, however, let us consider some examples which will further illustrate the point. The following section, especially the examples, has been heavily based on section 2 of~\cite{Hoult:2024qph}. 

\subsection{Examples}
\subsubsection{Example 1.}
\label{ch3:equiv:example_1}
Let us begin by considering the following example PDE:
\begin{equation}~\label{ch3:equiv:ex1-modified-wave-eq}
    u^\mu u^\nu \de_\mu \de_\nu \phi - c^2\lr{\phi} \Delta^{\mu\nu}\de_\mu \de_\nu \phi = 0\,,
\end{equation}
where $\phi(x)$ is a real scalar field, $u^\mu$ is some (constant) background timelike unit vector, and $\Delta^{\mu\nu} = u^\mu u^\nu + g^{\mu\nu}$ is the projector orthogonal to $u^\mu$. The equation~\eqref{ch3:equiv:ex1-modified-wave-eq} is just the wave equation in a reference frame specified by $u^\mu$, with a propagation speed $c(\phi)$ that depends on the local value of the field $\phi$. 

The characteristic equation associated with equation~\eqref{ch3:equiv:ex1-modified-wave-eq} is given by
\begin{equation}\label{ch3:equiv:ex1-full-char}
    Q = \lr{u{\cdot}\xi}^2 - c^2(\phi) \Delta^{\mu\nu}\xi_\mu \xi_\nu = 0\,.
\end{equation}
The equations will be causal so long as $0 \leq c^2(\phi) \leq 1$, such as if $c^2(\phi) =1/(1+\phi^2)$. Now, let us consider the linearization of equation~\eqref{ch3:equiv:ex1-modified-wave-eq} about some constant solution, i.e. $\phi(x) = \phi_0 + \delta \phi(x)$:
\begin{equation}\label{ch3:equiv:ex1-linearized-wave-eq}
    u^\mu u^\nu \de_\mu \de_\nu \delta \phi - c^2(\phi_0) \Delta^{\mu\nu} \de_\mu \de_\nu \delta \phi = 0\,.
\end{equation}
The characteristic equation associated with the linearized PDE~\eqref{ch3:equiv:ex1-linearized-wave-eq} is
\begin{equation}\label{ch3:equiv:ex1-lin-char}
    Q = \lr{u{\cdot}\xi}^2 - c^2(\phi_0) \Delta^{\mu\nu}\xi_\mu \xi_\nu = 0\,.
\end{equation}
Both characteristic equations are algebraic equations. It is clear from direct comparison that equation~\eqref{ch3:equiv:ex1-lin-char} is equivalent to equation~\eqref{ch3:equiv:ex1-full-char} under the operation $\phi_0 \to \phi(x)$. Therefore, so long as the equation~\eqref{ch3:equiv:ex1-linearized-wave-eq} is causal (as judged by the solutions of equation~\eqref{ch3:equiv:ex1-lin-char} satisfying the causality conditions~\eqref{ch2:hydro:real-space-causality}) for all real $\phi_0$, the non-linear PDE~\eqref{ch3:equiv:ex1-modified-wave-eq} will also be causal. Therefore, for equation~\eqref{ch3:equiv:ex1-modified-wave-eq}, showing (local) causality of the linearized system of equations is enough to show (local) causality of the full system of equations.

\subsubsection{Example 2.}
\label{ch3:equiv:example_2}
Next, let us consider a somewhat modified version of Example 1:
\begin{equation}\label{ch3:equiv:ex2-modified-wave-eq}
    u^\mu u^\nu \de_\mu\de_\nu \phi - c^2(\de\phi) \Delta^{\mu\nu}\de_\mu \de_\nu \phi = 0\,,
\end{equation}
with $c^2(\de \phi) = 1 + \Delta^{\mu\nu}\de_\mu \phi \de_\nu \phi$. The associated characteristic equation is given by
\begin{equation}
    \lr{u^\mu \xi_\mu}^2 - c^2(\de \phi) \Delta^{\mu\nu}\xi_\mu \xi_\nu = 0\,.
\end{equation}
Let us now again linearize about the constant solution $\phi_0$. Then equation~\eqref{ch3:equiv:ex2-modified-wave-eq} becomes
\begin{equation}
    u^\mu u^\nu \de_\mu \de_\nu \phi - \Delta^{\mu\nu} \de_\mu \de_\nu \phi = g^{\mu\nu}\de_\mu \de_\nu \phi = 0\,,
\end{equation}
and the characteristic equation is
\begin{equation}
    Q = \lr{u{\cdot}\xi}^2 - \Delta^{\mu\nu}\xi_\mu \xi_\nu = g^{\mu\nu} \xi_\mu \xi_\nu = 0\,.
\end{equation}
The characteristics for the linearized theory lie along the light cone, and are causal for all values of $\phi_0$. For the full non-linear theory, however, causality is broken for any spatially-varying $\phi(x)$ profile. The difference between the two cases is due to the derivatives in $c^2(\de \phi)$, i.e. due to derivatives in the principal part. In this example, the linearized analysis cannot be used to show causality of the non-linear theory.

\subsubsection{Example 3.}
Let us now consider another modification to Example 1:
\begin{equation}\label{ch3:equiv:ex3-modified-wave-eq}
    u^\mu u^\nu \de_\mu \de_\nu \phi - c^2(\phi) \Delta^{\mu\nu} \de_\mu \de_\nu \phi + \phi = 0\,,
\end{equation}
with $c^2(\phi) = 1 + \phi^2$. Unlike in Example 2, the principal part doesn't have any derivatives. Nevertheless, the equivalence between linearized and non-linear causality will still fail. The characteristic equation associated with equation~\eqref{ch3:equiv:ex3-modified-wave-eq} is given by
\begin{equation}
    \lr{u{\cdot}\xi}^2 - c^2(\phi) \Delta^{\mu\nu}\xi_\mu \xi_\nu = 0\,.
\end{equation}
Now, we would like to linearize about a constant solution, $\phi(x) = \phi_0 + \delta \phi(x)$. However, the zeroth-order term $\phi$ in equation~\eqref{ch3:equiv:ex3-modified-wave-eq} means that the only valid constant solution is $\phi_0 = 0$. Therefore, linearizing about $\phi_0 = 0$, i.e. $\phi(x) = 0 + \delta \phi$, equation~\eqref{ch3:equiv:ex3-modified-wave-eq} becomes
\begin{equation}
    u^\mu u^\nu \de_\mu \de_\nu \phi - \Delta^{\mu\nu} \de_\mu\de_\nu \phi = g^{\mu\nu} \de_\mu \de_\nu \phi = 0\,,
\end{equation}
and the characteristic equation is just again that of the lightcone,
$    g^{\mu\nu}\xi_\mu \xi_\nu = 0\,.$
The linearized theory is therefore causal. However, the non-linear theory is acausal for any $\phi(x) \neq 0$, and the connection between linear and non-linear is once again broken. In this case, the disconnect was due to constraints on the principal part imposed by the zeroth-order term. A more interesting example of this type is the MIS theory of hydrodynamics. Considering equations~\eqref{ch2:hydro:canonical-2}, we see that there is a similar zeroth-order term. 

\subsubsection{Example 4.}
Let us now consider the following slightly more general equation,
\begin{equation}\label{ch3:equiv:ex4-original}
    {\cal A}^{\mu\nu}[\lambda,\phi] \de_\mu \de_\nu \phi + {\cal C}[\phi] = 0\,.
\end{equation}
In the above, $\lambda$ is some free parameter in the theory. The characteristic equation is given by
\begin{equation}\label{ch3:equiv:ex4-char-eq}
    {\cal A}^{\mu\nu}[\lambda,\phi]\xi_\mu \xi_\nu = 0\,.
\end{equation}
``Causality constraints", as we have previously described them, amount to conditions on the parameter $\lambda$ and on $\phi$ such that the solutions $\xi_\mu$ to the equation~\eqref{ch3:equiv:ex4-char-eq} are always causal. 

As a simple example, if ${\cal A}^{\mu\nu} = u^\mu u^\nu- \lr{1 + \phi^2}^\lambda \Delta^{\mu\nu}$, then causality conditions would amount to $\lambda \leq 0$. Suppose, however, that ${\cal C}[\phi] = \phi$. Then, linearizing about the only constant solution ($\phi_0 = 0$) with $\phi = 0 + \delta \phi$, we arrive once again at
$g^{\mu\nu} \de_\mu \de_\nu \delta \phi = 0\,.$
This is causal for any value of $\lambda$ (since the equation doesn't depend on $\lambda$ at all), in stark contrast with the original non-linear equation. We find, once again, that the linearized causality conditions do not enforce the non-linear causality conditions.

There is, however, a workaround. Let us consider the ``partner system"
\begin{equation}\label{ch3:equiv:ex4-partner}
    {\cal A}^{\mu\nu}[\lambda,\chi] \de_\mu \de_\nu \chi = 0\,,
\end{equation}
which has been obtained from~\eqref{ch3:equiv:ex4-char-eq} by removing ${\cal C}[\phi]$ by hand. The equations~\eqref{ch3:equiv:ex4-char-eq} and~\eqref{ch3:equiv:ex4-partner} do not, generically, have the same solutions. However, the partner system~\eqref{ch3:equiv:ex4-partner} admits perturbations about any constant, $\chi = \chi_0 + \delta \chi$. The associated characteristic equation is given by
\begin{equation}\label{ch3:equiv:ex4-partner-characteristic}
    {\cal A}^{\mu\nu}[\lambda,\chi_0]\xi_\mu \xi_\nu = 0\,.
\end{equation}
This is an algebraic equation, and so the characteristic equation of the full system~\eqref{ch3:equiv:ex4-char-eq} can be obtained from~\eqref{ch3:equiv:ex4-partner-characteristic} via the replacement $\chi_0 \to \phi(x)$. Therefore, the (generally $\phi$-dependent) causality conditions of the original system~\eqref{ch3:equiv:ex4-original} may be obtained from the (generally $\chi_0$-dependent) causality conditions of the linearized partner system~\eqref{ch3:equiv:ex4-partner}.

This may be immediately generalized to the case of $n$ fields. The system of equations becomes
\begin{equation}\label{ch3:equiv:ex4-multin-gen}
    {\cal A}^{\mu\nu}_{AB}[\lambda, \phi^A] \de_\mu \de_\nu \phi^B + {\cal C}_A[\phi^B] = 0\,.
\end{equation}
In equation~\eqref{ch3:equiv:ex4-multin-gen}, the constant solutions $\phi^B = \phi^B_0$ are not, in general, independent. Instead, they are related by the constraint
${\cal C}_A\lr{\phi^1, \phi^2, ..., \phi^n} = 0\,.$
Therefore, the linearized theory obtained by expanding around $\phi_0^B$ cannot in general be used to determine the causality conditions of the full theory~\eqref{ch3:equiv:ex4-multin-gen}. However, the causality conditions of the full theory may be obtained by linearizing the partner system
\begin{equation}
    {\cal A}_{AB}^{\mu\nu}[\lambda,\chi^B] \de_\mu \de_\nu \chi^B = 0
\end{equation}
instead. The MIS theory is amenable to such a treatment.

\subsubsection{Example 5.}
In the previous examples, the failure of the linearized theory to reproduce the causality constraints of the full theory were due to either
\begin{enumerate}
    \item the presence of derivatives in the principal part, or
    \item constraints on the constant solutions $\phi^B$, usually due to the presence of zeroth-order terms.
\end{enumerate}
Therefore, let us consider the system of equations
\begin{equation}\label{ch3:equiv:ex5-PDE-gen}
    {\cal A}^{\mu\nu}_{AB}[\phi] \de_\mu \de_\nu ^B + {\cal B}^\mu_{AB}[\phi,\de\phi] \de_\mu \phi^B = 0\,.
\end{equation}
This system of equations has no derivative dependence in the principal part, and there are no zeroth-order contributions to the system of PDEs. The Landau-frame relativistic Navier-Stokes equations~\eqref{ch2:hydro:ns} and the BDNK theory of hydrodynamics~\eqref{ch2:hydro:conrel-BDNK} are of this form -- the MIS theory of hydrodynamics is not. The characteristic equation for this system of equations is given by
\begin{equation}
    \det\lr{{\cal A}^{\mu\nu}_{AB} \xi_\mu \xi_\nu} = 0\,.
\end{equation}
Suppose we now linearize the system of equations~\eqref{ch3:equiv:ex5-PDE-gen}. So long as we linearize about the most general constant solution, with all $\phi_0^B$ independent, the causality of the non-linear system of equations may be assured from the linearized analysis.

This was not what we did in Chapter~\ref{chapter:background}. We linearized about $u_0^\mu = \delta^\mu_0$, setting the spatial parts of $u^\mu_0$ zero -- and the non-linear causality conditions were still enforced by the demands of linearized causality. This must be squared with the previous examples where linearizing about $\phi_0 = 0$ ruined the equivalence; therefore, we would like to determine when linearization about a constrained subset $\tilde{\phi}^B$ of $\phi^B$ does not ruin the equivalence. Given the linear transformation $\phi^B \to G^B_{\,\,\,\,A} \phi^B$, the principal part has the symmetry $G$ if
\begin{equation}
    G {\cal A}(\phi) G^{-1} = {\cal A}(G \phi)
\end{equation}
It turns out that one can linearize about $\tilde{\phi}_0^B$ so long as the transformation that takes $\tilde{\phi}_0^B \to \phi_0^B$ is a symmetry of the principal part. Consider a system of equations where the principal part has such a symmetry. Then, linearizing about the constrained solution $\tilde{\phi}_0^B$, the characteristic equation is given by
\begin{equation}
\begin{split}
    \det\lr{{\cal A}^{\mu\nu}[\tilde{\phi}_0]\xi_\mu \xi_\nu}  &= \det\lr{{\cal A}^{\mu\nu}[\phi_0]\xi_\mu \xi_\nu} = 0\,.
\end{split}
\end{equation}
Therefore, the characteristic equation of the system linearized about $\tilde{\phi}_0^B$ is the same as that of the system of equations linearized about an arbitrary set of constants $\phi^B_0$.

Two quick explicit examples may be given. First, consider a vector $\phi_0^B$ in ${\mathbb R}^n$, where $B \in \{1, .., n\}$. Starting with a particular non-zero vector $\tilde{\phi}_0^B = (c, 0, ..., 0)$ with arbitrary constant $c$, any vector $\phi_0^B$ may be generated via an orthogonal transformation $G$ applied to $\tilde{\phi}_0^B$. Therefore, for any theory where ${\cal A}_{AB}$ is of the form ${\cal A}_{AB}^{\mu\nu} = f^{\mu\nu}(\Phi^2) \delta_{AB} + g^{\mu\nu}(\Phi^2) \phi_A \phi_B$, where $\Phi^2 = \delta_{AB} \phi^A \phi^B$, the non-linear causality constraints can be obtained from the theory linearized about $\tilde{\phi}_0^B$.

Another example is that of a timelike vector field $\beta^\mu$ and a scalar $\alpha$: $\phi^B = (\beta^\mu, \alpha)$. Let us take the equation~\eqref{ch3:equiv:ex5-PDE-gen} to be Lorentz covariant; then the characteristic equation is a Lorentz scalar of the form $Q(\beta^2, \beta{\cdot}\xi, \xi^2, \alpha) = 0$. Since the characteristic equation is a Lorentz scalar, it is invariant under Lorentz transformations, and so we can simply linearize about $\tilde{\phi}_0^B = ( \beta ,0, ..., 0, \alpha)$. This is exactly the case for the $U(1)$-charged BDNK fluid~\eqref{ch2:hydro:conrel-BDNK}.

\subsubsection{Example 6.}
Let us now consider the following system of PDEs in two fields, $U^B = (\chi, \psi)$:
\begin{equation}\label{ch3:equiv:ex6-equation}
    \begin{bmatrix}
        u^\mu u^\nu - c_1(\chi,\psi) \Delta^{\mu\nu} & f(\de\chi, \de\psi) u^\mu u^\nu\\
        0 & u^\mu u^\nu - c_2(\chi,\psi) \Delta^{\mu\nu} 
    \end{bmatrix}\de_\mu \de_\nu \begin{pmatrix}
        \chi\\
        \psi
    \end{pmatrix} + {\cal B}^\mu_{AB}[U,\de U]\de_\mu U^B = 0\,,
\end{equation}
where $c_1,\, c_2$ are real functions of $\chi$ and $\psi$. Note that derivatives have entered the principal part; however, they do not enter the characteristic equation
\begin{equation}\label{ch3:equiv:ex6-nonlin-char}
    Q = \lr{\lr{u{\cdot}\xi}^2 - c_1(\chi,\psi)\Delta^{\mu\nu}\xi_\mu\xi_\nu}\lr{\lr{u{\cdot}\xi}^2 - c_2(\chi,\psi)\Delta^{\mu\nu}\xi_\mu\xi_\nu}\,.
\end{equation}
Suppose we now linearize about the constant solution $U_0^B = (\chi_0,\psi_0)$. The principal part will become
\begin{equation}
    {\cal A}^{\mu\nu}_{AB} = \begin{bmatrix}
        u^\mu u^\nu - c_1(\chi_0,\psi_0)\Delta^{\mu\nu}& 0\\
        0 & u^\mu u^\nu - c_2(\chi_0,\psi_0)\Delta^{\mu\nu}
    \end{bmatrix}\,,
\end{equation}
and the characteristic equation of the linearized theory is given by
\begin{equation}\label{ch3:equiv:ex6-lin-char}
    Q = \lr{\lr{u{\cdot}\xi}^2 - c_1(\chi_0,\psi_0)\Delta^{\mu\nu}\xi_\mu\xi_\nu}\lr{\lr{u{\cdot}\xi}^2 - c_2(\chi_0,\psi_0)\Delta^{\mu\nu}\xi_\mu\xi_\nu}\,.
\end{equation}
It is clear from inspection that the linearized causality conditions arising from equation~\eqref{ch3:equiv:ex6-lin-char} given the full non-linear causality conditions~\eqref{ch3:equiv:ex6-nonlin-char} upon taking $\chi_0 \to \chi(x)$, $\psi_0 \to \psi(x)$, despite the derivatives appearing in the principal part of equation~\eqref{ch3:equiv:ex6-equation}. This makes it clear that the quantity affecting the equivalence is not, strictly speaking, the principal part -- rather, it is the characteristic equation.

\subsubsection{Example 7.}
Let us now return to Example 1:
\begin{equation}\label{ch3:equiv:ex7-original-equation}
    u^\mu u^\nu \de_\mu \de_\nu \phi - c^2(\phi) \Delta^{\mu\nu}\de_\mu \de_\nu \phi = 0\,.
\end{equation}
This is a second-order PDE, which may be generically cast down to a system of first-order PDEs. Let us introduce the auxiliary parameter $\chi_\mu = \de_\mu \phi$. The second-order PDE~\eqref{ch3:equiv:ex7-original-equation} may then be written
\begin{subequations}\label{ch3:equiv:ex7-first-order-system}
    \begin{align}
        u^\mu u^\nu \de_\mu \chi_\nu - c^2(\phi) \Delta^{\mu\nu} \de_\mu \chi_\nu &= 0\,,\\
        \de_\mu \chi_\nu - \de_\nu \chi_\mu &= 0\,,\\
        \de_\mu \phi - \chi_\mu &= 0\,.
    \end{align}
\end{subequations}
The first equation is simply equation~\eqref{ch3:equiv:ex7-original-equation}, the second equation is the statement $\de_\mu \de_\nu \phi - \de_\nu \de_\mu \phi = 0$, and the third equation is just the definition of $\chi_\mu$. In $d+1$ spacetime dimensions, there are then $d+2$ fields ($\chi_\mu, \phi)$ that serve as unknown variables in the theory. However, there are $2 + d (d+3)/2$ equations, which is more that $d+2$. This is not a problem, as not all of the equations in~\eqref{ch3:equiv:ex7-first-order-system} are dynamical equations. Only $(d+2)$ of the equations describe the evolution of the fields; the remaining $d(d+1)/2$ equations are constraint equations which serve solely to constrain initial data. The system of equations is therefore, in the parlance of Chapter~\ref{chapter:background}, ``closed".

In order to isolate the dynamical equations, let us prescribe initial conditions on some initial spacelike hypersurface, which we will denote by $\Sigma_0$. At each point $p$ on $\Sigma_0$, there exists a normal timelike unit covector $n_\mu$ such that $P_{\mu\nu} \equiv  g_{\mu\nu} + n_\mu n_\nu$ projects onto $\Sigma_0$. We may then split the system of equations~\eqref{ch3:equiv:ex7-first-order-system} at any particular point $p$ into dynamical equations (those with a derivative along $n_\mu$) and constraint equations (those which only have derivatives interior to $\Sigma_0$). The equations are given by
\begin{subequations}
    \begin{align}
        &\textbf{Dynamical }\begin{cases}
        u^\mu u^\nu \de_\mu \chi_\nu - c^2(\phi) \Delta^{\mu\nu} \de_\mu \chi_\nu = 0\,,\\
        n^\mu P^{\alpha \nu} \lr{\de_\mu \chi_\nu - \de_\nu \chi_\mu} = 0\,,\\
        n^\mu \de_\mu \phi - n^\mu \chi_\mu = 0\,,
    \end{cases}\label{ch3:equiv:ex7-dynamical-eqs}\\
    &\textbf{Constraint }\begin{cases}
        P^{\alpha\mu}P^{\beta\nu}\lr{\de_\mu \chi_\nu - \de_\nu \chi_\mu} = 0\,,\\
        P^{\alpha\mu}\lr{\de_\mu \phi - \chi_\mu} = 0\,,
    \end{cases}\label{ch3:equiv:ex7-constraint-eqs}
    \end{align}
\end{subequations}
where we note that $P^{\mu\nu}$ projects onto a $d$-dimensional subspace. The original equation is a dynamical equation because $u^\mu$ is timelike, and may be decomposed along $n^\mu$ according to $u^\mu = -\lr{u{\cdot}n} n^\mu + P^{\mu\nu}u_\nu$. The dynamical equations~\eqref{ch3:equiv:ex7-dynamical-eqs} may be written in the matrix form
\begin{equation}
    {\cal M}^\mu_{AB}(U) \de_\mu U^B + {\cal N}_A(U) = 0\,,
\end{equation}
where $U^B = (\chi^\nu,\phi)$. The characteristic equation can be shown to be given by
\begin{equation}
    Q = \det\lr{{\cal M}^\mu \xi_\mu} = \lr{n{\cdot}\xi}^{d} \lr{\lr{u{\cdot}\xi}^2 - c^2(\phi) \Delta^{\mu\nu}\xi_\mu \xi_\nu} = 0\,.
\end{equation}
We can see that in addition to the characteristic equation~\eqref{ch3:equiv:ex1-full-char} of the original second-order system, there are also $d$ factors of $(n{\cdot}\xi)$ which were not present in the characteristic equation of the original PDE. The characteristics which solve $n{\cdot}\xi=0$ have normal covectors entirely contained within $\Sigma_0$, and so they correspond to non-propagating solutions.

It is important to note that the separation of the equations into ``dynamical" and ``constraint" equations is not an invariant concept -- it depends on the choice of the hypersurface $\Sigma_0$, and therefore the choice of $n_\mu$. A different choice $\Sigma'_0$ and $n'_\mu$ would lead to a different set of dynamical equations, and a different set of constraint equations. In particular, the set of extra characteristics given by $\lr{n'{\cdot}\xi}=0$ will differ. Therefore, the first-order system~\eqref{ch3:equiv:ex7-first-order-system}, ignoring the constraint equations, is not covariant. More concretely, for initial spacelike hyperplanes, a change of $n_\mu \to n'_\mu = \Lambda_{\mu}^{\,\,\,\,\nu}n_\nu$, where $\Lambda_\mu^{\,\,\,\,\nu}$ is a Lorentz boost, is equivalent to leaving $n_\mu$ unchanged, and boosting
\begin{equation}
    u^\mu \to u'^\mu = \Lambda_\nu^{\,\,\,\,\mu} u^\nu, \quad \xi_\mu \to \xi'_\mu = \Lambda^\nu_{\,\,\,\,\mu}\xi_\nu\,.
\end{equation}
The solutions to $n{\cdot}\xi = 0$ will not be solutions to $n{\cdot}\xi' = 0$ after applying the boost, but the solutions to the original second-order characteristic equation $\lr{u{\cdot}\xi}^2 - c^2(\phi) \Delta^{\mu\nu}\xi_\mu \xi_\nu = 0$ are still solutions after applying the boost. Since the non-propagating characteristics depend on the choice of $n_\mu$, neglecting the constraint equations in the equations~\eqref{ch3:equiv:ex7-first-order-system} leads to a loss of boost covariance.

\subsection{General case}
Let us consider a general quasilinear second-order\footnote{We assume here that the equations are such that the variables all require both the value of the variable and its first time derivative to be specified to solve the initial value problem. We do not consider the case of mixed-order systems.} system of partial differential equations,
\begin{equation}
\label{ch3:equiv:geneq}
    L[U] = {\cal A}^{\mu\nu}_{AB}[U,\de U] \de_\mu \de_\nu U^B + {\cal B}^\mu_{AB}[U, \de U] \de_\mu U^B + {\cal C}_A[U] = 0\,.
\end{equation}
Based on the examples that were previously laid out, we can lay out the following sufficient conditions for the causality of the theory to be assured by linearized causality constraints. Given the linearization $U^B(x) = U_0^B + \delta U^B(x)$ of the system of equations~\eqref{ch3:equiv:geneq},
\begin{enumerate}
    \item The principal part ${\cal A}^{\mu\nu}_{AB}$ does not depend on derivatives of the fields, $\de U$, and
    \item The constant solutions $U_0^B$ are not subject to any algebraic constraints, except for those which also apply to $U^B(x)$.
\end{enumerate}
These conditions are sufficient, rather than necessary, as we have already seen examples where they are violated, and the equivalence still holds. The primary takeaway is that, given a system of quasilinear PDEs of the very general form~\eqref{ch3:equiv:geneq}, one can assure the non-linear causality of the system of equations from causality conditions on the linearized system if conditions 1. and 2. hold.

The first condition is a property that the system of equations either does or does not have; however, the second can be a matter of how the linearization is set up. One aspect of condition 2. above is that the constant solutions $U_0^B$ are all independent and non-zero. However, as we saw in Example 5, it is sometimes possible to violate this condition and still have the equivalence hold.

\subsection{Constraint equations}
In the previous subsection, we had a baked-in assumption that all of the equations in~\eqref{ch3:equiv:geneq} were dynamical equations. This is certainly not always true, such as in Maxwell's equations with the constraint $\nabla{\cdot}B=0$. Let us consider a foliation of spacetime by spacelike hypersurfaces $\Sigma_t$, where the parameter labelling the hypersurfaces $t$ could be time, or some other parameter. At each point $p$ on $\Sigma_t$, the normal to the hypersurface is given by a timelike unit covector $n_\mu$, and the projector onto the hypersurface is given by $P_{\mu\nu} = n_\mu n_\nu + g_{\mu\nu}$. We can define the initial hypersurface $\Sigma_0$ to be the hypersurface given by $t=0$. We prescribe initial conditions on the hypersurface $\Sigma_0$; for a second order system of PDEs, the initial conditions are comprised of the value of the field on the initial hypersurface $U^B\vert_{\Sigma_0}$, and the first derivative off the hypersurface $n^\mu \de_\mu U^B \vert_{\Sigma_0}$.

With these definitions in place, we may now define a ``dynamical equation" as one which contains $n^\mu n^\nu \de_\mu \de_\nu U^B$, and a ``constraint equation" as an equation which does not. Constraint equations constrain the form of the initial data $U^B\vert_{\Sigma_0}$ and $n^\mu \de_\mu U^B \vert_{\Sigma_0}$. Let us suppose that the full system of $N$ PDEs in $n$ variables $U^B$ is covariant. The dynamical equations can be given by
\begin{equation}\label{ch3:equiv:dynamical-eqs}
    \tilde{{\cal A}}^{\mu\nu}_{AB}[U,\de U] \de_\mu \de_\nu U^B + \tilde{{\cal B}}^\mu_{AB}[U, \de U] \de_\mu U^B + \tilde{{\cal C}}_A[U] = 0\,,
\end{equation}
where $\det\lr{\tilde{{\cal A}}^{\mu\nu}_{AB}n_\mu n_\nu}$ must be non-vanishing, and $A,B \in \{1,...,n\}$. The dynamical equations~\eqref{ch3:equiv:dynamical-eqs} are supplemented by $N_c = N-n$ constraint equations of the form
\begin{equation}\label{ch3:equiv:constraint-eqs}
    \tilde{{\cal D}}^{\mu\nu}_{a B}[U,\de U] \de_\mu \de_\nu U^B + \tilde{{\cal E}}^\mu_{a B}[U,\de U] \de_\mu U^B + \tilde{{\cal F}}_a[U] = 0\,,
\end{equation}
where $a \in \{1,...,N_c\}$ labels the constraint equations, and $\tilde{{\cal D}}^{\mu\nu}_{aB}n_\mu n_\nu = 0$. The full system of equations~\eqref{ch3:equiv:dynamical-eqs} and~\eqref{ch3:equiv:constraint-eqs} is assumed covariant, but the decomposition of the system into dynamical and constraint equations breaks covariance, and instead depends on how one defines the hypersurfaces $\Sigma_t$ and the normal vector $n_\mu$. In particular, this means that the dynamical equations~\eqref{ch3:equiv:dynamical-eqs} by themselves are not covariant, and the characteristics of the dynamical equations will be sensitive to a choice of $n$. The only exception to this is if $\tilde{{\cal D}}^{\mu\nu}_{AB} = 0$.

Let us now specialize to a case where both the dynamical and constraint equations satisfy the sufficient conditions 1. and 2. of the previous section, i.e. independence of the principal part from derivatives of the state vector, and no zeroth-order term. In other words,
\begin{subequations}~\label{ch3:equiv:dyn_con_equiv}
    \begin{align}
        \tilde{{\cal A}}^{\mu\nu}_{AB}[U] \de_\mu \de_\nu U^B + \tilde{{\cal B}}^{\mu}_{AB}[U,\de U]\de_\mu U^B &= 0\,,\label{ch3:equiv:dyneq-conditions}\\
        \tilde{{\cal D}}^{\mu\nu}_{aB}[U]\de_{\mu}\de_\nu U^B + \tilde{{\cal E}}^{\mu}_{a B}[U,\de U] \de_\mu U^B &= 0\,.
    \end{align}
\end{subequations}
Since covariance is lost if we consider just the dynamical equations alone, let us take a slightly more straightforward view of causality. Let us define a system of partial differential equations to be causal if the Monge conoid at a point $p$ lies inside the lightcone for all $p$ and all choices of timelike unit covector $n_\mu$. With this definition the equivalence will still hold for the system of dynamical equations~\eqref{ch3:equiv:dyneq-conditions}. However, because the dynamical equations by themselves are not covariant, Lorentz transformations are not a symmetry of the principal part of~\eqref{ch3:equiv:dyneq-conditions}. Letting $G$ correspond to a Lorentz transformation, requiring Lorentz symmetry would require the principal part to transform as
$    G \tilde{{\cal A}}[\phi,n]G^{-1} = \tilde{A}(G \phi,n)\,,$
which is not the case. In the specific context of relativistic hydrodynamics, this implies that it would not be sufficient to ensure causality for the fluid at rest; instead, one must ensure causality for all reference frames, and look at linearized fluctuations about a general $U_0^B$.

Let us finish off this section by noting that this (somewhat frustrating) situation is mitigated in the specific case where the characteristic equation for the dynamical equations is of the form
\begin{equation}\label{ch3:equiv:special-Q}
    Q = \lr{n{\cdot}\xi}^{\ell} \tilde{Q} = 0\,,
\end{equation}
where $\tilde{Q}$ is a Lorentz scalar independent of $n$, and $\ell > 0$ is a positive integer. The characteristic equation in Example 7 was of this form, and in general an arbitrary unconstrained second-order system of PDEs cast down to first order will be of the form~\eqref{ch3:equiv:special-Q}. The non-covariance of the characteristic equation~\eqref{ch3:equiv:special-Q} is entirely contained within $n{\cdot}\xi$. Since the solutions of $n{\cdot}\xi=0$ are always causal for any timelike unit covector $n_\mu$, the causality of the system of equations comes down to determining the solutions to $\tilde{Q} = 0$. However, $\tilde{Q}$ is a Lorentz scalar, and so may be evaluated with any choice of $n_\mu$.

In a hydrodynamic theory, one can choose to work in the local fluid rest frame. Since $n_\mu$ defines the coordinate time, the fluid rest frame is given by $n_\mu u^\mu = -1$ at a given point $p$. Once one has fixed this choice of $n_\mu$, the only non-zero component of $u^\mu$ at $p$ is the component along $n_\mu$; in other words, $v_j = 0$. This means that linearizing about the solution $u^\mu(x) = \delta^\mu_0 + \delta u^\mu(x)$ no longer violates the sufficient condition 2., because the condition $u^\mu = \delta^\mu_0$ was inherited from the non-linearized variable $u^\mu(x)$. The sufficient conditions laid out for the equivalence are then fully satisfied, and enforcing that the roots of $\tilde{Q} = 0$ are causal in the linearized analysis in the rest frame is sufficient to enforce causality of the full non-linear theory for any choice of $n_\mu$. 
\section{Covariant stability}
\label{ch3:sec_covarstable}
Let us now consider what is actually required for a theory to be ``covariantly stable". No attempt shall be made here to be comprehensive; regarding my own contribution, this section (specifically in the large-$k$ limit) primarily contains results from my paper~\cite{Hoult:2023clg}.

\subsection{Stability and causality}
In Chapter~\ref{chapter:background}, when investigating the causality of the theory of a $U(1)$ charged fluid, we made use of a result that may be roughly stated as follows:
\begin{center}
    \textit{If a causal hydrodynamic system is stable in the local fluid rest frame, it is stable in all Lorentz reference frames.}
\end{center}
This result was rigorously shown in~\cite{Bemfica:2020zjp} for the $U(1)$ charged fluid subject to the assumption that the system of equations were strongly hyperbolic, formalizing the intuition of Israel~\cite{Israel2009}. In~\cite{Gavassino:2021owo}, this result was extended, and it was shown that
\begin{enumerate}
    \item If a hydrodynamic theory is acausal and dissipative, there exists a frame in which it is unstable. This was illustrated by the acausality of the diffusion equation in a boosted frame in Chapter~\ref{chapter:background}.
    \item If a deviation from equilibrium uniformly decays over time in one Lorentz reference frame, and its support does not leave the lightcone, it decays uniformly over time in all reference frames.
    \item If a causal hydrodynamic theory is \textit{unstable} in one reference frame, it is unstable in all reference frames.
\end{enumerate}
Taken all together, we arrive at the unified picture that stability and causality are intricately tied together: a causal theory which is stable in one reference frame is stable in all reference frames. This is the condition of ``covariant stability", one which we will demand of all causal hydrodynamic theories.

\subsection{The covariant stability condition}
In what follows, for a complex variable $z$, we will denote the real and imaginary parts by $z = z' + i z''$. The demands of covariant stability can be packaged into a single convenient condition. Let us simultaneously consider a complex frequency $\omega = \omega' + i\,\omega''$ \textit{and} a complex wavevector $k_i = k'_i + i \,k''_i $. Note that $k'_i$ and $k''_i$ are both real vectors. The covariant stability condition may then be written as~\cite{Heller:2022ejw}
\begin{equation}\label{ch3:covar:main-condition}
     \omega'' \leq | k''_i|\,.
\end{equation}
This condition is known in the QFT literature~\cite{Itzykson:1980rh}, but its application to quantum many-body systems in~\cite{Heller:2022ejw} was novel. Roughly speaking, the condition arises by demanding that the dispersion relations of the theory be consistent with the \textit{microscopic} causality of the system. Let us consider the origin of this condition, following the arguments of~\cite{Heller:2022ejw}. The retarded two-point function of a local operator ${\cal O}(x^\mu)$ in Minkowski space is given by
\begin{equation}
    G^R_{{\cal O}{\cal O}}(x,y) = - i \theta(x^0 - y^0) \braket{[{\cal O}(x), {\cal O}(y)]}\,,
\end{equation}
where $\theta(x^0 - y^0)$ is the Heaviside step function, and $\braket{[\cdot,\cdot]}$ refers to the expectation value of the commutator taken in some ensemble. We have previously taken causality as a condition on coupled systems of partial differential equations. However, quantum field theories are not described by coupled PDEs, but rather by operators; causality instead amounts to the demand that spacelike-separated operators have a vanishing commutator. With this consideration in mind, we can note that $G^{R}_{{\cal O}{\cal O}}(0,x)$ is only non-zero when $x^\mu$ is in the closed\footnote{i.e. in or on the past light-cone.} past light-cone. 
We also impose an additional condition -- that the two-point function is a tempered distribution (see e.g.~\cite{Streater-Wightman} Ch.~2), which simply means that the two-point function doesn't grow ``too fast" as it goes to infinity.

Consider now the Fourier transform
\begin{equation}
    \tilde{G}^R(K^\mu) = \int d^{d+1}x G^{R}_{{\cal O}{\cal O}}(0,x) e^{i K_\mu x^\mu}\,.
\end{equation}
Let us take $K_\mu$ to be complex, i.e. $K_\mu = K'_\mu + i K''_\mu = (-\omega', k'_i) + i (-\omega'', k''_i)$. Then
\begin{equation}\label{ch3:covar:fourier-transform-G}
       \tilde{G}^R(K^\mu) = \int d^{d+1}x G^{R}_{{\cal O}{\cal O}}(0,x) e^{i K'_\mu x^\mu} e^{- K''_\mu x^\mu}\,,
\end{equation}
Recall however that $G^{R}_{{\cal O}{\cal O}}(0,x)$ is only non-zero for $x^\mu$ in the closed past lightcone. Therefore, the integrand is exponentially suppressed when $K''_\mu$ is in the open\footnote{i.e. in, but not on, the future lightcone.} future lightcone, since then $K''_\mu x^\mu > 0$. This combined with $G^R_{{\cal O}{\cal O}}(0,x)$ being a tempered distribution means that when $K''_\mu$ is in the open future lightcone ($\omega'' > |k''_i|$), $\tilde{G}^R(K^\mu)$ is an analytic function~\cite{Streater-Wightman}.

This is a problem, however, because (see Appendix~\ref{app:linear response}) dispersion relations are given by the singularities of the Fourier-transformed retarded two-point function. Therefore, any dispersion relation which is compatible with microscopic causality must be such that $K''_\mu$ lies \textit{outside} the open future-directed lightcone. We therefore arrive at the condition~\eqref{ch3:covar:main-condition}.

The original derivation of the condition~\eqref{ch3:covar:main-condition} in~\cite{Heller:2022ejw} only (explicitly) depended on the causality of the theory; the condition was then re-interpreted in~\cite{Gavassino:2023myj} as a condition on stability in a hydrodynamic system. Specifically,~\cite{Gavassino:2023myj} showed that the violation of condition~\eqref{ch3:covar:main-condition} led immediately to the existence of a reference frame in which the equilibrium state was unstable. Additionally, it was shown that if one has a stable and causal theory, then the condition~\eqref{ch3:covar:main-condition} must hold. The relation arises due to the close connection (though not equivalence) between stability and the demand that $G^{R}_{{\cal O}{\cal O}}(0,x)$ be a tempered distribution.

Now that we have investigated the origin and the interpretation of the condition~\eqref{ch3:covar:main-condition}, we will discover what it can do. We will first consider the small-$k$ limit, which was investigated in~\cite{Heller:2022ejw,Gavassino:2023myj,Heller:2023jtd}. We will then consider the large-$k$ limit, which was investigated in~\cite{Hoult:2023clg,Wang:2023csj}. In all of the following, we will assume an isotropic equilibrium state, such that the system enjoys an $SO(d)$ rotation symmetry, and we can consider $K^\mu = \{\omega,k,0,0,...\}$, where $k_i$ has been rotated to align with the $x$-axis.

\subsection{The small-$k$ limit}
The following techniques arise from~\cite{Heller:2022ejw}. Let us consider a dispersion relation which admits an expansion in small $k$ about the origin $k=0$. In other words, let us write
\begin{equation}
    \omega = \sum_{n=0}^\infty a_n k^n\,,
\end{equation}
where $a_n$ and $k$ are both complex. Let us further assume that this expansion has a radius of convergence $R$, which is set by the nearest singularity to the origin. Within the disk of radius $R$, $\omega$ is taken to be analytic. Now, let us set $k = r \exp(i \theta)$, where $0 \leq r < R$, and $\theta$ is the argument of $k$\footnote{We will no longer be making use of the Heaviside function, and so hereafter $\theta$ will exclusively refer to an angle.}. Recall that both $a_n$ and $k^n$ are complex; we can then equivalently write
\begin{equation}
    \omega = U(r,\theta) + i V(r,\theta)
\end{equation}
Let us write $a_n \equiv \alpha_n + i \beta_n = |a_n| \exp\lr{i \theta_n}$, where $\theta_n$ is the argument of $a_n$. Using deMoivre's identity, we can write
\begin{equation}
    \omega = \lr{ \sum_{n=0}^\infty r^n \lr{
    \alpha_n \cos(n \theta) - \beta_n \sin(n \theta)}} + i \lr{\sum_{n=0}^\infty r^n \lr{\alpha_n \sin(n \theta) + \beta_n \cos(n \theta)}}\,,
\end{equation}
leading immediately to the identification
\begin{subequations}
    \begin{align}
        U(r,\theta) &=  \sum_{n=0}^\infty r^n \lr{
    \alpha_n \cos(n \theta) - \beta_n \sin(n \theta)}\,,\\
    V(r,\theta) &= \sum_{n=0}^\infty r^n \lr{\alpha_n \sin(n \theta) + \beta_n \cos(n \theta)}\,.
    \end{align}
\end{subequations}
The condition~\eqref{ch3:covar:main-condition} then amounts to the demand
\begin{equation}\label{ch3:covar:main-condition-smallk}
    V(r,\theta) \leq r |\sin(\theta)|\,.
\end{equation}
Now, we can use the orthogonality of trigonometric functions to note that for $n>0$,
\begin{subequations}
    \begin{align}
        r^n \alpha_n &= \frac{1}{\pi} \int_{0}^{2\pi}d\theta\, V(r,\theta) \sin(n \theta)\,,\quad r^n \beta_n = \frac{1}{\pi} \int_0^{2\pi} d\theta\, V(r,\theta) \cos(n \theta)\,,
    \end{align}
\end{subequations}
Therefore, 
    \begin{equation}
    \begin{split}
        |a_n| r^n &= \lr{\alpha_n + i  \beta_n} e^{- i \theta_n} r^n= \frac{1}{\pi} \int_0^{2\pi} d\theta\, V(r,\theta) \sin(n \theta + \theta_n)\,,
    \end{split}
    \end{equation}
    where we have used the fact that $|a_n|r^n$ is real. We would now like to make use of the constraint~\eqref{ch3:covar:main-condition-smallk}, but we must ensure the quantity multiplying $V(r,\theta)$ is non-negative. Let us therefore consider the expression $|a_n| r^n + 2 \beta_0$, and write
    \begin{equation}
        \begin{split}
            |a_n| r^n  + 2 \beta_0&= \frac{1}{\pi} \int_0^{2\pi} d\theta\, V(r,\theta) \lr{\sin(n\theta + \theta_n) + 1}\\
            &\leq \frac{r}{\pi} \int_0^{2\pi} d\theta\, |\sin(\theta)| \lr{\sin(n\theta+\theta_n)+1}\,.
        \end{split}
    \end{equation}
    Evaluating the integral in the last line (recalling that $n \in \mathbb{Z}$), we arrive at the condition for $n>0$
    \begin{equation}
        |a_n| r^n + 2 \beta_0 = \frac{2 r}{\pi} \lr{\frac{2 (n^2-1) - \lr{1+(-1)^n} \sin(\theta_n)}{n^2-1}}\,.
    \end{equation}
    Taking $r$ to $R$ yields the following constraint on the magnitude of the coefficients of the small-$k$ expansion:
    \begin{equation}\label{ch3:covar:theorem1-heller}
        |a_n| \leq \frac{2}{\pi R^{n-1}} \lr{\frac{2(n^2-1) - (1+(-1)^n)\sin(\theta_n)}{n^2-1}} - \frac{2}{R^n} a''_0\,,
    \end{equation}
   where we have switched to the notation of the previous section, $a''_0 = \beta_0$. This is the constraint in Theorem 1 of~\cite{Heller:2022ejw}. Setting $|k_i| = 0$ in condition~\eqref{ch3:covar:main-condition} yields that $a''_0\leq 0$. We should note here that the condition~\eqref{ch3:covar:theorem1-heller} constrains the coefficients of the small-$k$ expansion from \textit{above}. This differs significantly from the constraints we found in Chapter~\ref{chapter:background}, in which coefficients of small-$k$ dispersion relations were constrained from \textit{below} (e.g. $\sigma \geq 0$). The bound is in terms of the radius of convergence $R$, which is assumed to be a microscopically computable quantity\footnote{This is indeed the case in theories which admit a holographic dual~\cite{Grozdanov:2019kge,Cartwright:2024rus}.}.

    We now investigate hydrodynamic dispersion relations. Consider a purely diffusive mode (e.g. the shear mode). Then the mode goes as $\omega = - i D k^2 + \oser{k^4}$. Taking the condition~\eqref{ch3:covar:theorem1-heller} for $n=2$ and setting $\theta_2 = -\pi/2$ (which maximizes the bound), we find that
    \begin{equation}\label{ch3:covar:diffusion_bound}
        0 \leq D \leq \frac{16}{3 \pi} \frac{1}{R}\,.
    \end{equation}
    Similarly, let us consider a sound mode. Then $\omega(k) = \pm v_s k - i D k^2 + \oser{k^3}$. For the linear term, let us just consider $\omega = \pm v_s k$. Then $\omega'' = \pm v_s r \sin(\theta)$. Inserting this dispersion relation into the covariant stability constraint~\eqref{ch3:covar:main-condition} yields
    \begin{equation}\label{ch3:covar:linear-small-k}
        \pm v_s \sin(\theta) \leq |\sin(\theta)|
    \end{equation}
Demanding that the condition~\eqref{ch3:covar:linear-small-k} hold for all $\theta$ leads immediately to the condition $-1 \leq v_s \leq 1$. The diffusive term also obeys the constraint~\eqref{ch3:covar:diffusion_bound}. Let us now take the limit $R\to \infty$ in the constraint~\eqref{ch3:covar:theorem1-heller}. If the Taylor expansion of a complex function $w(z)$ has an infinite radius of convergence (i.e. the function is analytic over the entire complex plane except infinity), the function is said to be ``entire". We see that the condition~\eqref{ch3:covar:theorem1-heller} forces $a_n \to 0$ for all $n\geq 2$. Therefore, if $\omega$ is an entire function, it must be a polynomial of at most linear order in $k$. 
    
This immediately rules out the Landau theory of relativistic hydrodynamics; as we saw in Chapter~\ref{ch2:sec_uncharged}, the shear mode was given by the entire function $\omega = - i \lr{\frac{\eta}{p+\epsilon}} k^2$, which is of quadratic order in $k^2$. On the other hand, the modes in the BDNK and MIS theories of hydrodynamics are not entire, as they contain branch points due to the square root in their dispersion relations. Therefore, they do not immediately violate the constraint~\eqref{ch3:covar:main-condition}. Finally, one can show that the condition~\eqref{ch3:covar:main-condition} enforces that $\omega = \omega(k)$ have no poles or essential singularities. Branch points and branch cuts are not ruled out, as evidenced by the fact that branch points appear in causal theories of hydrodynamics.

\subsubsection{The hydrohedron}
Before moving on to large-$k$, we briefly discuss the results of~\cite{Heller:2023jtd}. The existence of two-sided bounds on coefficients for dispersion relations implies the existence of a polyhedron in coefficient space describing the region of permitted values. The authors of~\cite{Heller:2023jtd} referred to this polyhedron as the Hydrohedron. While we do not present the full results here (the interested reader may refer directly to the paper), we note that one can derive bounds on coefficients in terms of lower-order terms. For example, let us consider a diffusion-type dispersion relation, i.e.
$    \omega(k) = -i \sum_{n=1}^{\infty} \beta_{2n} k^{2n}\,.$
Let us define the quantity $\bar{\beta}_{2n} \equiv R^{2n-1}\beta_{2n}$. Then the first two bounds are given by~\cite{Heller:2023jtd}
\begin{equation}
        0 \leq \bar{\beta}_2 \leq \frac{16}{3 \pi}\,,\quad \frac{15 \pi \bar{\beta}_2 \lr{8 + 3 \pi \bar{\beta}_2}-256}{90\pi} \leq \bar{\beta}_4 \leq \frac{64}{15\pi}\,.
\end{equation}
If we continue on for all $\bar{\beta}_{2n}$, we find an infinite-dimensional polyhedron. One can repeat a similar exercise for sound modes $\omega(k) = \sum_{n=0}^{\infty} \alpha_{2n+1} k^{2n+1} -i \sum_{n=0}^\infty \beta_{2n} k^{2n}$, finding bounds as well involving the infinite set of parameters. 

\subsection{The large-$k$ limit}
Let us turn now to the large-$k$ limit, which was investigated in~\cite{Hoult:2023clg,WangPu}. As a preface, in the following we will only be considering dispersion relations which arise from classical systems of partial differential equations. What we mean here by a ``classical" theory is a theory in $d+1$ dimensions which has a finite number of partial differential equations in a set of fields which depend on $d+1$ coordinate variables\footnote{Looking ahead to Chapter~\ref{chapter:micro}, the governing equations for both kinetic theory and holographic theories fail the criterion. In kinetic theory, the Boltzmann equation is not a PDE, but rather a integro-differential equation. In holography, there are a finite number of partial differential equations, but one typically considers the dynamics of the boundary theory which depends only on boundary coordinates $x^\mu$, and not the AdS radial coordinate $r$. This leads to holographic systems having an infinite number of dispersion relations~\cite{Grozdanov:2019uhi,Berti:2009kk} (``quasinormal modes"), rather than the finite number one finds in a classical theory.}. One consequence of demanding that the theory be classical is that the large-$k$ expansion of $\omega(k)$ goes as the same power of $k$, regardless of the phase of $k$. This will be a core assumption going forward -- a result that is certainly not true for quantum systems~\cite{Grozdanov:2023txs}. In what follows, we find constraints on the form of large-$k$ dispersion relations. These constraints are quite general, and dispersion relations in both BDNK and MIS-type theories satisfy the conditions. Let us consider a linearized system of partial differential equations
\begin{equation}
    L_{AB}[\de] \delta U^B = 0\,,
\end{equation}
where $L_{AB}$ is a differential operator. We assume that the system of equations is Lorentz covariant. If we insert the plane-wave solution $\delta U^B = \delta \tilde{U}^B \exp(i K_\mu x^\mu)$, where $K_\mu = (-\omega, k_j)$ as before, then non-trivial solutions are given by the spectral curve
\begin{equation}
    F(\omega,k_j) = \det\lr{L[i K]} = 0\,.
\end{equation}
Assuming a rotationally invariant equilibrium state as in the small-$k$ case, we fix $k_j = (k, 0,...)$ along the $x$-axis. The spectral curve $F(\omega,k)$ is then a finite-order polynomial in $\omega$ and $k$, which also describes an algebraic curve\footnote{Hence the name.} in $\mathbb{C}^2$. We would like to constrain dispersion relations in the limit of large-$k$; thanks to the Puiseux theorem~\cite{Wall-singular-points}, an expansion of $\omega$ in terms of $k$ about $1/k = 0$ is guaranteed to exist, and may be written in the form
\begin{equation}\label{ch3:covar:large-k-puiseux}
    \omega(k) = \sum_{m=m_0}^\infty c_m \zeta^m\,.
\end{equation}
The series~\eqref{ch3:covar:large-k-puiseux} is called a Puiseux series. A number of new things have been introduced in the above. First of all, the parameter $\zeta$ is defined by $\zeta^r = 1/k$.  In the sum, $m_0$ is an integer which depends on the details of the spectral curve $F(\omega,k)$. Secondly, if the order of the polynomial $F(\omega,k)$ in $\omega$ is given by $M$, then there will be $M$ expansions~\eqref{ch3:covar:large-k-puiseux}. These $M$ expansions come in $N$ sets, and each set comes with $r_a$ branches, such that $\sum_{a=1}^N r_a = M$. As a consequence, for each set $\zeta$ is given by
\begin{equation}
    \zeta = e^{2 \pi i \ell/r_a}k^{-1/r_a}
\end{equation}
where $\ell = 0, 1, ..., r_a-1$. Suppose there were a set of expansions with $r=3$. Then there would be three modes of the form
\begin{equation}
        \omega_0(k) = \sum_{m=m_0}^\infty c_m k^{-m/3}\,,\,\,\, \omega_1(k) = \sum_{m=m_0}^\infty c_m e^{2 \pi i/3} k^{-m/3}\,,\,\,\,\omega_2(k) = \sum_{m=m_0}^\infty c_m e^{4 \pi i/3} k^{-m/3}\,.
\end{equation}
As an immediate simple example of a Puiseux series, let us consider a sound mode in the limit of small-$k$ (the expansion would then be about $k=0$, instead of $1/k = 0$). By isotropy, the spectral curve is a function of $k^2$; one then finds that $\zeta^2 = k^2$ for the sound mode. Therefore $\zeta = e^{\pi i \ell} k = (-1)^{\ell} k$, where $\ell = 0,1$, and
\begin{equation}
    \omega_{\pm}(k) = \sum_{m=1}^\infty (-1)^{\ell m} c_m k^{m} = \pm c_1 k + c_2 k^2 \pm c_3 k^3 + ...\,,
\end{equation}
which reveals the origin of the $\pm$ present in the sound mode. Returning now to large-$k$, we combine the expansion~\eqref{ch3:covar:large-k-puiseux} with the covariant stability condition~\eqref{ch3:covar:main-condition}. Consider the leading-order term of the expansion~\eqref{ch3:covar:large-k-puiseux}; we denote this by $\omega(k) \sim c_p k^p$, where $p$ is a real, rational exponent, and $c_p \in \mathbb{C}$, i.e. $c_p = c'_p + i c''_p$. Let us now complexify $k = r e^{i \theta}$; then the condition~\eqref{ch3:covar:main-condition} becomes
\begin{equation}\label{ch3:covar:main-condition-largek-leading}
     \lr{c''_p \cos(p \theta) + c_p' \sin(p\theta)} \leq \frac{1}{r^{p-1}}|\sin(\theta)|\,.
\end{equation}
Taking the $r\to \infty$ limit, setting $\theta = 0$, $\theta = \pi/p$, and $\theta = \pm\pi/(2p)$ yields the following constraints for $p>1$:
\begin{equation}\label{ch3:covar:no-large-k}
    c''_p \leq 0, \quad c''_p \geq 0, \quad c'_p \leq 0, \quad c'_p \geq 0\,,
\end{equation}
which of course just amounts to $c_p = 0$ for $p>1$. We have immediately recovered that which we already knew, namely the condition that $\omega(k)$ may not grow faster than linearly with $k$ in the limit of large $k$. Referring back to the Puiseux series~\eqref{ch3:covar:large-k-puiseux}, we see that the condition~\eqref{ch3:covar:no-large-k} means that we must have $m_0 \geq - r$ for all modes to be causal. Let us now set $p=1$, and consider $\theta = 0,\pi, \pm \frac{\pi}{2}$. This yields
\begin{equation}
    c''_1 \leq 0, \quad c''_1 \geq 0, \quad  c'_p \leq 1, \quad c'_p \geq -1\,,
\end{equation}
which amounts to the first two conditions of~\eqref{ch2:hydro:full_momentum_causcon}. 

We can now try to fix the subleading terms in the expansion beyond linear order. First of all, the next subleading term after linear order must be another integer term\footnote{The reason for this may be simply stated; the imaginary component of the next term will be the dominant imaginary contribution to the mode in the large-$k$ limit. However, if the power of $k$ is non-integer, then there is an accompanying phase factor from~\eqref{ch3:covar:large-k-puiseux}, and the mode will be generically unstable.}. The constraints one can impose at next-to-leading order are much weaker than at leading order, as one can see from the factor of 1/$r^{p-1}$ in the right-hand side of~\eqref{ch3:covar:main-condition-largek-leading}. One can only impose constraints when $\theta = 0,\, \pm \pi$, as in that limit the right-hand side vanishes, as does the linear contribution $c_1 k$. Investigating these limits, we find that if the next-to-leading order term is even in $k$, it must obey the condition $c''_n \leq 0$, and we cannot constrain any further terms (at least with the same method). If the next-to-leading order term is odd in $k$, the coefficient is constrained to be real ($c''_n = 0$), and we are now free to constrain the next-to-next-to-leading order term. If the next-to-next-to-leading term is even in $k$, then it must have negative imaginary part, and we cannot constrain any further terms. If it is odd, we can continue the process.

Summarizing the above, non-integer contributions to the large-$k$ dispersion relations may appear, but only after the first even-integer term in the expansion. Therefore, the large-$k$ expansion of dispersion relations in a covariantly stable system must generically be of the form
\begin{equation}\label{ch3:covar:generic-large-k}
    \omega(k) = \sum_{n=0}^{n_0} c_{1-2n} k^{1-2n} + c_{-2n_0} k^{-2n_0} + ...\,,
\end{equation}
where $n_0$ is a non-negative integer, the coefficients $c_{1-2n}$ are all real (and not necessarily non-zero), $|c_1| \leq 1$, and $c''_{-2n_0}\leq 0$. The dots in~\eqref{ch3:covar:generic-large-k} denote further subleading terms, which are allowed to have fractional powers of $1/k$. Some examples of allowed expansions presented in~\cite{Hoult:2023clg} are
\begin{subequations}
    \begin{align}
        \omega(k) &= c_1 k + c_0 + c_{-1/2} k^{-1/2} + ...\,,\\
        \omega(k) &= c_0 + c_{-1/2} k^{-1/2} + ...\,,\\
        \omega(k) &= c_1 k + c_{-1} k^{-1} + c_{-3} k^{-3} + c_{-4} k^{-4} + c_{-9/2} k^{-9/2} + ...\,,.
    \end{align}
\end{subequations}

Finally, we note~\cite{Hoult:2023clg} that one can also extract the third condition of~\eqref{ch2:hydro:full_momentum_causcon} from the covariant stability constraint~\eqref{ch3:covar:main-condition}. As an addendum, one may use the third condition of~\eqref{ch2:hydro:full_momentum_causcon} to prove that unless $F(\omega,k)$ is independent of $k$, all covariantly stable fluids must have at least one mode which has a non-zero phase velocity at large-$k$. We will not present the proof here, but the interested reader may refer to Appendix B of~\cite{Hoult:2023clg} for more details.

	\startchapter{Causal Hydrodynamics from Microscopics}
\label{chapter:micro}
In this chapter, we will be discussing the emergence of causal theories of hydrodynamics from microscopic theories (or at least, microscopic relative to the scales at which hydrodynamics is well-defined). There are two frameworks that we will consider in this chapter. The first is kinetic theory, which is a relevant theory when the underlying microscopics admits either a particle or quasi-particle description. The other is holography, which is relevant for microscopic theories which admit a gravitational dual such as ${\cal N} = 4$ Super-Yang-Mills at large $N$ and large 't Hooft coupling. In both frameworks, we will discuss how causal hydrodynamics arises. My contributions to both subjects are contained in~\cite{Hoult:2021gnb}.

\section{Kinetic Theory}
\label{ch4:sec:KT}
Kinetic theory is an extremely rich and well-developed branch of theoretical physics~\cite{Rezzolla-Zanotti,DeGroot:1980dk,Cercignani1988,Cercignani2002}. It describes the many-body dynamics of interacting theories via the evolution of distribution functions on phase space. One can, in principle, include all $n$-particle distribution functions (this is referred to as the BBGKY hierarchy~\cite{Yvon:1935,Bogoliubov:1946,Kirkwood1946,Born:1946}); however, in practice, one often truncates the infinite coupled system of equations down to one equation containing only the one-particle distribution function: the Boltzmann equation. In this section, we will investigate the emergence of causal hydrodynamics from the Boltzmann equation. In the following, we always assume a Minkowski background.

\subsubsection{Setup}
Let us denote the one-particle distribution function by $f_p(x^\mu) \equiv f(x^\mu, p^\mu)$, where the momentum 4-vector $p^\mu = (E, p^j)$ is on-shell, so that $E^2 - p_j p^j = m^2$, where $E$ is the particle energy and $p^j$ is the 3-momentum. Put differently, the condition reads $p^0 = \sqrt{p_j p^j + m^2}$. The distribution function gives the number of particles in a region of phase space; the normalization of the distribution function is such that the number density is given by
\begin{equation}
    n(x^\mu) = \int \frac{d^3 p}{(2\pi)^3} f_p(x^\mu) = \int_p p^0 f_p(x^\mu)\,,
\end{equation}
where $\int_p = \int d^3p/\lr{(2\pi)^3 p^0}$ is shorthand for an integral over the Lorentz-invariant phase space measure. We will make use of this shorthand throughout the remainder of the section. The distribution function depends on time, and therefore describes the worldlines of particles in phase space; in the absence of any external forces, their evolution is given by the Liouville operator $L[f_p] = p^\mu \de_\mu f_p(x^\mu)$. In a collisionless gas (i.e. in free streaming), the evolution equation would then be
\[
L[f_p] = p^\mu \de_\mu f_p = 0\,.
\]
However, in real gases, collisions certainly happen, and the equation describing the evolution of the particles in phase space is appropriately modified. The collisions are in principle quite complicated, and depend on the underlying microscopic theory. One can contain all of the effects of the collisions in an integral operator known as the collision operator, $C[f_p]$. The equation of motion for the distribution function is then the relativistic version of the Boltzmann equation:
\begin{equation}\label{ch4:kinetic:Boltzmann}
    p^\mu \de_\mu f_p = C[f_p]\,.
\end{equation}
The form of the collision operator depends on which interactions occur, though we demand that it be at least quadratic in $f$; in the case of $2-$to$-2$ elastic collisions, the collision operator takes the form~\cite{DeGroot:1980dk}
\begin{equation}
    C[f_p] = \frac{1}{2} \int_{p'} \int_{k} \int_{k'}  W(p,p'|k,k') [f_k f_{k'} (1\pm f_p)(1\pm f_{p'}) - f_p f_{p'}(1\pm f_k)(1\pm f_{k'})]\,,
\end{equation}
where $+$ corresponds to bosons, and $-$ corresponds to fermions. In the following, whenever a $\pm$ appears, the top sign will always refer to bosons, and the bottom for fermions. The transition rates obey $W(p,p'|k,k') = W(k,k'|p,p') = W(p',p|k,k') = W(p,p'|k',k)$, and are proportional to $\delta(p+p'-k-k')$. One can then show that, for $2-$to$-2$ elastic collisions, that
\begin{equation}
    \int_p [a(x) + b_\mu(x) p^\mu]C[f_p] = 0\,,
\end{equation}
for any arbitrary functions $a(x)$ and $b_\mu(x)$. Let us now define the $U(1)$ charge-current and a stress-energy tensor:
\begin{equation}\label{ch4:kt:def_J_T}
    J^\mu \equiv \int_p p^\mu f_p, \qquad   T^{\mu\nu} \equiv \int_p p^\mu p^\nu f_p\,.
\end{equation}
One can straightforwardly see that these two quantities are conserved by multiplying both sides of the Boltzmann equation by $(a(x) + b_\mu(x) p^\mu)$ and integrating over $p$:
\begin{align*}
    \int_p\lr{a(x) + b_\mu(x) p^\mu} \lr{p^\nu \de_\nu f_p} &= \int_p \lr{a(x) + b_\mu(x) p^\mu} C[f_p]\\
    a(x) \de_\nu J^\nu + b_\mu(x) \de_\nu T^{\nu\mu} &= 0\,.
\end{align*}
Since $a(x)$ and $b_\mu(x)$ are arbitrary functions, this implies $J^\mu$ and $T^{\mu\nu}$ must be independently conserved. 

\subsubsection{An irreducible representation}
We will now take advantage of a very powerful decomposition; this follows the presentation of~\cite{DeGroot:1980dk}. Let us consider the rank-$n$ tensor $\pi^{\mu_1...\mu_n} \equiv p^{\mu_1} ... p^{\mu_n}$, and a unit timelike vector $u^\mu$. Under the little group of $SO(3,1)$ with respect to $u^\mu$, i.e. the set of Lorentz transformations such that $\Lambda^\mu_{\,\,\,\,\nu} u^\nu = u^\mu$, this tensor is reducible. Let us define $E_p = - p_\mu u^\mu$, and denote orthogonality to $u^\mu$ by an angular bracket, e.g. $p^{\braket{\mu}} = \Delta^{\mu\nu}p_\nu$. Then we can see the reducibility immediately by decomposing $\pi^{\mu_1...\mu_n}$ with respect to $u^\mu$ in its first index:
\begin{equation}
    \pi^{\mu_1...\mu_n} = \pi^{\braket{\mu_1}\mu_2...\mu_n} + E_p u^\mu p^{\mu_2} ... p^{\mu_n}\,,
\end{equation}
where $\pi^{\braket{\mu_1}\mu_2...\mu_n} = \Delta^{\mu_1}_{\,\,\,\,\,\nu_1} \pi^{\nu_1\mu_2...\mu_n}$. The second term transforms as a rank-$(n-1)$ tensor under the action of the little group, while $\pi^{\braket{\mu_1} \mu_2 ... \mu_n}$ still transforms as a rank-$n$ tensor. We can continue in this fashion until we construct the fully symmetric, transverse tensor $\pi^{\braket{\mu_1}\braket{\mu_2}...\braket{\mu_n}}$. We can also further reduce the representation by taking the trace of $\pi$; therefore, the fully irreducible set of tensors with respect to the little group are given by the set of transverse, symmetric, traceless tensors
\begin{equation}
    \pi^{\braket{\mu_1...\mu_n}} \equiv \Delta^{\mu_1 ... \mu_n}_{\nu_1 ... \nu_n} p^{\nu_1} ... p^{\nu_n}\,,
\end{equation}
where $\Delta^{\mu_1 ... \mu_n}_{\nu_1 ... \nu_n}$ is the transverse traceless projector of rank $n$. For $n=1$ and $n=2$, these projectors are by now familiar constructs: 
\begin{equation}
        \Delta^{\mu_1}_{\nu_1} = \delta^{\mu_1}_{\nu_1} + u^{\mu_1} u_{\nu_1}\,,\quad \Delta^{\mu_1\mu_2}_{\nu_1\nu_2} = \frac{1}{2}\lr{\Delta^{\mu_1}_{\nu_1} \Delta^{\mu_2}_{\nu_2} + \Delta^{\mu_2}_{\nu_1} \Delta^{\mu_1}_{\nu_2} - \frac{2}{d} \Delta^{\mu_1 \mu_2} \Delta_{\nu_1 \nu_2}}\,.
\end{equation}
For higher-order constructs, the interested reader may refer to Chapter VI, Section 2.a of~\cite{DeGroot:1980dk}. The tensors $\pi^{\braket{\mu_1... \mu_n}}$ are orthogonal, and therefore can then be shown to have the very useful property~\cite{Denicol:2012cn}
\begin{equation}\label{ch4:kt:orthogonality}
    \int_p \pi^{\braket{\mu_1 ... \mu_n}} \pi_{\braket{\nu_1... \nu_m}} g(E_p) = N_m \delta_{mn}\Delta^{\mu_1...\mu_m}_{\nu_1...\nu_m} \int_p g(E_p) \lr{p_{\braket{\lambda}} p^{\braket{\lambda}}}^m\,.
\end{equation}
In the above, $g(E_p)$ is an arbitrary (albeit requiring that the integrals converge) function of $E_p$, and $N_m$ is an $m$-dependent normalization constant. For $m=1$, $N_m = 1/d$, while for $m=2$, $N_m = 2/d(d+2)$. 

\subsubsection{Equilibrium}
Let us now consider functions of $p^\mu$ that set the collision operator to zero. Given the conservation of energy, momentum, and $U(1)$ charge, we can define the function
\begin{equation}\label{ch4:kinetic:equilibrium}
    \feq_p(x^\mu) \equiv \biggl[\exp\lr{-\beta_\mu(x) p^\mu - \alpha(x)} \mp 1\biggr]^{-1}\,,
\end{equation}
which will set $C[\feq_p] = 0$. This is the Fermi-Dirac/Bose-Einstein ($+/-$) distribution. Demanding that the distribution function~\eqref{ch4:kinetic:equilibrium} satisfy the Boltzmann equation~\eqref{ch4:kinetic:Boltzmann} yields the conditions
\begin{equation}\label{ch4:kinetic:killing}
    \de_\mu \beta_\nu + \de_\nu \beta_\mu  =0, \qquad \de_\mu \alpha = 0\,.
\end{equation}
In flat space\footnote{In curved space, this would instead amount to the demand that $\beta_\mu$ be a Killing vector.}, $\alpha$ and $\beta_\mu$ must therefore be constants, meaning $\feq_p$ is a function of only $p$ (more specifically, $E_p$). These quantities, familiar from the previous chapters, may once again be written as $\beta^\mu = \frac{u^\mu}{T},$ $\alpha = \frac{\mu}{T}\,,$ where $u^2 = -1$ and $T = 1/\sqrt{-\beta^2}$. Let us now decompose the definitions~\eqref{ch4:kt:def_J_T} of the charge current and the stress-energy tensor with respect to the fluid velocity $u^\mu$. One can write
\begin{subequations}\label{ch4:kt:gen_decomp}
    \begin{align}
        J^\mu  &= \lr{\int_p E_p f_p} u^\mu + \Delta^{\mu\nu} \int_p p_{\braket{\nu}} f_p \equiv {\cal N} u^\mu + {\cal J}^\mu\,,\label{ch4:kt:Jmu-general}\\
        T^{\mu\nu}  &= \lr{\int_p E_p^2 f_p} u^\mu u^\nu + \lr{\frac{1}{d} \Delta^{\alpha\beta}\int_p p_{\braket{\alpha}} p_{\braket{\beta}} f_p}\Delta^{\mu\nu} \nonumber\\
        &+ 2 u^{(\mu} \lr{\Delta^{\nu)\lambda} \int_p E_p p_{\braket{\lambda}} f_p} + \lr{\Delta^{\mu\nu\alpha\beta} \int_p p_{\langle \alpha} p_{\beta\rangle} f_p}\nonumber\\
        &\equiv {\cal E} u^\mu u^\nu + {\cal P} \Delta^{\mu\nu} + 2 {\cal Q}^{(\mu} u^{\nu)} + {\cal T}^{\mu\nu}\,,\label{ch4:kt:Tmunu-general}
    \end{align}
\end{subequations}
where $p_{\braket{\mu}} = \Delta_{\mu\nu}p^\nu$ and $p_{\langle \mu} p_{\nu\rangle} = \Delta_{\mu\nu\alpha\beta}p^\alpha p^\beta$ is the transverse traceless projection of $p^\alpha p^\beta$. We have implicitly defined the curly variables by equations~\eqref{ch4:kt:gen_decomp}. If we now take the distribution function to be the equilibrium distribution function $\feq_p = \feq_p(E_p)$, then by the orthogonality of the irreducible representation of the little group~\eqref{ch4:kt:orthogonality}, the general decompositions~\eqref{ch4:kt:gen_decomp} reduce to the familiar equilibrium forms
\begin{equation}
    T^{\mu\nu}_{\rm eq.} = \epsilon u^\mu u^\nu + p \Delta^{\mu\nu}\,, \quad J^{\mu}_{\rm eq.} = n u^\mu\,,
\end{equation}
where we can define the equilibrium energy density, isotropic pressure, and charge density by
\begin{equation}
\label{ch4:kinetic:eq_thermo_vars}
    \epsilon = \int_p E_p^2 \feq_p, \quad p = \frac{1}{d} \Delta^{\alpha\beta} \int_p p_{\braket{\alpha}}p_{\braket{\beta}} \feq_p, \quad n = \int_p E_p \feq_p\,.
\end{equation}
In particular, because $\feq_p$ does not depend on $p_{\braket{\mu}}$ nor $p_{\langle\mu}p_{\nu\rangle}$, the orthogonal basis leads to the condition that ${\cal Q}^{\mu} = {\cal T}^{\mu\nu}={\cal J}^\mu = 0$ in equilibrium.

\subsection{BDNK from kinetic theory}
We would now like to write down a solution to the Boltzmann equation~\eqref{ch4:kinetic:Boltzmann} apart from the equilibrium solution~\eqref{ch4:kinetic:equilibrium}. The solution should be in some sense ``near" equilibrium, as is the case in hydrodynamics. We will therefore assume spacetime derivatives are small, and try to write down a solution in a derivative expansion\footnote{This is the Hilbert series approach~\cite{Cercignani2002}, later modified into the Chapman-Enskog approach~\cite{DeGroot:1980dk,Cercignani2002}}
\begin{equation}\label{ch4:kinetic:Hilbert_expansion}
    f_p(x) = \feq_p(x) + \ce f_p^{(1)}(x) + \ce^2 f_p^{(2)}(x) + \oser{\ce^3}\,,
\end{equation}
where $\ce$ is a parameter which keeps track of derivatives (i.e. $\ce \sim \oser{\de}$, $\ce^2 \sim \oser{\de^2}$, etc.), to eventually be set to unity. One can also think of $\ce$ as tracking the order of the ratio of the mean free path and the macroscopic scale in the theory (in other words, the Knudsen number Kn)~\cite{Cercignani2002}. In the following, we will usually suppress the distribution function's $x$-dependence for notational brevity. Note that $\feq_p$ is now a function of $x$; in other words (note that we have re-labelled $\beta_\mu \to \beta_\mu^{(0)}$ and $\alpha \to \alpha^{(0)}$ for later convenience),
\begin{equation}\label{ch4:kinetic:ooe_f0}
    \feq_p(x) = \left[\exp\lr{-\beta^{(0)}_\mu(x) p^\mu - \alpha^{(0)}(x)} \mp 1\right]^{-1}\,.
\end{equation}
which will no longer exactly solve the Boltzmann equation due to the violation of the conditions~\eqref{ch4:kinetic:killing}. This will not be a problem: referring back to the Boltzmann equation~\eqref{ch4:kinetic:Boltzmann}, there was a derivative with respect to the spacetime variables which also gains a factor of $\ce$:
\begin{equation}\label{ch4:kinetic:scaled_Boltzmann}
\ce p^\mu \de_\mu \lr{\feq_p + \ce f_p^{(1)} + \ce^2 f_p^{(2)} + \oser{\ce^3}} = C[\feq_p + \ce f_p^{(1)} + \ce^2 f_p^{(2)} + \oser{\ce^3}]
\end{equation}
We see then that $\feq_p(x)$ solves equation~\eqref{ch4:kinetic:scaled_Boltzmann} in the limit $\ce\to0$. In equilibrium the parameters $\beta^\mu$ and $\alpha$ could be identified with the fluid velocity, temperature, and chemical potential. Out of equilibrium, however, no such identification exists, and the parameters $\beta_\mu^{(0)}$, $\alpha^{(0)}$ in~\eqref{ch4:kinetic:ooe_f0} are just that: parameters.

Inside the collision operator, it will be convenient to instead write the corrections to the equilibrium distribution function in terms of variables $\phi_p^{(n)} \equiv f_p^{(n)}/(\feq_p (1\mp \feq_p)$. The collision operator $C[f]$ may be expanded as follows:
\begin{equation}\label{ch4:kinetic:C_exp}
\begin{split}
    C[f_p]=& \ce \feq_p \LC[\phi^{(1)}_p] + \,... \,+ \ce^n \lr{\feq_p \LC[\phi_p^{(n)}] - C^{(n)}[f_p^{(n-1)}, f_p^{(n-2)}, ...]} 
    \end{split}
\end{equation}
where $\LC$ is the ``linearized collision operator", and $C^{(n)}$ is the remaining non-linear parts of the collision operator at $n^{\rm th}$ order. The non-linear part depends only on terms of order lower than $n$. The forms of both $\LC$ and $C^{(n)}$ depend on the details of the collision operator, and $n$ for $C^{(n)}$. For $2$-to-$2$ elastic collisions, $\LC$ is of the form
\begin{equation}\label{ch4:kinetic:lincoll_explicit}
    \LC[\phi_p] = \frac{1}{2} \int_{p'} \int_{k} \int_{k'} W(p,p'|k,k') \feq_{p'} \lr{1 \pm \feq_{k}}\lr{1\pm \feq_{k'}}(\phi_{k} + \phi_{k'} - \phi_p - \phi_{p'})\,.
\end{equation}
Inserting the expansion~\eqref{ch4:kinetic:C_exp} into the scaled Boltzmann equation~\eqref{ch4:kinetic:scaled_Boltzmann} gives an infinite tower of equations. Solving order-by-order, we find at $n^{\rm th}$-order that
\begin{equation}\label{ch4:kt:scaled_Boltz_OBO}
    p^\mu \de_\mu f_p^{(n-1)} + C^{(n)}[f_p^{(n-1)}, f_p^{(n-2)},...] = \feq_p \LC[\phi_p^{(n)}]\,.
\end{equation}
We see then the interesting result that the form of the equation repeats order-by-order; broadly speaking, the equation may be written as $\feq_p\LC[\phi_p^{(n)}] = S_n\lr{f_p^{(n-1)}, f_p^{(n-2)},...}$ where $S_n$ is the $n^{\rm th}$-order source term which depends only on lower-order terms. We would like to be able to simply invert $\LC$ at each order and get the solution $f_p$ -- however, we are stymied by a small wrinkle, which is that the operator $\LC$ is not formally invertible. To see this, we quickly elucidate some properties of the linearized collision operator. 

\subsubsection{Properties of the linearized collision operator}

First of all, as the name suggests, the linearized collision operator is linear in momentum, i.e. $\LC[c_1(x) a(p) + c_2(x) b(p)] = c_1(x) \LC[a(p)] + c_2(x) \LC[b(p)]$ for arbitrary functions $a,\,b,\,c_1,\,c_2$. Secondly, it is self-adjoint, in the sense that
\begin{equation}\label{ch4:kinetic:self-adjoint}
    \int_p \feq_p g(p) \LC[h(p)] = \int_p \feq_p h(p) \LC[g(p)]
\end{equation}for arbitrary functions $g$, $h$. Finally, and most importantly, in a system with conserved energy-momentum and charge, the linearized collision operator inherits the conservation of these quantities:
\begin{equation}\label{ch4:kinetic:lincoll_conservation}
    \int_p \feq_p (a(x) + b_\mu(x) p^\mu) \LC[\phi_p^{(n)}] = 0
\end{equation}
However, by the self adjoint property~\eqref{ch4:kinetic:self-adjoint}, this implies that
\begin{equation}
    \int_p \feq_p \phi_p^{(n)} \lr{a(x)\LC[1] + b_\mu(x) \LC[p^\mu]} = 0 \,.
\end{equation}
Since $\phi_p^{(n)}$ is unknown and $a(x)$, $b_\mu(x)$ are arbitrary functions, this implies that
\begin{equation}
    \LC[1] = \LC[p^\mu] = 0\,.
\end{equation}
However, this means that $\LC$ has zero modes\footnote{Morally speaking, one can think of the integral operator $\LC$ having zero modes in the same way as one might think of a matrix having a zero eigenvalue.}, and is therefore not formally invertible. One may only invert the operator $\LC$ if the source term is orthogonal to the zero modes, i.e.
\begin{equation}
    \int_p \feq_p (a(x) + b_\mu(x) p^\mu) \LC[\phi_p^{(n)}] =\int_p\lr{a(x) + b_\mu(x) p^\mu} S_n = 0
\end{equation}
However, the source term is entirely composed of lower-order terms which have already been solved for. Therefore, in order to solve the equation at $n^{\rm th}$-order, one must constrain the $(n-1)^{\rm th}$-order data. Let us now investigate how this works at first order.

\subsubsection{First order}

At first order, the Boltzmann equation~\eqref{ch4:kt:scaled_Boltz_OBO} reads
\begin{equation}\label{ch4:kinetic:first-order}
    p^\mu \de_\mu \feq_p = \feq_p \LC[\phi_p^{(1)}]
\end{equation}
We can then identify the source term as $S_1 = p^\mu \de_\mu \feq_p$. In order to invert equation~\eqref{ch4:kinetic:first-order}, the source term must be projected orthogonal to the zero modes:
\begin{equation}
\label{ch4:kinetic:orthogonality_condition}
    \int_p \lr{a(x) + b_\mu(x) p^\mu(x)} \lr{p^\nu \de_\nu \feq_p} = a(x) \de_\nu J_{(0)}^\nu(x) + b_\mu(x) \de_\nu T^{\mu\nu}_{(0)}(x) =  0
\end{equation}
where $T^{\mu\nu}_{(0)} = \int_p p^\mu p^\nu \feq_p$ and $J_{(0)}^\mu = \int_p p^\nu \feq_p$ are the zeroth-order stress-tensor and charge current respectively. Since $a(x)$ and $b_\mu(x)$ are arbitrary, these must independently vanish, and so the constraint that the source term be orthogonal to the zero modes amounts to the demand that the zeroth-order stress-tensor and charge current are independently conserved.

This in turn imposes constraints on the parameters $\beta^\mu_{(0)}$ and $\alpha_{(0)}$. In particular, defining $\epsilon(\beta_{(0)}(x),\alpha_{(0)}(x))$, $p(\beta_{(0)}(x),\alpha_{(0)}(x))$, $n(\beta_{(0)}(x),\alpha_{(0)}(x))$ analogously to the equilibrium case~\eqref{ch4:kinetic:eq_thermo_vars}, as well as $\beta_{(0)}^\mu = u^\mu_{(0)}/T_{(0)}$, $\alpha_{(0)} = \mu_{(0)}/T_{(0)}$, the condition~\eqref{ch4:kinetic:orthogonality_condition} imposes that
\begin{subequations}\label{ch4:kinetic:euler-constraint}
    \begin{align}
        u^\mu_{(0)} \de_\mu n + n \de_\mu u^\mu_{(0)} &= 0\,,\\
        u^\mu_{(0)} \de_\mu \epsilon + \lr{\epsilon + p} \de_\mu u^\mu_{(0)} &= 0\,,\\
        \lr{\epsilon + p}\Delta^{\alpha\nu}_{(0)} u^\mu_{(0)} \de_\mu u_{\nu,(0)} + \Delta^{\alpha\mu}_{(0)}\de_\mu p &= 0\,,
    \end{align}
\end{subequations}
where $\Delta^{\mu\nu}_{(0)} = u^\mu_{(0)} u^\nu_{(0)} + \eta^{\mu\nu}$. These are, of course, simply the relativistic Euler equations~\eqref{ch2:hydro:euler_eqs} that we saw in Chapter~\ref{chapter:background}. Let us now suppose that the conditions~\eqref{ch4:kinetic:euler-constraint} have been imposed, such that we can invert the linearized collision operator. The source term may be broken up into scalar, vector, and tensor contributions with respect to the little group:
\begin{equation}
    \begin{split}
        \frac{p^\mu \de_\mu \feq_p}{\feq_p \lr{1\pm \feq_p}} &= \biggl[ E_p^2 u^\mu u^\nu \de_\mu \beta_{\nu,(0)} + \frac{1}{d} p_{\braket{\alpha}} p^{\braket{\alpha}} \Delta^{\mu\nu}_{(0)} \de_\mu \beta_{\nu,(0)} + E_p u^\mu_{(0)} \de_\mu \alpha_{(0)}\biggr]_{\rm scalar} \\
        &+ \biggl[\lr{2 E_p p_{\braket{\lambda}}} \Delta^{\lambda(\mu}_{(0)} u^{\nu)}_{(0)} \de_\mu \beta_{\nu,(0)} + \lr{p_{\braket{\lambda}}}\Delta^{\lambda\mu}_{(0)} \de_\mu \alpha_{(0)}\biggr]_{\rm vector}\\
        &+ \biggl[p_{\langle \alpha}p_{\beta\rangle} \Delta^{\mu\nu\alpha\beta}_{(0)} \de_\mu \beta_{(0),\nu} \biggr]_{\rm tensor}
    \end{split}
\end{equation}
We now invert $\LC$. Due to the presence of the zero modes, the solution to equation~\eqref{ch4:kinetic:first-order} may be written in the form $\phi_p^{(1)} = \phi_{h} + \Phi_p$, where $\phi_h$ is the homogeneous solution, and $\Phi_p$ is the inhomogeneous solution found from inverting the operator $\LC$. The inhomogeneous solution is given by $\Phi_p = \LC^{-1}\left[p^\mu \de_\mu \feq_p / \feq_p \right]$, i.e.
\begin{align}
\label{ch4:kinetic:inhomogeneous_sol}
    \Phi_p &= \biggl[\LC^{-1}[(1\pm \feq_p)E_p^2] u^\mu_{(0)} u^\nu_{(0)} \de_\mu \beta_{\nu,(0)} + \frac{1}{d}\LC^{-1}[(1\pm \feq_p)p_{\braket{\alpha}}p^{\braket{\alpha}}] \Delta^{\mu\nu}_{(0)} \de_\mu \beta_{\nu,(0)}\\
    &+ \LC^{-1}[(1\pm \feq_p)E_p] u^\mu_{(0)} \de_\mu \alpha_{(0)} \biggr]_{\rm scalar}+ \biggl[2\LC^{-1}[ (1\pm \feq_p) E_p p_{\braket{\lambda}}]\Delta^{\lambda(\mu}_{(0)} u^{\nu)}_{(0)} \de_\mu \beta_{\nu,(0)}\nonumber\\
    &+ \LC^{-1}[(1\pm \feq_p) p_{\braket{\lambda}}] \Delta^{\lambda \mu}_{(0)} \de_\mu \alpha_{(0)}\biggr]_{\rm vector}+ \biggl[ \LC^{-1}[(1\pm \feq_p)p_{\langle \alpha} p_{\beta\rangle}]\Delta^{\mu\nu\alpha\beta}_{(0)}\de_\mu \beta_{(0),\nu}\biggr]_{\rm tensor}\nonumber
\end{align}
However, the terms in the solution~\eqref{ch4:kinetic:inhomogeneous_sol} are not all independent, due to the imposition of the ideal-order equations of motion. We can use the ideal-order equations to eliminate\footnote{We note that this decision for how to apply the ideal-order equation is a choice; one could just as easily choose to use the constraint equations to eliminate some other combination of two scalars and one vector.} $u^\mu u^\nu \de_\mu \beta_{\nu,(0)}$, $u^\mu_{(0)} \de_\mu \alpha_{(0)}$ and $\Delta^{\lambda(\mu}_{(0)} u^{\nu)}_{(0)} \de_\mu \beta_{\nu,(0)}$ in favour of $\Delta^{\mu\nu}_{(0)} \de_\mu \beta_{\nu,(0)}$ and $\Delta^{\lambda\mu}_{(0)} \de_\mu \alpha_{(0)}$, thereby removing all time-derivatives from the solution in the rest frame. With this choice, the inhomogeneous solution may be written
\begin{equation}
    \Phi_p = {\cal K}_{\zeta}(x,E_p) \Delta^{\mu\nu}_{(0)} \de_\mu \beta_{\nu,(0)} + {\cal K}_{\sigma}(x,E_p) p^{\braket{\mu}}\Delta^{\,\,\nu}_{\mu,(0)} \de_\nu \alpha_{(0)} + {\cal K}_{\eta}(x,E_p) p^{\langle \mu} p^{\nu\rangle}\Delta_{\mu\nu,(0)}^{\quad \alpha \beta} \de_\alpha \beta_{\beta,(0)}
\end{equation}
where ${\cal K}_{\zeta}(x, E_p)$, ${\cal K}_{\sigma}(x, E_p)$, and ${\cal K}_{\eta}(x, E_p)$ are functions of $\beta^\mu_{(0)}(x)$, $\alpha_{(0)}(x)$, and $E_p$ which depend on the details of the microscopic theory by way of ${\cal L}^{-1}$. The factors of $p^{\mu}$ in the solution above are determined based on symmetry grounds. The homogeneous solution is proportional to the zero modes of the linearized collision operator, and so $\phi_h = \mathfrak{a}^{(1)}(x) + \mathfrak{b}^{(1)}_\mu(x) p^\mu$, where $\mathfrak{a}^{(1)}(x)$ and $\mathfrak{b}^{(1)}_\mu(x)$ are integration functions. We then have the solution for $f_p$ up to first order in derivatives:
\begin{equation}\label{ch4:kinetic:1o-sol-zerovar}
\begin{split}
    f_p &= \feq_p + \ce \feq \lr{1\pm \feq} \biggl[\mathfrak{a}^{(1)}(x) + \mathfrak{b}^{(1)}_\mu(x) p^\mu +  {\cal K}_{\zeta}(x,E_p) \Delta^{\mu\nu}_{(0)} \de_\mu \beta_{\nu,(0)} \\
    &+ {\cal K}_{\sigma}(x,E_p) p^{\braket{\mu}}\Delta^{\,\,\nu}_{\mu,(0)} \de_\nu \alpha_{(0)} + {\cal K}_{\eta}(x,E_p) p^{\langle \mu} p^{\nu\rangle}\Delta_{\mu\nu,(0)}^{\quad \alpha \beta} \de_\alpha \beta_{\beta,(0)} \biggr] + \oser{\ce^2}
    \end{split}
\end{equation}
We will not go to second order\footnote{If we were to proceed to second order, we would find that demanding the invertibility of $\LC$ would fix the form of $\mathfrak{a}^{(1)}(x)$ and $\mathfrak{b}^{(1)}_\mu(x)$ entirely in terms of $\Phi_p$, and therefore in terms of $\beta^\mu_{0}$ and $\alpha_0$. Correspondingly, a new set of integration functions, $\mathfrak{a}^{(2)}(x)$ and $\mathfrak{b}_\mu^{(2)}(x)$ would be introduced at second order.}; we will instead do something which at first will seem somewhat opaque. Let us define new degrees of freedom $\alpha$, $\beta^\mu$ which are implicitly defined via:
\begin{subequations}\label{ch4:kinetic:frame-choice}
\begin{align}
    \beta^\mu + \ce b_1 \lr{\Delta^{\alpha\beta}\de_\alpha \beta_\beta} \beta^\mu + \ce b_2 \Delta^{\mu\nu}\de_\nu \alpha - \ce \mathfrak{b}^\mu_{(1)}(x) + \oser{\ce^2}&= \beta^\mu_{(0)} \,,\\
    \quad \alpha + a_1 \ce  \Delta^{\alpha\beta}\de_\alpha \beta_\beta - \ce \mathfrak{a}^{(1)}(x) + \oser{\ce^2}&= \alpha_{(0)} \,.
\end{align}
\end{subequations}
In the above, we have temporarily left $a_1$, $b_1$, $b_2$ undefined -- we will fix them momentarily. Let us now substitute the expressions~\eqref{ch4:kinetic:frame-choice} for $\beta_{(0)}^\mu$, $\alpha_{(0)}$ into the solution~\eqref{ch4:kinetic:1o-sol-zerovar}. Anywhere that $\alpha_{(0)}$ and $\beta_{(0)}^\mu$ appear in the first-order contribution $f_p^{(1)}$ (including inside ${\cal K}_{\zeta,\sigma,\eta}$), this substitution simply replaces $\beta^\mu_{(0)} \to \beta^\mu$, $\alpha_{(0)} \to \alpha$ up to an $\oser{\ce^2}$ error, which we neglect.

In the zeroth-order contribution $\feq_p$, this replacement is slightly more involved. Let us define $f'^{\rm eq.}_p$ as $\feq_p$ with the simple replacements $\beta_{(0)}^\mu\to\beta^\mu$ and $\alpha_{(0)} \to \alpha$. Then, to first order in $\ce$, we find
\begin{equation}
\begin{split}
    \feq_p &= f'^{\rm eq.}_p - \ce f'^{\rm eq.}_p (1\pm f'^{\rm eq.}_p) \biggl(\mathfrak{a}^{(1)}(x) + \mathfrak{b}_\mu^{(1)}(x)p^\mu \\
    &- b_1 \lr{\Delta^{\alpha\beta}\de_\alpha \beta_\beta} \beta_\mu p^\mu - b_2 \Delta^{\mu\nu} \de_\nu \alpha p_\mu - a_1 \Delta^{\alpha\beta}\de_\alpha\beta_\beta\biggr)+ \oser{\ce^2}\,.
    \end{split}
\end{equation}
However, this first-order term completely cancels out the integration constants, and we are left with a solution of the form
\begin{align}\label{ch4:kinetic:Landau_sol}
    f_p &= f'^{\rm eq.}_p + \ce f'^{\rm eq.}_p\lr{1 \pm f'^{\rm eq.}_p} \biggl[b_1 \lr{\Delta^{\alpha\beta}\de_\alpha \beta_\beta} \beta_\mu p^\mu + b_2 p_\mu\Delta^{\mu\nu} \de_\nu \alpha  + a_1 \Delta^{\alpha\beta}\de_\alpha\beta_\beta\\
    &+{\cal K}_{\zeta}(x,E_p) \Delta^{\mu\nu} \de_\mu \beta_{\nu} + {\cal K}_{\sigma}(x,E_p) p^{\braket{\mu}}\Delta^{\,\,\nu}_{\mu} \de_\nu \alpha + {\cal K}_{\eta}(x,E_p) p^{\langle \mu} p^{\nu\rangle}\Delta_{\mu\nu}^{\quad \alpha \beta} \de_\alpha \beta_{\beta} \biggr] + \oser{\ce^2}\,.\nonumber
\end{align}
Let us now fix $a_1$, $b_1$, $b_2$. Defining the shorthand $\braket{...} \equiv \int_p f'^{\rm eq.}_p \lr{1 \pm f'^{\rm eq.}_p} \lr{...}$, we set
\begin{subequations}\label{ch4:kt:defs_a1b1b2}
    \begin{alignat}{4}
        &b_1 &&= \left[\frac{\braket{E_p} \braket{E_p^2 {\cal K}_\zeta} -\braket{E_p^2} \braket{E_p {\cal K}_\zeta} }{\braket{E_p^2}^2 - \braket{E_p} \braket{E_p^3}}  \right]\,,\quad &&b_2 &&= -\left[\frac{\braket{E_p p_{\braket{\lambda}} p^{\braket{\lambda}} {\cal K}_{\sigma}}}{\braket{E_p p_{\braket{\lambda}} p^{\braket{\lambda}}}} \right]\,,\\
        &a_1 &&=- \left[\frac{\braket{E_p^2} \braket{E_p^2 {\cal K}_\zeta} -\braket{E_p^3} \braket{E_p {\cal K}_\zeta} }{\braket{E_p^2}^2 - \braket{E_p} \braket{E_p^3}} \right]\,.
    \end{alignat}
\end{subequations}
Inserting the solution~\eqref{ch4:kinetic:Landau_sol} along with the definitions~\eqref{ch4:kt:defs_a1b1b2} into the stress-energy tensor and charge current and decomposing with respect to $u^\mu = \beta^\mu/\sqrt{-\beta^2}$, we find
\begin{subequations}
    \begin{align}
        T^{\mu\nu} &= \lr{\epsilon' + \ce{\cal E}_1} u^\mu u^\nu + \lr{p' + \ce {\cal P}_1} \Delta^{\mu\nu} + 2 \ce {\cal Q}_1^{(\mu}u^{\nu)} + {\cal T}^{\mu\nu}_1\\
        J^\mu &= \lr{n'+\ce {\cal N}_1} u^\mu + \ce {\cal J}_1^\mu
    \end{align}
\end{subequations}
where
\begin{equation}
    \begin{split}
        \epsilon' &= \int_p  E_p^2 f'^{\rm eq.}_p, \quad p' = \frac{1}{d} \Delta^{\alpha\beta} \int_p p_{\braket{\alpha}} p_{\braket{\beta}} f'^{\rm eq.}_p, \quad n' = \int_p E_p f'^{\rm eq.}_p\,,
    \end{split}
\end{equation}
and
\begin{subequations}
    \begin{align}
{\cal E}_1 &= \int_p E_p^2 f'^{\rm eq.}_p \lr{1\pm f'^{\rm eq.}_p} \biggl[b_1 \frac{E_p}{T}  + a_1   + {\cal K}_\zeta(x,E_p)  \biggr]\Delta^{\mu\nu}\de_\mu \beta_\nu = 0\,,\\
{\cal P}_1 &= \frac{1}{d}\int_p p_{\braket{\lambda}}p^{\braket{\lambda}} f'^{\rm eq.}_p \lr{1\pm f'^{\rm eq.}_p}\biggl[b_1 \frac{E_p}{T}  + a_1   + {\cal K}_\zeta(x,E_p)  \biggr]\Delta^{\mu\nu}\de_\mu \beta_\nu \equiv - \zeta \de_\mu u^\mu\,, \\
{\cal N}_1 &=\int_p E_p f'^{\rm eq.}_p \lr{1\pm f'^{\rm eq.}_p} \biggl[b_1 \frac{E_p}{T}  + a_1   + {\cal K}_\zeta(x,E_p)  \biggr]\Delta^{\mu\nu}\de_\mu \beta_\nu = 0\,,
\end{align}
\begin{align}
{\cal Q}_1^\mu &= \Delta^{\mu\lambda} \int_p E_p p_{\braket{\lambda}}f'^{\rm eq.}_p \lr{1\pm f'^{\rm eq.}_p}\biggl[b_2  + K_\sigma(x,E_p) \biggr]p_{\braket{\alpha}}\Delta^{\alpha\beta}\de_\beta \alpha = 0 \,,\\
{\cal J}_1^\mu &= \Delta^{\mu\nu} \int_p p_{\braket{\nu}}  f'^{\rm eq.}_p \lr{1\pm f'^{\rm eq.}_p} \biggl[b_2 + K_\sigma(x,E_p)\biggr] p_{\braket{\alpha}}\Delta^{\alpha\beta}\de_\beta \alpha \equiv - \sigma T \Delta^{\mu\nu}\de_\nu \alpha\,,\\
{\cal T}_1^{\mu\nu} &= \Delta^{\mu\nu\alpha\beta} \int_p p_{\langle\alpha} p_{\beta\rangle}f'^{\rm eq.}_p \lr{1\pm f'^{\rm eq.}_p} \lr{K_{\eta}(x,E_p) p_{\langle \rho}p_{\sigma\rangle}}\Delta^{\rho\sigma\lambda\xi}\de_\lambda \beta_\xi \equiv - \eta \sigma^{\mu\nu}\,, 
    \end{align}
\end{subequations}
where $T = 1/\sqrt{-\beta^2}$. Inserting these expressions into the stress-energy tensor and charge current, we find (dropping the primes and scaling $\ce \to 1$)
\begin{equation}
        T^{\mu\nu} = \epsilon u^\mu u^\nu + \lr{p -  \zeta \de_\mu u^\mu} \Delta^{\mu\nu} - \eta  \sigma^{\mu\nu}\,,\quad J^\mu = n u^\mu - \sigma \Delta^{\mu\nu}\de_\nu \lr{\frac{\mu}{T}}\,.
\end{equation}
We have arrived at the Landau-frame formulation of hydrodynamics~\eqref{ch2:hydro:ns}, a result of the definitions of $\beta^\mu$ and $\alpha$ in equations~\eqref{ch4:kinetic:frame-choice},~\eqref{ch4:kt:defs_a1b1b2}. These definitions ultimately serve to fix\footnote{An alternative (equivalent) approach would have been to send $(\beta_{(0)}^\mu,\alpha_{(0)}) \to (\beta^\mu,\alpha)$, i.e. some other functions of $x^\mu$ which are ``off-shell" with respect to the ideal-order equations, with no comment on how ($\beta^\mu,\alpha$) are related to $(\beta^{\mu}_{(0)}, \alpha_{(0)}$), and then fix the integration functions in terms of $\alpha$, $\beta^\mu$ to set the frame. This approach is more direct, but the assumptions being made are much less clear, and so we have avoided it.} the integration functions $\mathfrak{a}^{(1)}(x)$, $\mathfrak{b}^{(1)}_\mu(x)$ in $\phi_h(x)$. With an understanding of how the Landau frame arises in kinetic theory, let us now consider how to arrive at a general fluid frame. The integration functions $\mathfrak{a}^{(1)}$, $\mathfrak{b}_\mu^{(1)}$ are first order in the derivative expansion; they can therefore be written in terms of a basis of one-derivative functions of $\beta^\mu$, $\alpha$. We will broadly write
\begin{subequations}\label{ch4:kinetic:betaalpha_genframe}
    \begin{align}
        \beta^\mu + \ce \biggl[ \lr{b_1 u^\alpha u^\beta \de_\alpha \beta_\beta + b_2 \Delta^{\alpha\beta} \de_\alpha \beta_\beta + b_3 u^\alpha \de_\alpha \alpha} \beta^\mu&\nonumber \\
        + \lr{ b_4 \Delta^{\lambda(\alpha}u^{\beta)}\de_\alpha\beta_\beta + b_5 \de_\lambda \alpha}\Delta^{\mu}_{\,\,\,\,\lambda} \biggr] - \ce \mathfrak{b}^\mu_{(1)} + \oser{\ce^2}&= \beta_{(0)}^\mu\\
        \alpha + \ce \biggl[a_1 u^\alpha u^\beta \de_\alpha \beta_\beta + a_2 \Delta^{\alpha\beta} \de_\alpha \beta_\beta + a_3 u^\alpha \de_\alpha \alpha\biggr] - \ce \mathfrak{a}_{(1)}+ \oser{\ce^2} &= \alpha_{(0)}
    \end{align}
\end{subequations}
Substituting the definitions~\eqref{ch4:kinetic:betaalpha_genframe} into the solution~\eqref{ch4:kinetic:1o-sol-zerovar} and keeping only terms up to first order in $\ce$ yields a solution in terms of $\alpha$, $\beta^\mu$, and eight free parameters $b_{1,2,3,4,5}$, $a_{1,2,3}$. This is surprising: in Chapter~\ref{chapter:background}, we saw that a general BDNK theory for a $U(1)$ charged fluid had thirteen free parameters (and the shear viscosity, which was frame invariant). When the hydrodynamic description arises from kinetic theory, those thirteen transport parameters become related. This is because the form of the inhomogeneous term was fixed by an application of the constraint equations (i.e. we chose to remove all time derivatives). This fixed the form of the frame-invariants $f_{1,2,3}$, $\ell_{1,2}$, leaving $13-5 = 8$ free parameters in the theory. Let us insert the definition~\eqref{ch4:kinetic:betaalpha_genframe} into the solution~\eqref{ch4:kinetic:1o-sol-zerovar} and keep to only first order in $\ce$. Then, using the shorthand $\braket{...}$ again, the scalar viscous corrections to the stress-energy tensor and charge current may be given by in terms of the building blocks~\eqref{ch2:hydro:the-basis}:
\begin{subequations}\label{ch4:kt:scalar_contribution}
    \begin{align}
        {\cal E}' &=  \braket{E_p^2(b_1 \frac{E_p}{T} + a_1)} \lr{\frac{s_1}{T}} + \braket{E_p^2(b_2 \frac{E_p}{T} + a_2 + {\cal K}_\zeta)} \lr{\frac{s_2}{T}} + \braket{E_p^2(b_3 \frac{E_p}{T} + a_3)}s_3 \,,\\
        {\cal P}' &= \braket{\frac{1}{d} p_{\braket{\alpha}}p^{\braket{\alpha}}(b_1 \frac{E_p}{T} + a_1)} \lr{\frac{s_1}{T}}+ \braket{\frac{1}{d} p_{\braket{\alpha}}p^{\braket{\alpha}}(b_2 \frac{E_p}{T} + a_2 + {\cal K}_\zeta)} \lr{\frac{s_2}{T}} \nonumber\\
        &+ \braket{\frac{1}{d} p_{\braket{\alpha}}p^{\braket{\alpha}}(b_3 \frac{E_p}{T} + a_3)}s_3 \,,\\
        {\cal N}' &= \braket{E_p(b_1 \frac{E_p}{T} + a_1)} \lr{\frac{s_1}{T}} + \braket{E_p(b_2 \frac{E_p}{T} + a_2 + {\cal K}_\zeta)} \lr{\frac{s_2}{T}}+ \braket{E_p(b_3 \frac{E_p}{T} + a_3)}s_3 \,,
    \end{align}
\end{subequations}
the vector corrections are given by
\begin{subequations}\label{ch4:kt:vector_contribution}
    \begin{align}
        {\cal Q}'^\mu &= \frac{1}{d}\braket{E_p p_{\braket{\lambda}} p^{\braket{\lambda}} b_4 } \lr{\frac{V_1^\mu}{T}} + \frac{1}{d}\braket{E_p p_{\braket{\lambda}} p^{\braket{\lambda}} \lr{b_5 + {\cal K}_\sigma} } V_2^\mu\,,\\
        {\cal J}'^\mu &= \frac{1}{d}\braket{ p_{\braket{\lambda}} p^{\braket{\lambda}} b_4 } \lr{\frac{V_1^\mu}{T}} + \frac{1}{d}\braket{ p_{\braket{\lambda}} p^{\braket{\lambda}} \lr{b_5 + {\cal K}_\sigma} } V_2^\mu\,,
    \end{align}
\end{subequations}
while the tensor correction is unchanged. The transport parameters $\{\ce_i, \pi_i, \nu_i, \theta_i, \gamma_i\}$ may then be straightforwardly read off from equations~\eqref{ch4:kt:scalar_contribution},~\eqref{ch4:kt:vector_contribution}. It may be shown using the definition of $p$, $\epsilon$, and $n$ that $f_1 = f_3 = \ell_1 = 0$. We may set the frame $b_{1,2,3,4,5}$, $a_{1,2,3}$, tuning the transport parameters so as to ensure a stable and causal theory. In this manner, we have arrived at the BDNK theory of hydrodynamics from kinetic theory.

\subsection{The DNMR theory}
In the above discussion of extracting causal hydrodynamics from kinetic theory, we used a Hilbert expansion. There exists a philosophically distinct method of extracting equations of hydrodynamics from kinetic theory, which makes use of Grad's method of moments~\cite{Grad1949}. Originally written down in a more limited form by Israel and Stewart~\cite{Israel:1979wp}, the DNMR theory~\cite{Denicol:2012cn} extracts equations of motion for transient hydrodynamics, i.e. M\"uller-Israel-Stewart. For the sake of brevity, we here only outline the procedure. In the following, we restrict to $d=3$.

Let us consider deviations of the distribution function away from equilibrium, i.e. $f_p = f_p^{\rm eq.} + \delta f_{p} = f_{p}^{\rm eq.}\lr{1 +(1\pm \feq_p) \phi_p}$, which we take as the definition of $\phi_p$. Here, $\feq_p$ is the $x^\mu$-dependent ``equilibrium" distribution function~\eqref{ch4:kinetic:ooe_f0}; we will drop the superscript $(0)$ on $\beta^\mu$ and $\alpha$, so as to disambiguate with quantities we later define. We will now adopt the following shorthand:
\begin{equation}
    \braket{...} = \int_p \lr{...} f_p, \quad \braket{...}_0 = \int_p \lr{...} \feq_p, \quad \braket{...}_\delta = \braket{...}-\braket{...}_0\,.
\end{equation}
Note that this shorthand differs from that used at the end of the previous section. Note also that $x^\mu$-dependence is baked into the various $\braket{...}$ by way of the distribution function. Let us consider the tensor decomposition of $T^{\mu\nu}$ and $J^\mu$ in the Landau frame. Then (note that $p$ with no index is the isotropic pressure)
\begin{subequations}\label{ch4:dnmr:conrel-Landau}
    \begin{align}
        T^{\mu\nu}&= \epsilon u^\mu u^\nu + \lr{p + \Pi} \Delta^{\mu\nu} + \pi^{\mu\nu}\,,\\
        J^\mu &= n u^\mu + n^\mu\,,
    \end{align}
\end{subequations}
where $u^\mu = \beta^\mu/\sqrt{-\beta_\mu \beta^\mu}$. We can then make the connection
\begin{equation}
\label{ch2:dnmr:defs_vars}
    \begin{gathered}
        \epsilon = \braket{E_p^2}_0, \quad p = \frac{1}{3} \Delta^{\alpha\beta}\braket{p_{\braket{\alpha}} p_{\braket{\beta}}}_0, \quad n = \braket{E_p}_0\,,\\
        \Pi = \frac{1}{3} \Delta^{\alpha\beta}\braket{p_{\braket{\alpha}} p_{\braket{\beta}}}_\delta, \quad \pi^{\mu\nu} = \braket{p^{\langle \mu}p^{\nu\rangle}}_\delta, \quad n^\mu = \braket{p^{\braket \mu}}_\delta\,.
    \end{gathered}
\end{equation}
We would like expressions for the corrections $\Pi$, $\pi^{\mu\nu}$, and $n^\mu$. The method of moments provides these for us from the transient (i.e. non-hydrodynamic) modes in the system. In order to find the viscous corrections, we must first find $\phi_p$, which is a function of both $p^\mu$ and $x^\mu$. To begin with, let us expand $\phi_p$ in the irreducible representation of the little group of $SO(3,1)$ with respect to $u^\mu$, i.e.
\begin{equation}
    \phi_p(x) = \sum_{\ell=0}^\infty \lambda^{\braket{\mu_1\mu_2...\mu_\ell}}p_{\langle \mu_1} p_{\mu_2}...p_{\mu_\ell\rangle}\,,
\end{equation}
where $\lambda = \lambda(E_p,x)$. Next, let us expand the coefficients $\lambda$ in a basis set of orthogonal polynomials in $E_p$:
\begin{equation}\label{ch4:dnmr:truncatable}
    \lambda^{\braket{\mu_1\mu_2...\mu_\ell}} = \sum_{n=0}^\infty c_n^{\braket{\mu_1\mu_2...\mu_{\ell}}}(x) P_{n\ell}(E_p)\,,
\end{equation}
where the polynomials $P_{n\ell}$ are a set of orthogonal polynomials in $E_p$, the details of which may be found in Appendix E of~\cite{Denicol:2012cn}. Note that while the sum is in principle infinite, in practice it must be truncated at some ${\cal N}_{\ell}$. We will return to using this truncation. Note as well that, because of the orthogonality of the polynomials, the expansion coefficients $c_n^{\braket{\mu_1...\mu_{\ell}}}$ may be immediately obtained.

Rather than using the polynomials $P_{n\ell}$, it turns out to be more convenient to write coefficients in terms of irreducible moments of $\delta f_p = f_p - \feq_p$:
\begin{equation}
\label{ch4:dnmr:irrep_mom}
    \rho^{\mu_1 ... \mu_\ell}_n = \braket{E_p^n p^{\langle \mu_1} p^{\mu_2} ... p^{\mu_{\ell}\rangle}}_\delta\,.
\end{equation}
where we recall that $\ell$ is a non-negative integer; $\ell=0$ corresponds to the scalar case $\rho_n = \braket{E_p^n}_\delta$. 
Imposing Landau frame, as was done in equations~\eqref{ch4:dnmr:conrel-Landau}, would correspond to demanding that $\rho_1 = \rho_2 = \rho_1^{\mu} = 0$. Assuming this to be the case allows us to immediately make the identification (by equations~\eqref{ch2:dnmr:defs_vars} and the fact $p_{\braket{\mu}}p^{\braket{\mu}} = E_p^2 - m^2$)
\begin{equation}~\label{ch4:dnmr:fix-rho-0}
\rho_0 = - \frac{3}{m^2} \Pi, \quad \rho_0^\mu = n^\mu, \quad \rho_0^{\mu\nu} = \pi^{\mu\nu}\,. 
\end{equation}
After obtaining the coefficients $c_n^{\braket{\mu_1...\mu_{\ell}}}$ in terms of the irreducible moments~\eqref{ch4:dnmr:irrep_mom}, we can straightforwardly write the complete distribution function:
\begin{equation}
\label{ch4:dnmr:full_fp}
    f_p = \feq_p + \feq_p (1\pm \feq_p) \sum_{\ell=0}^\infty \sum_{n=0}^\infty {\cal H}_n^{(\ell)} \rho_n^{\mu_1 ... \mu_{\ell}} p_{\langle \mu_1} ... p_{\mu_{\ell}\rangle}\,,
\end{equation}
where ${\cal H}_{n}^{(\ell)}$ is a particular linear combination of the $P_{n\ell}$~\cite{Denicol:2012cn}. Now, let us define the irreducible projection of the comoving derivative of the moments as
$\dot{\rho}^{\braket{\mu_1...\mu_{\ell}}} = \Delta^{\mu_1...\mu_{\ell}}_{\nu_1...\nu_{\ell}} u^\lambda \de_\lambda \rho^{\nu_1 ... \nu_\ell}\,.$
In the following, we will use the shorthand $\dot{A} = u^\lambda \de_\lambda A$. We can then re-write the Boltzmann equation~\eqref{ch4:kinetic:Boltzmann} in the form (defining $\dperp_\mu = \Delta_\mu^{\,\,\,\,\nu}\de_\nu$)
\begin{equation}\label{ch4:dnmr:rewrite-Boltz}
    \dot{\delta f}_p = - \dot{\feq_p} - E_p^{-1} p^{\braket{\mu}} \dperp_\mu \feq_p - E_p^{-1} p^{\braket{\mu}}\dperp_\mu \delta f_p + E_p^{-1} C[f_p]\,, 
\end{equation}
which follows from inserting $f_p = \feq_p + \delta f_p$ and decomposing $p^\mu$ with respect to $u^\mu$. We can then take irreducible moments of the re-written Boltzmann equation~\eqref{ch4:dnmr:rewrite-Boltz}, in other words $\langle E_p^r$~\eqref{ch4:dnmr:rewrite-Boltz} $p_{\langle \mu_1} ... p_{\mu_\ell \rangle}\rangle$. The full equations are quite lengthy even only considering the first few moments; the full details may be found in~\cite{Denicol:2012cn}. Schematically however, we find
\begin{subequations}\label{ch4:dnmr:moment-eqs}
    \begin{align}
        \dot{\rho_r} - C_{r-1} &= \alpha_r^{(0)} \theta + ...\,,\\
        \dot{\rho_r}^{\braket{\mu}} - C_{r-1}^{\braket{\mu}} &=\alpha_r^{(1)} I^\mu + ...\,,\\
        \dot{\rho_r}^{\braket{\mu \nu}} - C_{r-1}^{\braket{\mu\nu}} &= \alpha_r^{(2)} \sigma^{\mu\nu} + ...\,.
    \end{align}
\end{subequations}
A significant amount of new notation has been introduced here. First of all, the $...$ contain the remaining complexities of the equations; in particular, they contain couplings between the viscous corrections $\Pi$, $\pi^{\mu\nu}$, and $n^\mu$, as well as with the other moments, including higher moments that do not appear in the equations above. The viscous corrections have been included via~\eqref{ch4:dnmr:fix-rho-0}. The terms appearing in equations~\eqref{ch4:dnmr:moment-eqs} are included because, with the power-counting scheme we will later introduce, they are in some sense ``small". The moment of the collision operator is defined by by
\begin{equation}
    C_r^{\braket{\mu_1...\mu_\ell}} = \int_p E_p^r p^{\langle \mu_1}...p^{\mu_\ell\rangle} C[f]\,.
\end{equation}
Note that $C_r^{\braket{\mu_1...\mu_\ell}}$ is a function of $x^\mu$ by way of $u^\mu$ and $f$.
Finally, the shorthand $\theta = \de_\mu u^\mu$, $I^\mu = D^\mu_\perp \alpha$, and $\sigma^{\mu\nu} = 2 \Delta^{\mu\nu\alpha\beta}\de_\alpha u_\beta$ have been used; the final is, of course, the shear tensor. Note that the full equations also contain dependence on the vorticity tensor $\omega_{\mu\nu} = \frac{1}{2} \lr{\dperp_\mu u_\nu - \dperp_\nu u_\mu}$. One could in principle include higher moments in the irreducible representation of the little group; however, for our purposes, this will not be necessary. Regardless, every single moment is either directly or indirectly coupled to every other moment. We must find some way to truncate this infinite set of coupled integro-differential equations. 

\subsubsection{Truncation scheme}
To truncate the system of equations, one must introduce a power-counting scheme. One obvious candidate for a small parameter is the Knudsen number Kn that we used in the Hilbert expansion. This will be useful for characterizing derivatives of the hydrodynamic variables, but it is not particularly helpful for handling the corrections $\Pi$, $n^\mu$, $\pi^{\mu\nu}$. With this in mind, the DNMR method uses the following additional parameters in the theory, which they call inverse Reynolds numbers:
\begin{equation}
    \Re^{-1}_{\Pi} \equiv \frac{|\Pi|}{p}, \quad \Re^{-1}_n \equiv \frac{|n^\mu|}{n}, \quad \Re^{-1}_\pi \equiv \frac{|\pi^{\mu\nu}|}{p}\,,
\end{equation}
where $|{\cdot}|$ denotes the ``size" of the variable in question, and $p$ (with no index) is the isotropic pressure. The inverse Reyonolds numbers involve ratios between quantities defined in terms of moments of $\delta f_p = f_p - \feq_p$, and equilibrium quantities. Therefore, when the system is near equilibrium, we expect these inverse Reynolds numbers to be small. We would like to write equations which describe the evolution of $\Pi$, $\pi^{\mu\nu}$, and $n^\mu$ which are accurate up to $\oser{\text{Kn}^2}$, $\oser{(\text{Re}_i^{-1})^2}$, and $\oser{\text{Kn Re}_i^{-1}}$. On a practical level, one may determine this by employing the scaling
\begin{equation}
    \Pi \to \eta_{\Pi}\, \Pi, \quad n^\mu \to \eta_n \, n^\mu, \quad \pi^{\mu\nu} \to \eta_{\pi}\, \pi^{\mu\nu}, \quad D_\mu^\perp \to \ce D_\mu^\perp\,,
\end{equation}
where $\eta_{\Pi}$, $\eta_n$, $\eta_\pi$, and $\ce$ are used as counting parameters for the various inverse Reynolds' numbers and the Knudsen number\footnote{Note that a different power-counting scheme is being used here compared to the previous section; before, we scaled $\de_\mu \to \ce \de_\mu$. Now, it is only the transverse derivatives which are scaled. Longitudinal derivatives $u^\mu \de_\mu$ of the hydrodynamic variables $\beta^\mu$, $\alpha$ can be consistently removed using the conservation equations.} respectively.

To begin with let us expand the moments of the collision operator in deviations from equilibrium $\delta f$:
\begin{equation}
    C_{r-1}^{\braket{\mu_1...\mu_\ell}} = - \sum_{n=0}^{\infty} {\cal A}_{rn}^{(\ell)} \rho_n^{\mu_1...\mu_\ell} + ...\,,
\end{equation}
where the $...$ are higher-order terms in $\delta f$. The term ${\cal A}_{rn}^{(\ell)}$ is the $(rn)$ coefficients of an, in principle, infinite matrix ${\cal A}^{(\ell)}$. In practice, one must truncate; therefore, we truncate~\eqref{ch4:dnmr:truncatable} to ${\cal N}_{\ell}$ as described below that equation. One can write the matrix ${\cal A}_{rn}^{(\ell)}$ in terms of integrals over the ${\cal H}_{n}^{(\ell)}$ of equation~\eqref{ch4:dnmr:full_fp}, and therefore truncating~\eqref{ch4:dnmr:truncatable} turns ${\cal A}_{rn}^{(\ell)}$ into the element of a finite $({\cal N}_{\ell}+1)\times({\cal N}_{\ell} + 1)$ matrix. Due to the use of Landau frame, ${\cal A}^{(0)}$ has rows and columns of zeroes in the second and third row/column $(n,r) = 1,2$, while ${\cal A}^{(1)}$ has a row/column of zeroes in the second row and column $(n,r)=1$. If we wish to invert ${\cal A}^{(\ell)}$, we must exclude those rows and columns.

Let us now assume that ${\cal A}^{(\ell)}$ may be diagonalized, i.e.
    $P^{-1}_{(\ell)} {\cal A}^{(\ell)} P_{(\ell)} = D_{(\ell)}\,,$
where $D_{(\ell)}$ is a diagonal matrix with the eigenvalues of ${\cal A}^{(\ell)}$; we denote the $n^{\rm th}$ eigenvalue by $\chi^{(\ell)}_n$. These eigenvalues are fixed by the microscopics by way of the collision operator; they are in general functions of $\beta^\mu$ and $\alpha$. We can then define
\begin{equation}\label{ch4:dnmr:diagonalization-rho}
    X_n^{\mu_1...\mu_{\ell}} =  (P_{(\ell)}^{-1})_{nm} \rho_m^{\mu_1...\mu_\ell}\,,
\end{equation}
where we recall that since $\rho_r^{\mu_1...\mu_\ell} = \rho_r^{\braket{\mu_1...\mu_\ell}}$, the same holds for $X_r^{\mu_1...\mu_\ell}$. Let us now apply a factor of $(P_{(\ell)}^{-1})_{ir}$ from the left to~\eqref{ch4:dnmr:moment-eqs}, with $\ell$ set as appropriate. This gives
\begin{subequations}~\label{ch4:dnmr:moment-relax}
    \begin{align}
        \dot{X}_i + \chi_i^{(0)} X_i &= \lr{P_{ir}^{(0)} \alpha_r^{(0)}} \theta + ...\,,\\
        \dot{X}_i^{\braket{\mu}} + \chi_i^{(1)} X_i^{\mu} &= \lr{P_{ir}^{(1)} \alpha_r^{(1)}} I^\mu + ...\,,\\
        \dot{X}_i^{\braket{\mu\nu}} + \chi_i^{(2)} X_i^{\mu\nu} &= \lr{P_{ir}^{(2)} \alpha_r^{(2)}} \sigma^{\mu\nu} + ...\,,
    \end{align}
\end{subequations}
where the $...$ are the same as those in equations~\eqref{ch4:dnmr:moment-eqs}, plus the non-linear contributions for $C_{r-1}^{\langle\mu_1...\mu_\ell\rangle}$. We can denote the terms in brackets above by $\beta_i^{(\ell)}$. Note as well that there is no implied summation over $i$ in the second term of each line. The moment equations~\eqref{ch4:dnmr:moment-relax} have now separated into individual relaxation equations (though there is still coupling in the non-linear regime, contained within the $...$). Let us impose an ordering on the eigenvalues such that $\chi_{n}^{(\ell)} < \chi_{n+1}^{(\ell)}$. We can also recall that for a relaxation equation of the form
\begin{equation}
    \dot{y}(x) + \chi y(x) = f(x) 
\end{equation}
the variable $y(x)$ relaxes towards $(1/\chi) f(x)$ with a characteristic time of $\tau=\chi^{-1}$. Therefore, the most significant modes at late times are those with the smallest value of $\chi$. Referring back to equations~\eqref{ch4:dnmr:moment-relax}, we see that the most relevant mode for each equation will be $\chi_0^{(\ell)}$. 

We therefore arrive at the central approximation of the DNMR method:
\begin{center}
    \textit{Let all $X_r^{\mu_1...\mu_\ell}$ with $r>0$ be approximated by their asymptotic values.}
\end{center}
In other words, for all $r>0$,
\begin{equation}\label{ch4:dnmr:main_approx}
    X_r \simeq \lr{\frac{\beta_r^{(0)}}{\chi_r^{(0)}}} \theta\,, \quad X_r^\mu \simeq \lr{\frac{\beta_r^{(1)}}{\chi_r^{(1)}}} I^\mu\,, \quad X_r^{\mu\nu} \simeq \lr{\frac{\beta_r^{(2)}}{\chi_r^{(2)}} }\sigma^{\mu\nu}\,.
\end{equation}
Finally, let us return back to the moments $\rho_r^{\mu_1...\mu_{\ell}}$, i.e. the quantities which actually appear in the distribution function. Inverting the relation~\eqref{ch4:dnmr:diagonalization-rho} and making use of the approximations~\eqref{ch4:dnmr:main_approx} yields
    \begin{alignat}{4}\label{ch4:dnmr:rho-X-eqs-1}
        &\,\, \rho_r &&\simeq \Omega_{r0}^{(0)} X_0 + \sum_{j=1}^{{\cal N}_0} \Omega_{rj}^{(0)} \lr{\frac{\beta_j^{(0)}}{\chi_j^{(0)}}} \theta\,,&&\rho_r^{\mu} &&\simeq \Omega_{r0}^{(1)}X_0^\mu + \sum_{j=1}^{{\cal N}_1} \Omega_{rj}^{(1)} \lr{\frac{\beta_j^{(1)}}{\chi_j^{(1)}}} I^\mu\,,\\
        &\rho_r^{\mu\nu} &&\simeq \Omega_{r0}^{(2)} X_0^{\mu\nu} + \sum_{j=1}^{{\cal N}_2} \Omega_{rj}^{(2)}\lr{\frac{\beta_j^{(2)}}{\chi_j^{(2)}}} \sigma^{\mu\nu}\,.\nonumber
    \end{alignat}
This may also be used to eliminate higher-order moments. Now, let us set $r=0$ above, and take $\Omega_{00}^{(\ell)} = 1$ without loss of generality. We know what $\rho_0$, $\rho_0^\mu$, and $\rho_0^{\mu\nu}$ are by equation~\eqref{ch4:dnmr:fix-rho-0}. The equations~\eqref{ch4:dnmr:rho-X-eqs-1} therefore become
\begin{subequations}
    \begin{alignat}{4}
        &-\frac{3}{m^2} \Pi &&\simeq X_0 + \sum_{j=1}^{{\cal N}_0} \Omega_{0j}^{(0)} \lr{\frac{\beta_j^{(0)}}{\chi_j^{(0)}}} \theta\,,&&n^\mu &&\simeq X_0^\mu + \sum_{j=1}^{{\cal N}_1} \Omega_{0j}^{(1)} \lr{\frac{\beta_j^{(1)}}{\chi_j^{(1)}}} I^\mu\,,\\
        &\qquad\,\pi^{\mu\nu} &&\simeq  X_0^{\mu\nu} + \sum_{j=1}^{{\cal N}_2} \Omega_{0j}^{(2)}\lr{\frac{\beta_j^{(2)}}{\chi_j^{(2)}}} \sigma^{\mu\nu}\,.
    \end{alignat}
\end{subequations}
Inverting these to solve for $X_0^{\mu_1...\mu_\ell}$, we therefore find that we can eliminate all $\rho_{r}^{\mu_1...\mu_{\ell}}$ for $r>0$ in favour solely of quantities that appear in the hydrodynamic constitutive relations. Let us define by $\tau^{(\ell)}$ the inverse of ${\cal A}^{(\ell)}$, with elements given by $\tau^{(\ell)}_{rn}$. Then we can write
\begin{subequations}\label{ch4:dnmr:eliminate-rho}
    \begin{alignat}{4}
        &\frac{m^2}{3} \rho_r &&\simeq - \Omega_{r0}^{(0)} \Pi - \lr{\zeta_r - \Omega_{r0}^{(0)} \zeta_0} \theta\,,&&\rho_r^\mu &&\simeq \Omega_{r0}^{(1)} n^\mu + \lr{\kappa_{r} - \Omega_{r0}^{(1)} \kappa_0} I^\mu\,,\\
        &\quad\,\rho_r^{\mu\nu} &&\simeq \Omega_{r0}^{(2)} \pi^{\mu\nu} +  \lr{\eta_r - \Omega_{r0}^{(2)} \eta_0} \sigma^{\mu\nu}\,,
    \end{alignat}
\end{subequations}
where the coefficients $\zeta_r$, $\kappa_r$, and $\eta_r$ are defined by
\begin{equation}
    \zeta_r \equiv \frac{m^2}{3} \sum_{n=0}^{{\cal N}_0} \tau_{rn}^{(0)}\alpha_{n}^{(0)}, \quad \kappa_r \equiv \sum_{n=0}^{{\cal N}_1} \tau_{rn}^{(1)} \alpha_n^{(1)}\,,\quad \eta_{r} = \sum_{n=0}^{{\cal N}_2} \tau_{rn}^{(2)} \alpha_{n}^{(2)}\,.
\end{equation} 
In the above, we have made the implicit assumption that the sum in $\zeta_r$ skips over $n=1,2$, and that the sum in $\kappa_r$ skips over $r=1$. Though we will not show it here, one can additionally show that $\rho^{\mu_1...\mu_\ell}_r$ for $\ell > 2$ are higher order in Kn and Re$^{-1}_i$ than the desired order for all $r$. 

We are now almost done. The final step is as follows:  Recall that before, we had $C_{r}^{\braket{\mu_1...\mu_{\ell}}} = - {\cal A}_{rn}^{(\ell)} \rho_{n}^{\mu_1...\mu_{\ell}} + ...$. Let us now invert ${\cal A}^{(\ell)}$; then
    $\tau_{mr}^{(\ell)} C_{r}^{\braket{\mu_1...\mu_{\ell}}} = - \rho_m^{\mu_1...\mu_\ell} + ...\,.$
We therefore apply $\tau^{(\ell)}$ to the left of the equations~\eqref{ch4:dnmr:moment-eqs}, and then use the relations~\eqref{ch4:dnmr:eliminate-rho} to find the relations~\cite{Denicol:2012cn}
\begin{subequations}
    \begin{align}
        \tau_{\Pi} \dot{\Pi} + \Pi &= - \zeta \theta + {\cal J} + {\cal K} + {\cal R}\,,\\
        \tau_{n} \dot{n}^\mu + n^\mu &= \kappa I^\mu + {\cal J}^\mu + {\cal K}^\mu + {\cal R}^\mu\,,\\
        \tau_{\pi} \dot{\pi}^{\mu\nu} + \pi^{\mu\nu} &=  \eta' \sigma^{\mu\nu} + {\cal J}^{\mu\nu} + {\cal K}^{\mu\nu} + {\cal R}^{\mu\nu}\,.
    \end{align}
\end{subequations}
In the above, the ${\cal J}$ terms denote terms of order $\oser{\text{Kn Re}^{-1}_i}$, the ${\cal K}$ denote terms of order $\oser{\text{Kn}^2}$, and the ${\cal R}$ denote terms of order $\oser{(\text{Re}_i^{-1})^2}$. The parameters $\zeta, \kappa, \eta'$ are simple $\zeta_0$, $\kappa_0$, and $\eta_0$. Finally, $\tau_{\Pi} = \sum_{r=0}^\infty \tau_{0r}^{(0)} \Omega_{r,0}^{(0)} = 1/\chi_{0}^{(0)}$, $\tau_{n} = \sum_{r=0}^\infty \tau_{0r}^{(1)} \Omega_{r,0}^{(1)}= 1/\chi_{0}^{(1)}$, and $\tau_{\pi} = \sum_{r=0}^\infty \tau_{0r}^{(2)} \Omega_{r,0}^{(2)}= 1/\chi_{0}^{(2)}$. We are now essentially done -- we have ``closed" the hydrodynamic equations via the method of moments, leading to equations that can in principle be solved. One may see from inspection that, as promised, the DNMR method leads to MIS-type equations. We can put the equations into a form more familiar from Chapter~\ref{chapter:background} via the substitution
\begin{equation}
    \kappa = - \sigma T, \qquad \eta' = - \eta\,,
\end{equation}
where $\sigma$ is the charge conductivity and $\eta$ is the shear viscosity, while $T^{-1} = \sqrt{-\beta^\mu \beta_\mu}$ is the inverse temperature associated with the ``equilibrium" solution $\feq_p(x)$.

The DNMR theory represents the fourth formulation of the MIS equations that we have discussed in this dissertation. Often in practice the $\oser{\text{Kn}^2}$ terms are omitted, as they cause problems for the solvability of the equations. One method proposed for handling these terms is the so-called ``Inverse Reynolds Dominance" (IReD) approach~\cite{Wagner:2022ayd}. In the IReD approach, terms in ${\cal K}$ end up absorbed into the transport coefficients inside of ${\cal J}$, leading to solvable equations.

\section{Holography}
\label{ch4:sec:holography}
The Anti-de Sitter/Conformal Field Theories correspondence (AdS/CFT correspondence), also called the gauge/gravity duality or holography, is one of the most exciting tools in modern high-energy physics. Originally conjectured by Maldacena~\cite{Maldacena:1997re} in 1997, the AdS/CFT correspondence subsequently exploded in popularity, making it now one of the most active areas of theoretical physics~\cite{Natsuume2015}. The simplest, one line description of the duality is that for some $D$-dimensional quantum field theories, the strong-coupling dynamics of the theory can be encoded in the (classical) dynamics of a $D+1$ dimensional gravitational theory. In particular, the prototypical example is that there is a duality between ${\cal N}=4$ Super-Yang-Mills theory in $D=4$ spacetime dimensions and type IIB string theory compactified of \AdS5$\times S^5$~\cite{Maldacena:1997re,Aharony:1999ti}; because type IIB string theory contains supergravity in the low-energy limit, certain observables can be well-described by classical gravity in the strong-coupling\footnote{The $\lambda \to \infty$ limit, where $\lambda$ is the t'Hooft coupling, after previously taking the large $N$ limit, where $N$ is the degree of the gauge group.} limit.

\begin{figure}[t]
    \centering
    \includegraphics[width=0.49\linewidth]{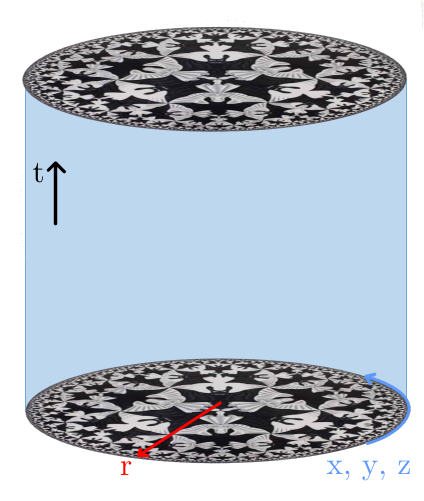}
    \caption{A schematic representation of anti-de Sitter spacetime. The spacetime has a boundary at $r=\infty$. The spacetime is hyperbolic; the tessellated image which serves to represent this type of space is ``Circle Limit IV (Heaven and Hell)" by M. C. Escher.}
    \label{fig:holo:escher}
\end{figure}

Anti-de Sitter (AdS) spacetimes have a fundamentally different nature to Minkowski, Schwarzschild, or other asymptotically flat spacetimes. $(D+1)$-dimensional AdS, often written as AdS$_{D+1}$ (e.g. AdS$_5$), has a boundary at infinity~\cite{Aharony:1999ti}. Due to the structure of AdS, it takes massive particles an infinite amount of time to reach this boundary; however, massless particles (e.g. photons) can reach this boundary in a finite amount of time and return to the interior~\cite{Aharony:1999ti}. It is then almost inevitable\footnote{It has been found that there exist ``islands of stability" of parameter space in which a perturbation of small enough amplitude will not collapse~\cite{Deppe:2015qsa,Cownden:2017bog,Cownden:2020oge}.} that, given an initial energy distribution, a horizon will eventually form~\cite{Bizon:2011gg}. Spacetimes of this form are called ``hyperbolic spacetimes"; for an illustration, refer to Figure~\ref{fig:holo:escher}.
The AdS boundary is crucial to the duality. Given the AdS theory is $D+1$-dimensional, the boundary is a $D$-dimensional surface. The dual quantum field theory is said to ``live" on this asymptotic boundary; therefore, in common parlance, the dual field theory is often referred to as the ``boundary theory", and the AdS interior is referred to as the ``bulk". We will adopt this terminology in the remainder of this chapter.

Holography provides a ``dictionary" relating quantities in the bulk and quantities on the boundary. Fields in the bulk correspond to gauge-invariant operators on the boundary. Geometries in the bulk correspond to states on the boundary. A black hole (or black brane, i.e. a solution where the horizon has translational symmetry) solution in the bulk corresponds to a thermal state on the boundary. As our interest in this dissertation is hydrodynamics, we will be particularly interested in this final entry in the dictionary.

\subsubsection{Foundations}
To begin with, let us consider the simplest possible case: Einstein gravity with a negative cosmological constant. There is a natural scale in AdS, appropriately referred to as the AdS scale $L_{\rm AdS}$; it arises because AdS$_{D+1}$ can be embedded as a negatively curved surface in a $D+2$ dimensional space the same way a 2-sphere can be embedded into three-dimensional space. We know that for a 2-sphere, the embedding is given by $x_1^2 + x_2^2 + x_3^2 = r^2$, where $r$ is the radius of the sphere. For AdS$_{D+1}$, the embedding is given by
\begin{equation}\label{ch4:holo:AdS-embedding}
    x_1^2 + x_2^2 + ... + x_D^2 - x_{D+1}^2 - x_{D-2}^2 = -L_{\rm AdS}^2\,.
\end{equation}
The AdS$_{D+1}$ metric is then given in Poincar\'e coordinates by~\cite{Natsuume2015}
\begin{equation}
    ds^2 = \lr{\frac{r}{L_{\rm AdS}}}^2 \lr{-dt^2 + \delta_{ij}dx^i dx^j} + L_{\rm AdS}^2\frac{dr^2}{r^2}\,,
\end{equation}
where $i,j \in (1,...,D)$. One can write the Einstein-Hilbert action as
\begin{equation}
    S_{\rm EH} = \frac{1}{16 \pi G_N} \int d^{D+1} x \sqrt{-g} \lr{ R^{(D+1)} - 2 \Lambda}\,,
\end{equation}
where $R^{(D+1)}$ is the Ricci scalar, $G_N$ is the $(D+1)$-dimensional Newton's constant, $g$ is the metric, and $\Lambda$ is the cosmological constant. The cosmological constant is negative, and is given by $\Lambda = - D(D-1)/(2 L_{\rm AdS}^2)$. Let us now scale the variables in the theory such that $L_{\rm AdS} = 1$; then
\begin{equation}\label{ch4:holo:EH-action}
    S_{\rm EH} = \frac{1}{16 \pi G_N} \int d^{D+1}x \sqrt{-g} \lr{R^{(D+1)} + D(D-1)}\,.
\end{equation}
Varying this action with respect to the metric gives rise to the standard vacuum Einstein equations:
\begin{equation}\label{ch4:holo:vac_Ein}
    R_{MN}^{(D+1)} - \frac{1}{2} R^{(D+1)} g_{MN} - \lr{\frac{D\lr{D-1}}{2}}g_{MN} = 0
\end{equation}
where $R_{MN}^{(D+1)}$ is the Ricci tensor, and capital latin letters $M,N$ run from $0$ to $D$. In the following, we will use capital latin letters for bulk spacetime indices, greek letters for boundary spacetime indices, and lowercase latin letters for boundary spatial indices. 

Let us now specialize to the case $D=4$, i.e. a 5-dimensional bulk spacetime. There are many solutions to the Einstein equations~\eqref{ch4:holo:vac_Ein}. One solution is the previously mentioned black brane solution, which is a \textit{planar} black hole with a horizon which is translationally invariant in the boundary coordinates $x^\mu$. In AdS-Schwarzschild coordinates, the black brane solution is given by~\cite{Berti:2009kk,Bhattacharyya:2008jc}
\begin{equation}\label{ch4:holo:black-brane-schwarz}
    ds^2 = - r^2 f(r) dt^2 + \frac{1}{r^2 f(r)} dr^2 +r^2 \lr{dx^2+dy^2+dz^2}
\end{equation}
where $t,x,y,z$ are boundary coordinates, and $r \in [0, \infty)$ is the bulk radial coordinate from the origin to the AdS boundary at $r=\infty$. The factor $f(r)$ is defined by
    $f(r) = 1 -\frac{r_0^4}{r^4}\,.$
 The parameter $r_0$ is the location of the black-brane horizon. The solution~\eqref{ch4:holo:black-brane-schwarz} has the somewhat inconvenient property shared by all Schwarzschild-like solutions, in that there is a coordinate singularity at $r=r_0$. It would be more convenient to write this solution in a coordinate system which is regular on the horizon. As such, we will transform the metric into infalling Eddington-Finkelstein (EF) coordinates~\cite{Carroll:2004st}.

EF coordinates are adapted to null trajectories. We can define a new coordinate $v = t + r^*$, where $r^*$ is the tortoise coordinate defined by $dr^*/dr = 1/(r^2 f(r))$, or
\begin{equation}
    r^*= \frac{1}{4 r_0} \lr{2\arctan\lr{\frac{r}{r_0}} + \ln\lr{\frac{r-r_0}{r+r_0}}}
\end{equation}
for $r\geq r_0$. As $r \to r_0$, $r^* \to -\infty$, as expected of a singular coordinate transformation. Inserting $t = v-r^*$ into the metric, we find
\begin{align}
    ds^2 &= - r^2 f(r) dv^2 + 2 dv dr + r^2 \lr{dx^2 + dy^2 + dz^2}\label{ch4:holo:EF-no-boost}
\end{align}
where we have used the fact that $dr^* = (dr^*/dr) dr = \lr{r^2 f(r)}^{-1} dr$. We therefore have arrived at the black brane solution in EF coordinates; these are the coordinates we will use going forward. The coordinate singularity at $r=r_0$ has now been removed.

One thing we can consider is how the solution~\eqref{ch4:holo:EF-no-boost} changes if one acts on it with the isometries of \AdS5. The isometries of \AdS5, i.e. the transformations that leave the line element of \AdS5 invariant, act on the boundary as the Poincar\'e transformations of $x^\mu$ (the ``boundary" coordinates, though the transformation acts in the bulk), scaling, and the special conformal transformation. In other words, the group of isometries is isomorphic to the group $SO(4,2)$ which leaves the embedded surface~\eqref{ch4:holo:AdS-embedding} invariant. The black-brane solution~\eqref{ch4:holo:EF-no-boost} is not invariant under boosts, and under dilations, and we use these to generate a family of solutions.

Let us consider the scaling $(r,v,x,y,z) \to  (\lambda  r, \lambda^{-1} v,\lambda^{-1} x,\lambda^{-1} y,\lambda^{-1} z)$; then $f(r) \to f(\lambda r) = 1 - \frac{r_0^4}{\lambda^4 r^4} \neq f(r)$. Dilations therefore shift the horizon location, taking $r_0 \to r_0/\lambda$. Similarly, if one takes $dx^\mu \to dx^{\mu'} = \Lambda^{\mu'}_{\,\,\,\,\nu}dx^{\nu}$, where $dx^\mu = \{dv,dx,dy,dz\}$ is raised and lowered with the Minkowski metric, we find that
    $dv \to -u_\mu dx^\mu$, and $\lr{dx^2 + dy^2 +dz^2} \to \Delta_{\mu\nu} dx^\mu dx^\nu$,
where $u^\mu$ is the future-directed timelike unit four-vector $(u^2=1)$ generated by the boosts, and $\Delta_{\mu\nu} = u_\mu u_\nu + \eta_{\mu\nu}$ is the projector orthogonal to $u_\mu$. Since $f(r) \neq 1$, this does not leave the metric~\eqref{ch4:holo:EF-no-boost} invariant. We may therefore write the full four-parameter family of solutions generated by dilations and boosts as
\begin{equation}\label{ch4:holo:boosted-black-brane-eq}
    ds^2 = -2 u_\mu dx^\mu dr - r^2 f(r) u_\mu u_\nu dx^\mu dx^\nu + r^2 \Delta_{\mu\nu} dx^\mu dx^\nu\,.
\end{equation}
We will hereafter work with the ``mass parameter" $b=1/r_0$ instead of $r_0$. The four-parameter $(b,u^\mu)$ family of solutions~\eqref{ch4:holo:boosted-black-brane-eq} is called the boosted black brane solution. 

One can repeat this process for the Einstein-Maxwell equations, including a $U(1)$ gauge field in the bulk (the gauge field associated with an ``R charge"\footnote{An R charge is a charge associated with a type of global symmetry which may be present in supersymmetric theories (such as ${\cal N} = 4$ Super-Yang-Mills). We know from the holographic dictionary that global symmetries on the boundary correspond to gauge fields in the bulk; therefore, we consider the gauge field associated with the boundary $R$ charge. In this particular case, we consider the case of a $U(1)$ subgroup of an $SO(6)$ $R$-symmetry~\cite{Erdmenger:2008rm}.}). This would correspond to a field theory on the boundary which has a conserved $U(1)$ charge current. The action is given (with $L_{\rm AdS}$ again scaled to one) by
\begin{equation}
    S_{\rm EHM} = \frac{1}{16 \pi G_N} \int d^5x \sqrt{-g} \lr{R^{(5)} +12 - \frac{1}{4} F_{MN}F^{MN}}
\end{equation}
where $F_{MN} = \de_M A_N - \de_N A_M$ is the five-dimensional field strength tensor for the $U(1)_{R}$ gauge field $A_M$. Varying this action with respect to the metric and the gauge field, we arrive at the Einstein-Maxwell equations:
\begin{subequations}\label{ch4:holo:EM-eq}
    \begin{align}
        R_{MN}^{(5)} - \frac{1}{2} R^{(5)} g_{MN} - 6 g_{MN} + \frac{1}{2} \biggl[F_{MK}F^{K}_{\,\,\,\,N} + \frac{1}{4} g_{MN} F_{KL}F^{KL} \biggr]&= 0\,,\\
        \nabla_M F^{MN} &= 0\,.
    \end{align}
\end{subequations}
In analogy with the previous case, one can find a boosted black brane solution to the equations~\eqref{ch4:holo:EM-eq}~\cite{Banerjee:2008th,Erdmenger:2008rm}:
\begin{subequations}\label{ch4:holo:charge_sol_full_eq}
    \begin{align}
        ds^2 &= -2 u_\mu dx^\mu dr - r^2 f(r) u_\mu u_\nu dx^\mu dx^\nu + r^2 \Delta_{\mu\nu} dx^\mu dx^\nu\,,\label{ch4:holo:bbb-charged}\\
        A_M dx^M &= \frac{\sqrt{3} Q}{2 r^2} u_\mu dx^\mu 
        \,,\label{ch4:holo:gaugefield-eq}
    \end{align}
\end{subequations}
where
$f(r) = 1 - \frac{1}{b^4 r^4}+ \frac{Q^2}{r^6}\,,$
once again $b$ is the mass parameter, and $Q$ is related to the black brane charge. The solution~\eqref{ch4:holo:bbb-charged} has two horizons; we will exclusively consider the outer horizon, which we denote by $R$. We have used the radial gauge, $A_r = 0$. 
The Hawking temperature of the black brane is given by the surface gravity $\kappa$ according to the canonical formula~\cite{Carroll:2004st}
\begin{equation}
    T_{\rm H} = \frac{\kappa}{2\pi}
\end{equation}
The surface gravity may be found by~\cite{Carroll:2004st}
\begin{equation}
    \kappa^2 = - \frac{1}{2} \nabla_M \xi_N \nabla^M \xi^N\,\biggl\vert_{r = R},
\end{equation}
where $\xi^M$ is a Killing vector, the norm of which becomes zero on the (Killing) horizon, and $\nabla_M$ is the covariant derivative. We note\footnote{More clearly,  $\xi^M = (0)\, \de_r + u^\mu \de_\mu$.} that $u^\mu\de_\mu$ satisfies this condition. The norm of $u^\mu\de_\mu$ is given by $- r^2 f(r)$, which vanishes when $r \to R$. For the charged boosted black brane~\eqref{ch4:holo:bbb-charged},\footnote{In the process of this computation we took advantage of auxiliary variables~\cite{Banerjee:2008th} $M = 1/(b R)^4$, $q = Q/R^3$, and $q^2 =M - 1$ to write the solution in terms of the outer radius $R$.}
\begin{equation}\label{ch4:holo:hawkingtemp_charged}
    T_{\rm H} = \frac{3 R^4 b^4 - 1}{2 \pi R^3 b^4}\,.
\end{equation}
In the uncharged limit, $R \to 1/b$, and~\eqref{ch4:holo:hawkingtemp_charged} collapses to $T_H = 1/(\pi b)$. Since $R=R(b,Q)$, the temperature depends on both the mass parameter and the charge, as one might expect. 

Given the presence of a charge in the dual theory, just as we asked about the temperature of the dual spacetime, we might equivalently ask about the chemical potential in the boosted black brane spacetime. The chemical potential may be given by the gauge-invariant difference of gauge fields~\cite{Erdmenger:2008rm}
\begin{equation}\label{ch4:holo:chemical_potential}
    \mu = u^\mu A_\mu(r=R) - u^\mu A_\mu (r=\infty) = \frac{\sqrt{3} Q}{2 R^2}\,.
\end{equation}
Just like the Hawking temperature, the chemical potential depends on both $Q$ and $b$ by way of $R=R(b,Q)$. These relations can be inverted to find expressions for $Q$ and $b$ in terms of $T_{\rm H}$ and $\mu$.

The last thing we would like to do is extract details of the relevant boundary operators from the gravitational solutions~\eqref{ch4:holo:charge_sol_full_eq}. Following the prescription of~\cite{Iqbal:2008by}, we will extract the one-point function of the dual operator to the metric (the stress-energy tensor, $\braket{T^{\mu\nu}}$) and to the gauge field (the $U(1)$ charge current, $\braket{J^\mu}$). Let us consider a generic field in the bulk $\phi(r,x^\mu)$; this could be the gauge field, the metric, or something else. The general prescription is that
\begin{equation}
    \braket{{\cal O}(x^\mu)}_{\phi_0} = \lim_{r \to \infty} (\Pi (r,x^\mu) - \text{Counter terms})
\end{equation}
where $\braket{{\cal O}(x^\mu)}_{\phi_0}$ is the expectation value of the operator ${\cal O}$ in the presence of a classical source $\phi_0 = \lim_{r \to \infty} \phi(r,x^\mu)$, and $\Pi(r,x^\mu)$ is the conjugate momentum to $\phi$ with respect to $r$-foliation. For the gauge field $A_\mu$, no counterterms are required~\cite{Iqbal:2008by}, and the expression is
\begin{equation}
    \braket{J^\mu}_{A_\mu^{\rm bdry}} = \lim_{r \to \infty} j^\mu = \frac{1}{16 \pi G_N}\lim_{r \to \infty} \sqrt{-g} F^{r\mu}\,,
\end{equation}
where $j^\mu$ is the conjugate momentum to the gauge field. For the metric, things are somewhat more complicated~\cite{Balasubramanian:1999re}. After taking counter-terms into account and re-writing, the stress-energy tensor can be given by~\cite{Bhattacharyya:2008jc}
\begin{equation}
    \braket{T^{\mu\nu}}_{g} = -\frac{2\eta^{\mu\rho}}{16 \pi G_N} \lim_{r \to \infty} r^4\lr{K^\nu_{\,\,\,\,\rho} - \delta^\nu_{\,\,\,\,\rho}}\,.
\end{equation}
where $K^\nu_{\,\,\,\,\rho}$ is the extrinsic curvature of a surface of constant $r$. The solutions~\eqref{ch4:holo:charge_sol_full_eq} therefore yield (dropping the $\braket{\cdot}$)
\begin{equation}
    J^\mu = \lr{\frac{\sqrt{3} Q}{16 \pi G_N}} u^\mu\,,\quad T^{\mu\nu} = \lr{\frac{3}{16 \pi b^4 G_N}}u^\mu u^\nu + \lr{\frac{1}{16 \pi b^4 G_N}} \Delta^{\mu\nu}\,.
\end{equation}
Matching with the equilibrium expressions for $T^{\mu\nu}$ and $J^\mu$~\eqref{ch2:hydro:eq-const-first}, we find that
\begin{equation}
        \epsilon = \frac{3}{16 \pi \,b(T,\mu)^4 \,G_N}\,,\quad p = \frac{1}{16 \pi \,b(T,\mu)^4 \,G_N}\,,\quad n = \frac{\sqrt{3} Q(T,\mu)}{16 \pi G_N}\,.
\end{equation}

Note that $\epsilon$ and $p$ obey the conformal relation $\epsilon = 3 p$; in other words, the stress-energy tensor is traceless. With the equilibrium properties of the boosted black brane and the dual fluid resolved, we now consider deviations away from equilibrium, i.e. hydrodynamics.

\subsection{The fluid/gravity correspondence}
There have long been hints that there was a connection between black holes and hydrodynamics, beginning with attempts in the 1970s to move beyond the black hole thermodynamics of Bekenstein~\cite{Bekenstein:1973ur} and Hawking~\cite{Hawking:1975vcx} using the so-called ``membrane paradigm"~\cite{Thorne:1986iy}. To an external observer, a black hole appears to behave like a dynamical fluid membrane; however, asymptotically flat black holes do not behave like normal fluids. For one, Schwarzschild black holes have a negative heat capacity~\cite{Witten:2024upt}, growing colder as they gain energy. This stymied the usefulness of the paradigm for a while.

The membrane paradigm gained a second life in the context of the gauge/gravity duality, and anti-de Sitter spacetimes. As discussed in the previous section, one can construct a fluid living on the \textit{boundary} of AdS; some of the earlier calculations done with the AdS/CFT correspondence were the calculation of the shear viscosity and the R-charge diffusion constant of a ${\cal N}=4$ Super-Yang-Mills plasma~\cite{Policastro:2001yc,Policastro:2002se}. This led to the conjecture~\cite{Kovtun:2004de} of a universal lower bound on shear viscosity\footnote{Note that $\hbar=1$.}, $\eta/s \geq 1/(4 \pi)$.

It was also discovered that the long-wavelength limit of quasinormal modes for black branes displayed hydrodynamic behaviour~\cite{Kovtun:2003wp,Kovtun:2005ev,Iqbal:2008by}. All of these things together led eventually to a methodology for determining hydrodynamic transport coefficients \textit{non-linearly} for fluids describing microscopic theories with holographic duals. This method, the fluid/gravity correspondence~\cite{Bhattacharyya:2008jc,Banerjee:2008th,Erdmenger:2008rm}, is what we shall discuss from this point on.

Let us begin by considering the solutions~\eqref{ch4:holo:charge_sol_full_eq} with the parameters $Q$, $b$, and $u^\mu$ now promoted to dynamical functions of the boundary spacetime, i.e.
\begin{subequations}\label{ch4:holo:charged_zeroth_order}
    \begin{align}
        ds^2 &= - 2 u_\mu(x) dx^\mu dr - r^2 f(r) u_\mu(x) u_\nu(x) dx^\mu dx^\nu + r^2 \Delta_{\mu\nu}(x) dx^\mu dx^\nu\,,\\
        A_M dx^M &= \frac{\sqrt{3} Q(x)}{2 r^2} u_\mu(x) dx^\mu\,,
    \end{align}
\end{subequations}
where
$    f(r) = 1 - \frac{1}{b(x)^4 r^4} + \frac{Q(x)^2}{r^6}\,,$
and we have made a gauge choice to set $A_\mu^{\rm bdry}$ to zero, a choice we will consistently make throughout the remainder of this chapter. The expressions~\eqref{ch4:holo:charged_zeroth_order} are no longer solutions to the equations of motion~\eqref{ch4:holo:EM-eq}. If we assume, however, that $b(x)$, $Q(x)$, and $u^\mu(x)$ are very slowly varying functions of the boundary spacetime $x^\mu$, then we may ``tubewise" approximate a boosted black brane solution~\cite{Bhattacharyya:2008jc}. In practice, we attempt to solve the Einstein-Maxwell equations via a derivative expansion in derivatives with respect to boundary coordinates $\de_\mu$ for $g_{MN}$ and $A_M$.

Let us define the following order-by-order expansion in derivatives
\begin{subequations}\label{ch4:holo:order-by-order}
    \begin{align}
        g_{MN} &= g_{MN}^{(0)} + \ce g_{MN}^{(1)} + \ce^2 g_{MN}^{(2)} + \oser{\ce^3}\,,\\
        A_M &= A_M^{(0)} + \ce A_M^{(1)} + \ce^2 A_M^{(2)} + \oser{\ce^3}\,,
    \end{align}
\end{subequations}
where $\ce$ is once again a derivative-counting parameter ($\de_\mu \to \ce \de_\mu$) which can be taken to unity at the end of the procedure, as in the previous section. The zeroth-order pieces $g_{MN}^{(0)}$ and $A_M^{(0)}$ are simply given by equations~\eqref{ch4:holo:charged_zeroth_order}. In the following, we will only consider solving to first order in the expansion; the interested reader may refer to~\cite{Bhattacharyya:2008jc,Banerjee:2008th,Erdmenger:2008rm} for a derivation at second order. Taking the expansions~\eqref{ch4:holo:order-by-order} and inserting them into the Einstein-Maxwell equations~\eqref{ch4:holo:EM-eq} yields equations order-by-order in $\ce$. The equations, like the order-by-order equations in kinetic theory, have the same form at each order. At $n^{\rm th}$ order, the equations are of the form
\begin{equation}\label{ch4:holo:H-OBO}
    {\cal H}[g^{(n)}, A^{(n)}] = S_n[A^{(n-1)}, g^{(n-1)}, A^{(n-2)}, g^{(n-2)},..., A^{(0)}, g^{(0)}]\,.
\end{equation}
The operator ${\cal H}$ is a derivative operator solely\footnote{Since boundary derivatives $\de_\mu$ come with a factor of $\ce$, any boundary derivatives necessarily act on a lower-order term in the expansion, contributing only to the source term $S_n$.} in $r$ which repeats order-by-order just as the linearized collision operator ${\cal L}$ did in kinetic theory. The right-hand side of equation~\eqref{ch4:holo:H-OBO} $S_n$ is the source term, which depends only on lower-order solutions in $\ce$ which have (in principle) already been obtained. To solve for $g^{(n)}_{MN}$, $A_{M}^{(n)}$, all one need do in principle is invert ${\cal H}$ at each order.

As the astute reader may have been able to foresee, this is na\"ive. Just as in the case of kinetic theory, the operator ${\cal H}$ has zero modes which make it formally non-invertible unless the source term is constrained to be orthogonal to the zero modes. Demanding the invertibility of ${\cal H}$ leads to imposing
\begin{equation}
    \de_\mu T^{\mu\nu}_{(n-1)} =0, \quad \de_\mu J^\mu_{(n-1)} = 0\,,
\end{equation}
where $T^{\mu\nu}_{(n-1)}$ is the $(n-1)^{\rm th}$ order contribution to $T^{\mu\nu}$, and $J^\mu_{(n-1)}$ is the $(n-1)^{\rm th}$ order contribution to $J^\mu$. Note that the contributions to $T^{\mu\nu}$, $J^\mu$ are individually conserved order-by-order. We also note that the parameters of the equilibrium solution $b(x)$, $Q(x)$, and $u_\mu(x)$ are corrected order-by-order as well:
\begin{subequations}
    \begin{align}
        b(x) &= b_{(0)}(x) + \ce b_{(1)}(x) + \ce^2 b_{(2)}(x) + \oser{\ce^3}\,,\\
        Q(x) &= Q_{(0)}(x) + \ce Q_{(1)}(x) + \ce^2 Q_{(2)}(x) + \oser{\ce^3}\,,\\
        u^\mu(x) &= u^\mu_{(0)}(x) + \ce u^\mu_{(1)}(x) + \ce^2 u^\mu_{(2)}(x) + \oser{\ce^3}\,.
    \end{align}
\end{subequations}
Let us now consider the expansion taken to first order. Because of the zero-modes of ${\cal H}$, the solution is comprised of of the particular solution found by inverting ${\cal H}$, and the homogeneous solution which is proportional to the zero modes, i.e.
\begin{subequations}\label{ch4:holo:first-order-sol}
    \begin{align}
        g_{MN} &= g_{MN}^{(0)} + \ce \lr{g_{MN}^{(1),p} + g_{MN}^{(1),h}} + \oser{\ce^2}\,,\\
        A_M &= A_M^{(0)} + \ce \lr{A_M^{(1),p} + A_M^{(1),h}} + \oser{\ce^2}\,.
    \end{align}
\end{subequations}
We will not focus on the particular solution here; instead, we shall consider only the homogeneous solution. We can write more specifically that
\begin{subequations}\label{ch4:holo:homogeneous-part}
    \begin{align}
        g_{MN}^{(1),h}dx^M dx^N &= b_{(1)} \lr{-\frac{4}{b_{(0)}^5 r^2} u_\mu^{(0)} u_\nu^{(0)}} dx^\mu dx^\nu +Q_{(1)} \lr{- \frac{2 Q_{(0)}}{r^4} u_\mu^{(0)} u_\nu^{(0)}} dx^\mu dx^\nu \nnl
        &+ u^\lambda_{(1)} \lr{r^2 (1-f^{(0)}(r)) \lr{\Delta_{\lambda\mu}^{(0)}u_\nu^{(0)} + \Delta_{\lambda\nu}^{(0)} u_{\mu}^{(0)}}} dx^\mu dx^\nu\,,\\
        A_M^{(1),h} dx^M &= Q_{(1)} \lr{\frac{\sqrt{3}}{2 r^2} u_\mu^{(0)}} dx^\mu + u_{(1)}^\lambda \lr{\frac{\sqrt{3} Q_{(0)}}{2 r^2} \Delta_{\lambda\mu}}dx^\mu\,,
    \end{align}
\end{subequations}
where $f^{(0)}(r) = 1 - \frac{1}{b_{(0)}^4 r^4} + \frac{Q_{(0)}^2}{r^6}$, and $\Delta_{\mu\nu}^{(0)} = \eta_{\mu\nu} + u_\mu^{(0)} u_\nu^{(0)}$. That the coefficients $b_{(1)}, Q_{(1)}, u_{(1)}^\mu$ in~\eqref{ch4:holo:homogeneous-part} are in fact the first-order corrections to $b$, $Q$, and $u^\mu$ may be straightforwardly seen by inserting the order-by-order corrections~\eqref{ch4:holo:order-by-order} into the ideal-order solution~\eqref{ch4:holo:charged_zeroth_order} and expanding to first order in $\ce$. 

Note that, in order to write down the solution~\eqref{ch4:holo:first-order-sol} in the first place, one had to invert ${\cal H}$, which implies that $b_{(0)}$, $Q_{(0)}$, and $u^\mu_{(0)}$ are all constrained to obey the ideal-order conservation equations~\eqref{ch2:hydro:euler_eqs}. Since the hydrodynamic parameters $\beta^\mu = u^\mu/T$ and $\alpha = \mu/T$ are defined in terms of $b$, $Q$, and $u^\mu$, the hydrodynamic parameters also receive order-by-order corrections. At first order, the corrections are given by
\begin{subequations}
    \begin{align}
        \alpha_{(1)} &= \pder{\alpha_{(0)}}{b_{(0)}} b_{(1)} + \pder{\alpha_{(0)}}{Q_{(0)}} Q_{(1)}\,,\\
        \beta^\mu_{(1)} &= - \frac{1}{T_{(0)}^2} \lr{\pder{T_{(0)}}{b_{(0)}} b_{(1)} + \pder{T_{(0)}}{Q_{(0)}} Q_{(1)}} u^\mu_{(0)} + \frac{1}{T_{(0)}} \Delta^{\mu\lambda}_{(0)} u_{\lambda}^{(1)}\,.
    \end{align}
\end{subequations}
Now, finally, let us repeat the same trick as in kinetic theory. We define new degrees of freedom $b'$, $Q'$, and $u'^\mu$ implicitly via the relations
\begin{subequations}\label{ch4:holo:replacement}
    \begin{align}
        b' -  4 \pi G_N b'^5\ce\,\biggl[\pi_1 \lr{\frac{u'^\lambda \de_\lambda T'}{T'} - \frac{1}{3} \de_\mu u'^\mu} + \pi_3 u^\mu \de_\mu \lr{\frac{\mu'}{T'}}\biggr] - \ce b_{(1)} + \oser{\ce^2} &= b_{(0)}\,,\\
        Q' + \frac{16\pi G_N}{\sqrt{3}}\ce\, \biggl[\nu_1 \lr{\frac{u'^\lambda \de_\lambda T'}{T'} - \frac{1}{3} \de_\mu u'^\mu} + \nu_3 u^\mu \de_\mu \lr{\frac{\mu'}{T'}} \biggr] - \ce Q_{(1)}+ \oser{\ce^2} &= Q_{(0)}\,,\\
        u'^\mu + \ce\, \biggl[a_1\, \Delta'^{\mu\nu} \lr{\frac{\de_\nu T'}{T'} + u^\lambda \de_\lambda u'_\nu} + a_2 \Delta'^{\mu\nu}\de_\nu \lr{\frac{\mu'}{T'}} \biggr] - \ce u^\mu_{(1)}+ \oser{\ce^2} &= u^\mu_{(0)}\,,
    \end{align}
\end{subequations}
where $T', \mu'$ above are taken to be functions of $b',Q'$. We then replace $b_{(0)}$, $Q_{(0)}$, and $u^\mu_{(0)}$ everywhere they appear. In all first-order contributions, $\{b_{(0)}, Q_{(0)}, u^\mu_{(0)}\} \to \{b', Q', u^\mu\}$ in a straightforward fashion, as the further corrections are all $\oser{\ce^2}$. However, in the zeroth-order part of the solution, using equations~\eqref{ch4:holo:replacement} leads to a $\oser{\ce}$ correction which fixes the homogeneous solution. In this manner, analogously to in kinetic theory, one may set a choice of hydrodynamic frame. In particular, one can choose either a ``good frame", where causality and stability are present after truncating the expansion at $\oser{\ce}$, or a ``bad frame" (such as Landau frame) where the truncated hydrodynamic equations are unstable and acausal. Regardless of frame choice, the calculation proceeds completely analogously to the kinetic theory case, with one major difference: the number of parameters. 

We see that $\pi_1,\pi_3,\nu_1,\nu_3$ appear directly in~\eqref{ch4:holo:replacement}, as opposed to the parameters $a_{1,2,3}$, $b_{1,2,3}$ which appeared in the analogous kinetic theory procedure. The reason for this is the conformal symmetry of the boundary theory. The reduction in free parameters in the kinetic theory case was due to $f_{1,2,3}$ having been fixed. However, in a conformal theory, $f_{1,2,3}$ are generically fixed to be zero in any fluid frame, and so no reduction in the number of free parameters occurs. Note, however, that $\ell_{1,2}$ are not fixed by conformal symmetry, and so there a reduction of degrees of freedom still occurs in the vector sector; $\theta_1$ and $\gamma_1$ both depend on $a_1$, while $\theta_2$ and $\gamma_2$ both depend on $a_2$.

One final point to make in this chapter, which applies to both kinetic theory and holography, is that the acausal nature of the macroscopic theory we extract from the microscopics in Landau frame is in no way a symptom of anything ``wrong" with the microscopic theory. It is the \textit{truncation} of the derivative expansion at finite order that leads to the acausality. The only way to prevent this acausality, as far as we know, is to either break Lorentz symmetry~\cite{Armas:2020mpr,Bhambure:2024axa,Gavassino:2025hwz}, or introduce some kind of regulation. In BDNK, this regulation takes the form of the additional transport parameters. In MIS-type theories, these are the additional relaxational modes in the theory.

	\startchapter{Extensions of Causal Hydrodynamics}
\label{chapter:extensions}
In this chapter, we will discuss a number of extensions of the causal formulations of hydrodynamics that have thus-far been discussed in the dissertation. The number of possible extensions are vast; as such, we will limit to only those extensions on which I have personally worked. The section on one-form magnetohydrodynamics is based on my paper~\cite{Hoult:2024qph}, while the section on relativistic superfluids is based on my paper~\cite{Hoult:2024cyx}. Both sections primarily focus on the BDNK procedure for rendering hydrodynamics causal; however, the theories under consideration are amenable to an MIS-type procedure as well (see e.g.~\cite{Biswas:2020rps,Gavassino:2021crz} for MIS-type constructions for related theories).
\section{Magnetohydrodynamics}
\label{ch5:sec:mhd}
Magnetohydrodynamics is the effective theory describing the interactions between fluids and electromagnetic fields. The subject of interest is usually a plasma which is electrically neutral on hydrodynamic length scales -- electric charges in the plasma adjust on a microscopic level to screen the electric field. The magnetic field is not screened, and so remains a relevant degree of freedom even in equilibrium.

The standard formulation of relativistic MHD is textbook material at ideal order~\cite{Anile}; plasmas in the standard formulation are described by the conservation of the stress-energy tensor, as well as those of Maxwell's equations in matter that control the magnetic field. At higher order in a derivative expansion, the dynamics of the electric field become relevant, as do viscous effects -- see~\cite{Hernandez:2017mch}. In astrophysical applications, there is also often an additional global $U(1)$ charge on top of the gauge $U(1)$ for electromagnetism. This global $U(1)$ symmetry corresponds to the conservation of particle number. The constituents are often massive neutral particles, and so the current associated with this global $U(1)$ is often called the ``mass current" $J_m^\mu = m J^\mu$ (where $J^\mu$ is the actual conserved current for the global $U(1)$ charge). We will denote this global $U(1)$ symmetry by $U(1)_m$.

Let us neglect the dynamics of the electric field entirely. Then the relevant equations of motion are given by~\cite{Hoult:2024qph}
\begin{equation}
        \nabla_\mu T^{\mu\nu} = 0\,,\quad \nabla_\mu J^{\mu\nu} =0\,,\label{ch5:mhd:Jmunu-cons}\quad \nabla_\mu J^\mu = 0\,,    
\end{equation}
where $J^{\mu\nu} = (1/2) \epsilon^{\mu\nu\rho\sigma} F_{\rho\sigma}$ is proportional to the dual field-strength tensor. The hydrodynamic degrees of freedom are the temperature $T$, the fluid velocity $u^\mu$, the $U(1)_m$ chemical potential $\mmu$, and the magnetic field $B^\mu$. 

We could at this point write constitutive relations for the conserved quantities; however, we will instead take a short detour to talk about symmetry. The conservation equations~\eqref{ch5:mhd:Jmunu-cons} are not all on the same footing. The conservation of the stress-energy tensor is a manifestation of diffeomorphism invariance, while the conservation of the $U(1)_m$ charge current is a manifestation of the global $U(1)_m$ symmetry. The conservation of the dual field-strength tensor, on the other hand, arises from the $U(1)$ gauge symmetry -- which is not, in truth, a symmetry at all, but rather simply a redundancy in our description of the theory~\cite{Elitzur1975,TongGT}. There \textit{is} in fact a global symmetry which is responsible for the conservation of the dual field-strength tensor; however, it is not a traditional symmetry in the sense that we are used to thinking of them. 

The symmetry responsible for the conservation of the dual field-strength tensor is a global one-form symmetry, an example of what is known as a ``higher-form symmetry"~\cite{Schafer-Nameki:2023jdn}. A discussion of all the details behind higher-form symmetries is beyond the scope of this dissertation. However, a brief summary is included hereafter.

\subsection{One-form MHD}
\subsubsection{Higher-form symmetries}
Let us begin by discussing a normal global $U(1)$ symmetry. Under this symmetry, a local charged field $\phi(x)$ transforms by a phase
$    \phi \to \phi' = e^{i q \alpha} \phi\,.$
By Noether's theorem, whenever there is a continuous global symmetry, there exists a conserved current ${\cal J}^\mu$. In $d+1$-dimensional Minkowski spacetime, the charge associated with the symmetry is given by integrating the time component of the current over $d$-dimensional space
$    Q = \int d^dx {\cal J}^0\,.$
This can be generalized and re-written in a more suggestive way. First of all, let us instead consider the integral over a $d$-dimensional spatial manifold ${\cal M}_{d}$ with a time-directed volume element:
\begin{equation}
    Q = \int_{{\cal M}_d} d\Sigma_\mu {\cal J}^\mu\,.
\end{equation}
We take ${\cal M}_d$ to be such that it divides the spacetime into two disconnected parts, and has itself no boundary. Next, let us note (or recall) that the general volume element $\sqrt{-g}d^{d+1}x$ is, in fact, a $(d+1)$-form
$    \sqrt{-g}d^{d+1}x = \sqrt{-g}dx^0 \wedge dx^1 \wedge ... \wedge dx^{d}.$
We are considering the directed volume element for a spacelike $d$-dimensional hypersurface. Let us denote the internal coordinates of the hypersurface by $y_1,...,y_d$; then the volume element for the hypersurface is given by
$    \sqrt{\gamma} d^dy = \sqrt{\gamma} dy^1 \wedge ...\wedge dy^d =  \frac{\sqrt{\gamma}}{d!} n^\mu \epsilon_{\mu\nu_1\nu_2...\nu_d} dx^{\nu_1}\wedge...\wedge dx^{\nu_d}\,,$
where $n^\mu$ is the timelike future-directed unit normal to the spacelike hypersurface ${\cal M}_d$, $\gamma$ is the determinant of the induced metric on ${\cal M}_d$, and $\epsilon$ is the Levi-Civita symbol (as opposed to tensor). We can see then that $d\Sigma_\mu$ may be written in terms of $dx^\mu$ by writing
\begin{equation}
    d\Sigma_\mu =  -n_\mu \frac{\sqrt{\gamma}}{d!}\,n^\lambda\epsilon_{\lambda\nu_1\nu_2...\nu_d}dx^{\nu_1}\wedge dx^{\nu_2}\wedge...\wedge dx^{\nu_d}\,,
\end{equation}
and the charge is given by
\begin{equation}\label{ch5:mhd:charge-no-hodge}
    Q =  -\int_{{\cal M}_d} d^dy \sqrt{\gamma}\, n_\mu {\cal J}^\mu
\end{equation}
Now, let us define ${\bf J} = {\cal J}_\alpha dx^\alpha$. The Hodge dual $\star {\bf J}$ is defined on the hypersurface ${\cal M}_d$ by
\begin{equation}
    \star {\bf J} = \frac{\sqrt{\gamma}}{d!} {\cal J}_\lambda g^{\mu\lambda} \epsilon_{\mu\nu_1\nu_2...\nu_d} dx^{\nu_1} \wedge dx^{\nu_2}\wedge  ... \wedge dx^{\nu_d}\,.
\end{equation}
We may now define the projector orthogonal to $n^\mu$, $P^{\mu\nu} = g^{\mu\nu} + n^\mu n^\nu$. Decomposing ${\cal J}_\mu$ with respect to $n_\mu$ 
$    ({\cal J}_{\mu} = - n_\mu n^\lambda {\cal J}_\lambda + P_{\mu}^{\,\,\,\,\lambda}{\cal J}_\lambda)$
allows us to write $\star {\bf J}$ as
$    \star{\bf J} = \frac{\sqrt{\gamma}}{d!} \lr{- n^\lambda {\cal J}_{\lambda} n^\mu + P^{\mu\lambda}{\cal J}_\lambda} \epsilon_{\mu\nu_1...\nu_d} dx^{\nu_1} \wedge ... \wedge dx^{\nu_d}\,.$
The integral over the orthogonal projection will vanish due to the Levi-Civita symbol, meaning we can finally write~\eqref{ch5:mhd:charge-no-hodge} as
\begin{equation}
    Q = \int_{{\cal M}_d} \star {\bf J}\,.
\end{equation}
This is the standard definition of the charge in terms of differential forms\footnote{Note the differing sign in~\cite{Carroll:2004st}.}~\cite{Carroll:2004st,Schafer-Nameki:2023jdn,Shao:2023gho}. Given the spacetime was $d+1$ dimensional, the $d$-dimensional surface ${\cal M}_d$ is said to be a codimension one surface, ``codimension" referring to the number of dimensions lower than the total dimension of the spacetime. If the codimension-one surface is a time slice, we can think of the surface as counting the amount of charge associated with the particle world-lines that intersect the time slice.

We now generalize the above. Suppose that instead of finding the charge associated with a local operator, we want to find the charge associated with an extended operator such as a line operator (e.g. Wilson line) or a surface operator. The procedure above may be generalized to find the charge associated with these extended operators. The symmetry from which the charge arises is called a ``higher form symmetry". For a zero-form symmetry, we integrated over a codimension-one surface. In general, for a $p$-form symmetry, we will need to integrate over a codimension-$(p+1)$ surface. 

We will focus now on so-called ``one-form" symmetries. These are the symmetries associated with line operators, and we can obtain the charge by integrating over a codimension-two surface. One can think of the codimension-two surface as counting the amount of charge associated with the worldsheets of the line operators that intersect the surface. In analogy with the above, the charge is given by
\begin{equation}
    Q_1 = \int_{{\cal M}_2} \star\Jform_1
\end{equation}
where $\Jform_1 = J_{\mu\nu} dx^{\mu} \wedge dx^\nu$. The current associated with the global $U(1)$ one-form symmetry, $J^{\mu\nu}$, is antisymmetric and conserved, $\nabla_\mu J^{\mu\nu} = 0$. 

\subsubsection{Ideal-Order Magnetohydrodynamics}

In the context of MHD, the quantity being conserved is the number of magnetic field lines. The formulation of MHD in terms of this one-form symmetry was initiated by~\cite{Schubring:2014iwa,Grozdanov:2016tdf}, and has been shown~\cite{Hernandez:2017mch,Armas:2018atq,Armas:2018zbe} to be entirely equivalent to the standard formulation of MHD. However, the one-form formulation of MHD has the benefit of being based entirely on global symmetries. In the following, we will frequently use the terms ``one-form MHD" and ``dMHD" interchangeably. The name ``dMHD" was introduced in~\cite{Hoult:2024qph} as a shorthand for ``dual MHD", as the one-form formulation represents a dual formulation of magnetohydrodynamics.

We would like to now determine the equilibrium constitutive relations for one-form MHD with an additional global zero-form $U(1)_m$ symmetry. We will do so using the equilibrium generating functional~\cite{Armas:2018atq}. First of all, we will introduce background source fields to couple to the conserved quantities. For the stress tensor, we bring in a background metric $g_{\mu\nu}$ as before. For the current $J^\mu$, we will introduce a background $U(1)_m$ gauge field $A_\mu$. Finally, for the two-form current $J^{\mu\nu}$, we introduce a background two-form gauge field $b_{\mu\nu}$. Under the combination of diffeomorphisms, zero-form gauge transformations, and one-form gauge transformations generated by arbitrary parameters $G = \{\xi_\mu, \lambda^{(1)}_\mu, \lambda^{(0)}\}$ respectively, these background fields transform as
\begin{subequations}
    \begin{align}
        g_{\mu\nu} \to g'_{\mu\nu} &= g_{\mu\nu} + \Lied_{\xi} g_{\mu\nu}\,,\\
        b_{\mu\nu} \to b'_{\mu\nu} &= b_{\mu\nu} + \Lied_{\xi} b_{\mu\nu} + \lr{\de_\mu \lambda^{(1)}_\nu -  \de_\nu \lambda^{(1)}_\mu}\,,\\
        A_\mu \to A'_\mu &= A_\mu + \Lied_\xi A_\mu + \de_\mu \lambda^{(0)}\,.
    \end{align}
\end{subequations}

One could now be tempted to write down the generating functional in terms of solely these background source fields,
$    W[g,A,b] = \int d^{d+1}x \sqrt{-g} {\cal F}[g,A,b]\,.$
However, this is not quite right. Let us begin by trying to write down gauge-invariant combinations of variables as we did for the $U(1)$ zero-form charged fluid back in Chapter~\ref{chapter:background}. Let us introduce a timelike Killing vector $K^\mu$, a zero-form gauge parameter $\Lambda^{(0)}$, and a one-form gauge parameter $\Lambda_{\nu}^{(1)}$. We then demand the equilibrium conditions
\begin{subequations}\label{ch5:mhd:equil-cons-Lie}
    \begin{align}
        \Lied_{K} g_{\mu\nu} &= 0\,,\quad \Lied_{K} b_{\mu\nu} + \de_\mu \Lambda_{\nu}^{(1)}-\de_\nu \Lambda_{\mu}^{(1)} = 0\,,\quad \Lied_{K} A_\mu + \de_\mu \Lambda^{(0)} = 0\,.
    \end{align}
\end{subequations}
which we take to define $\Lambda^{(0)}$ and $\Lambda_\mu^{(1)}$. Let us now demand that these conditions are gauge-invariant. This tells us that under a gauge transformation given by $G = \{\lambda^{(1)}_\mu, \lambda^{(0)}\}$, the gauge parameters must transform as
\begin{equation}
        \Lambda^{(0)} \to \Lambda'^{(0)} = \Lambda^{(0)} - \Lied_{K}\lambda^{(0)}\,,\quad \Lambda^{(1)}_\mu \to \Lambda'^{(1)}_\mu = \Lambda^{(1)}_\mu -  \Lied_K \lambda^{(1)}_\mu\,.
\end{equation}
Now, let us define equilibrium variables in analogy to the $U(1)$ zero-form case (i.e. as in Section~\ref{ch2:sec_uncharged}):
\begin{subequations}
    \begin{alignat}{6}
        &T&&= \frac{T_0}{\sqrt{-K^2}}\,,\quad &&u^\mu &&= \frac{K^\mu}{\sqrt{-K^2}}\,,\quad &&\mmu &&= \frac{K^\mu A_\mu + \Lambda}{\sqrt{-K^2}}\,,\label{ch5:mhd:rest_of_defs}\\
        & && &&\mu^\Phi_\nu &&= \frac{K^\mu b_{\mu\nu} + \Lambda_\nu}{\sqrt{-K^2}}\label{ch5:mhd:bad-muphi}\,,
    \end{alignat}
\end{subequations}
where the subscript $\Phi$ on the vector $\mu_\nu^{\Phi}$ denotes the magnetic flux, and $\mu^\Phi_\nu$ is the vector serving as the source for the magnetic flux. However, we run into an immediate issue -- while $\mmu$ is gauge invariant, $\mu_\nu^{\Phi}$ is not! Under a $U(1)$ one-form gauge transformation,
\begin{equation}
    \mu_\nu^{\Phi} \to \mu'^{\Phi}_\nu = \mu^{\Phi}_\nu - \frac{\de_\nu \lr{K^\mu \lambda_\mu^{(1)}}}{\sqrt{-K^2}} \neq \mu_\nu^{\Phi}\,.
\end{equation}
This is not an ideal situation; we would like to have a truly gauge-invariant definition of $\mu_\nu^{\Phi}$. The solution was pointed out in~\cite{Armas:2018atq,Armas:2018zbe}. Let us consider the case where the one-form symmetry is \textit{partially} spontaneously broken. When the one-form symmetry is fully broken, a vector Goldstone boson $\varphi_\mu$ arises~\cite{Gaiotto:2014kfa,Lake:2018dqm,Hofman:2018lfz} which transforms as $\varphi_\mu \to \varphi'_\mu = \varphi_\mu + \lambda_\mu^{(1)}$. The symmetry being only partially broken\footnote{For more detail on what it means for a symmetry to be ``partially broken", the interested reader may refer to~\cite{Armas:2018zbe}.} leads to there being only a scalar Goldstone boson $\varphi \equiv K^\mu \varphi_\mu$, which transforms as
\begin{equation}
    \varphi \to \varphi' = \varphi + K^\mu \lambda_\mu^{(1)}
\end{equation}
under a $U(1)$ one-form gauge transformation. We may then treat $g_{\mu\nu}$, $b_{\mu\nu}$, $A_\mu$, and $\varphi$ as the fields upon which the generating functional depends. We note that $\varphi$ is a dynamical field rather than a background source; however, we still demand the condition
\begin{equation}
    \Lied_K \varphi  + K^\mu \Lambda_\mu^{(1)} = 0\,.
\end{equation}
We can then modify the definition of $\mu_\nu^{\Phi}$ to
\begin{equation}
\label{ch5:mhd:good_muphi}
    \mu_\nu^{\Phi} = \frac{K^\mu b_{\mu\nu} + \Lambda_\nu^{(1)} + \de_\nu \varphi}{\sqrt{-K^2}}\,.
\end{equation}
This is now a gauge-invariant quantity. The generating functional is given by
$    W[g,b,A,\varphi] = \int d^{d+1}x \sqrt{-g} {\cal F}(g,A,b,\varphi)$. 
Variation of $W$ yields
\begin{equation}
\label{ch5:mhd:gen_func_sources_var}
    \delta W = \int d^{d+1}x \sqrt{-g} \biggl[\frac{1}{2} T^{\mu\nu}\delta g_{\mu\nu} + \frac{1}{2}J^{\mu\nu}\delta b_{\mu\nu} + J^\mu \delta A_\mu + E \delta \varphi \biggr]
\end{equation}
where $E = 0$ is the equation of motion for the Goldstone. The equations of motion may be found by setting the perturbations to be due to diffeomorphisms, one-form gauge transformations, and zero-form gauge transformations, yielding
\begin{subequations}\label{ch5:mhd:mhd-eqs-wsources}
    \begin{align}
    \nabla_\mu T^{\mu\nu} &= \bar{F}^{\nu\lambda}J_\lambda + H^{\nu\lambda\sigma}J_{\lambda\sigma} 
    \,,\\
        \nabla_\mu J^{\mu\nu} &= E K^\nu\,,\\
        \nabla_\mu J^\mu &= 0\,.
    \end{align}
\end{subequations}
where $\bar{F}_{\mu\nu} = \de_\mu A_\nu - \de_\nu A_\mu$ is the field-strength for the zero-form $U(1)$ gauge field $A_\mu$ associated with the mass current, and $H_{\mu\nu\rho} = \de_{[\mu} b_{\nu\rho]}$ is the (completely antisymmetric) field strength associated with the gauge field $b_{\mu\nu}$. It is therefore clear that the equilibrium equation of motion for the Goldstone $E=0$ is given by
$    u_\nu \nabla_\mu J^{\mu\nu} = 0\,.$
Turning back temporarily to the definition of $J^{\mu\nu} = \frac{1}{2} \epsilon^{\mu\nu\rho\sigma}F_{\rho\sigma}$ (recalling that $F_{\rho\sigma}$ is the field strength tensor for the electromagnetic $U(1)$), we can see that in the rest frame of the fluid $u^\mu = \delta^\mu_0$, this corresponds to the no-monopole condition on $B$. In order to fully evaluate the equation of motion for the Goldstone, we will need to know what $J^{\mu\nu}$ is in more detail.

Let us now compute the ideal-order equilibrium constitutive relations associated with one-form MHD. It is useful to separate $\mu_\nu^{\Phi}$ into a magnitude and a direction according to $\mu_\nu^{\Phi} = \mu_{\Phi} h_\nu$. Since $\mu_\nu^{\Phi}$ acts as a (vector) source for the magnetic flux, we take the magnitude $\muphi$ as the scalar source, and the direction $h^\mu$ as giving the local direction of the magnetic flux, with $h^2 = 1$. Generically, $h^\mu u_\mu \neq 0$; however, we would like $\mu^{\Phi}_\nu$ to align with magnetic field lines, and so we can use a redefinition freedom~\cite{Armas:2018zbe,Armas:2022wvb} to take $h^\mu u_\mu = 0$. This is a choice we make going forward. As with the other hydrodynamic variables~\eqref{ch5:mhd:rest_of_defs}, the definition~\eqref{ch5:mhd:good_muphi} only makes sense in equilibrium. At zeroth order, we can write
\begin{equation}
    W = \int d^{d+1}x \sqrt{-g} \lr{p(T,\muphi,\mmu) + ...}
\end{equation}
where the $...$ contain higher-derivative hydrostatic terms. The Goldstone $\varphi$ is present in the generating functional by way of $\mu_{\Phi}$ and equation~\eqref{ch5:mhd:good_muphi}. Let us also supplement this expression with the Gibbs-Duhem relation
$    dp = s\, dT + \rho_{\Phi} \,d\muphi + n d\mmu\,,$
where $s$ is the entropy density, $\rhophi$ is the scalar magnetic flux density, and $n$ is the $U(1)$ charge density. The energy density is given by
$    \epsilon = - p + s T+ n \mmu + \rho_{\Phi} \muphi\,.$
Varying the generating functional as in~\eqref{ch5:mhd:gen_func_sources_var}, we find the ideal-order constitutive relations
\begin{subequations}\label{ch5:mhd:eq-con-rel}
    \begin{align}
        T^{\mu\nu}_{\rm eq.} &= \epsilon u^\mu u^\nu + p \Delta_{\perp}^{\mu\nu} + \lr{p - \muphi \rhophi} h^\mu h^\nu\,,\\
        J^{\mu\nu}_{\rm eq.} &= \rhophi \lr{u^\mu h^\nu - u^\nu h^\mu}\,,\label{ch5:mhd:eq-con-rel-Jmunu}\\
        J^\mu_{\rm eq.} &= n u^\mu\,,
    \end{align}
\end{subequations}
where $\delperp^{\mu\nu} = u^\mu u^\nu + g^{\mu\nu} - h^\mu h^\nu$ is the projector orthogonal to both $u^\mu$ and $h^\mu$. 

Let us now determine which one-derivative quantities vanish in equilibrium. From the demands~\eqref{ch5:mhd:equil-cons-Lie} and the demand that $E=0$, we can get (by contracting with $u^\mu$, $h^\mu$, and $\delperp^{\mu\nu}$, and turning off the background gauge fields) the following vanishing scalars~\cite{Armas:2022wvb}\cite{Hoult:2024qph}
\begin{subequations}\label{ch5:mhd:basis-set}
    \begin{alignat}{6}
       &s_1 &&\equiv \frac{u^\mu \nabla_\mu T}{T}, \quad &&s_2&& \equiv \nabla_\mu u^\mu, \quad &&s_3&& \equiv u^\lambda \nabla_\lambda \lr{\frac{\mmu}{T}}\,,\nonumber\\       &s_4&& \equiv h^\mu h^\nu \nabla_\mu u_\nu\,, \quad &&s_5 &&\equiv \frac{T}{\muphi} u^\mu\nabla_\mu \lr{\frac{\muphi}{T}}\,,\label{ch5:mhd:basis-scalar}\\
       &p_1 &&\equiv h^\mu \lr{\frac{\nabla_\mu T}{T} + u^\lambda \nabla_\lambda u_\mu}, \quad &&p_2&& \equiv \frac{1}{T \rhophi} \nabla_\mu \lr{T \rhophi h^\mu}, \quad &&p_3&& \equiv h^\lambda \nabla_\lambda \lr{\frac{\mmu}{T}}\,,\label{ch5:mhd:basis-pseudoscalar}
       \end{alignat}
       vectors,
    \begin{alignat}{4}
       &Y_1^\mu &&\equiv T \delperp^{\mu\lambda} \lr{\nabla_\lambda \lr{\frac{\muphi}{T}} - \frac{\muphi}{T} h^\alpha \nabla_\alpha h_{\lambda}}, \quad &&Y_2^\mu&& \equiv T \delperp^{\mu\alpha} \lr{\frac{\nabla_\alpha T}{T} + u^\lambda \nabla_\lambda u_\alpha}\,,\nonumber\\
       &Y_3^\mu &&\equiv T \delperp^{\mu\lambda} \nabla_\lambda \lr{\frac{\mmu}{T}}\,,\label{ch5:mhd:basis-vector}\\
       &\Sigma_1^{\mu} &&\equiv \delperp^{\mu\alpha} h^{\nu} \lr{\nabla_\alpha u_\nu + \nabla_\nu u_\alpha},\quad &&\Sigma_2^\mu &&\equiv \delperp^{\mu\alpha}u^\nu \lr{\nabla_\alpha h_\nu - \nabla_\nu h_\alpha}\,,\label{ch5:mhd:basis-pseudovector}
    \end{alignat}
    and tensors,
    \begin{align}
        \sigma^{\mu\nu}_\perp &\equiv \lr{\delperp^{\mu\alpha} \delperp^{\nu\beta} + \delperp^{\mu\beta} \delperp^{\nu\alpha} - \frac{2}{d-1} \delperp^{\mu\nu} \delperp^{\alpha\beta}}\nabla_\alpha u_\beta\,,\label{ch5:mhd:basis-tensor}\\
       Z^{\mu\nu} &\equiv \muphi \delperp^{\mu\rho} \delperp^{\nu\sigma} \lr{\nabla_\rho h_\sigma - \nabla_\sigma h_\rho}\,,\label{ch5:mhd:basis-pseudotensor}
    \end{align}
\end{subequations}
where~\eqref{ch5:mhd:basis-scalar} is the set of parity-even scalars that vanish in equilibrium,~\eqref{ch5:mhd:basis-pseudoscalar} is the set of parity-odd scalars (pseudoscalars),~\eqref{ch5:mhd:basis-vector} is the set of parity-odd vectors,~\eqref{ch5:mhd:basis-pseudovector} is the set of parity-even vectors (pseudovectors),~\eqref{ch5:mhd:basis-tensor} is the parity-even rank two traceless symmetric tensor, and~\eqref{ch5:mhd:basis-pseudotensor} is the parity-odd rank two antisymmetric tensor that vanishes in equilibrium.
Note that in the above, we have $p_2 \equiv \frac{1}{T \rhophi} \nabla_\mu \lr{T \rhophi h^\mu}$; the vanishing of $p_2$ in equilibrium did not, strictly speaking, arise from the conditions~\eqref{ch5:mhd:equil-cons-Lie}, but rather from the demand that the equation of motion for $\varphi$ be satisfied. The equation of motion was given by $u_\nu \nabla_\mu J^{\mu\nu}_{\rm eq.} = 0$; along with the equilibrium constitutive relation~\eqref{ch5:mhd:eq-con-rel-Jmunu}, the vanishing of $p_2$ (making use of $p_1 = 0$) subsequently follows.

Let us now take the fluid slightly out of equilibrium. For notational brevity, we will suppress subscripts on thermodynamic derivatives; it is assumed we are working in a basis of $T, \muphi, \mmu$. At ideal-order, the constitutive relations are given by~\eqref{ch5:mhd:eq-con-rel}; together with equations~\eqref{ch5:mhd:mhd-eqs-wsources} (with the sources turned off, and $E=0$), they give us the equations of perfect-fluid MHD. There are eight dynamical equations in equations~\eqref{ch5:mhd:mhd-eqs-wsources}, and so to close the equations, an equation of state $p = p(T,\mu_{\Phi}, \mmu)$ is required.

We will now write down the perfect-fluid equations explicitly. Let us define the shorthand ${\cal D}_\chi = (\pder{\chi}{T} T + \pder{\chi}{\muphi} \muphi + \pder{\chi}{\mmu} \mmu)$. Then the three scalar equations of motion are given by
\begin{subequations}
    \begin{align}
-u_\nu \nabla_\mu T^{\mu\nu} &= {\cal D}_\epsilon\frac{u^\mu \nabla_\mu T}{T} + \pder{\epsilon}{\muphi} T u^\mu \nabla_\mu \lr{\frac{\muphi}{T}} \nonumber\\
&+ \pder{\epsilon}{\mmu} T u^\mu \nabla_\mu \lr{\frac{\mmu}{T}} + \lr{\epsilon + p} \nabla_\mu u^\mu - \muphi \rhophi h^\mu h^\nu \nabla_\mu u_\nu =0\,,\\
h_\nu \nabla_\mu J^{\mu\nu} &= {\cal D}_{\rhophi} \frac{u^\mu \nabla_\mu T}{T} + \pder{\rhophi}{\muphi} T u^\mu \nabla_\mu \lr{\frac{\muphi}{T}} \nonumber\\
&+ \pder{\rhophi}{\mmu}T u^\mu \nabla_\mu \lr{\frac{\mmu}{T}}+ \rhophi \nabla_\mu u^\mu - \rhophi h^\mu h^\nu \nabla_\mu u_\nu=0\,,\\
\nabla_\mu J^\mu &= {\cal D}_n\frac{u^\mu \nabla_\mu T}{T} + \pder{n}{\muphi} T u^\mu \nabla_\mu \lr{\frac{\muphi}{T}}\nonumber\\
&+ \pder{n}{\mmu} T u^\mu \nabla_\mu \lr{\frac{\mmu}{T}} + n \nabla_\mu u^\mu=0\,.
    \end{align}
\end{subequations}
The two pseudoscalar equations of motion are given by
\begin{subequations}
    \begin{align}
h_\nu \nabla_\mu T^{\mu\nu} &= {\cal D}_{(p-\muphi\rhophi)}\frac{h^\mu \nabla_\mu T}{T}+ \pder{(p-\muphi \rhophi)}{\muphi} T h^\mu \nabla_\mu \lr{\frac{\muphi}{T}}\\
&+ \pder{(p-\muphi \rhophi)}{\mmu} T h^\mu \nabla_\mu \lr{\frac{\mmu}{T}}+(p+\epsilon) h^\nu u^\mu \nabla_\mu u_\nu - \muphi \rhophi \nabla_\mu h^\mu = 0\,,\nonumber\\ 
u_\nu \nabla_\mu J^{\mu\nu} &= {\cal D}_{\rhophi} \frac{h^\mu \nabla_\mu T}{T} + \pder{\rhophi}{\muphi} T h^\mu \nabla_\mu \lr{\frac{\muphi}{T}}\nonumber\\
&+ \pder{\rhophi}{\mmu}T h^\mu \nabla_\mu \lr{\frac{\mmu}{T}}+ \rhophi \nabla_\mu h^\mu - \rhophi h^\nu u^\mu \nabla_\mu u_\nu = 0\,.
    \end{align}
\end{subequations}
Finally, the (pseudo)vector equations of motion are given by
\begin{subequations}
    \begin{align}
        \Delta^{\alpha}_{\perp\nu}\nabla_\mu T^{\mu\nu} &= (p+\epsilon)\Delta_{\perp}^{\alpha\mu} \lr{\frac{\nabla_\mu T}{T} + u^\nu \nabla_\nu u_\mu} + \rhophi T \Delta^{\alpha\mu}_{\perp} \nabla_\mu \lr{\frac{\muphi}{T}} \nonumber\\
        &+ n T \Delta^{\alpha\mu}_{\perp} \nabla_\mu \lr{\frac{\mmu}{T}}- \muphi \rhophi \delperp^{\alpha\nu} h^\mu \nabla_\mu h_\nu = 0\,,\\
        \Delta^{\alpha}_{\perp\nu} \nabla_\mu J^{\mu\nu} &= \rhophi \delperp^{\alpha\nu} \lr{u^\mu \nabla_\mu h_\nu - h^\mu \nabla_\mu u_\nu} = 0 \,.
    \end{align}
\end{subequations}
We now specialize to $d=3$. Let us consider perturbations about a homogeneous rest-frame equilibrium state in flat space, i.e.
\begin{equation}\label{ch5:mhd:equilibrium-state}
    \begin{split}
        T(x) &= T_0 + \delta T(x), \quad \mmu(x) = \mmu_0 + \delta \mmu(x), \quad \muphi(x) = \mu_{\Phi,0} + \delta \muphi\,,\\
        u^\mu(x) &= \delta_0^\mu + \delta u^\mu(x), \quad h^\mu(x) = \delta^\mu_z + \delta h^\mu(x)\,,
    \end{split}
\end{equation}
where $u^\mu \delta u_\mu = h^\mu \delta h_\mu = 0 + \oser{\de^2}$, and demanding $h_\mu(x)u^\mu(x) = 0$ hold to quadratic order in the perturbations implies that $\delta u^z =  \delta h^0$. We have fixed $h^\mu$ to point in the $z$-direction, meaning that the equilibrium state has an $SO(2)$ symmetry corresponding to rotations in the $xy$-plane. All scalars, vectors, and tensors are categorized with respect to the equilibrium $SO(2)$ symmetry. We can, more specifically, consider plane-wave perturbations, e.g.
\begin{equation}
T = T_0 + \delta T(\omega,k_j) \exp\lr{- i \omega t + i k \lr{x \sin(\theta) +  z \cos(\theta)}}\,,
\end{equation}%
where $\theta$ is the angle between the wavevector $k_j$ and the $z$-axis (i.e. the equilibrium direction of $h^\mu$). The $SO(2)$ symmetry has been used to align $k_j$ with the $xz$-plane. Once again, $k = \sqrt{k_j k^j}$. We can compute the spectral curve $F(\omega,k)$ in the same way as in Chapter~\ref{chapter:background}; by demanding that the coefficient matrix for the perturbations in momentum space be singular. Note that in the following, we only consider the dynamical equations\footnote{It is possible to take the constraints into account in the linearized analysis in a straightforward way; we neglect to do so because we wish to make a direct comparison to the non-linear theory, in accordance with the discussion in Chapter~\ref{ch3:sec_equivalence}. We will take constraint equations into account in the linearized analysis at the end of this section, as well as in the next section when we discuss relativistic superfluids.}; having specialized to $d=3$, there are eight dynamical equations, and one constraint equation given by $ \nabla_\mu J^{\mu0} = 0$. This matches the eight degrees of freedom in~\eqref{ch5:mhd:equilibrium-state}.

Due to the demand of the $SO(2)$ symmetry, along with the theory respecting parity $P$ and charge conjugation symmetry $C$ for the charge associated with the one-form symmetry\footnote{The discrete symmetries $P$ and $C$ are not quite the same symmetries as are usually considered in MHD (e.g.~\cite{Hernandez:2017mch}); see appendix A.2 of~\cite{Armas:2022wvb} for a discussion of discrete symmetry and one-form charge.} $C$, the spectral curve factorizes to all orders:
\begin{equation}
    F(\omega,k) = F_{\rm Alfv\acute{e}n}(\omega,k)F_{\rm d-ms}(\omega,k) = 0\,,
\end{equation}
where the transverse factor $F_{\rm Alfv\acute{e}n}$ is called the Alfv\'en channel. Named after Swedish physicist Hannes Alfv\'en, the Alfv\'en channel describes (in traditional MHD) the mixing of the components of the magnetic field and the fluid velocity perturbations transverse to both the background magnetic field and the wave vector. In our case, this corresponds to the mixing of $\delta u^y$ and $\delta h^y$. For his work on Alfv\'en waves~\cite{ALFVN1942}, Alfv\'en was awarded the Nobel prize in 1970. The remaining factor, $F_{\rm d-ms}$, describes the mixing of $\delta \mmu$, $\delta T$, and the remaining components of the fluid velocity and magnetic field perturbations. For this reason, we refer to this factor as the ``diffusion-magnetosonic" channel.

For the Alfv\'en channel, there are two hydrodynamic modes. They are given by
\begin{equation}
    \omega_{\rm A} = \pm \sqrt{\frac{\muphio \rhophio}{\epsilon_0 + p_0}} \cos(\theta) k \equiv \pm {\cal V}_A \cos(\theta) k \,.
\end{equation}
In the above, ${\cal V}_A$ is the so-called ``Alfv\'en speed"; by the demands of causality, it must be such that $|{\cal V}_A| \leq 1$. In the diffusion-magnetosonic channel, there are six hydrodynamic modes. Two of them have $\omega = 0$ identically at ideal order; based on their behaviour at first order, we will differentiate between the two, referring to one as the ``diffusion mode" $\omega_q$, and the other as the ``spurious mode" $\omega_0$. The other four modes correspond to magnetosonic waves, and are given by
\begin{equation}
    \omega_{\rm ms} = \pm {\cal V}_{\pm} k\,,
\end{equation}
where ${\cal V}_{\pm}$ is a complicated function of thermodynamic parameters and the angle $\theta$; note that the $\pm$ in the label of ${\cal V}$ is independent of the $\pm$ in the coefficient of $k$, hence leading to four types of magnetosonic waves. The ${\cal V}_{+}$ are the ``fast" magnetosonic modes, while the ${\cal V}_{-}$ are the ``slow" magnetosonic modes. For the expression of ${\cal V}_{\pm}$ in the absence of the $U(1)_m$ charge, the interested reader may refer to~\cite{Armas:2022wvb,Grozdanov:2016tdf}. By the demands of causality, we must again have $ |{\cal V}_{\pm}| \leq 1$.

\subsubsection{First-Order dMHD}
To determine the constitutive relations at first order, we must consider both hydrostatic and non-hydrostatic contributions. The hydrostatic contributions come from the generating functional, while the non-hydrostatic contributions are built out of the building blocks~\eqref{ch5:mhd:basis-set}. In particular, we demand that the equations respect parity ($P$) symmetry, charge conjugation symmetry ($C$) for the one-form charge, and charge conjugation symmetry ($C_m$) for the $U(1)_m$ charge, which we denote by $C_m$.

Under the assumption of $P$ and $C$ symmetries for the microscopic theory, one can show~\cite{Armas:2022wvb} that the first-order contributions to the generating functional vanish. Therefore, the only contributions to the constitutive relations are non-hydrostatic contributions. Let us write down a decomposition of $T^{\mu\nu}$, $J^{\mu\nu}$, and $J^\mu$ with respect to $u^\mu$ and $h^\mu$:
\begin{subequations}
    \begin{align}
        T^{\mu\nu} &= {\cal E} u^\mu u^\nu + {\cal P}\delperp^{\mu\nu} + {\cal S}h^\mu h^\nu + 2{\cal Q}_{\parallel} h^{(\mu}u^{\nu)} + 2 {\cal Q}_{\perp}^{(\mu} u^{\nu)} + {\cal T}_{\perp}^{(\mu} h^{\nu)} + {\cal T}^{\mu\nu}_{\perp}\,,\\
        J^{\mu\nu} &= 2{\cal B}_{\parallel} u^{[\mu} h^{\nu]} + 2u^{[\mu}{\cal B}_{\perp}^{\nu]} + 2{\cal D}^{[\mu}_{\perp}h^{\nu]} + {\cal D}_{\perp}^{\mu\nu}\,,\\
        J^\mu &= {\cal N} u^\mu + {\cal J}_{\parallel} h^\mu + {\cal J}_{\perp}^\mu\,,
    \end{align}
\end{subequations}
where
\begin{subequations}
\begin{alignat}{6}
      &\,\,\,\,\,\,{\cal E} &&\equiv T^{\mu\nu} u_\mu u_\nu\,, \quad &&\,\,\,\,{\cal P} &&\equiv \frac{1}{d-1} T^{\mu\nu} \Delta_{\mu\nu}^{\perp}\,,\quad  && \,\,\,\,\,{\cal S} &&\equiv T^{\mu\nu}h_\mu h_\nu\,,\\
&\,\,\,\,{\cal B}_{\parallel} &&\equiv J^{\mu\nu}h_\mu u_\nu, \quad &&\,\,\,\,{\cal N}&&\equiv - J^\mu u_\mu\,,\\
       &\,\,\,{\cal Q}_{\parallel} &&\equiv -T^{\mu\nu} u_\mu h_\nu \,, \quad &&\,\,{\cal J}_{\parallel} &&\equiv J^\mu h_\mu\,,\\
       &\,{\cal Q}_{\perp}^\alpha &&\equiv - T_{\mu\nu} u^\mu \delperp^{\nu\alpha}\,,\quad &&{\cal D}_{\perp}^\alpha &&\equiv J_{\mu\nu} \delperp^{\mu\alpha}h^\nu\,, \quad &&{\cal J}_{\perp}^\alpha &&\equiv \delperp^{\alpha\mu}J_\mu\,,\\
       &\,{\cal T}_{\perp}^\alpha &&\equiv \delperp^{\alpha\mu}h^\nu T_{\mu\nu}\,,\quad &&{\cal B}_{\perp}^\alpha &&\equiv \delperp^{\alpha\mu}u^\nu J_{\mu\nu}\,,\\
        &{\cal T}^{\mu\nu}_{\perp} &&\equiv \Delta^{\mu\nu\alpha\beta}T_{\alpha\beta}, \qquad&&{\cal D}^{\mu\nu}_{\perp} &&\equiv \delperp^{\mu[\alpha}\delperp^{\nu\beta]}J_{\alpha\beta}\,.
    \end{alignat}
\end{subequations}
Comparing to the ideal-order constitutive relations, we have (listing only the componenents which are non-zero are ideal order)
\begin{equation}
\begin{gathered}
        {\cal E} = \epsilon + {\cal E}_1\,, \quad {\cal P} = p + {\cal P}_1\,,\quad  {\cal S} = p - \muphi \rhophi + {\cal S}_1\,,\\
        {\cal B}_{\parallel} = \rhophi + {\cal B}_{\parallel,1}\,,\quad {\cal N} = n + {\cal N}_1\,,
    \end{gathered}
\end{equation}
where the subscript $1$ denotes the first-order (and beyond) corrections to the constitutive relations. In the notation\footnote{Note that, in comparison to~\cite{Hoult:2024qph}, we use slightly different definitions of ${\cal P}$ and ${\cal S}$. The final results will be the same.} of~\cite{Hoult:2024qph}, the first-order derivative corrections are then given by
\begin{subequations}\label{ch5:mhd:transport-params}
    \begin{alignat}{6}
        &\,\,\,\,\,\,\,{\cal E} &&= \epsilon + \sum_{n=1}^5 \ce_n s_n\,, \quad &&\,\,\,\,\,\,{\cal P} &&= p + \sum_{n=1}^5 \pi_n s_n\,, \quad &&\,\,\,\,{\cal S}&& = p - \muphi \rhophi + \sum_{n=1}^5 \sigma_n s_n\,,\\
        &\,\,\,\,{\cal B}_{\parallel} &&= \rhophi + \sum_{n=1}^5 \beta_{\parallel n} s_n\,, \quad &&\,\,\,\,\,\,{\cal N} &&= n + \sum_{n=1}^5 \nu_n s_n\,,\\
        &\,\,\,{\cal Q}_{\parallel} &&= \sum_{n=1}^3 \theta_{\parallel n} p_n\,,\quad &&\,\,\,\,{\cal J}_{\parallel} &&= \sum_{n=1}^3 \gamma_{\parallel n} p_n\,,\\
        &\,\,{\cal Q}^\mu_{\perp} &&= \sum_{n=1}^3 \theta_{\perp n} Y_n^\mu\,,\quad &&\,\,\,{\cal D}^\mu_{\perp} &&= \sum_{n=1}^3 \rho_{\perp n} Y_n^\mu\,,\quad &&{\cal J}^\mu_{\perp} &&= \sum_{n=1}^3 \gamma_{\perp n} Y_n^\mu\,,\\
        &\,\,{\cal T}^\mu_{\perp} &&= \sum_{n=1}^2 \tau_n \Sigma_n^\mu\,,\quad &&\,\,\,{\cal B}^\mu_{\perp} &&= \sum_{n=1}^2 \beta_{\perp n} \Sigma_n^\mu\,,\\
        &{\cal T}^{\mu\nu}_{\perp} &&= - \eta_{\perp} \sigma^{\mu\nu}_{\perp}\,,\quad&&{\cal D}^{\mu\nu}_{\perp} &&= - r_{\parallel} Z^{\mu\nu}\,.
    \end{alignat}
\end{subequations}
In alignment with the notation of Chapter~\ref{chapter:background}, we will refer to the 46 functions $\{\ce_n, \pi_n, \sigma_n, \beta_{\parallel n}, \nu_n, \theta_{\parallel n}, \gamma_{\parallel n}, \theta_{\perp n}, \rho_{\perp n}, \gamma_{\perp n}, \tau_n, \beta_{\perp n}, \eta_{\perp}, r_{\parallel}\}$ of $(T,\mmu,\muphi)$ as ``transport parameters". Those quantities amongst them which have Kubo formulae may be promoted to the more distinguished status of physical transport coefficients. The transverse shear viscosity $\eta_{\perp}$ and the longitudinal resistivity $r_{\parallel}$ are two examples of genuine physical transport coefficients.

There are far more transport parameters in dMHD than in the case of a $U(1)$ charged fluid. We would like to identify which (combinations of) parameters are physical transport coefficients. The first step in this process is to identify frame-invariant quantities. Let us consider an arbitrary first-order frame redefinition
\begin{equation}
\label{ch5:mhd:frame-redef-mhd}
    \begin{split}
        T \to T' &= T + \delta T, \quad \mmu \to \mmu' = \mmu + \delta \mmu\,, \quad \muphi \to \muphi' = \muphi + \delta \muphi\,\\
        u^\mu \to u'^\mu &= u^\mu + \delta u_{\parallel} h^\mu + \delta u_{\perp}^\mu, \quad h^\mu \to h'^\mu = h^\mu + \delta u_{\parallel} u'^\mu + \delta h^\mu_{\perp}\,.
    \end{split}
\end{equation}
The redefinitions are such that $u^2 = -1$, $h^2 = 1$, and $u{\cdot}h = 0$ are respected up to second order in derivatives. Since the redefinitions are first order and vanish in equilibrium, they may be generically decomposed with respect to the basis set~\eqref{ch5:mhd:basis-set}:
\begin{subequations}
    \begin{align}
        \delta T &= \sum_{i=1}^5 \delta \mathfrak{a}_i s_i, \quad \delta \muphi = \sum_{i=1}^5 \delta \mathfrak{b}_i s_i, \quad \delta \mmu = \sum_{i=1}^5 \delta \mathfrak{c}_i s_i\,,\\
        \delta u_{\parallel} &= \sum_{j=1}^3 \delta \mathfrak{d}_j p_j, \quad \delta u^\mu_{\perp} = \sum_{n=1}^3 \delta \mathfrak{e}_n Y_n^\mu, \quad \delta h^\mu_{\perp} = \sum_{m=1}^2 \delta \mathfrak{f}_m \Sigma_m^\mu\,.
    \end{align}
\end{subequations}
Let us now define the the notation $\delta \chi_i \equiv \chi_i' - \chi_i$ for any transport parameter $\chi_i$, where $\chi'_i$ denotes the transport parameter in the new hydrodynamic frame. Demanding the frame-invariance of the one-point functions $T^{\mu\nu}$, $J^{\mu\nu}$, $J^\mu$ up to second order, we find that the transport parameters must transform (defining the notation $\delta \chi_i = \chi'_i - \chi_i$ for transport parameter $\chi_i$) as
\begin{subequations}
    \begin{align}
        \delta \ce_i &= \lr{\pder{\epsilon}{T} \delta \mathfrak{a}_i + \pder{\epsilon}{\muphi} \delta \mathfrak{b}_i + \pder{\epsilon}{\mmu} \delta \mathfrak{c}_i}\,,\qquad (i \in \{1,..,5\})\\
        \delta \pi_i &=\lr{\pder{p}{T} \delta \mathfrak{a}_i + \pder{p}{\muphi} \delta \mathfrak{b}_i + \pder{p}{\mmu} \delta \mathfrak{c}_i}\,,\\
        \delta \sigma_i &= \lr{\pder{(p - \muphi \rhophi)}{T} \delta \mathfrak{a}_i + \pder{(p - \muphi \rhophi)}{\muphi} \delta \mathfrak{b}_i + \pder{(p - \muphi \rhophi)}{\mmu} \delta \mathfrak{c}_i}\,,\\
        \delta \beta_{\parallel i} &= \lr{\pder{\rhophi}{T} \delta \mathfrak{a}_i + \pder{\rhophi}{\muphi} \delta \mathfrak{b}_i + \pder{\rhophi}{\mmu} \delta \mathfrak{c}_i}\,,\\
        \delta \nu_i &= \lr{\pder{n}{T} \delta \mathfrak{a}_i + \pder{n}{\muphi} \delta \mathfrak{b}_i + \pder{n}{\mmu} \delta \mathfrak{c}_i}\,,\\
        \delta \theta_{\parallel j} &= \lr{\epsilon + p +\muphi\rhophi} \delta \mathfrak{d}_j, \quad \delta \gamma_{\parallel j} = n \delta \mathfrak{d}_j\,,\quad (j \in \{1,...,3\})\\
        \delta \theta_{\perp n} &= \lr{\epsilon + p} \delta \mathfrak{e}_n\,,\quad \delta \gamma_{\perp n} = n \delta \mathfrak{e}_n, \quad \delta \rho_{\perp n} = \rhophi \delta \mathfrak{e}_n\,, \quad (n \in \{1,...,3\})\\
        \delta \tau_m &= - \muphi \rhophi \delta \mathfrak{f}_m, \quad \beta_{\perp m} = \rhophi \delta \mathfrak{f}_m, \quad (m \in \{1,2\})\,.
    \end{align}
\end{subequations}
The transport parameters for the tensors $(\eta_{\perp} ,r_{\parallel})$ are invariant under frame transformations. From these transport parameters, we can form 23 frame-invariant quantities:
\begin{subequations}
\label{ch5:mhd:frame-invar}
    \begin{align}
        f_i &\equiv \pi_i - \lr{\pder{p}{\epsilon}}_{(\rhophi,n)} \ce_i - \lr{\pder{p}{\rhophi} }_{(\epsilon,n)} \beta_{\parallel i} - \lr{\pder{p}{n}}_{(\epsilon,\rhophi)}\nu_i\,,\\
        g_i &\equiv \sigma_i - \lr{\pder{(p-\muphi \rhophi)}{\epsilon}}_{(\rhophi,n)} \ce_i  -\lr{\pder{(p-\muphi \rhophi)}{\rhophi}}_{(\epsilon,n)} \beta_{\parallel i}\nonumber\\
        &- \lr{\pder{(p-\muphi \rhophi)}{n}}_{(\epsilon,\rhophi)} \nu_i\,,\\
        h_j &\equiv \gamma_{\parallel j} - \lr{\frac{n}{\epsilon + p - \muphi \rhophi}} \theta_{\parallel j}\,,\quad k_n \equiv \rho_{\perp n} - \lr{\frac{\rhophi}{\epsilon + p}} \theta_{\perp n}\,,\\
        \ell_n &\equiv \gamma_{\perp n} - \lr{\frac{n}{\epsilon + p}} \theta_{\perp n}\,,\quad m_m \equiv\tau_m + \muphi \beta_{\perp m}\,,\\
        &\eta_{\perp}, \qquad r_{\parallel}\,.
    \end{align}
\end{subequations}
Let us now define an equivalent of the Landau frame for dMHD. Note by looking at the redefinitions~\eqref{ch5:mhd:frame-redef-mhd} that there are three scalar parameters, one pseudoscalar parameter, one vector parameter, and one pseudovector parameter. The redefinition freedom may therefore be used to set three scalar corrections, one pseudoscalar correction, one vector correction, and one pseudovector correction to the perfect-fluid constitutive relations to zero. We define the Landau frame as
\begin{equation}
    {\cal E} = \epsilon, \quad {\cal B}_{\parallel} = \rhophi, \quad {\cal N} = n, \quad {\cal Q}_{\parallel} = {\cal Q}_{\perp}^\mu = {\cal B}_{\perp}^\mu = 0\,.
\end{equation}
The constitutive relations are then given by
\begin{subequations}\label{ch5:mhd:frame-invar-conrel}
    \begin{align}
        T^{\mu\nu} &= \epsilon u^\mu u^\nu + \lr{ p + \sum_{i=1}^5 f_i s_i} \delperp^{\mu\nu} + \lr{p - \rhophi \muphi + \sum_{i=1}^5 g_i s_i} h^\mu h^\nu \nonumber\\
        &+ 2\sum_{m=1}^2 m_m \Sigma_m^{(\mu} h^{\nu)} - \eta_{\perp} \sigma^{\mu\nu}_{\perp}\,,\\
        J^{\mu\nu} &= 2\rhophi u^{[\mu}h^{\nu]} + 2\sum_{n=1}^3 k_n V_n^{(\mu}h^{\nu)} - r_{\parallel} Z^{\mu\nu}\,,\\
        J^\mu &= n u^\mu  + \sum_{j=1}^3 h_j p_j h^\mu + \sum_{n=1}^3 \ell_n V_n^\mu\,.
    \end{align}
\end{subequations}
There are now 23 parameters in the theory. As in Section~\ref{ch2:sec_uncharged}, we can go further by once again applying the ideal-order equations of motion. There are three scalar equations of motion ($u_\nu \nabla_\mu T^{\mu\nu} = 0$, $h_\nu \nabla_\mu J^{\mu\nu} = 0$, and $\nabla_\mu J^\mu = 0$), two pseudoscalar equations of motion ($h_\nu \nabla_\mu T^{\mu\nu} = 0$ and $u_\nu \nabla_\mu J^{\mu\nu} = 0$), one vector equation of motion ($\Delta^{\alpha}_{\perp \nu} \nabla_\mu T^{\mu\nu} =0$), and one pseudovector equation of motion ($\Delta^{\alpha}_{\perp \nu} \nabla_\mu J^{\mu\nu} =0$). We choose to eliminate $s_1, s_3, s_5, p_1,p_2,V_2^\mu,$ and $\Sigma_2^\mu$. After applying the equations of motion, we can write
\begin{subequations}\label{ch5:mhd:mhd_Landau_conrel}
    \begin{align}
        T^{\mu\nu} &= \epsilon u^\mu u^\nu + \lr{p - \muphi \rhophi - \zeta_\parallel s_4 - \zeta_{\times}(s_2-s_4)}h^\mu h^\nu + \lr{ p - \zeta'_{\times} s_2 - \zeta_{\perp} (s_2 - s_4)}\delperp^{\mu\nu}\nonumber\\
        &- 2\eta_{\parallel} \Sigma_1^{(\mu} h^{\nu)} - \eta_{\perp} \sigma^{\mu\nu}_{\perp}\,,\\
        J^{\mu\nu} &= 2\rhophi u^{[\mu}h^{\nu]} - 2r_{\perp} Y_1^{[\mu} h^{\nu]} - 2 \tilde{\sigma}' Y_3^{[\mu} h^{\nu]}-r_{\parallel} Z^{\mu\nu}\,,\\
        J^\mu &= n u^\mu  - \lr{\sigma_{\parallel} p_3}h^\mu - \tilde{\sigma} Y_1^\mu - \sigma_{\perp} Y_3^\mu\,.
    \end{align}
\end{subequations}
Due to their complexity, the expressions for the transport coefficients above in terms of the frame invariants~\eqref{ch5:mhd:frame-invar} are not reproduced here; they may be straightforwardly (albeit tediously) obtained by applying the ideal-order equations of motion to the constitutive relations~\eqref{ch5:mhd:frame-invar-conrel}. Now, let us consider the following ``canonical" formulation for the entropy current:~\cite{Grozdanov:2016tdf}
\begin{equation}
    S^\mu = p \beta^\mu - T^{\mu\nu}\beta_\nu - \frac{\muphi}{T} J^{\mu\nu}h_\nu - \frac{\mmu}{T} J^\mu\,.
\end{equation}
The divergence of the entropy current can then be shown (via an application of the equations of motion) to be
\begin{equation}
    \nabla_\mu S^\mu = - T_1^{\mu\nu} \nabla_\mu \beta_\nu - \nabla_\mu \lr{\frac{\muphi}{T}} J^{\mu\nu}_1 h_\nu - \nabla_\mu \lr{\frac{\mmu}{T}} J_1^\mu + \oser{\de^2}\,,
\end{equation}
where the subscript $1$ refers to the $\oser{\de}$ contributions to the conserved currents. Inserting the constitutive relations~\eqref{ch5:mhd:mhd_Landau_conrel} and demanding the non-negativity of the divergence of the entropy current in an analogous manner to Chapter~\ref{chapter:background}, we find ten inequality-type conditions. There are also two Onsager relations:~\cite{Grozdanov:2016tdf,Hoult:2024qph}\footnote{The parameter $\tilde{\sigma}$ was missing in previous formulations of dMHD; it was first written down in~\cite{Hoult:2024qph}.} 
\begin{subequations}
    \begin{align}
        \zeta_\times &= \zeta'_{\times} \quad \tilde{\sigma}' = \tilde{\sigma}\,,\\
        \eta_{\parallel} &\geq 0, \quad \eta_{\perp} \geq 0, \quad \sigma_{\parallel} \geq 0, \quad r_{\parallel} \geq 0\,,\\
        \zeta_{\parallel} &\geq 0, \quad \zeta_{\perp} \geq 0, \quad \zeta_{\parallel} \zeta_{\perp} \geq \zeta_{\times}^2\,,\\
        \sigma_{\perp} &\geq 0, \quad r_{\perp} \geq 0, \quad \sigma_{\perp} r_{\perp} \geq \tilde{\sigma}^2\,.
    \end{align}
\end{subequations}
After fixing a frame, applying the ideal-order equations of motion, and obtaining Onsager relations between the parameters (see Appendix~\ref{app:linear response} for details on Onsager relations), the number of independent transport coefficients has dropped from 46 to 23 to 10. With the constitutive relations~\eqref{ch5:mhd:mhd_Landau_conrel} in hand, let us discuss dispersion relations. We once again consider perturbations of the form~\eqref{ch5:mhd:equilibrium-state}. We will investigate the diffusion-magnetosonic channel and the Alfv\'en channel separately\footnote{Recall that the spectral curve factorizes to all orders due to the discrete symmetries we impose.}. Since the diffusion-magnetosonic channel is more complicated than the Alfv\'en channel, we will consider that channel only in broad strokes. Once again, we only take into account the dynamical equations.

\paragraph{Alfv\'en channel.} The Alfv\'en channel has two hydrodynamic modes, as before. The spectral curve can be written 
\begin{align}\label{ch5:mhd:viscous-Alfven-spectral}
    F_{\rm Alfv\acute{e}n}(\omega,k,\theta) &=\omega^2 + i k^2\lr{\sin^2(\theta) \lr{\frac{\eta_{\perp}}{p_0+\epsilon_0} + \frac{r_{\parallel} \muphio}{\rhophio}} + \cos^2(\theta) \lr{\frac{\eta_{\parallel}}{p_0+\epsilon_0} + \frac{r_{\perp} \muphio}{\rhophio}}} \omega\nonumber\\
    &-\frac{\muphio}{ (p_0+ \epsilon_0) \rhophio} k^4\biggl[\cos^4(\theta) r_{\perp} \eta_{\parallel} + \sin^4(\theta) r_{\parallel} \eta_{\perp} \\
    &+ \sin^2(\theta) \cos^2(\theta) \lr{r_{\parallel} \eta_{\parallel} + r_{\perp} \eta_{\perp} } \biggr] - {\cal V}_A^2 \cos^2(\theta) k^2 \,.\nonumber
\end{align}
The dispersion relations for these modes, $\omega_A$, are given in the limit of small $k$ by 
\begin{equation}\label{ch5:mhd:Alfven-dispersion-FO}
    \omega_A = \pm {\cal V}_A \cos(\theta) k - \frac{i}{2} \lr{\Gamma_{A1} \sin^2(\theta) + \Gamma_{A2} \cos^2(\theta)} k^2 + \oser{k^3}
\end{equation}
where
\begin{equation}
    \Gamma_{A1} = \frac{\eta_{\perp}}{p_0 + \epsilon_0} + \frac{r_{\parallel}\muphio}{\rhophio}, \quad \Gamma_{A2} = \frac{\eta_{\parallel}}{p_0+\epsilon_0} + \frac{r_{\perp} \muphio}{\rhophio}\,.
\end{equation}
Note that $\omega_A$ is a function of both $k$ and $\theta$. The dispersion relations above were given in the limit of $k \ll 1$ for arbitrary $\theta$. In particular, setting $\theta = \pi/2$ in the dispersion relation~\eqref{ch5:mhd:Alfven-dispersion-FO} yields
\begin{equation}
    \omega_A = - \frac{i}{2} \lr{\frac{\eta_{\perp}}{p + \epsilon} + \frac{r_{\parallel}\muphi}{\rhophi}} k^2 + \oser{k^4}\,.
\end{equation}
Both modes are identical. Let us now instead reverse the ordering, and first set $\theta =  \pi/2$ in equation~\eqref{ch5:mhd:viscous-Alfven-spectral}.  We then solve for $\omega$ in the limit of small $k$, and find that the two modes differ:
\begin{equation}
    \omega_{A1} = - i \lr{\frac{r_{\parallel} \muphio}{\rhophio}} k^2 + \oser{k^4}\,, \quad \omega_{A2} =- i \lr{\frac{\eta_{\perp}}{p_0 + \epsilon_0}}k^2 + \oser{k^4}\,.
\end{equation}
We see therefore that there is a non-commutativity between the $k \to 0$ limit, and the $\theta \to \pi/2$ limit~\cite{Hernandez:2017mch,Fang:2024hxa,Fang:2024skm}. While this non-commutativity may appear surprising at first glance, another way to think about the limits involved may prove more enlightening. The background magnetic field was fixed to point in the $z$-direction. There are therefore two scalars one can form with respect to the $SO(2)$ symmetry of the background equilibrium state as well as $P$:
\[
k_z^2, \quad k_{\perp}^2 = k_i k_j \lr{\delta^{ij} - \delta^i_z \delta^j_z}\,.
\]
In the above, we used the $SO(2)$ symmetry to fix $k_j$ to lie in the $xz$-plane. The two scalars $k_z^2$ and $k_{\perp}^2$ may be independently taken to zero; the two limits are not necessarily commutative, and this is indeed what we find in this case.

\begin{figure}[t]
    \centering
    \begin{subfigure}[t]{0.49\linewidth}
    \includegraphics[width=\linewidth]{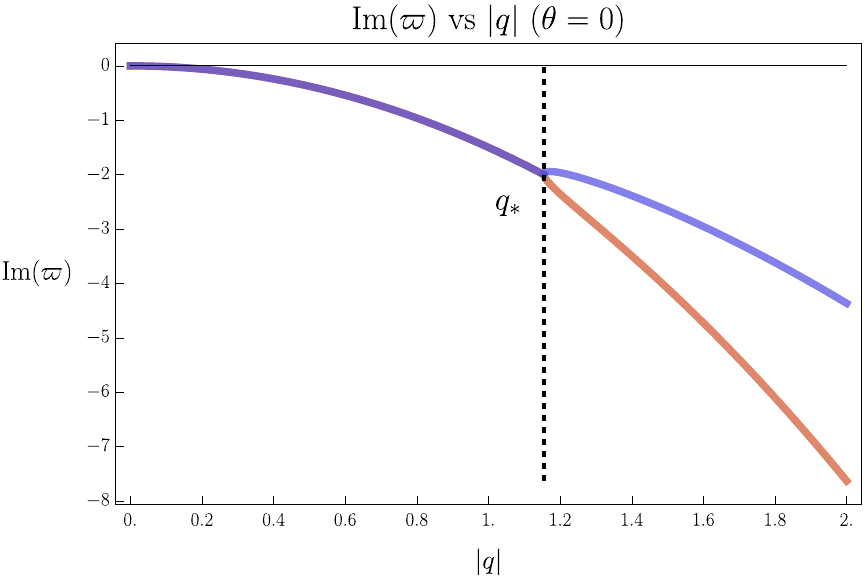}
    \end{subfigure}
    \begin{subfigure}[t]{0.49\linewidth}
    \includegraphics[width=\linewidth]{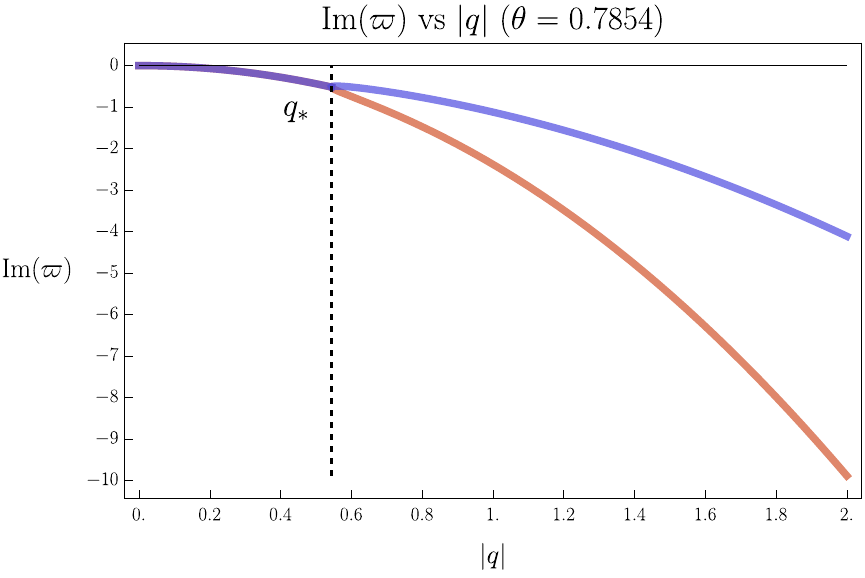}
    \end{subfigure}
    \hfill
    \vspace{0.5em}
    \begin{subfigure}[t]{0.49\linewidth}
    \includegraphics[width=\linewidth]{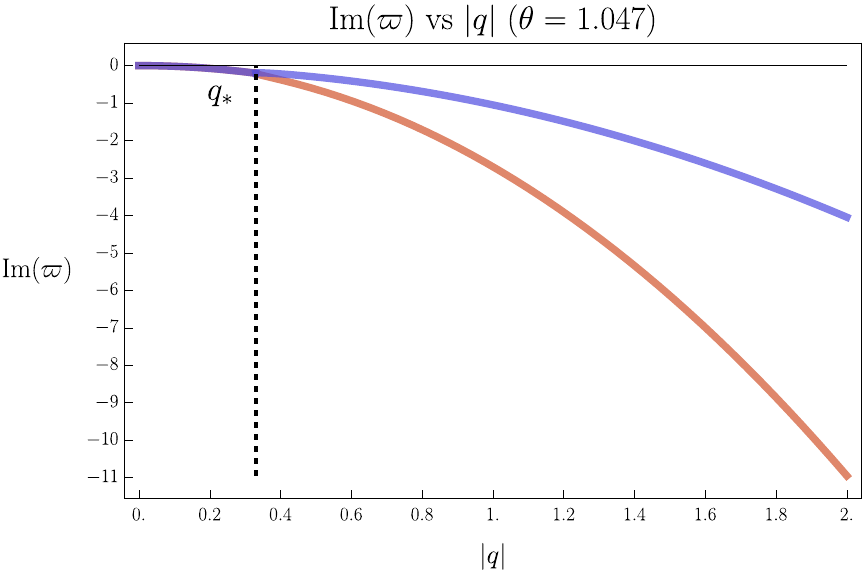}
    \end{subfigure}
    \begin{subfigure}[t]{0.49\linewidth}
    \includegraphics[width=\linewidth]{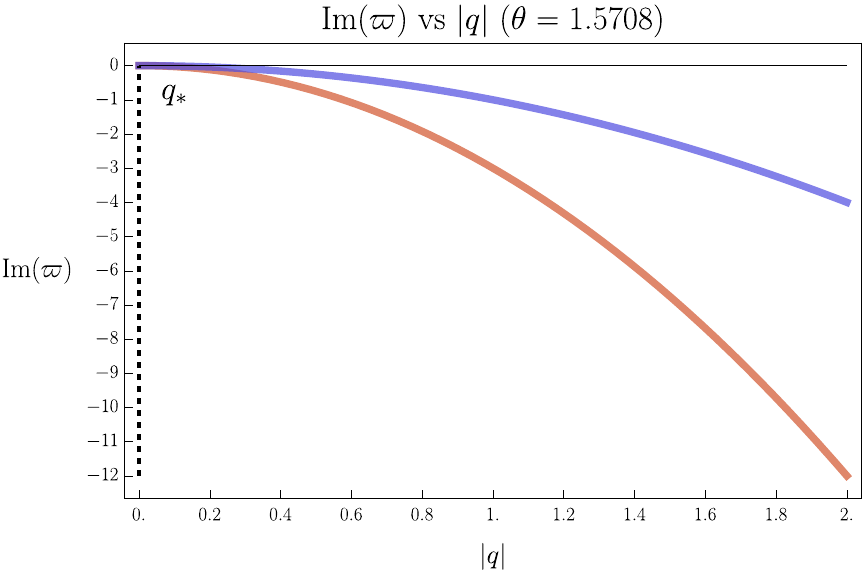}
    \end{subfigure}
    \caption{The imaginary parts of the solutions~\eqref{ch5:mhd:alfven-full-sol} with $\eta_{\parallel}/(p+\epsilon) = 2 T$, $r_{\perp} \muphi/\rhophi = T$, $r_{\parallel} \muphi/\rhophi = T$, $\eta_{\perp}/(p+\epsilon) = 3 T$, and ${\cal V}_A = 1/\sqrt{3}$ for $\theta \in \{0,\frac{\pi}{4}, \frac{\pi}{3}, \frac{\pi}{2}\}$. As $\theta \to \pi/2$, the branch point of the solution at $q_*$ goes towards zero. We plot the unitless quantities $\varpi \equiv \omega/T$ and $|q| \equiv |k|/T$.}
    \label{ch5:fig:Alfven-1}
\end{figure}

The natural follow-up question is: why do the two limits not commute? This may be answered by considering the full solution to the equation $F_{\rm Alfv\acute{e}n} = 0$:
\begin{equation}\label{ch5:mhd:alfven-full-sol}
\begin{split}
    \omega_A &= - \frac{i}{2}\lr{\frac{r_{\perp} \muphio}{\rhophio} + \frac{\eta_{\parallel}}{(p_0+\epsilon_0)}}k_z^2  -\frac{i}{2} \lr{\frac{r_{\parallel} \muphio}{\rhophio} + \frac{\eta_{\perp}}{(p_0+\epsilon_0)}}k_x^2\\
    &\pm \frac{i}{2}\sqrt{\lr{k_x^2 \lr{\frac{r_{\parallel} \muphio}{\rhophio} - \frac{\eta_{\perp}}{(p_0+\epsilon_0)}} + k_z^2 \lr{\frac{r_{\perp} \muphio}{\rhophio} - \frac{\eta_{\parallel}}{(p_0+\epsilon_0)}}}^2 - 4 {\cal V}_A^2 k_z^2}\,.
\end{split}
\end{equation}
We see that the solution has a square root, and therefore a branch point when the discriminant is zero. The branch point sets the radius of convergence of the small-$|k|$ expansion. In Figure~\ref{ch5:fig:Alfven-1}, the imaginary part of the full solution~\eqref{ch5:mhd:alfven-full-sol} has been plotted for varying values of $\theta$; one can visually see that as $\theta \to \pi/2$, the radius of convergence of the small-$|k|$ expansion goes to zero. We also note that in the fine-tuned case where
\begin{equation}
    \frac{r_{\parallel} \muphio}{\rhophio} = \frac{\eta_{\perp}}{(p_0+\epsilon_0)}
\end{equation}
the non-commutativity vanishes, and the radius of convergence goes to infinity at $\theta = \pi/2$. As the values of the transport coefficients are microscopically determined, we should expect such an equivalence to only hold if there were secretly an underlying symmetry to make it so. There is, to the best of my knowledge, no reason to think such a symmetry exists.

Finally, it should come as no surprise at this point that the Alfv\'en channel in Landau frame in acausal. It inherits all of the issues that the Landau frame has in the regular $U(1)$ charged fluid of Chapter~\ref{chapter:background}. Just as in that case, there are a number of ways to resolve the acausality. We will only consider the BDNK formulation here, though one can also write down an MIS-type formulation of one-form MHD. 

\paragraph{Diffusion-magnetosonic channel.} The diffusion-magnetosonic channel in Landau frame has six hydrodynamic modes, as before. One of these is the ``spurious mode" $\omega_0 = 0$, which remains identically zero. Another is the diffusion mode $\omega_q$, which is now of the form
\begin{equation}
    \omega_q = - i \Gamma_q k^2 + \oser{k^4}\,,
\end{equation}
where $\Gamma_q>0$ is a (complicated) function of the transport coefficients and the thermodynamic quantities in the theory. The remaining four modes are the magnetosonic modes, and are given by
\begin{equation}
    \omega_{ms} = \pm {\cal V}_{\pm} k - i \Gamma_{\pm} k^2 \pm \oser{k^3}\,,
\end{equation}
where the $\Gamma_{\pm}>0$ are again complicated functions of the transport coefficients and the thermodynamic quantities in the theory. The diffusive coefficients $\Gamma_q$, $\Gamma_{\pm}$ must be positive to ensure stability. Just like the Alfv\'en channel, the diffusion-magnetosonic channel is also acausal, and therefore unstable after boosting. To rectify this, we consider the BDNK theory of one-form MHD.

\subsection{BDNK}
The BDNK theory for one-form MHD was originally written down in~\cite{Armas:2022wvb}. It was later extended to include the mass current in my paper~\cite{Hoult:2024qph}; there, it was also shown that both the original BDNK dMHD and the extended version with the mass current enjoyed the equivalence between linear and non-linear causality discussed in Chapter~\ref{ch3:sec_equivalence}. In the case of dMHD, the constitutive relations for the BDNK formulation are given by equation~\eqref{ch5:mhd:transport-params}. Of the forty-six parameters, only ten are physical; the remaining thirty-six act as regulators in the theory to ensure stability and causality. Let us once again consider plane-wave perturbations about the equilibrium state as in equations~\eqref{ch5:mhd:equilibrium-state}. The spectral curve again factorizes into the Alfv\'en channel and the diffusion-magnetosonic channel; we will briefly consider each in turn.

\paragraph{Alfv\'en channel.} 
In the Alfv\'en channel, there are now two hydrodynamic modes, and two non-hydrodynamic (gapped) modes. In the limit of small $k$, they are given by
\begin{subequations}
    \begin{align}
        \omega_A &= \pm {\cal V}_A \cos(\theta) k - \frac{i}{2} \lr{\Gamma_{A1} \sin^2(\theta) + \Gamma_{A2} \cos^2(\theta)} k^2 + \oser{k^3}\,,\\
        \omega &= -\lr{\frac{p_0+\epsilon_0}{\theta_{\perp 2}}} + \oser{k^2}\,,\qquad  \omega = i \lr{\frac{\rhophio}{\beta_{\perp 2}}} + \oser{k^2}\,,
    \end{align}
\end{subequations}
where ${\cal V}_A$, $\Gamma_{A1}$, and $\Gamma_{A2}$ are as before. Stability of the gaps leads to the demands
\begin{equation}
\label{ch5:mhd:Alfven-gap}
    \frac{p_0+\epsilon_0}{\theta_{\perp 2}} \geq 0, \quad \frac{\rhophio}{\beta_{\perp 2}} \leq 0\,.
\end{equation}
It is clear then the issue with Landau frame -- it sets both $\theta_{\perp 2}$ and $\beta_{\perp 2}$ to zero. This is completely analogous to how the Landau frame in regular charged hydro set $\theta_1 = 0$, which killed the non-hydrodynamic mode in the shear sector. Let us move on to causality. Thanks to the presence of the two non-hydrodynamic modes, one can show that the third causality condition of~\eqref{ch2:hydro:full_momentum_causcon} is satisfied. In the limit of large $k$, the dispersion relations are linear (as desired), $\omega \sim c k$. The solutions for $c$ are given by the roots of the biquadratic
\begin{equation}\label{ch5:mhd:Alfven-large-k-poly}
    P_{\rm Alfv\acute{e}n}(c^2) \equiv \lr{- \beta_{\perp 2} \theta_{\perp 2}} c^4 + a  c^2 + b = 0\,,
\end{equation}
where
\begin{subequations}
    \begin{align}
        a &= \frac{1}{T_0} \lr{\beta_{\perp 1} \theta_{\perp 1} \muphio + \theta_{\perp 2} \rho_{\perp 1} \muphio - \theta_{\perp 1} \rho_{\perp 2} \muphio - T_0 \beta_{\perp 2} \tau_1 + T_0 \lr{\beta_{\perp 1} - \rho_{\perp2}} \tau_2} \cos(\theta)^2\nonumber \\
        &+ \lr{\beta_{\perp 2} \eta_{\perp} - r_{\parallel} \theta_{\perp 2} \muphi} \sin(\theta)^2\,,\\
        b &= \frac{\muphio}{4 T_0} \lr{ - r_{\parallel} T_0 + \rho_{\perp 1} + \lr{r_{\parallel} T_0 + \rho_{\perp 1}}\cos(2\theta)}\lr{-\eta_{\perp} + \tau_1 + \lr{\eta_{\perp} + \tau_1} \cos(2\theta)}\,.
    \end{align}
\end{subequations}
One can ensure the causality of the Alfv\'en channel by applying a sequence of criteria to the polynomial~\eqref{ch5:mhd:Alfven-large-k-poly}. Firstly, we can use the Schur-Cohn criterion (refer to Chapter~\ref{ch3:sec_RHandSC} for a reminder) to ensure that the roots in $c^2$ lie in the open unit disk in the complex plane. Next, we can enforce reality of the roots by enforcing that the discriminant of~\eqref{ch5:mhd:Alfven-large-k-poly} in $c^2$ is positive. Finally, we can ensure that $0\leq c^2<1$ by imposing the Routh-Hurwitz criterion (refer back to Chapter~\ref{ch3:sec_RHandSC} as well) on $P_{\rm Alfv\acute{e}n}(-c^2)$. As the RH-criterion forces roots to lie in the left-hand complex plane, this condition forces roots in $c^2$ to lie in the right-hand complex plane.

After applying these constraints, sufficient causality conditions are given by~\cite{Armas:2022wvb}
\!\begin{subequations}\label{ch5:mhd:causality_con_Alfven}
    \begin{align}
        \theta_{\perp 1} = - \frac{\tau_2}{\muphio} \quad \text{or}&\quad \beta_{\perp 1} = T_0 \rho_{\perp 2}\,,\\
        T_0 \theta_{\perp 2} \rho_{\perp 1} - \frac{\tau_1}{\muphio} \beta_{\perp 2} + \rho_{\perp 1} \tau_1 > 0, \quad &r_{\parallel} \eta_{\perp} - T_0 r_{\parallel} \theta_{\perp 2} + \frac{\eta_{\perp}}{\muphio} \beta_{\perp 2} > 0\,,\\
        r_{\parallel} \eta_{\perp} + \eta_{\perp} \rho_{\perp 1} + r_{\parallel} \tau_1& + \rho_{\perp 1} \tau_1 < 0\,.
    \end{align}
\end{subequations}
These conditions are not empty.

\paragraph{Diffusion-magnetosonic channel.} With the inclusion of all the transport parameters, the spectral curve of the diffusion-magnetosonic channel becomes an order-12 polynomial in $\omega$ which satisfies the third condition of~\eqref{ch2:hydro:full_momentum_causcon}. In the limit of small-$k$, there are the six previously discussed hydrodynamic modes:
    \begin{align}
        \omega_0 &= 0\,,\quad  \omega_q = - i \Gamma_q k^2 + \oser{k^4}\,,\quad \omega_{\pm} = \pm {\cal V}_{\pm} k - i \Gamma_{\pm} k^2 + \oser{k^3}\,.
    \end{align}
In addition to these modes, there are six non-hydrodynamic modes. Three of them are straightforward to write down in the small-$k$ limit:
\begin{equation}
\begin{split}
   \omega = - \lr{\frac{p_0 + \epsilon_0}{\theta_{\perp 2}}}& + \oser{k^2}\,, \quad \omega = i \lr{\frac{\rhophio}{\beta_{\perp 2}}}  + \oser{k^2}\,,\\ \omega = -i&\lr{\frac{p_0 + \epsilon_0 - \muphio \rhophio}{\theta_{\parallel 1} + \theta_{\parallel 2}}} + \oser{k^2}\,.
\end{split}
\end{equation}
Two of these modes have the same $k\to0$ limit as the non-hydrodynamic modes of the Alfv\'en channel. The reason for this is that in the limit $k \to 0$, the $SO(2)$ symmetry is restored, and a second factor of the Alfv\'en channel factors out of the diffusion-magnetosonic channel. The remaining three non-hydrodynamic modes are given at leading order in small $k$ by the roots of
\begin{align}
    P_{3g}(\omega) &= \det\left| \begin{pmatrix}
        \pder{\epsilon_0}{T_0}&\pder{\epsilon_0}{\muphio}&\pder{\epsilon_0}{\mmu_0}\\
        \pder{\rhophio}{T_0}&\pder{\rhophio}{\muphio}&\pder{\rhophio}{\mmu_0}\\
        \pder{n_0}{T_0}&\pder{n_0}{\muphio}&\pder{n_0}{\mmu_0}
    \end{pmatrix} + \lr{- i \omega} \begin{pmatrix}
        \frac{(\ce_1 - \ce_5) T_0 - \ce_3 \mmu}{T^2_0} &\frac{\ce_5}{\mmu_0} & \frac{\ce_3}{T_0}\\
        \frac{\lr{\beta_{\parallel 1} - \beta_{\parallel 5}}T_0 - \beta_{\parallel 3} \mmu_0}{T_0^2} & \frac{\beta_{\parallel 5}}{\mmu_0} & \beta_{\parallel 3}{T_0}\\
        \frac{\lr{\nu_1 - \nu_5} T_0 - \nu_3 \mmu_0}{T^2_0} & \frac{\nu_5}{\mmu_0} & \frac{\nu_3}{T_0}
    \end{pmatrix} \right| \nonumber\\
    &= a \lr{- i \omega}^3 + b \lr{-i \omega}^2 + c \lr{- i\omega} + d\,,\label{ch5:mhd:cubic-gap}
\end{align}
where $a,b,c,d$ are functions of both the transport parameters and the thermodynamic parameters of the equilibrium state. Imposing the Routh-Hurwitz criteria on the polynomial $P_{3g}(i \Delta)$ for variable $\Delta = - i \omega$, we find that the modes are stable so long as (see Appendix~\ref{app:RH-criteria})
\begin{equation}
    a>0,\quad  b>0,\quad  d>0,\quad  bc - ad >0\,.
\end{equation}
Let us now consider the large-$k$ limit. Due to the third condition of~\eqref{ch2:hydro:full_momentum_causcon} being satisfied, in this limit all of the modes must go as $\omega \sim c \,k$ (allowing for the case that $c=0$). The controlling equation for the phase velocity $c$ is then given by
\begin{equation}
    c^2 P_{\rm d-ms}(c^2) = 0\,.
\end{equation}
In addition to the spurious mode $\omega_0$, there is another mode which is non-propagating in the limit of large $k$. This is one of the modes in the cubic gap~\eqref{ch5:mhd:cubic-gap}. We will also refer to this mode as a ``spurious mode" for reasons that will soon become apparent. The remaining ten modes are controlled by $P_{\rm d-ms}(c^2)$, which is a quintic polynomial in $c^2$. There exists a particular hydrodynamic frame, which we refer to as the ``decoupled frame" in analogy to~\cite{Hoult:2020eho}, in which the diffusion mode factors out of $P_{\rm d-ms}$. This frame is given by
\begin{equation}
\label{ch5:mhd:decoupled_frame}
    \ce_3 =\theta_{\parallel 2} = \theta_{\parallel 3} = \theta_{\perp 3} = \pi_3 = \sigma_3 = \rho_{\perp 3} = \beta_{\parallel 3} = 0\,.
\end{equation}
If one chooses to put the equations into the decoupled frame, the controlling equation factorizes according to
\begin{equation}
    P_{\rm d-ms}(c^2) = \lr{c^2 \nu_3 \rhophio + \lr{\pder{\rhophio}{\mmu_0} T_0 \gamma_{\parallel 2} + \gamma_{\parallel 3} \rhophio} \cos^2(\theta) + \gamma_{\perp 3} \rhophio \sin^2\theta} P_{\rm ms}(c^2)\,,
\end{equation}
where $P_{\rm ms}(c^2)$ is a quartic polynomial in $c^2$. One can demand the causality of the modes in the same manner as in the Alfv\'en channel, recalling that positivity of the roots for a polynomial with real coefficients requires more than positivity of the discriminant for polynomials of order $n\geq 4$.

We now circle back to the two non-propagating modes, which we referred to as ``spurious". The reason for the name is that the modes are removed from the spectrum entirely once the constraint equation $\de_\mu J^{\mu 0} = 0$ is fully taken into account.
Let us momentarily return to ideal order. The ideal-order constraint may be solved by writing the perturbations in the form~\cite{Hoult:2024qph}
\begin{equation}\label{ch5:mhd:constraint-sol}
    \begin{split}
        \delta \mu_{\Phi} &= \pder{\rhophio}{T_0} \delta \phi(\omega, k_j), \quad \delta \mmu = \pder{\rhophio}{T_0} \delta \psi(\omega,k_j), \quad \delta h^x = \pder{\rhophio}{T_0} \cos(\theta) \delta \chi(\omega,k_j)\,,\\
        \delta T &= - \pder{\rhophio}{\muphio} \delta \phi(\omega, k_j) - \pder{\rhophio}{\mmu_0} \delta \psi(\omega,k_j) - \rhophio \sin(\theta) \delta \chi(\omega,k_j)\,,
    \end{split}
\end{equation}
with the remaining perturbations still independent, and $\delta \phi$, $\delta \psi$, and $\delta \chi$ being some new degrees of freedom. On this solution, the spurious mode $\omega_0$ vanishes from the spectral curve even at ideal order.

The solution~\eqref{ch5:mhd:constraint-sol} to the ideal-order constraint equations will \textit{also} be a solution to the first-order constraint equations (recall that the constraint equation receives viscous corrections) in the following hydrodynamic frame which we term the ``constraint frame":
\begin{equation}\label{ch5:mhd:constraint-frame}
    \begin{split}
        \beta_{\parallel 1} = - \frac{{\cal D}_{\rhophio}}{\rhophio} \beta_{\parallel 4}, \quad \beta_{\parallel 2} = - \beta_{\parallel 4}, \quad \beta_{\parallel 3} &= - \frac{T_0}{\rhophio} \pder{\rhophio}{\mmu_0} \beta_{\parallel 4}, \quad \beta_{\parallel 5} = - \frac{\muphio}{\rhophio} \pder{\rhophio}{\muphi} \beta_{\parallel 4}\,,\\
        \beta_{\perp 1} = \beta_{\parallel 4}, \quad &\beta_{\perp 2} = \beta_{\parallel 4}\,.
    \end{split}
\end{equation}
In this frame, introducing the solution~\eqref{ch5:mhd:constraint-sol} leads to both of the ``spurious modes" vanishing from the spectral curve entirely. The other modes are not affected. If the hydrodynamic frame~\eqref{ch5:mhd:constraint-frame} is used without introducing the solution~\eqref{ch5:mhd:constraint-sol}, then the spurious modes do not vanish from the spectrum; however, they \textit{do} factorize out of the spectral curve. We note here that the constraint frame~\eqref{ch5:mhd:constraint-frame} and the decoupled frame~\eqref{ch5:mhd:decoupled_frame} introduced above are not consistent with one another unless $\beta_{\parallel 4} = 0$.

We have not applied the constraint~\eqref{ch5:mhd:constraint-sol} in the analysis above for a simple reason -- we would like to be able to apply the equivalence discussed in Chapter~\ref{ch3:sec_equivalence}, and it is more straightforward to do so if one does not take the constraint equation into account directly. Neglecting a constraint is, however, very unsatisfying. Various means of handling constraint equations on an operational level exist in the literature such as ``divergence cleaning"~\cite{Porth:2016rfi} and the CCZ4 formulation of general relativity~\cite{Rezzolla-Zanotti}; the situation with regard to including constraint equations in the non-linear causality analysis is murkier. We will not attempt to tackle the subject here.

Finally, speaking of the equivalence, we can note that the spectral curve of the system is of the form~\eqref{ch3:equiv:special-Q} with $\ell=1$. This is convenient, as it means that the equations are amenable to the equivalence between linear and non-linear causality, and the causality constraints in the linearized theory (e.g.~\eqref{ch5:mhd:causality_con_Alfven}) can be promoted to constraints which ensure the non-linear causality of the equations. It is somewhat surprising, however, that $\ell=1$; there are two ``spurious" modes in the spectrum which appear non-propagating in the rest frame. However, upon boosting, one of the modes remains non-propagating, and the other propagates along with the fluid velocity. In other words, the characteristic equation is given by
\begin{equation}
    Q = \lr{n{\cdot}\xi} \lr{u{\cdot}\xi} \tilde{Q} = 0,
\end{equation}
where $\tilde{Q}$ is a Lorentz scalar. Since $u{\cdot}\xi$ is a Lorentz-invariant factor, it does not contribute to the non-invariance of the characteristic equation.

With our investigation of a causal theory of one-form magnetohydrodynamics complete, let us now move on to the second extension to BDNK hydrodynamics that we will consider in this dissertation: relativistic superfluids.

\section{Superfluids}
\label{ch5:sec:superfluids}
Superfluids describe thermal matter in a spontaneously broken phase, where an operator has acquired a thermal expectation value and a collective massless degree of freedom (a Goldstone boson) has subsequently emerged. The prototypical example of a superfluid in a non-relativistic context is that of liquid helium, where the spontaneous breaking of a $U(1)$ symmetry (specifically, particle number) at the lambda point ($\sim 2.17$K for He-4) gives rise to a superfluid. 

Superfluids were first discovered experimentally in He-4 by Kapitza~\cite{Kapitza1938}, as well as by Allen and Misener~\cite{ALLEN1938}. Almost immediately thereafter, an explanation was put forward by Tisza~\cite{TISZA1,tisza:1938:cr1,tisza:1940:vol1,tisza:1940:vol2} in a series of papers, which postulated that superfluids could be thought of as two fluids coexisting independently on top of one another, with independent flows. This model was put on more rigorous footing by Landau three years later in 1941~\cite{Landau:1941,Landau:1941ussr}, explaining superfluidity in terms of collective excitations, rather than individual atoms. Landau emphasized that one should not literally consider two fluids on top of one another; rather, the effective description of the flow is that of two fluids. It was also at that time that Landau proposed what would come to be called the Landau criterion, whereby if the relative velocity between the superfluid and normal components grows too large, superfluidity begins to decay. The cores of neutron stars have also been postulated to be superfluidic in nature~\cite{MIGDAL:1959655,BAYM1969,Haskell:2017lkl}. The non-relativistic theory is discussed in e.g.~\cite{LL6}; for an interesting history of the discovery of the theory, one may refer to~\cite{Balibar:2017}. The relativistic theory has been worked on by many~\cite{ClarkThesis,putterman1974}, in particular Carter~\cite{Carter:1992} for the viscous theory. The modern formulation of the viscous theory of relativistic superfluid hydrodynamics was written down in~\cite{Herzog:2011ec,Bhattacharya:2011eea,Bhattacharya:2011tra}. In this section, we will consider the causal theory of relativistic, viscous superfluid hydrodynamics, which was first written down in my paper~\cite{Hoult:2024cyx}.

We will consider an equilibrium state in which an operator which is charged under a global $U(1)$ symmetry gains an thermal expectation value. We denote the phase of this expectation value by $\varphi$. This dynamical variable characterizes the equilibrium state along with the usual temperature, chemical potential\footnote{There is some subtlety in how to think about the chemical potential in terms of the transition from the unbroken phase to the broken phase. We will consider here systems sufficiently far enough away from the critical point so as to not worry about the transition. In addition, this also means we consider systems which are well-separated enough from the critical point to neglect the dynamics of the so-called ``amplitude mode"\cite{Pekker-Varma,Donos:2022xfd}, the gapped mode describing the (massive) fluctuations of the radial part of the operator.}, and fluid velocity.

\subsection{Thermodynamics}
To begin with, let us describe the equilibrium state via the use of the generating functional. The conserved quantities in the theory are the stress-energy tensor and the $U(1)$ charge current\footnote{We consider here only zero-form $U(1)$ symmetries.}. There is also the Goldstone boson $\varphi$, which is dynamical. Therefore, we will couple the fluid to a background metric $g_{\mu\nu}$ and background $U(1)$ gauge field $A_\mu$, such that
\begin{equation}
    W[g,A,\varphi] = \int d^{d+1}x \sqrt{-g} {\cal F}(g,A,\varphi)\,.
\end{equation}
Even though $\varphi$ is dynamical and not a background field, we will imprecisely refer to the collection $\{g,A,\varphi\}$ as ``source fields". Arbitrary variation of the generating functional gives
\begin{equation}
    \delta W[g,A,\varphi] = \int d^{d+1} x \sqrt{-g} \biggl[ \frac{1}{2} T^{\mu\nu} \delta g_{\mu\nu} + J^\mu \delta A_\mu + E \,\delta \varphi\biggr]\,.
\end{equation}
The symmetries of the theory with the background sources $(g,A)$ turned on are diffeomorphisms and a $U(1)$ gauge symmetry; with the background sources $(g,A)$ turned off, these reduce to the (physical) spacetime symmetries and global $U(1)$ symmetry. Under diffeomorphisms generated by a parameter $\chi_\mu$ and $U(1)$ gauge transformations generated by a gauge parameter $\lambda$, the source fields transform as
\begin{equation}
    g_{\mu\nu} \to g'_{\mu\nu} = g_{\mu\nu} + \Lied_\chi g_{\mu\nu}, \quad A_\mu \to A'_\mu = A_\mu + \Lied_\chi A_\mu + \de_\mu \lambda, \quad \varphi \to \varphi' = \varphi + \Lied_\chi \varphi + \lambda\,.
\end{equation}
i.e. $\varphi$ transforms like a phase. Demanding that variations of the generating functional vanish under diffeomorphisms and gauge transformations respectively, we find the conservation equations to be
    \begin{align}
        \nabla_\mu T^{\mu\nu} &= F^{\nu\lambda} J_\lambda\,,\qquad \nabla_\mu J^\mu = E\,.
    \end{align}
The equation for $\varphi$ is $E=0$; therefore, the equilibrium equation of motion for $\varphi$ is $\nabla_\mu J^\mu=0$.

Now, we wish to impose that the fluid is in an equilibrium state, and so let us once again repeat the procedure of Section~\ref{ch2:sec_uncharged}. Let us introduce a timelike Killing vector $K^\mu$, and a $U(1)$ gauge parameter $\Lambda$ which transforms as $\Lambda \to \Lambda' = \Lambda - \Lied_K \lambda$ under a $U(1)$ gauge transformation. We may then demand the following gauge-invariant equilibrium conditions:
    \begin{align}\label{ch5:super:equib-conditions}
        \Lied_K g_{\mu\nu} &= 0\,,\qquad \Lied_K A_\mu + \de_\mu \Lambda = 0\,,\qquad\Lied_K \varphi + \Lambda = 0\,.
    \end{align}
Let us now define the thermodynamic parameters that we use to parametrize the equation of state. First of all, we can define the temperature, chemical potential, and fluid velocity:
\begin{equation}\label{ch5:super:thermo-defs}
    T = \frac{T_0}{\sqrt{-K^2}}, \quad \mu = \frac{K^\mu A_\mu + \Lambda}{\sqrt{-K^2}}, \quad u^\mu = \frac{K^\mu}{\sqrt{-K^2}}\,.
\end{equation}
Next, while we could parametrize the equation of state via $\varphi$ directly, it is a bit inconvenient to do so. After all, $\varphi$ is not a gauge-invariant quantity. Let us therefore instead work with the following gauge-invariant parameter:
\begin{equation}
\label{ch5:super:defn-xi}
    \xi_\mu \equiv - \de_\mu \varphi + A_\mu\,.
\end{equation}
It follows immediately from this definition that
$\nabla_\mu \xi_\nu - \nabla_\nu \xi_\mu = F_{\mu\nu}\,.$
We note that $\xi_\mu$ has a microscopic definition in terms of the Goldstone boson $\varphi$. Therefore, in the following, we will take $\xi_\mu$ to stand on the same footing as $T^{\mu\nu}$ and $J^\mu$. In terms of derivative counting, we take $\xi_\mu \sim \oser{1}$; therefore, $\varphi \sim \oser{\de^{-1}}$. The vector $\xi^\mu$ is called the ``superfluid velocity", in connection with the Landau-Tisza two-fluid formulation. It also follows directly from the definition of $\xi_\mu$ in equation~\eqref{ch5:super:defn-xi} that
\begin{equation}
\label{ch5:super:Josephson}
    u^\mu \xi_\mu = \mu\,.
\end{equation}
The chemical potential $\mu$ is then (in the rest frame) the time-component of $\xi_\mu$, rather than an independent variable. The relationship~\eqref{ch5:super:Josephson} is called the Josephson equation. The superfluid velocity may be decomposed with respect to $u_\mu$ to yield
\begin{equation}
    \xi_\mu = - \mu\, u_\mu + \zeta_\mu\,.
\end{equation}
The new parameter $\zeta_\mu \equiv \Delta_{\mu}^{\,\,\,\,\nu}\xi_\nu$ is the transverse superfluid velocity, or the relative superfluid velocity. The equilibrium state of the superfluid may then be characterized by $T$, $\mu$, $u^\mu$, and $\zeta^\mu$. At zeroth order in the derivative expansion, the generating functional is given by
\begin{equation}
    W[g,A,\varphi] = \int d^{d+1}x \sqrt{-g} \lr{p(T,\mu,\zeta) + \oser{\de}}\,,
\end{equation}
where $p$ is the isotropic pressure, $\zeta = \sqrt{\zeta_\mu \zeta^\mu}$. Let us supplement this generating functional with the Gibbs-Duhem relation
$    dp = s \, dT + \rho\, d\mu + \tilde{\rho}\,d\zeta\,,$
where $s$ is the entropy density, $\rho$ is the total $U(1)$ charge density (comprising both the normal and superfluid components), and $\tilde{\rho}$ is the flux density for the superfluid component. Varying the generating functional leads to the following equilibrium constitutive relations at zeroth-order in the derivative expansion.
\begin{subequations}\label{ch5:super:ideal-order-conrel}
    \begin{align}
        T^{\mu\nu} &= \epsilon u^\mu u^\nu + \tilde{\epsilon} z^\mu z^\nu + p \delperp^{\mu\nu} + 2 \mu \tilde{\rho} u^{(\mu} z^{\nu)}\,,\\
        J^\mu &= \rho u^\mu  + \tilde{\rho} z^\mu\,,\\
        \xi_\mu &= - \mu u_\mu + \zeta_\mu\,,
    \end{align}
\end{subequations}
where $z^\mu = \zeta^\mu/\zeta$ is the unit vector in the direction of $\zeta$, $\delperp^{\mu\nu} = u^\mu u^\nu + g^{\mu\nu} - z^\mu z^\nu$ is the projector perpendicular to both $u^\mu$ and $z^\mu$, as in the case of dMHD.
We also find
$        \epsilon \equiv - p + s T + \rho \mu\,,$
$        \tilde{\epsilon} \equiv p - \zeta \tilde{\rho}\,,$
where $\epsilon$ is the total energy density, and $\tilde{\epsilon}$ is the anisotropic contribution to the pressure in the $z^\mu$ direction due to the superfluid velocity. Turning off the background gauge field, the equations of motion are the conservations equations and the (definitional) antisymmetric equation for $\xi_\mu$:
    \begin{align}\label{ch5:super:EoM}
        \nabla_\mu T^{\mu\nu} = 0\,,\qquad \nabla_\mu J^\mu = 0\,,\qquad 2 \nabla_{[\mu} \xi_{\nu]} = 0\,.
    \end{align}
Let us now account for which one-derivative quantities vanish in equilibrium. We are interested in a superfluid for which the equilibrium state respects $P$ and $T$. With some work, we find that the following scalars
\begin{equation}
    \begin{gathered}
        s_1 =\frac{u^\alpha \nabla_\alpha T}{T}, \quad s_2 = u^\alpha \nabla_\alpha \lr{\frac{\mu}{T}}, \quad s_3 = u^\alpha \nabla_\alpha \lr{\frac{\zeta}{T}}, \quad s_4 =z^\alpha \nabla_\alpha \lr{\frac{\mu}{T}} \,,\\
        s_5 = z^\alpha \lr{\frac{\nabla_\alpha T}{T} + a_\alpha}, \quad s_6 =\nabla_\mu u^\mu , \quad s_7 = z^\alpha z^\beta \nabla_\alpha u_\beta, \quad s_8 = \frac{1}{T^{d-1}} \Delta^{\alpha\beta} \nabla_\alpha \lr{\frac{\tilde{\rho} z_\beta}{T}}\,,
    \end{gathered}
\end{equation}
vectors,
\begin{equation}\label{ch5:super:basis-vectors}
    \begin{gathered}
        V_1^\mu = \delperp^{\mu\nu} \lr{\frac{\nabla_\nu T}{T} + a_\nu}, \quad V_2^\mu = \delperp^{\mu\nu} \nabla_\nu \lr{\frac{\mu}{T}},\\
        V_3^\mu = 2 \delperp^{\mu\nu} z^\lambda \nabla_{(\lambda} u_{\nu)}, \quad V_4^\mu = \delperp^{\mu\nu}\lr{u^\lambda \nabla_\lambda z_\nu + z^\lambda \nabla_\nu u_\lambda},\\
        V_5^\mu = \delperp^{\mu\nu} \biggl[ \nabla_\nu \lr{\frac{\zeta}{T}} + \frac{\mu}{T} \lr{u^\lambda \nabla_\lambda z_\nu + z^\lambda \nabla_\lambda u_\nu} + \frac{\zeta}{T} \lr{\frac{\nabla_\nu T}{T} - z^\lambda \nabla_\lambda z_\nu} \biggr]\,,
    \end{gathered}
\end{equation}
and tensor
\begin{equation}\label{ch5:super:basis-tensor}
    \sigma^{\mu\nu}_u = \lr{\delperp^{\mu\alpha}\delperp^{\nu\beta} + \delperp^{\mu\beta} \delperp^{\nu\alpha} - \frac{2}{d-1} \delperp^{\mu\nu} \delperp^{\alpha\beta}}\nabla_\alpha u_\beta\,,
\end{equation}
vanish in equilibrium. Of these, the vanishing of $s_{1-7}$, $V_{1-4}^\mu$, and $\sigma_u^{\mu\nu}$ were obtained from the conditions~\eqref{ch5:super:equib-conditions}, while $s_8$ is the Goldstone equation of motion $\nabla_\mu J^\mu = 0$, and $V_5^\mu$ can be obtained from the demand (which must hold in equilibrium) that
\begin{equation}
    \delperp^{\rho\nu} \xi^\mu \lr{\nabla_\mu \xi_\nu - \nabla_\nu \xi_\mu} = 0\,.
\end{equation}

In the following, we will be restricting ourselves to conformal superfluids for the sake of simplicity. We therefore must impose the additional condition that under a Weyl transformation of the background metric $g_{\mu\nu} \to g'_{\mu\nu} = e^{-2\phi} g_{\mu\nu}$, the vanishing scalars, vectors, and tensors transform homogeneously. For a list of how all of the hydrodynamic variables transform under a Weyl transformation, refer to Table~\ref{ch5:super:conformal-table}.

\begin{table}[t!]
\centering
\begin{tabular}{ |c|c|c|c|c|c|c|c|c|c|c|c|c|c|c|c| } 
 \hline
 Quantity & $T$ & $\mu$ & $u^\mu$ & $u_\mu$ & $\xi^\mu$ & $\xi_\mu$ & $z^\mu$ & $z_\mu$ & $\zeta$ & $p$ & $\epsilon$ & $\tilde{\epsilon}$ & $s$ & $\rho$ & $\tilde{\rho}$ \\ 
 \hline
 $\boldsymbol{w}$ & 1 & 1 & 1 & $-1$ & 2 & 0 & 1 & $-1$ & 1 & $d+1$ & $d+1$ & $d+1$ & $d$ & $d$ & $d$ \\ 
 \hline
\end{tabular}
\caption{A table for the Weyl weights of many of the thermodynamic quantities appearing in this section. A quantity is said to transform with a Weyl weight $\boldsymbol{w}$ if it transforms as ${\cal O} \to e^{\boldsymbol{w} \phi} {\cal O}$ under the Weyl transformation $g_{\mu\nu} \to e^{-2 \phi} g_{\mu\nu}$. Reproduced from~\cite{Hoult:2024cyx}.}
\label{ch5:super:conformal-table}
\end{table}

In particular, $s_1$, $s_4$, and $s_5$ do not transform covariantly, and so we must work with covariant linear combinations of these scalars. The set of scalars for the conformal theory are given by~\cite{Hoult:2024cyx}
\begin{equation}\label{ch5:super:basis-scalars}
    \begin{gathered}
        \mbb{s}_1 \equiv s_1 + \frac{1}{d} s_6, \quad \mbb{s}_2 = s_2, \quad \mbb{s}_3 = s_3, \quad \mbb{s}_4 = s_4\,,\\
        \mbb{s}_5 = s_5, \quad \mbb{s}_6 = s_7 - \frac{1}{d} s_6, \quad \mbb{s}_7 = s_8\,. 
    \end{gathered}
\end{equation}
All of the scalars~\eqref{ch5:super:basis-scalars} transform with Weyl weight $\boldsymbol{w} = 1$. The vectors are the same as~\eqref{ch5:super:basis-vectors}, and transform with Weyl weight $\boldsymbol{w} = 2$. Finally, the transverse traceless tensor $\sigma^{\mu\nu}_u$ is also the same as~\eqref{ch5:super:basis-tensor}, and transforms with Weyl weight $\boldsymbol{w} = 3$. With these ``building blocks" in hand, we are now in a position to write down the constitutive relations for a conformal relativistic superfluid.

\subsection{Constitutive relations}
The constitutive relations for the superfluid relate the stress-energy tensor $T^{\mu\nu}$, the charge current $J^\mu$, and the superfluid velocity $\xi^\mu$ to the parameters characterizing the equilibrium state: the temperature $T$, the fluid velocity $u^\mu$, the $U(1)$ chemical potential $\mu$, and the transverse superfluid velocity $\zeta_\mu$. Now, let us consider taking the fluid slightly out of equilibrium. Then the equilibrium description is no longer valid; nevertheless, we may still use the fields $T(x)$, $\mu(x)$, $u^\mu(x)$, $\zeta_\mu(x)$ to characterize the system. In equilibrium, these variables reduce to their equilibrium values.

We may decompose the stress-energy tensor, charge current, and superfluid velocity with respect to $u^\mu$ and $z^\mu = \zeta^\mu/\zeta$. Then
\begin{subequations}\label{ch5:super:symmetry-decomposition}
    \begin{align}
        T^{\mu\nu} &= {\cal E} u^\mu u^\nu + {\cal V} z^\mu z^\nu + {\cal P} \delperp^{\mu\nu} + 2 {\cal U} u^{(\mu} z^{\nu)} + 2 {\cal Q}^{(\mu} u^{\nu)} + 2 {\cal R}^{(\mu} z^{\nu)} + {\cal T}^{\mu\nu}\,,\\
        J^\mu &= {\cal N} u^\mu + {\cal S} z^\mu + {\cal J}^\mu\,,\\
        \xi^\mu &= - {\cal M} u^\mu + {\cal Z} z^\mu + {\cal X}^\mu\,,
    \end{align}
    where the negative sign in front of ${\cal M}$ is a matter of definition. The curly variables are given by
    \begin{equation}
        \begin{gathered}
            {\cal E} \equiv T^{\mu\nu} u_\mu u_\nu, \quad {\cal V} \equiv T^{\mu\nu} z_\mu z_\nu, \quad {\cal P} \equiv \frac{1}{d-1} T^{\mu\nu}\Delta_{\mu\nu}^{\perp}, \quad {\cal U} \equiv - T^{\mu\nu} u_\mu z_\nu\,,\\
            {\cal N} \equiv - J^\mu u_\mu,\quad {\cal S} \equiv J^\mu z_\mu, \quad {\cal M} \equiv u^\mu \xi_\mu, \quad {\cal Z} \equiv \xi_\mu z^\mu\,,\\
            {\cal Q}^\alpha \equiv -\Delta^{\alpha}_{\perp\mu} u_\nu T^{\mu\nu}, \quad {\cal R}^\alpha \equiv \Delta^\alpha_{\perp\mu} z_\nu T^{\mu\nu},\quad {\cal J}^\alpha \equiv \delperp^{\alpha\mu}J_\mu, \quad {\cal X}^\alpha \equiv \delperp^{\alpha\mu}\xi_\mu\,,\\
            {\cal T}^{\mu\nu} \equiv\frac{1}{2}\lr{\delperp^{\mu\alpha} \delperp^{\nu\beta} + \delperp^{\mu\beta} \delperp^{\nu\alpha} - \frac{2}{d-1} \delperp^{\mu\nu} \delperp^{\alpha\beta}}T_{\alpha\beta}\,.
        \end{gathered}
    \end{equation}
\end{subequations}
In a conformal theory, the stress-energy tensor must be traceless. This leads to the relation
\begin{equation}\label{ch5:super:notrace}
    {\cal V} = {\cal E} - \lr{d-1} {\cal P}\,. 
\end{equation}
With the general form of the constitutive relations set, let us now investigate the theory at zeroth and first order in the derivative expansion.
\subsubsection{Ideal order}
In the following, we neglect subscripts on derivatives, and assume we work in a basis of $T$, $\mu$, $\zeta$. At ideal order, the constitutive relations are that of equilibrium, but with variables promoted to slowly varying functions of spacetime. In terms of the decomposition~\eqref{ch5:super:symmetry-decomposition}, the constitutive relations are given by
\begin{equation}
    \begin{gathered}
    {\cal E} = \epsilon, \quad {\cal V} = \tilde{\epsilon}, \quad {\cal P} = p, \quad {\cal U} = \mu \tilde{\rho}\,, \quad {\cal N} = n, \quad {\cal M} = \mu, \quad {\cal Z} = \zeta\,,\\
    {\cal Q}^\mu = {\cal R}^\mu ={\cal T}^{\mu\nu} = {\cal S} =  {\cal J}^\mu = {\cal X}^\mu = 0\,.
    \end{gathered}
\end{equation}
In a conformal theory, equation~\eqref{ch5:super:notrace} then becomes
\begin{equation}
    \tilde{\epsilon} = \epsilon - (d-1) \,p\,,
\end{equation}
an expression which also follows from the conformal equation of state
\begin{equation}
    p = a\, T^4 f\lr{\frac{\mu}{T}, \frac{\zeta}{T}}\,.
\end{equation}
Let us now consider the equations of motion. Inserting the ideal-order constitutive relations~\eqref{ch5:super:ideal-order-conrel} into the equations of motion~\eqref{ch5:super:EoM}, there are four scalar equations, and three vector equations~\cite{Hoult:2024cyx}. The scalar equations are given by
\begin{subequations}
    \begin{align}
        u_\nu \nabla_\mu T^{\mu\nu} = 0 \implies& \lr{d+1} \epsilon \,\mbb{s}_1 + T \pder{\epsilon}{\mu} \mbb{s}_2 + T \pder{\epsilon}{\zeta} \mbb{s}_3 + T \tilde{\rho} \mbb{s}_4 + 2 \mu \tilde{\rho} \mbb{s}_5\nnl
        &- \zeta \tilde{\rho}\, \mbb{s}_6 + \mu T^d \mbb{s}_7 = 0\,,\\
        z_\nu \nabla_\mu T^{\mu\nu} = 0 \implies& \lr{d+1} \mu \tilde{\rho} \mbb{s}_1 + T \lr{\tilde{\rho} + \mu \pder{\tilde{\rho}}{\mu}} \mbb{s}_2 + \mu T \pder{\tilde{\rho}}{\zeta} \mbb{s}_3\nnl
        &+ T \rho \mbb{s}_4 + \lr{\epsilon + \tilde{\epsilon}} \mbb{s}_5 + \mu \tilde{\rho} \mbb{s}_6 - \zeta T^d \mbb{s}_7 = 0\,,\\
        \nabla_\mu J^\mu = 0 \implies & d \rho \mbb{s}_1 + T \pder{\rho}{\mu} \mbb{s}_2 + T \pder{\rho}{\zeta} \mbb{s}_3 + \tilde{\rho} \mbb{s}_5 + T^d \mbb{s}_7 = 0\,,\\
        u^\mu z^\nu \lr{\de_\mu \xi_\nu - \de_\nu \xi_\mu} = 0 \implies& \zeta \mbb{s}_1 + T \mbb{s}_3 - T \mbb{s}_4 - \mu \mbb{s}_5 + \zeta \mbb{s}_6 = 0\,,
    \end{align}
\end{subequations}
while the vector equations are given by
\begin{subequations}
    \begin{align}
        \Delta^\alpha_{\perp\nu}\nabla_\mu T^{\mu\nu} = 0 \implies& \lr{p + \epsilon} V_1^\mu + T \rho V_2^\mu + T \tilde{\rho} V_5^\mu = 0\,,\\
        \delperp^{\alpha\mu}u^\nu \lr{\de_\mu \xi_\nu - \de_\nu \xi_\mu} = 0 \implies& \mu V_1^\mu + T V_2^\mu - \zeta V_4^\mu = 0\,,\\
        \delperp^{\alpha\mu} z^\nu \lr{\de_\mu \xi_\nu - \de_\nu \xi_\mu} = 0 \implies& \mu V_4^\mu - T V_5^\mu = 0\label{ch5:super:constraint-eq}\,.
    \end{align}
\end{subequations}
Note that $V_3^\mu$ does not appear in these equations, nor does $\sigma_u^{\mu\nu}$. 

We now restrict ourselves to $d=3$. We would like to consider linearized perturbations about equilibrium, to understand the mode structure present in the theory. In order to do so, there is one detail we must first consider. Of the $11$ independent equations in~\eqref{ch5:super:EoM}, only $8$ are dynamical. The remaining $3$ equations are constraint equations. Unlike in the case of one-form MHD, we will here take the constraint equations into account; the followup to this is that we will not compare to the non-linear causality constraints.

With this in mind, let us consider perturbations about the rest frame $u^\mu = \delta^\mu_0$ with $\zeta_\mu$ aligned along the $z$-axis. In other words, we consider
\begin{equation}
    \begin{gathered}
        T(x) = T_0 + \delta T(\omega,k) e^{-i \omega t + i k_j x^j}, \quad \mu(x) = \mu_0 + \delta \mu(\omega,k) e^{-i \omega t + i k_j x^j}\,,\\
        u^\mu(x) = \delta^\mu_t + \delta u^\mu(\omega,k) e^{-i \omega t + i k_j x^j}, \quad \zeta^\mu(x) = \zeta_0 \delta^\mu_z + \delta \zeta^\mu(\omega,k) e^{-i \omega t + i k_j x^j}\,. 
    \end{gathered}
\end{equation}
We demand that $u^2 = -1$ and $u^\mu \zeta_\mu = 0$ hold to quadratic order in the perturbations, as well as demanding that the constraint equations~\eqref{ch5:super:constraint-eq} are satisfied. This allows us to further decompose the perturbations to $u^\mu$ and $\zeta^\mu$ as
\begin{subequations}\label{ch5:super:constraint-eq-solution}
    \begin{align}
        \delta u^\mu(\omega,k) &= \delta^\mu_z \delta u_{\parallel}(\omega,k) + \delta u^\mu_{\perp}(\omega,k)\,,\\
        \delta \zeta^\mu &= \zeta_0 \delta^\mu_t \delta u_{\parallel}(\omega,k) + \mu_0 \delta u_{\perp}^\mu(\omega,k) + \delta \hat{\zeta}(\omega,k) \hat{k}^\mu\,, 
    \end{align}
\end{subequations}
where $\delta u_{\parallel} = \delta u^z$, $\delta u_\perp^\mu = \delperp^{\mu\nu} \delta u_{\nu}$, and $\hat{k}^\mu = \delta^\mu_i k^i/|k|$ is a vector pointing in the direction of the wave vector. There are only $3$ components to $\delta u^\mu$, and only $1$ component to $\delta \zeta^\mu$. That the perturbations to $\zeta^\mu$ have only one independent component is reflective of the true underlying degree of freedom, i.e. the Goldstone boson.

Let us now briefly enumerate the modes that appear in the spectral curve. We will only express the modes schematically. There are six hydrodynamic modes, four of which are propagating. Two of these are the sounds modes in the normal component of the fluid, with speed $v_s$; the other is the so-called ``second sound" with speed $v_2$, which describe the propagation of entropy in the fluid.\footnote{In~\cite{Gouteraux:2022qix}, an interesting observation was made that the Landau criterion, i.e. the condition that the transverse superfluid velocity not grow too large, could be observed in relativistic superfluid hydrodynamics. Specifically, when the transverse superfluid velocity grows large enough, one of the elements of the static susceptibility matrix $\chi_{ab}$ diverges, and then becomes negative. This leads directly to a hydrodynamic instability, with one of the sound modes moving into the upper-half complex plane. While we will not make use of this result here, it would be interesting to investigate this instability further.} Finally, there is a shear mode, and a charge diffusion mode, both of which are zero at ideal order. Schematically,
    \begin{align}
        \omega_s &= \pm v_s k \,,\quad \omega_2 = \pm v_2 k \,,\quad \omega_\sigma = 0\,,\quad \omega_\eta =0\,.
    \end{align}
The expressions for $v_s$, $v_2$ are fairly complex at arbitrary $\zeta$, and so we do not reproduce them here. Let us now turn to the first-order result.

\subsubsection{First order}
At first order, there are two types of corrections that can appear in the constitutive relations -- hydrostatic corrections which come from the generating functional, and non-hydrostatic corrections. For a conformal superfluid which respects parity and time-reversal symmetry of the equilibrium state, there are no first-order scalars that one can construct which are non-vanishing in equilibrium~\cite{Hoult:2024cyx}. Therefore, the only contributions to the constitutive relations at first order are non-hydrostatic contributions, which may be written down in terms of the first-order data that vanishes in equilibrium. 

With this in mind, the constitutive relations to first order in derivatives are given in terms of the decomposition~\eqref{ch5:super:symmetry-decomposition} by\footnote{The transport parameters $\varphi_n$ and the Goldstone $\varphi$ are unrelated.}
\begin{subequations}\label{ch5:super:first-order-corrections-gen}
    \begin{alignat}{4}
        &\,\,\,\,\,{\cal E} &&= \epsilon + \sum_{n=1}^7 \ce_n \mbb{s}_n, \quad &&\,\,\,\,\,{\cal V} &&= \tilde{\epsilon} + \sum_{n=1}^7 \lr{\ce_n - \lr{d-1} \pi_n} \mbb{s}_n\,,\\
        &\,\,\,\,\,{\cal U} &&= \mu \tilde{\rho} + \sum_{n=1}^7 \varphi_n \mbb{s}_n, \quad &&\,\,\,\,{\cal N} &&= n + \sum_{n=1}^7 \nu_n \mbb{s}_n,\\
        &\,\,\,\,\,{\cal S} &&= \sum_{n=1}^7 \lambda_n \mbb{s}_n\,, && \,\,\,\,{\cal P} &&= p + \sum_{n=1}^7 \pi_n \mbb{s}_n, \\
        &\,\,{\cal M} &&= \mu + \sum_{n=1}^7 \alpha_n \mbb{s}_n, \quad &&\,\,\,\,{\cal Z} &&= \zeta + \sum_{n=1}^7 \beta_n \mbb{s}_n\,,\\
        &\,{\cal Q}^\mu &&= \sum_{n=1}^5 \theta_n V_n^\mu, \quad &&{\cal R}^\mu &&= \sum_{n=1}^5 \varrho_n V_n^\mu,\\
        &\,{\cal X}^\mu &&= \sum_{n=1}^5 \varsigma_n V_n^\mu\,,\quad&&{\cal J}^\mu &&= \sum_{n=1}^5 \gamma_n V_n^\mu,\\
        &{\cal T}^{\mu\nu} &&= - \eta \sigma^{\mu\nu}_u\,.
    \end{alignat}
\end{subequations}
Looking at the above constitutive relations for a conformal superfluid, we see that there are seventy\footnote{This is after applying conformal symmetry; a non-conformal superfluid has eighty-five transport parameters.} transport parameters $\{\ce_n, \pi_n, \varphi_n, \nu_n, \lambda_n$, $ \alpha_n, \beta_n, \theta_n, \varrho_n, \gamma_n, \varsigma_n\}$, dwarfing even the dMHD case. The difference comes down to symmetry; the magnetic flux direction $h^\mu$ was a pseudovector, while the transverse fluid velocity $\zeta^\mu$ is a regular vector. We will now attempt to determine the physical transport coefficients. Let us begin by considering the following frame transformations:
\begin{equation}
    \begin{gathered}
        T \to T' = T + \delta T, \quad \mu \to \mu' = \mu + \delta \mu\,,\\
        u'^\mu = u^\mu + \delta u^\mu = u^\mu + \delta u_{\parallel} z^\mu + \delta u^\mu_{\perp}\,,\\
        \zeta^\mu \to \zeta'^\mu = \zeta^\mu + \delta \zeta^\mu = \zeta^\mu + \zeta \delta u_{\parallel} u^\mu + \delta \zeta^\mu_{\perp}\,,
    \end{gathered}
\end{equation}
where the perturbations are all $\oser{\de}$, and the perturbations $\delta u^\mu_{\perp}$ and $\delta \zeta^\mu_{\perp}$ lie in the plane transverse to both $u^\mu$ and $\zeta^\mu$. For future convenience, we can write down the transformations of $\zeta$ and $z^\mu$ as
\begin{equation}
\label{ch5:super:zeta-z-transform}
    \zeta \to \zeta' = \zeta + \delta \zeta, \quad z^\mu \to z'^\mu = z^\mu + \delta z^\mu = z^\mu + \delta u_{\parallel} u^\mu + \frac{1}{\zeta} \delta \zeta_{\perp}^\mu\,.
\end{equation}
In~\eqref{ch5:super:zeta-z-transform}, we assume $\zeta \neq 0$ (as otherwise the unit vector $z^\mu$ is not well-defined). With these transformations, the transformation of the scalar viscous corrections $\delta {\cal E} \equiv {\cal E}'-{\cal E}$ etc. for a generic (i.e. not necessarily conformal) superfluid are given by
\begin{subequations}\label{ch5:super:frame-transforms-scalar}
    \begin{align}
        \delta {\cal E} &= -\pder{\epsilon}{T} \delta T - \pder{\epsilon}{\mu} \delta \mu - \pder{\epsilon}{\zeta} \delta \zeta - 2 \mu \tilde{\rho} \delta u_{\parallel}\,,\\
        \delta {\cal V} &= - \pder{\tilde{\epsilon}}{T} \delta T - \pder{\tilde{\epsilon}}{\mu} \delta \mu - \pder{\tilde{\epsilon}}{\zeta} \delta \zeta - 2 \mu \tilde{\rho} \delta u_{\parallel}\,,\\
        \delta {\cal P} &= - \pder{p}{T} \delta T - \pder{p}{\mu} \delta \mu - \pder{p}{\zeta} \delta \zeta\,,\\
        \delta {\cal U} &= - \mu \pder{\tilde{\rho}}{T} \delta T - \lr{\tilde{\rho} + \mu \pder{\tilde{\rho}}{\mu}} \delta \mu - \mu \pder{\tilde{\rho}}{\zeta} \delta \zeta - \lr{\epsilon + \tilde{\epsilon}}\delta u_{\parallel}\,,\\
        \delta {\cal N} &= - \pder{\rho}{T} \delta T - \pder{\rho}{\mu} \delta \mu - \pder{\rho}{\zeta} \delta \zeta - \tilde{\rho} \delta u_{\parallel}\,,\\
        \delta {\cal S} &= - \pder{\tilde{\rho}}{T} \delta T - \pder{\tilde{\rho}}{\mu} \delta \mu - \pder{\tilde{\rho}}{\zeta} \delta \zeta - \rho \delta u_{\parallel}\,,\\
        \delta {\cal M} &= - \delta \mu + \zeta \delta u_{\parallel}\,,\\
        \delta {\cal Z} &= - \delta \zeta + \mu \delta u_{\parallel}\,,
    \end{align}
\end{subequations}
while the transformations of the vector viscous corrections $\delta {\cal Q}^\mu = {\cal Q}'^\mu - {\cal Q}^\mu$ etc. are given by
\begin{subequations}\label{ch5:super:frame-transforms-vector}
    \begin{alignat}{4}
        &\delta {\cal Q}^\mu &&= - \lr{p + \epsilon} \delta u^\mu_{\perp} - \frac{\mu\tilde{\rho}}{\zeta} \delta \zeta^\mu_{\perp}\,,\quad&&
        \delta {\cal R}^\mu &&=  - \mu \tilde{\rho} \delta u^\mu_{\perp} + \tilde{\rho} \delta \zeta_{\perp}^\mu\,,\\
        &\delta {\cal J}^\mu &&= - \rho \delta u_{\perp}^\mu - \frac{\tilde{\rho}}{\zeta} \delta \zeta_{\perp}^\mu\,,\quad &&
        \delta {\cal X}^\mu &&= \mu \delta u^\mu_\perp - \delta \zeta_\perp^\mu\,.
    \end{alignat}
\end{subequations}
Finally, the tensor contribution is invariant, with
$    \delta {\cal T}^{\mu\nu} = 0\,.$
Using the subscript $1$ to denote the first-order viscous contributions, there are four frame-invariant scalar viscous corrections we can write down. Schematically,
\begin{subequations}\label{ch5:super:frame-invars-scalar}
    \begin{align}
        {\cal F} &\equiv {\cal V}_1 - \mathbb{a}_1 {\cal E}_1 - \mathbb{a}_2 {\cal N}_1 - \mathbb{a}_3 {\cal S}_1 - \mathbb{a}_4 {\cal Z}_1\,,\\
        {\cal G} &\equiv {\cal P}_1 - \mathbb{b}_1 {\cal E}_1 - \mathbb{b}_2 {\cal N}_1 - \mathbb{b}_3 {\cal S}_1 - \mathbb{b}_4 {\cal Z}_1\,,\\
        {\cal L} &\equiv {\cal U}_1 - \mathbb{c}_1 {\cal E}_1 - \mathbb{c}_2 {\cal N}_1 - \mathbb{c}_3 {\cal S}_1 - \mathbb{c}_4 {\cal Z}_1\,,\\
        {\cal H} &\equiv {\cal M}_1 - \mathbb{d}_1 {\cal E}_1 - \mathbb{d}_2 {\cal N}_1 - \mathbb{d}_3 {\cal S}_1 - \mathbb{d}_4 {\cal Z}_1\,.
    \end{align}
\end{subequations}
The coefficients $\mathbb{a}_i, \mathbb{b}_i,\mathbb{c}_i,\mathbb{d}_i$, $d\in \{1,..,4\}$ are purely thermodynamic in nature. They may be obtained in detail by performing the frame transformations~\eqref{ch5:super:frame-transforms-scalar} on the frame-invariant quantities~\eqref{ch5:super:frame-invars-scalar}. There are also two vector frame invariants we can form:
\begin{subequations}
    \begin{align}
        {\cal L}^\mu &\equiv {\cal Q}^\mu_1 - \frac{(p+\epsilon) \zeta + \mu^2 \tilde{\rho}}{\mu \tilde{\rho} + \zeta \rho} {\cal J}_1^\mu + \frac{T s \tilde{\rho}}{\mu \tilde{\rho} + \zeta \rho} {\cal X}_1^\mu\,,\\
        {\cal K^\mu} &\equiv {\cal R}_1^\mu + \tilde{\rho} {\cal X}_1^\mu\,.
    \end{align}
\end{subequations}
We now fix the hydrodynamic frame to be the ``Landau-Lifshitz-Clark-Putterman" frame (LLCP frame), so called due to its use by Landau and Lifshitz~\cite{LL6}, Clark~\cite{ClarkThesis}, and Putterman~\cite{putterman1974}. The LLCP frame is defined by
\begin{equation}
    {\cal E}_1 = {\cal N}_1 = {\cal S}_1 = {\cal Z}_1 = {\cal J}_1^\mu = {\cal X}_1^\mu = 0\,.
\end{equation}
Let us now consider the conformal case; then ${\cal V}_1 = - (d-1) {\cal P}_1$ due to the tracelessness of the stress-energy tensor. Referring back to the scalar frame invariants~\eqref{ch5:super:frame-invars-scalar}, this imposes the frame-invariant relation
\begin{equation}
    {\cal F} = {\cal V}_1 = - \lr{d-1} {\cal P}_1 = - \lr{d-1} {\cal G}\,.
\end{equation}
There are therefore three scalar corrections (${\cal G}$, ${\cal L}$, ${\cal H}$), two vector corrections (${\cal L}^\mu$, ${\cal K}^\mu$), and one tensor correction (${\cal T}^{\mu\nu}$) for a conformal superfluid in the LLCP frame. These corrections may be decomposed with respect to the scalars~\eqref{ch5:super:basis-scalars}, vectors~\eqref{ch5:super:basis-vectors}, and tensor~\eqref{ch5:super:basis-tensor} that vanish in equilibrium:
\begin{subequations}\label{ch5:super:frame-invar-conrel}
    \begin{align}
        T^{\mu\nu} &= \epsilon u^\mu u^\nu + \lr{\tilde{\epsilon} - \lr{d-1} \sum_{n=1}^7 g_n \mbb{s}_n} z^\mu z^\nu + \lr{p + \sum_{n=1}^7 g_n \mbb{s}_n}\delperp^{\mu\nu} \\
        &+ 2 \lr{\mu \tilde{\rho} + \sum_{n=1}^7\ell_{\parallel n} \mbb{s}_n}u^{(\mu} z^{\nu)}+ 2 \lr{\sum_{n=1}^5 \ell_{\perp n} V_n^{(\mu}}u^{\nu)} + 2 \lr{ \sum_{n=1}^5 k_n V_n^{(\mu}}z^{\nu)} \nnl
        &- \eta \sigma^{\mu\nu}_u\,,\nnl
        J^\mu &= \rho u^\mu + \tilde{\rho} z^\mu\,,\\
        \xi_\mu &= - \lr{\mu + \sum_{n=1}^7 h_n \mbb{s}_n}u_\mu + \zeta z_\mu\,,
    \end{align}
\end{subequations}
where $g_n$, $\ell_{\parallel n}$, $h_n$, $\ell_{\perp n}$, and $k_n$ are defined as the coefficients of ${\cal G}$, ${\cal L}$, ${\cal H}$, ${\cal L}^\mu$, and ${\cal K}^\mu$ respectively decomposed with respect to the building blocks. Let us now apply the ideal-order equations of motion~\eqref{ch5:super:ideal-order-conrel} and eliminate four scalars and three vectors. For the scalars, we will choose to eliminate $\mbb{s}_{1,2,3,4}$, leaving just $\mbb{s}_{5,6,7}$. Of the vectors we choose to eliminate $V_{2,4,5}$, leaving only $V_1^\mu$ and the mandatory\footnote{Recall that the ideal-order equations of motion do not contain $V_3^\mu$.} choice $V_3^\mu$. We may then write the constitutive relations in the form
\begin{subequations}\label{ch5:hydro:LLCP-conrel-preonsager}
    \begin{align}
        T^{\mu\nu} &= \epsilon u^\mu u^\nu + \lr{\tilde{\epsilon} - \lr{d-1} \sum_{n=5}^7 \bar{\pi}_n \mbb{s}_n} z^\mu z^\nu + \lr{p + \sum_{n=5}^7 \bar{\pi}_n \mbb{s}_n}\delperp^{\mu\nu} \\
        &+ 2 \lr{\mu \tilde{\rho} + \sum_{n=5}^7 \bar{\varphi}_n \mbb{s}_n}u^{(\mu} z^{\nu)}+ 2 \lr{\sum_{n=1,3}\bar{\theta}_n V_n^{(\mu}}u^{\nu)} + 2 \lr{ \sum_{n=1,3} \bar{\varrho}_n V_n^{(\mu}}z^{\nu)} \nnl
        &- \eta \sigma^{\mu\nu}_u\,,\nnl
        J^\mu &= \rho u^\mu + \tilde{\rho} z^\mu\,,\\
        \xi_\mu &= - \lr{\mu + \sum_{n=5}^7 \bar{\alpha}_n \mbb{s}_n}u_\mu + \zeta z_\mu\,,
    \end{align}
\end{subequations}
where the barred quantities denote transport coefficients in the LLCP frame. For details of the relationships between the LLCP transport coefficients, the frame invariants, and the seventy transport parameters, the interested reader may refer to Appendix A of~\cite{Hoult:2024cyx}.

The canonical entropy current is modified for a superfluid, and is given in the LLCP frame by
\begin{equation}
S^\mu = s u^\mu - \frac{u_\nu}{T} T^{\mu\nu}_{(1)} - \frac{\tilde{\rho}}{T} {\cal M}_1 z^\mu\,,
\end{equation}
while the divergence is given by
\begin{equation}
    \de_\mu S^\mu = - T_{(1)}^{\mu\nu} \de_\mu \lr{\frac{u_\nu}{T}} - {\cal M}_1 \Delta^{\mu\nu} \de_\mu \lr{\frac{\tilde{\rho} z_\nu}{T}}\,.
\end{equation}
Inserting the constitutive relations~\eqref{ch5:hydro:LLCP-conrel-preonsager}, one can find the following Onsager relations:
\begin{equation}
    d\, \bar{\pi}_5 = - \bar{\varphi}_6, \quad d\, \bar{\pi}_7 = - T^d \bar{\alpha}_6, \quad \bar{\varphi}_7 = T^d \bar{\alpha_5}, \quad \bar{\theta}_3 = \bar{\varrho}_1\,.
\end{equation}
This brings the number of physical transport coefficients down to ten. These ten coefficients are further constrained by demanding positivity of entropy production:
\begin{equation}
    \begin{gathered}
        \bar{\varphi}_5 \leq 0, \quad \bar{\pi}_6 \geq 0, \quad \bar{\alpha}_7 \leq 0, \quad \bar{\pi}_5^2 \leq - \frac{1}{d} \bar{\varphi}_5 \bar{\pi}_6, \quad \bar{\pi}_7^2 \leq - \frac{T^d}{d} \bar{\pi}_6 \bar{\alpha}_7, \quad \bar{\varphi}_7^2 \leq T^d \bar{\varphi}_5 \bar{\alpha}_7\,,\\
        \bar{\theta}_1 \leq 0, \quad \bar{\varrho}_3 \leq 0, \quad \bar{\theta}_3^2 \leq \bar{\theta}_1 \bar{\varrho}_3, \quad \eta \geq 0\,.
    \end{gathered}
\end{equation}
The $\oser{k^2}$ viscous corrections to the modes we previously found may be written down in terms of the ten parameters $\{\bar{\pi}_5, \bar{\pi}_6, \bar{\pi}_7, \bar{\varphi}_5, \bar{\varphi}_7, \bar{\alpha}_7, \bar{\theta}_1, \bar{\theta}_3, \bar{\varrho}_3, \eta\}$. Additionally, in the LLCP frame, there are three non-hydrodynamic modes in the spectrum. Unsurprisingly, the LLCP frame is acausal, just like the Landau frames in both dMHD~\ref{ch5:sec:mhd} and the $U(1)$ charged fluid~\ref{ch2:sec_uncharged}. The spectral curve of the theory violates the third condition of~\eqref{ch2:hydro:full_momentum_causcon}, with
\begin{equation}
    9 = {\cal O}_{\omega}F(\omega,k,\theta) \neq {\cal O}_{k}F(\omega = a \,k, k_j = s_j \,k, \theta) = 12\,.
\end{equation}
Consequently, upon further investigation we find that one mode goes as $k^3$, while three others go as $k^{4/3}$. All of these clearly indicate acausality. 

\subsection{BDNK theory of superfluids}
Let us repeat the linearized analysis of the superfluid in a more general fluid frame. In a general frame, the constraint equations receive viscous corrections. However, as in dMHD, there exists a class of frames (the ``constraint frames") in which the solution ~\eqref{ch5:super:constraint-eq-solution} is still valid. The constraint frames are given by the conditions
\begin{equation}
    \begin{gathered}
        \beta_1= \frac{\zeta_0}{T_0} \beta_3, \quad \beta_2 = 0, \quad \beta_4 = \varsigma_2, \quad \beta_5 = - \frac{\mu_0}{T_0} \beta_3, \quad \beta_6 = \frac{\zeta_0}{T_0} \beta_3, \quad \beta_7 = 0\,,\\
        \varsigma_1 = - \frac{\mu_0}{T_0} \beta_3, \quad \varsigma_3 = 0, \quad \varsigma_4 = \lr{\frac{\zeta_0}{T_0}} \beta_3 - \lr{\frac{\mu_0}{T_0}}\varsigma_5\,.
    \end{gathered}
\end{equation}
We see then that in the constraint frames, there are only three\footnote{The LLCP frame was a member of the class of constraint frames, with $\beta_3 = \varsigma_2 = \varsigma_5 = 0$; hence, no modification was needed to the solution~\eqref{ch5:super:constraint-eq-solution}.} independent transport parameters $\{\beta_3, \varsigma_2, \varsigma_5\}$ that contribute to ${\cal Z}$, ${\cal X}^\mu$. Let us consider the modes evaluated in the constraint frames. A natural place to start is the $k \to 0$ limit. In this limit, the spectral curve takes the form
\begin{equation}\label{ch5:super:gapped-modes}
\begin{split}
    \lim_{k \to 0} F(\omega,k,\theta) &= \omega^6 \times G_3(\omega) \times\lr{i + \frac{\beta_3}{T_0} \omega}\\
    &\times \biggl[ i \zeta_0 (4 p - \zeta_0 \tilde{\rho}_0) + \lr{\zeta_0 \theta_1 + \mu_0 \lr{\theta_4 + \frac{\mu_0}{T_0} \theta_5}}\omega\biggr]^2 = 0\,.
\end{split}
\end{equation}
where $G_3$ is a complicated cubic polynomial which has a strong dependence on the equation of state. It may be written in the generic form
\begin{equation}
    G_3(\omega) = \mathbb{a} \lr{-i \omega}^3 + \mathbb{b} \lr{-i \omega}^2 + \mathbb{c} \lr{-i \omega} + \mathbb{d}\,,
\end{equation}
where the coefficients $\mathbb{a},\mathbb{b},\mathbb{c}$ depend on both the transport parameters and the thermodynamic parameters, while $\mathbb{d}$ depends solely on the thermodynamic parameters in the theory, and therefore the details of the equation of state. Let us make the assumption that the equation of state is such that $\mathbb{d} \neq 0$. Then, dividing through by $\mathbb{d}$ yields
\begin{equation}
    G'_3 = \frac{G_3}{\mathbb{d}} = \mathbb{a}' \lr{-i \omega}^3 + \mathbb{b}' \lr{-i \omega}^2 + \mathbb{c}' \lr{-i \omega} + 1\,,
\end{equation}
We may now ensure that the gapped modes are stable by applying the Routh-Hurwitz criterion (see Chapter~\ref{ch3:sec_RHandSC}). This yields the conditions
\begin{subequations}\label{ch5:super:gap-con-1}
    \begin{align}
        \beta_3 &> 0\,,\\
        (4p-\zeta_0 \tilde{\rho}_0)/ \lr{\theta_1 + \frac{\mu_0}{\zeta_0} \lr{\theta_4 + \frac{\mu_0}{T_0} \theta_5}} &> 0\,,\\
        \mathbb{a}'>0, \quad \mathbb{b}'>0, \quad \mathbb{b}'\mathbb{c}'-\mathbb{a}'&>0\,.
    \end{align}
\end{subequations}
We can immediately see that the condition $\beta_3 > 0$ is violated by the LLCP frame (and any other frame that sets ${\cal Z} =0$). These constraints may be simplified slightly if we introduce an additional class of frames to be layered on top of the constraint frames. We will refer to this class of frames as the decoupled frame due to its effect at large-$k$, rather than the $k\to0$ limit. The decoupled class of frames are defined by
\begin{equation}\label{ch5:super:decoupled-frame}
\ce_{2,3,4,7} = \pi_{2,3,4,7} = \varphi_{2,3,4,7} = \theta_{2,4,5} = \varrho_{2,4,5} = 0, \quad \alpha_4 = - \frac{1}{T_0^2} \pder{\tilde{\rho}}{\mu} \alpha_7\,.
\end{equation}
Note that the act of imposing the decoupled frames on top\footnote{Unlike in dMHD, the two classes of frames are not inconsistent for a relativistic superfluid.} of the constraint frames was a choice -- the two classes of frames are, in principle, independent. In this set of frames, the conditions~\eqref{ch5:super:gap-con-1} become
\begin{subequations}\label{ch5:super:gap-con-2}
    \begin{alignat}{4}
        &\qquad \,\,\,\frac{\nu_2 T_0^2}{\mathbb{d}} \lr{\ce_1 \varphi_5 - \ce_5 \varphi_1} &&> 0\,, \quad&&\,\,\lr{4 p - \zeta_0 \tilde{\rho}}/\theta_1 &&> 0\,,\\
      &\mathbb{b}' \mathbb{c}' - \frac{\nu T_0^2}{\mathbb{d}} \lr{\ce_1 \varphi_5 - \ce_5 \varphi_1} &&>0\,, \quad&&\,\,\,\,\,\beta_3 > 0\,, \quad\,\mathbb{b}' &&> 0\,.
    \end{alignat}
\end{subequations}
If the constraints~\eqref{ch5:super:gap-con-2} are satisfied, then the gapped modes will all be stable. Let us now turn to the large-$k$ limit. We know that to ensure causality, the modes must all go linearly in $k$, i.e.
\begin{equation}
    \omega \sim c\,k + ...
\end{equation}
where the $...$ denotes subleading terms in $k$ and $-1 \leq c \leq 1$. Substituting the linear ansatz for $\omega$, we find that the spectral curve $F(\omega,k)$ factorizes in the large-$k$ limit into an order-2 polynomial and an order-10 polynomial in $c$. The order-2 polynomial is easily handled, but the order-10 polynomial is analytically intractible. The situation is simplified, however, if one works in the decoupled frames~\eqref{ch5:super:decoupled-frame}; then the order-ten polynomial further factorizes into two quadratics and one order-six polynomial in $c$:
\begin{subequations}\label{ch5:super:large-k-modes}
    \begin{align}
        \theta_1 c^2 - \cos(\theta) \lr{\theta_3 + \varrho_1} c + \varrho_3 \cos^2(\theta) - \eta \sin^2(\theta) &= 0\label{ch5:super:shear-large}\,,\\
        \beta_3 c^2 - \alpha_3 \cos(\theta) c - \frac{\alpha_7}{2 T_0^2 \zeta_0} \lr{ \pder{\tilde{\rho}}{\zeta} \zeta_0 + \lr{\pder{\tilde{\rho}}{\zeta} \zeta_0 - 2 \tilde{\rho}_0}\lr{\cos^2(\theta) - \sin^2(\theta)}} &= 0\label{ch5:super:zeta-large}\,,\\
        \nu_2 c^2 - \lr{\lambda_2 + \nu_4 + \frac{1}{T_0^2} \pder{\tilde{\rho}}{\mu} \nu_7} \cos(\theta) c + \lr{\lambda_4 + \frac{1}{T_0^2} \pder{\tilde{\rho}}{\mu} \lambda_7} \cos^2(\theta) + \gamma_2 \sin^2(\theta) &= 0\label{ch5:super:charge-large}\,,\\
        6 \theta_1 \lr{\ce_1 \varphi_5 - \ce_5 \varphi_1} c^6 + \mathfrak{f}_1 \cos(\theta) c^5 + \mathfrak{f}_2 c^4 + \mathfrak{f}_3 \cos(\theta)c^3 + \mathfrak{f}_4 c^2 + \mathfrak{f}_5 \cos(\theta) c + \mathfrak{f}_6 &= 0\,,\label{ch5:super:sound-large}
    \end{align}
\end{subequations}
where $\theta$ is once again the angle between $k_j$ and the background direction along which $\zeta_0^\mu$ is aligned, in this case the $z$-direction. The coefficients $\mathfrak{f}_{1..6}$ depend on the transport parameters, thermodynamic quantities, and the angle $\theta$. The first three equations in~\eqref{ch5:super:large-k-modes} are straightforward to constrain, and correspond respectively to shear perturbations, the fluctuations of the transverse superfluid velocity $\delta \hat{\zeta}$, and the fluctuations of the chemical potential. The remaining mode~\eqref{ch5:super:sound-large} describes the mixing of the perturbations to $T$, $v_x$, and $v_z$.

Let us begin with the quadratics, which are easier to constraint. In order to ensure that the roots of a quadratic $A x^2 + B x + C $ with $A>0$ and $B,C\neq 0$ are such that $0 < x^2 < 1$ (we exclude the possibility of luminal propagation so that the Schur-Cohn criterion may be used), we find that the coefficients must obey the following conditions:
\begin{equation}
    \Delta \equiv B^2 - 4 A C > 0, \quad B< 0, \quad 0 < C < A\,, \quad A+B+C > 0\,.
\end{equation}
It is immediately clear that these conditions cannot possibly be satisfied by the quadratics in~\eqref{ch5:super:large-k-modes}. The linear term in all three equations is proportional to $\cos(\theta)$, and since the transport parameters are not functions of $\theta$, there will always exist a value of $\theta$ such that $B>0$. We therefore must have either $B$ or $C$ equal to zero.

Via a further frame choice (thereby further restricting the class of frames within which we work), the linear term is set to zero in~\eqref{ch5:super:shear-large},~\eqref{ch5:super:charge-large}. We also set the constant term to zero in~\eqref{ch5:super:zeta-large}, as the thermodynamics does not guarantee the term's sign. This leaves~\cite{Hoult:2024cyx}
\begin{subequations}
    \begin{align}
&\text{Shear mode:}\left\{
\begin{aligned}
    &v^2 - \lr{\frac{\eta \sin^2(\theta)  - \varrho_3 \cos^2(\theta)}{\theta_1}} = 0,\\
    &\theta_3 + \varrho_1 = 0, \label{eq:shear_mode}\\
\end{aligned}\right.\\
&\text{Transverse superfluid velocity mode:}\left\{
\begin{aligned}
     &v\lr{v - \frac{\alpha_3}{\beta_3} \cos(\theta)} = 0,\\
    &\alpha_7 = 0,\label{eq:ssm}\\
\end{aligned}\right.\\
&\text{Charge mode:}\left\{
\begin{aligned}
    &v^2 + \frac{1}{\nu_2} \left[\lr{\lambda_4 + \frac{1}{T_0^2} \pder{\tilde{\rho}}{\mu} \lambda_7} \cos^2(\theta) + \gamma_2 \sin^2(\theta)\right] = 0,\\
    &\lambda_2 + \nu_4 + \frac{1}{T_0^2} \pder{\tilde{\rho}}{\mu} \nu_7 = 0.\label{eq:charge_mode}\\
\end{aligned}\right.
\end{align}
\end{subequations}
We can achieve the frame in which the equations above hold by imposing
$    \theta_3 = - \varrho_1, \, \alpha_7 = 0, \, \nu_7 = 0, \, \lambda_2 = - \nu_4.$
We can therefore clearly see that the three large-$k$ modes~\eqref{ch5:super:shear-large},~\eqref{ch5:super:zeta-large},~\eqref{ch5:super:charge-large} may be rendered causal in this frame via satisfying the constraints
\begin{equation}
    \begin{gathered}
        0 \leq \frac{\eta}{\theta_1} \leq 1, \quad -1 \leq \frac{\varrho_3}{\theta_1} \leq 0\,,\quad -1 \leq \frac{\alpha_3}{\beta_3} \leq 1\,,\\
        -1 \leq \frac{\lambda_4 + \frac{1}{T_0^2} \pder{\tilde{\rho}}{\mu} \lambda_7}{\nu_2} \leq 0, \quad -1 \leq \frac{\gamma_2}{\nu_2} \leq 0\,.
    \end{gathered}
\end{equation}
Let us now turn our attention to the order-six polynomial. As in~\cite{Hoult:2024cyx}, we will only consider three cases -- the co-aligned limit $\theta = 0$, the transverse limit $\theta = \pi/2$, and the anti-aligned limit $\theta = \pi$.

\subsubsection{Co-aligned limit}
In the co-aligned limit, with $\theta=0$, the $SO(2)$ symmetry of the equilibrium state is restored, and the order-six polynomial factorizes into an quartic polynomial and a second copy of the shear mode~\eqref{ch5:super:shear-large} with $\theta = 0$. The quartic polynomial is of the form
\begin{equation}
    P_0(c) \equiv {\mathfrak a} c^4 + {\mathfrak b} c^3 + {\mathfrak c} c^2 + {\mathfrak d} c + {\mathfrak e} = 0\,,
\end{equation}
with
\begin{subequations}\label{ch5:super:aligned-coeff}
    \begin{align}
        {\mathfrak a} &= 3 \lr{\varepsilon_5 \varphi_1 - \varepsilon_1 \varphi_5},\\
    {\mathfrak b} &= 6 \lr{\varepsilon_5 \pi_1 - \varepsilon_1 \pi_5} + 2 \lr{\varepsilon_1 \varphi_6 - \varepsilon_6 \varphi_1},\\
    {\mathfrak c} &= 4 \lr{\varepsilon_1 \pi_6 - \varepsilon_6 \pi_1} + 4 \lr{\varepsilon_1 \varphi_5 - \varepsilon_5 \varphi_1} + 6 \lr{ \pi_5 \varphi_1 - \pi_1 \varphi_5} + 2 \lr{\varepsilon_6 \varphi_5 - \varepsilon_5 \varphi_6},\\
    {\mathfrak d} &= 2 \lr{ \varepsilon_1 \pi_5 - \varepsilon_5 \pi_1} + 4 \lr{\varepsilon_6 \pi_5 - \varepsilon_5 \pi_6} + 4 \lr{\pi_1 \varphi_6 - \pi_6 \varphi_1} + 2 \lr{\varepsilon_6 \varphi_1 - \varepsilon_1 \varphi_6},\\
    {\mathfrak e} &= 2 \lr{\pi_1 \varphi_5 - \pi_5 \varphi_1} +4 \lr{ \pi_6 \varphi_5 - \pi_5 \varphi_6}+ \lr{\varepsilon_5 \varphi_1 - \varepsilon_1 \varphi_5} + 2 \lr{\varepsilon_5 \varphi_6 - \varepsilon_6 \varphi_5}.
    \end{align}
\end{subequations}
We now must impose that the modes are causal. This involves imposing reality of the roots, as well as Schur-Cohn stability. These conditions may be found by
\begin{subequations}\label{ch5:super:theta-zero-conditions}
    \begin{align}
&\text{Real roots:}\left\{
\begin{aligned}
    &\Delta \equiv 256 s^3 - 128 q^2 s^2 + 144 q r^2 s + 16 q^4 s - 27 r^4 - 4 q^3 r^2 > 0,\\
    & \frac{q^2}{4} - s > 0, \label{eq:rr_1}\\
\end{aligned}\right.\\
&\text{Schur-Cohn stability:}\left\{
\begin{aligned}
    &\mc{S} \equiv \mf{a}+\mf{b}+\mf{c}+\mf{d}+\mf{e} >0,\\
    &2\mf{a} + \mf{b} - \mf{d} -2\mf{e} >0,\\
    &{\cal P} \equiv \lr{2 \mf{a} + \mf{b} - \mf{d} -2\mf{e}}\lr{6\mf{a}-2\mf{c}+6\mf{e}} - \mc{S} \lr{2\mf{a} -\mf{b} + \mf{d} -2 \mf{e}} > 0,\\
    &2\lr{2\mf{a} -\mf{b} +\mf{d} -2\mf{e}} {\cal P} - \lr{\mf{a} - \mf{b} + \mf{c} -\mf{d} +\mf{e}} \lr{6\mf{a} -2\mf{c}+6\mf{e}}^2 >0,\\
    &2\mf{a} - \mf{c} + 2\mf{e} > 0,\label{eq:schur_1}
\end{aligned}\right.
\end{align}
\end{subequations}
In the above, the coefficients $q,r,s$ are the coefficients of the depressed quartic associated with $P_0(c)$. They are given by
\begin{subequations}
    \begin{align}
        q &\equiv \frac{1}{\mf{a}^2} \lr{\mf{a} \mf{c} - \frac{3}{8} \mf{b}^2},\\
        r &\equiv\frac{1}{\mf{a}^3} \lr{\frac{\mf{b}^3 - 4 \mf{a} \mf{b} \mf{c}}{8} - \mf{d} \mf{a}^2},\\
        s &\equiv \frac{1}{\mf{a}^4} \lr{\mf{e} \mf{a}^3 - \frac{\mf{b}}{256} \lr{3 \mf{b}^3 + 16 \mf{abc} + 64 \mf{d} \mf{a}^2}},
    \end{align}
\end{subequations}
If the conditions~\eqref{ch5:super:theta-zero-conditions} are satisfied, then the fluid will be causal when $\theta = 0$.

\subsubsection{Transverse limit}
In the limit that $\theta \to \pi/2$, all of the odd-powered terms in~\eqref{ch5:super:sound-large} vanish. We are then left with a cubic polynomial in $c^2$, i.e.
\begin{equation}
    P_{\pi/2}(c^2) = \mf{a} x^3 + \mf{b} x^2  \mf{c} x + \mf{d} = 0\,, 
\end{equation}
where we have defined $x= c^2$. The coefficients $\mf{a}$, $\mf{b}$, $\mf{c}$, $\mf{d}$ are given by
\begin{subequations}
    \begin{align}
    {\mathfrak a} &= 3 \theta_1 \lr{\varepsilon_5 \varphi_1 - \varepsilon_1 \varphi_5},\\
    {\mathfrak b} &= -\biggl[3 \varrho_1^2 \varepsilon_1 + \varepsilon_1 \lr{3 \theta_1 \varrho_3 - \lr{\theta_1 + 3 \eta} \varphi_5 + \lr{\pi_5 \varphi_6 - \pi_6 \varphi_5}} \nonumber\\
    &\quad+ \theta_1 \lr{\varepsilon_6 \varphi_5 - \varepsilon_5 \varphi_6 - 3 \pi_1 \varphi_5} + \pi_1 \lr{\varepsilon_6 \varphi_5 - \varepsilon_5 \varphi_6}\nonumber\\
    &\quad+ \varphi_1 \lr{\varepsilon_5 \lr{3 \eta + \theta_1} + 3 \theta_1 \pi_5 + \varepsilon_5 \pi_6 - \varepsilon_6 \pi_5} \nonumber\\
    &\quad+ \varrho_1 \lr{3 \lr{ \varepsilon_5 \pi_1 - \varepsilon_1 \pi_5} + \lr{\varepsilon_6 \varphi_1 - \varepsilon_1 \varphi_6}}\biggr],\\
    {\mathfrak c} &= -\biggl[\varrho_1^2 \lr{\varepsilon_6 - 3 \pi_1 - \varepsilon_1} +  \varrho_3 \lr{\theta_1 \lr{\varepsilon_6 - 3 \pi_1 - \varepsilon_1} +\lr{\varepsilon_6 \pi_1 - \varepsilon_1 \pi_6} -  3\varepsilon_1 \eta}\nonumber \\
    &\quad+ \varrho_1 \lr{\lr{\pi_1 \varphi_6 - \pi_6 \varphi_1} + \lr{\varepsilon_1 \pi_5 - \varepsilon_5 \pi_1} + \lr{\varepsilon_5 \pi_6 - \varepsilon_6 \pi_5} + 3 \eta \lr{\varepsilon_5 - \varphi_1}}\nonumber\\
    &\quad+  \theta_1 \lr{ \lr{\pi_5 \varphi_6 - \pi_6 \varphi_5} + \lr{\pi_1 \varphi_5 - \pi_5 \varphi_1} - 3 \eta \varphi_5}\biggr],\\
    {\mathfrak d} &= \lr{3 \eta + \pi_6 - \pi_1} \lr{\varrho_1^2 + \theta_1 \varrho_3}.
    \end{align}
\end{subequations}
We now impose that the roots of $P_{\pi/2}(x)$ lie in the open unit disk in the complex plane, are real, and also lie in the right-hand complex plane; in other words, we impose that they lie on the real axis in the interval $x \in (0,1)$. This may be done by imposing the positivity of the discriminant, Schur-Cohn stability, and Routh-Hurwitz stability of $P_{\pi/2}(-x)$.
\begin{subequations}\label{ch5:super:theta-pi2-constraint}
    \begin{align}
&\text{Real roots:}
\quad \Delta \equiv \mf{b}^2 \mf{c}^2 - 4 \mf{b}^3 \mf{d} + 18 \mf{abcd} - \mf{a} \lr{4 \mf{c}^3 + 27 \mf{ad}^2} >0, \\
&\text{Schur-Cohn stability:}\left\{
\begin{aligned}
    &\mc{S}\equiv \mf{a}+\mf{b}+\mf{c}+\mf{d} >0,\\
    &3\mf{a} + \mf{b} -\mf{c} -3\mf{d} >0,\\
    &\mf{a}-\mf{b}+\mf{c}-\mf{d} >0,\\
    &\lr{3 \mf{a} + \mf{b} - \mf{c} - 3\mf{d}} \lr{3\mf{a} - \mf{b} -\mf{c} +3\mf{d}} - \mc{S} \lr{\mf{a}-\mf{b}+\mf{c}-\mf{d}} >0, \\
\end{aligned}\right.\\
&\text{Positivity:}\quad \mf{a}>0, \quad \mf{b}< 0, \quad \mf{ad}-\mf{bc}>0, \quad \mf{d}<0.
\end{align}
\end{subequations}
Imposing these constraints ensures causality of the superfluid in the limit $\theta = \pi/2$. Finally, let us consider the anti-aligned limit.

\subsubsection{Anti-aligned limit}
In the anti-aligned limit, when $\theta = \pi$, the $SO(2)$ symmetry of the equilibrium state is restored, and the mode~\eqref{ch5:super:sound-large} again factorizes into a quartic polynomial and a copy of the shear mode, this time with $\theta = \pi$. The quartic polynomial is given by
\begin{equation}
    P_\pi(c) \equiv {\mathfrak a} c^4 + {\mathfrak b} c^3 + {\mathfrak c} c^2 + {\mathfrak d} c + {\mathfrak e} = 0\,,
\end{equation}
with
\begin{subequations}\label{ch5:super:antialigned-coeff}
    \begin{align}
    {\mathfrak a} &= 3 \lr{\varepsilon_5 \varphi_1 - \varepsilon_1 \varphi_5},\\
    {\mathfrak b} &= 6 \lr{\varepsilon_1 \pi_5 - \varepsilon_5 \pi_1} + 2 \lr{\varepsilon_6 \varphi_1 - \varepsilon_1 \varphi_6},\\
    {\mathfrak c} &= 4 \lr{\varepsilon_1 \pi_6 - \varepsilon_6 \pi_1} + 4 \lr{\varepsilon_1 \varphi_5 - \varepsilon_5 \varphi_1} + 6 \lr{ \pi_5 \varphi_1 - \pi_1 \varphi_5} + 2 \lr{\varepsilon_6 \varphi_5 - \varepsilon_5 \varphi_6},\\
    {\mathfrak d} &= 2 \lr{ \varepsilon_5 \pi_1 - \varepsilon_1 \pi_5} + 4 \lr{\varepsilon_5 \pi_6 - \varepsilon_6 \pi_5} + 4 \lr{\pi_6 \varphi_1 - \pi_1 \varphi_6} + 2 \lr{\varepsilon_1 \varphi_6 - \varepsilon_6 \varphi_1},\\
    {\mathfrak e} &= 2 \lr{\pi_1 \varphi_5 - \pi_5 \varphi_1} +4 \lr{ \pi_6 \varphi_5 - \pi_5 \varphi_6}+ \lr{\varepsilon_5 \varphi_1 - \varepsilon_1 \varphi_5} + 2 \lr{\varepsilon_5 \varphi_6 - \varepsilon_6 \varphi_5}.
\end{align}
\end{subequations}
Note the sign differences between equations~\eqref{ch5:super:aligned-coeff} and~\eqref{ch5:super:antialigned-coeff}. The constraint equations are identical to~\eqref{ch5:super:theta-zero-conditions}, but with the definitions of $\mf{a,b,c,d,e}$ given by~\eqref{ch5:super:antialigned-coeff}.

\begin{figure}[!t]
    \centering
    \def\ImageScale{0.25}
    \includegraphics[scale=\ImageScale]{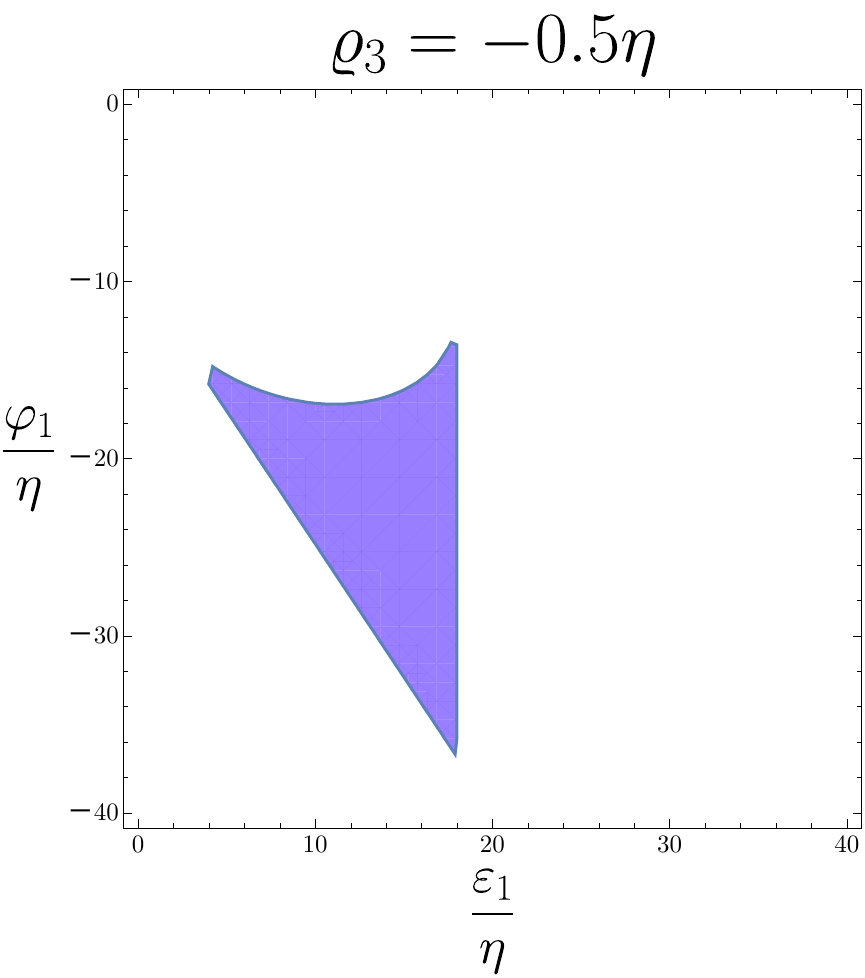}
    \includegraphics[scale=\ImageScale]{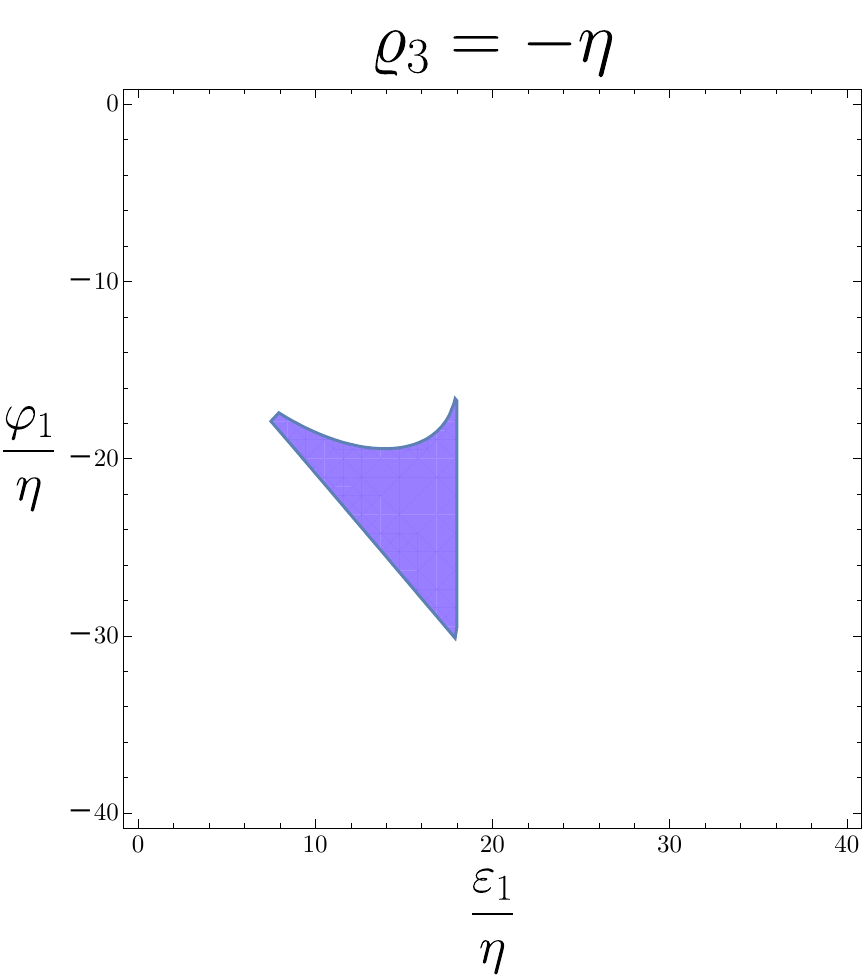}
    \includegraphics[scale=\ImageScale]{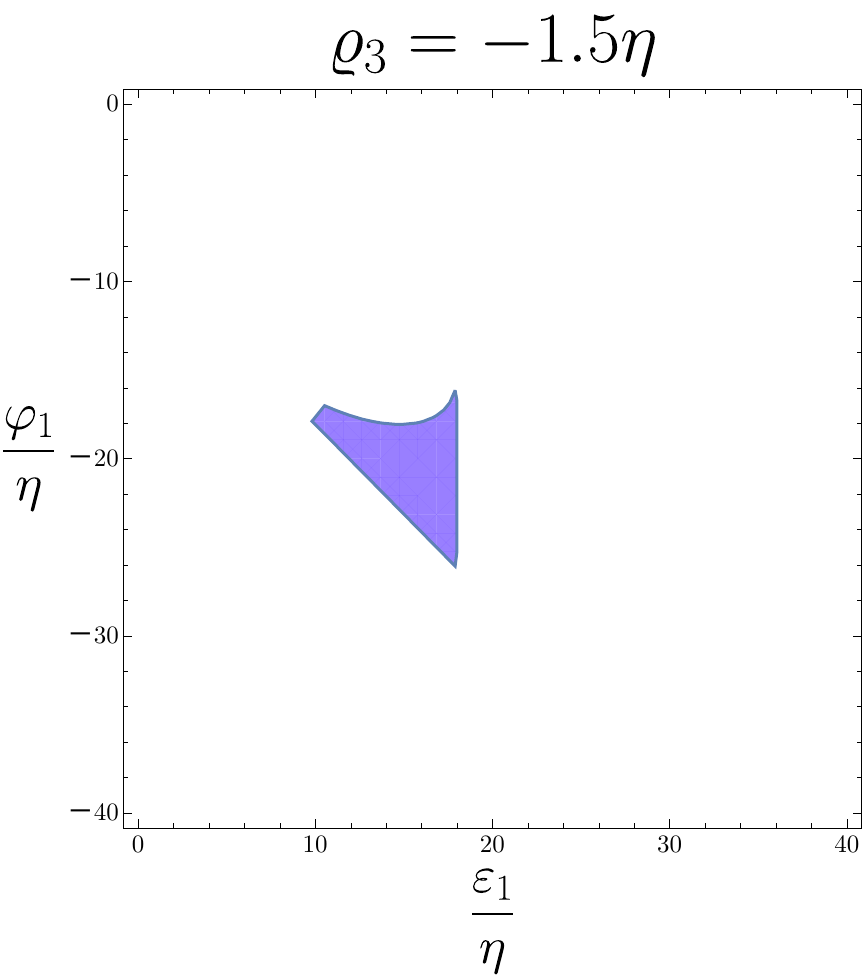}
    \includegraphics[scale=\ImageScale]{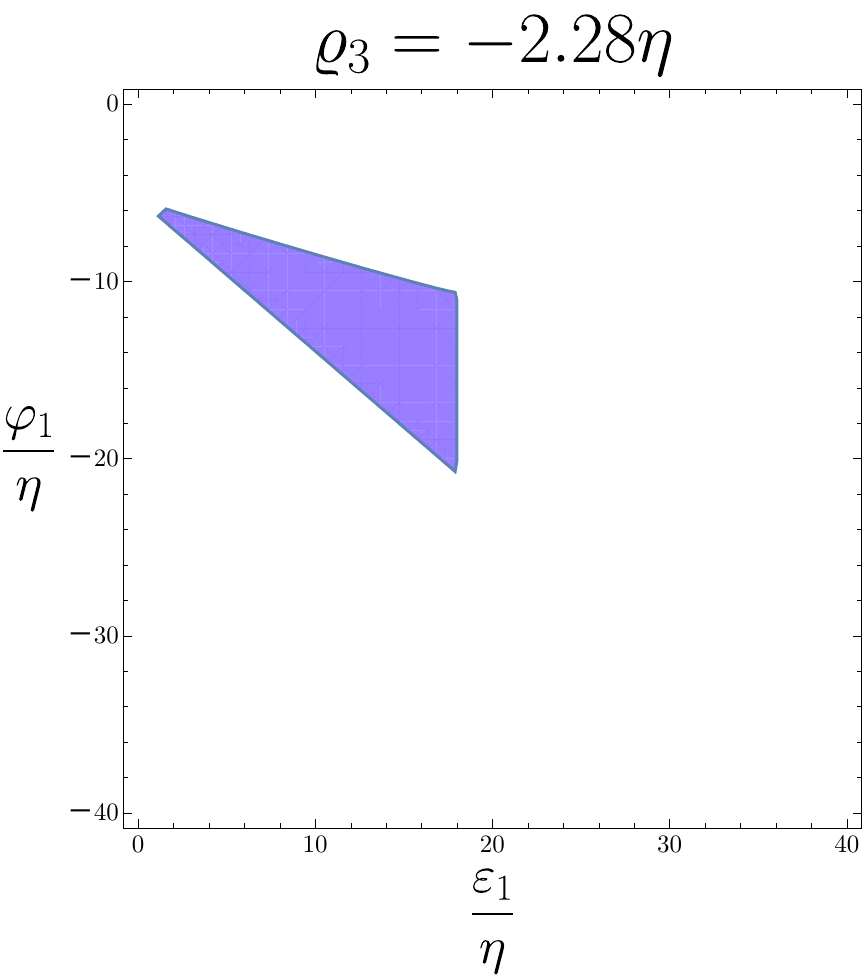}
    \includegraphics[scale=\ImageScale]{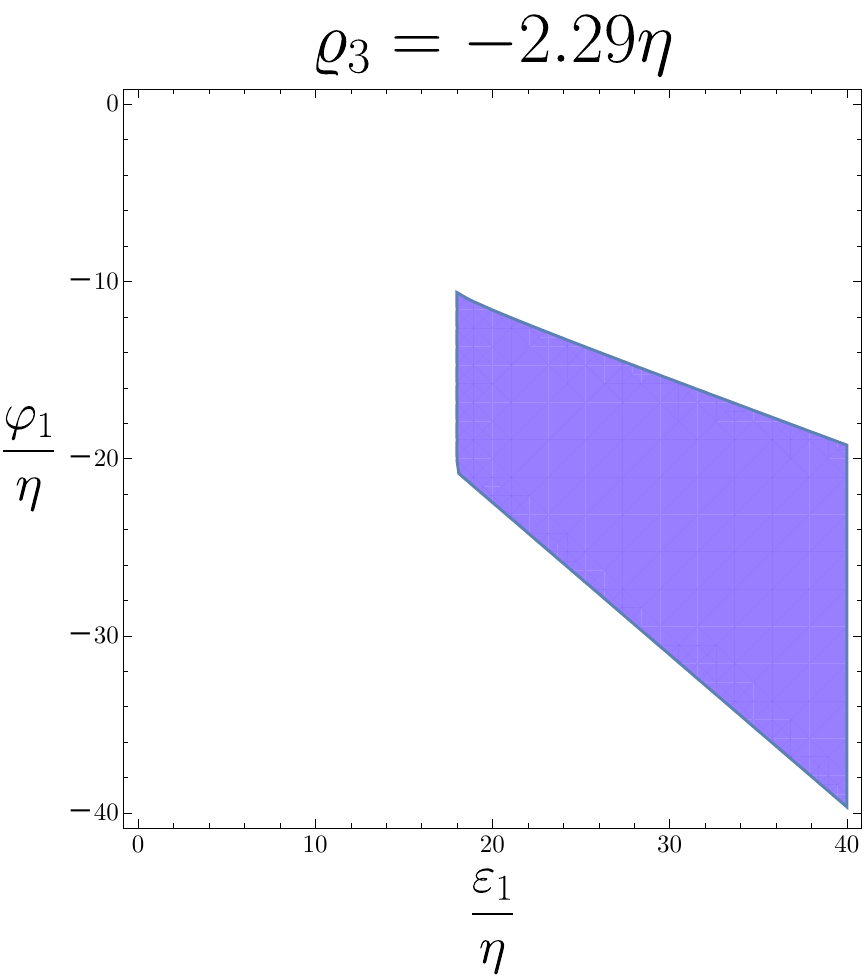}
    \includegraphics[scale=\ImageScale]{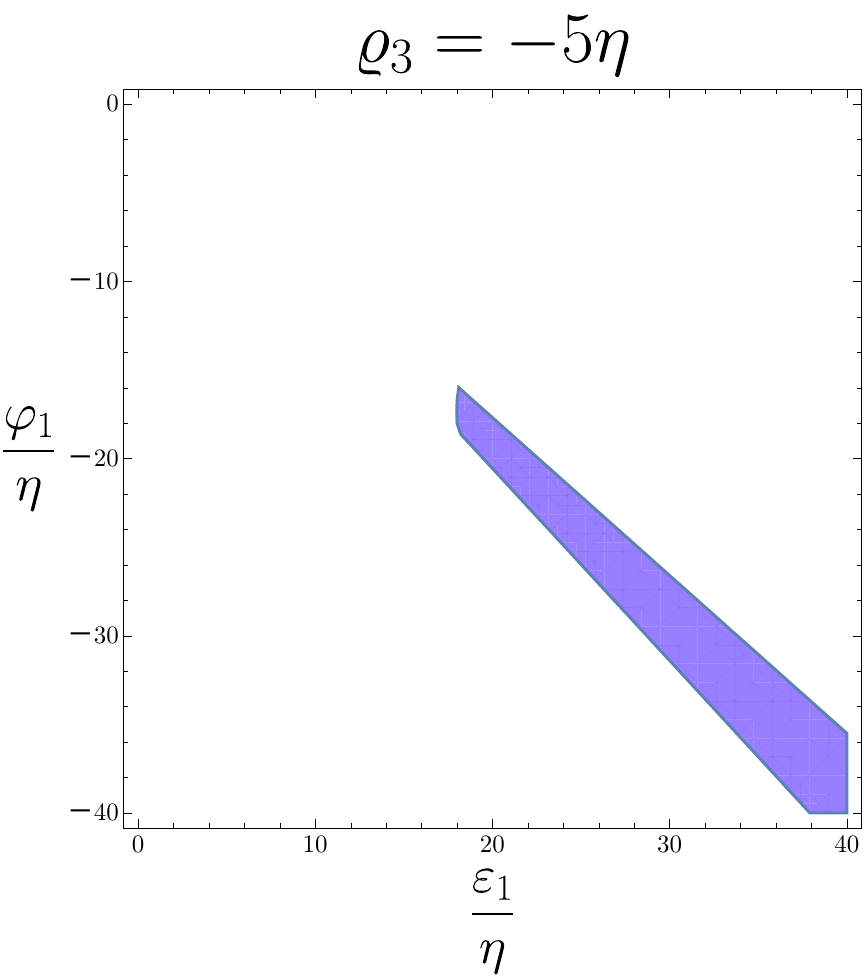}
    \caption{Regions in $(\ce_1/\eta, \varphi_1/\eta)$ parameter space for different values of the ratio $M= (-\varrho_3)/\eta$ where the large-$k$ modes are causal for $\theta = 0$, $\pi/2$, and $\pi$ for the demonstrative choice eq.~\eqref{ch5:super:plotregions} within the extended decoupled family of frames. Figure reproduced from~\cite{Hoult:2024cyx}.}
    \label{ch5:super:causal_sound}
\end{figure}

Putting all of these conditions together, we can ensure causality of the superfluid in the cases $\theta \in \{0,\pi/2,\pi\}$ by simultaneously imposing the constraints~\eqref{ch5:super:theta-zero-conditions} for both $\theta = 0, \pi$, as well as the $\pi/2$ constraints~\eqref{ch5:super:theta-pi2-constraint}. To illustrate that these conditions are not empty, let us consider the following ``test frame", which is written in terms of $M \equiv (-\varrho_3)/\eta \geq 0$:
\begin{align}\label{ch5:super:plotregions}
        \ce_5 &= 0,\quad \pi_5 = 0,\quad \varrho_5 = 0, \quad \ce_6 = \ce_1/2,\quad \pi_6 =  \ce_1/6,\quad \varphi_6 = \varphi_1/2,\quad \varrho_6 = 0,\nonumber\\
\pi_1 &= \ce_1/3, \quad \alpha_3 = \beta_3/2, \quad \gamma_2 = - \nu_2/2,  \quad \lambda_4 = - \lr{\pder{\tilde{\rho}}{\mu} \lambda_7/T_0^2 +\nu_2/2},\quad \varphi_5 = - \pi \eta\,,\nonumber\\
\theta_1 &= 2 \lr{\eta - \varrho_3},\quad 
\varrho_1 = -\frac{\sqrt{1+2M}}{2}\lr{\eta - \varrho_3}\,.
\end{align}
In this test frame, the regions in which the constraints all simultaneously hold are plotted in Figure~\ref{ch5:super:causal_sound}. As is clearly seen, these constraints are not empty. It is important to note that constraining causality for three discrete values of $\theta$ is not enough to ensure causality for all values of $\theta$. Such an undertaking is best taken with a particular equation of state in mind, via a numerical procedure. 

With these conditions determined, we have gone about as far as we can analytically with respect to the causality of the relativistic superfluid. For arbitrary $\theta$, the controlling equation is an order-six polynomial, and order-six polynomial equations do not generically have closed-form solutions.

	\startchapter{Conclusions}
\label{chapter:concl}

This dissertation investigated causal theories of relativistic hydrodynamics. In the seemingly disparate works that have preceded, a unifying programme existed: the development of effective theories of the macroscopic transport of conserved densities (hydrodynamics) which both incorporate viscous effects and respect the demands of stability and causality. In the various forms this programme took, whether in the mathematical structure of the hydrodynamic equations (Chapter~\ref{chapter:math}), the connection to microscopic descriptions (Chapter~\ref{chapter:micro}), or extensions of causal frameworks to new hydrodynamic theories (Chapter~\ref{chapter:extensions}), there was a shared goal: determining how covariant stability manifested itself, and determining its impact on the effective theories under consideration.

Chapter~\ref{chapter:math}, based on the results of my paper~\cite{Hoult:2024qph}, showed there can exist an equivalence between linearized and nonlinear causality conditions, and found sufficient conditions for the equivalence to hold. This equivalence will greatly ease the analysis of causality for any postulated theories going forward. So long as the sufficient conditions outlined in Section~\ref{ch3:sec_equivalence} hold, a linearized causality analysis will be enough to ensure non-linear causality, without needing to perform a separate investigation of the characteristics of the partial differential equations. 

The ways covariant stability constrains dispersion relations $\omega = \omega(k)$ were also investigated; in particular, based on the findings of my paper~\cite{Hoult:2023clg}, constraints were found on dispersion relations in the short-wavelength limit due to the demands of covariant stability. This requirement represents completely new conditions on dispersion relations, constraining beyond the previously known causality conditions at linear order in large-$k$.

Chapter~\ref{chapter:micro} demonstrated how causal theories of hydrodynamics arise from considerations of microscopic theories such as kinetic theory and theories with holographic duals. As I found in my paper~\cite{Hoult:2021gnb}, the BDNK formulation of hydrodynamics arises from a careful treatment of zero-modes in a recurring operator in a Hilbert expansion; fixing the associated integration functions also fixes the hydrodynamic frame. This result clarifies the relationship between transport parameters in the hydrodynamic theory and the solutions to the equations of the microscopic theories, providing a clear map for choosing a hydrodynamic frame.

Finally, Chapter~\ref{chapter:extensions} contained the BDNK theory of one-form MHD with a ``mass current" for the first time. Based on results from my paper~\cite{Hoult:2024qph}, constitutive relations in a general frame were written down, and frame-invariant combinations leading to physical transport coefficients were identified. In the process, a new physical transport coefficient was identified which had been previously missed. Finally, the controlling equations for large-$k$ dispersion relation were determined, and causality conditions to be imposed were identified. This lays the groundwork for astrophysical applications, work that has already begun for the theory without an additional $U(1)$ in~\cite{Lier:2025wfw}.

The BDNK theory of relativistic superfluids was also written down for the first time. Following the findings in my paper~\cite{Hoult:2024cyx}, constitutive relations were written down, frame-invariant quantities and physical transport coefficients were determined, and then finally the controlling equations for the large-$k$ dispersion relations at linear order were identified. Causality constraints were imposed, and the resulting inequalities were explicitly demonstrated to not be empty in a subset of parameter space. This work can pave the way towards applications in binary neutron star mergers~\cite{Baiotti:2016qnr}, due to the postulated superfluidity of neutron star cores~\cite{MIGDAL:1959655,BAYM1969}.

There are numerous paths forward prompted by the groundwork I have laid in this dissertation. Following Section~\ref{ch3:sec_equivalence}, an immediate follow-up question arises from the assumptions made. The analysis was performed under the assumption that all equations were of the same order; application to systems of equations of mixed order are an immediate extension. As well, a more complete accounting of the role of constraints such as $\nabla{\cdot}B = 0$ in the causality conditions would be very useful for application to the extensions discussed in Chapter~\ref{chapter:extensions}.

Another immediate consideration follows from Section~\ref{ch3:sec_covarstable}. In that section, an assumption was made of isotropy. However, many interesting physical systems have anisotropies in the equilibrium state. Investigating how anisotropy affects causality constraints at both small- and large-$k$ seems of immediate interest, and is something to which I intend to return.

In Chapter~\ref{chapter:micro} both kinetic theory and holography were investigated. In the case of kinetic theory, we wrote down a BDNK-type theory, and also discussed an MIS-type theory (DNMR). However, in the case of holography we only considered a BDNK-type theory. No holographic version of MIS exists in the non-linear regime as of yet. The development of such a theory is an endeavour upon which I will be actively working moving forward.

Finally, in Chapter~\ref{chapter:extensions} two extensions to the BDNK formulation of causal hydrodynamics were written down -- one-form MHD, and relativistic superfluids. In both cases, the theories were presented in Minkowski spacetime. Coupling the theories to general relativity would be the next important step in their use in astrophysical applications. Additionally, in Section~\ref{ch5:sec:superfluids} the effect of vortices were not included, which play a critical role in the theory of superfluidity. Adding in this effect is an immediate extension.

There is still a great deal left to understand. In this dissertation a programme focused on the development of causal theories of hydrodynamics was presented, and this programme continues. Refinement and development of these theories, in particular with a focus on the minimal necessary number of transport parameters, will lead to new applications. I hope that this dissertation will prove useful in service to the formulation of new causal theories of relativistic hydrodynamics, and further investigation into existing theories.
	\appendix
	\startappendix{Species of MIS}
\label{app:species_MIS}

In this appendix, we present in more detail the three M\"uller-Israel-Stewart-type theories discussed in Chapter~\ref{ch2:sec_uncharged}. They were the ``canonical" approach, resummed BRSSS, and the generating functional approach. We consider each in turn.

\subsubsection{``Canonical" Approach}
An approach using the entropy current was originally described by Israel and Stewart~\cite{Israel-Stewart}, and then was later refined by Hiscock and Lindblom~\cite{Hiscock:1983zz}. Here, we present the formalism as described in~\cite{Noronha:2021syv}. Let us consider the canonical entropy current~\eqref{ch2:hydro:canonical-entropy} with a potential second-order modification $Q^\mu$:
\begin{equation}
    S^\mu = p \beta^\mu - T^{\mu\nu}\beta_\nu - \alpha J^\mu - Q^\mu
\end{equation}
Given the constitutive relations~\eqref{ch2:hydro:MIS-gen-constitutive}, the entropy current can be re-written in the form
\begin{equation}
    S^\mu = s u^\mu - \alpha n^\mu - Q^\mu
\end{equation}
Let us now postulate a form for $Q^\mu$. Since we would like to be agnostic about the underlying microscopic theory, let us simply write down a general form based on symmetry:
\begin{equation}\label{app:hydro:second-order-Q}
\begin{split}
    Q^\mu &= \frac{1}{2T}\biggl[ \beta_\pi \pi^{\alpha\beta} \pi_{\alpha\beta} + \beta_{\Pi} \Pi^2 + \beta_n n_\alpha n^\alpha \biggr]u^\mu - \frac{1}{T} \biggl[ \alpha_{\Pi n} \Pi  n^\mu + \alpha_{\pi n} \pi^{\mu\alpha} n_\alpha\biggr]\,.
\end{split}
\end{equation}
Let us now take the divergence of the entropy current. Inserting the definition~\eqref{app:hydro:second-order-Q}, we find
\begin{align}
    \nabla_\mu S^\mu &= \nabla_\mu \lr{s u^\mu} - \nabla_\mu \lr{\alpha n^\mu} - \frac{\beta_\pi}{T} \pi_{\alpha\beta} u^\mu \nabla_\mu \pi^{\alpha\beta} - \frac{\beta_\Pi}{T} \Pi u^\mu \nabla_\mu \Pi - \frac{\beta_n}{T} n_\alpha u^\mu \nabla_\mu n^\alpha \nonumber\\
    &- \pi_{\alpha\beta}\pi^{\alpha\beta} \nabla_\mu \lr{\frac{\beta_\pi}{2T} u^\mu} - \Pi^2 \nabla_\mu \lr{\frac{\beta_\Pi}{2T} u^\mu} - n_\alpha n^\alpha \nabla_\mu \lr{\frac{\beta_n}{2T} u^\mu} \nonumber\\
    &+ \frac{\alpha_{\Pi n}}{T} \lr{\Pi \nabla_\mu n^\mu + n^\mu \nabla_\mu \Pi} + \frac{\alpha_{\pi n}}{T} \lr{\pi^{\mu\alpha} \nabla_\mu n_\alpha + n_\alpha \nabla_\mu \pi^{\mu\alpha}}\label{app:hydro:MIS1-div-entropy}\\
    &+ \Pi n^\mu \nabla_\mu \lr{\frac{\alpha_{\Pi n}}{T}} + \pi^{\mu\alpha}n_\alpha \nabla_\mu \lr{\frac{\alpha_{\pi n}}{T}}\nonumber
\end{align}
Let us now turn to the equations of motion. The scalar equations of motion are given by
\begin{subequations}
    \begin{align}
        -u_\nu \nabla_\mu T^{\mu\nu} &=  \nabla_\mu \lr{\epsilon u^\mu} + (p + \Pi) \nabla_\mu u^\mu - u_\nu \nabla_\mu \pi^{\mu\nu} = 0\,,\\
        \nabla_\mu J^\mu &= \nabla_\mu \lr{n u^\mu} + \nabla_\mu n^\mu\,.
    \end{align}
\end{subequations}
Combining these two equations gives an expression for $\nabla_\mu \lr{s u^\mu}$:
\begin{equation}\label{app:hydro:MIS-1-sumu}
    \nabla_\mu \lr{s u^\mu} = \alpha \nabla_\mu n^\mu + \beta_\nu \nabla_\mu \pi^{\mu\nu} - \Pi \Delta^{\mu\nu}\nabla_\mu \beta_\nu\,,
\end{equation}
where we have made use of the identity $\Delta^{\mu\nu}\nabla_\mu \beta_\nu = (1/T) \nabla_\mu u^\mu$. Inserting the expression~\eqref{app:hydro:MIS-1-sumu} into the divergence of the entropy current~\eqref{app:hydro:MIS1-div-entropy} yields (after some finessing)
\begin{equation}
    \begin{split}
        \nabla_\mu S^\mu =& - \frac{\Pi}{T} \biggl[T\Delta^{\mu\nu}\nabla_\mu \beta_\nu + \beta_\Pi u^\mu \nabla_\mu \Pi + T \Pi  \nabla_\mu \lr{\frac{\beta_\Pi}{2 T} u^\mu} \\
        &- \alpha_{\Pi n} \nabla_\mu n^\mu - \gamma_{\Pi n} T n^\mu  \nabla_\mu \lr{\frac{\alpha_{\Pi n}}{T}} \biggr]\\
    &- \frac{n^\mu}{T} \biggl[ T\nabla_\mu \alpha + \beta_n u^\lambda \nabla_\lambda n_\mu + T n_\mu \nabla_\alpha \lr{\frac{\beta_{n}}{2 T} u^\alpha}- \alpha_{\Pi n} \nabla_\mu \Pi  \\
    & -\alpha_{\pi n} \nabla_\alpha \pi^\alpha_{\,\,\,\,\mu}- (1-\gamma_{\Pi n}) T \Pi \nabla_\mu \lr{\frac{\alpha_{\Pi n}}{T}}  - \gamma_{\pi n} T \pi^{\alpha}_{\,\,\,\,\mu} \nabla_\alpha \lr{\frac{\alpha_{\pi n}}{T}}\biggr] \\
    &- \frac{\pi^{\mu\nu}}{T} \biggl[T \nabla_\mu \beta_\nu + \beta_\pi u^\lambda \nabla_\lambda \pi_{\mu\nu} + T\pi_{\mu\nu} \nabla_\alpha \lr{\frac{\beta_\pi}{2 T} u^\alpha} \\
    &- \alpha_{\pi n} \nabla_\mu n_\nu - \lr{1-\gamma_{\pi n}}T n_\nu \nabla_\mu \lr{\frac{\alpha_{\pi n}}{T}} \biggr]\,,
    \end{split}
\end{equation}
where the parameters $\gamma_{\Pi n}$, $\gamma_{\pi n}$ have been introduced to control the distribution of the $\nabla_\mu \lr{\alpha_{xx}/T}$ between the square brackets using the trivial identity $X = \gamma X + \lr{1-\gamma} X$. Now, we demand that the entropy current have non-zero divergence for all possible field configurations. This leads immediately to the identification
\begin{subequations}\label{app:hydro:MIS1-EoM}
    \begin{align}
        \Pi =& - \zeta\biggl[T\Delta^{\mu\nu}\nabla_\mu \beta_\nu + \beta_\Pi u^\mu \nabla_\mu \Pi + T \Pi  \nabla_\mu \lr{\frac{\beta_\Pi}{2 T} u^\mu} \\
        &- \alpha_{\Pi n} \nabla_\mu n^\mu - \gamma_{\Pi n} T n^\mu  \nabla_\mu \lr{\frac{\alpha_{\Pi n}}{T}} \biggr]\nonumber\,,\\
        n^\alpha =& - \sigma \Delta^{\alpha\mu}\biggl[ T\nabla_\mu \alpha + \beta_n u^\lambda \nabla_\lambda n_\mu + T n_\mu \nabla_\lambda \lr{\frac{\beta_{n}}{2 T} u^\lambda}- \alpha_{\Pi n} \nabla_\mu \Pi  \\
    & -\alpha_{\pi n} \nabla_\lambda \pi^\lambda_{\,\,\,\,\mu}- (1-\gamma_{\Pi n}) T \Pi \nabla_\mu \lr{\frac{\alpha_{\Pi n}}{T}}  - \gamma_{\pi n} T \pi^{\lambda}_{\,\,\,\,\mu} \nabla_\lambda \lr{\frac{\alpha_{\pi n}}{T}}\biggr]\,,\nonumber \\
    \pi^{\alpha\beta} =& - 2 \eta \Delta^{\alpha\beta\mu\nu} \biggl[T \nabla_\mu \beta_\nu + \beta_\pi u^\lambda \nabla_\lambda \pi_{\mu\nu} + T\pi_{\mu\nu} \nabla_\lambda \lr{\frac{\beta_\pi}{2 T} u^\lambda} \\
    &- \alpha_{\pi n} \nabla_\mu n_\nu - \lr{1-\gamma_{\pi n}}T n_\nu \nabla_\mu \lr{\frac{\alpha_{\pi n}}{T}} \biggr]\,,\nonumber
    \end{align}
\end{subequations}
With this identification in place, the divergence of the entropy current becomes
\begin{equation}
    \nabla_\mu S^\mu = \frac{\Pi^2}{\zeta T} + \frac{n_\mu n^\mu}{\sigma T} + \frac{\pi^{\mu\nu} \pi_{\mu\nu}}{2 \eta T}
\end{equation}
which is always non-negative so long as 
\[
\zeta \geq 0, \qquad \sigma \geq 0, \qquad \eta \geq 0\,.
\]
The identifications~\eqref{app:hydro:MIS1-EoM} may be re-written in the following more suggestive form:
\begin{subequations}\label{app:hydro:MIS-1-relaxEq}
    \begin{align}
        \tau_\Pi \dot{\Pi} + \Pi &= - \zeta \nabla_\mu u^\mu + \,...\,,\\
        \tau_n \dot{n}^\mu + n^\mu &= - \sigma T \Delta^{\mu\nu} \nabla_\nu \lr{\frac{\mu}{T}} + \,...\,,\\
        \tau_\pi \dot{\pi}^{\mu\nu} + \pi^{\mu\nu} &= - \eta \sigma^{\mu\nu} + \,...\,,
    \end{align}
\end{subequations}
where the $...$ contain all of the ``interaction terms" between $\{\Pi, n^\mu, \pi^{\mu\nu}\}$, as well as interactions with derivatives of the hydrodynamic variables. In the above, we have defined $\tau_\Pi = \zeta \beta_{\Pi}$, $\tau_n = \sigma \beta_n$, $\tau_\pi = 2 \eta \beta_\pi$, and $\dot{X} = u^\mu \nabla_\mu X$. The equations~\eqref{app:hydro:MIS-1-relaxEq} are relaxation-type equations, morally the same as equation~\eqref{ch2:diffusion:MC_additional_eq}. In the late-time limit, the corrections $\Pi$, $n^\mu$, and $\pi^{\mu\nu}$ reduce to their Navier-Stokes limits as in~\eqref{ch2:hydro:ns}, and so we identify $\zeta$, $\sigma$, and $\eta$ as the bulk viscosity, charge conductivity, and shear viscosity respectively.

Finally, we note here that non-linear causality constraints on the parameters appearing in the MIS theory have been derived~\cite{Bemfica:2020xym} for a system with zero chemical potential. The form of the MIS equations appearing in~\cite{Bemfica:2020xym} are those arising from the DNMR formalism, which we discussed in Chapter~\ref{chapter:micro}. These conditions were then later applied~\cite{Plumberg:2021bme} to realistic simulations of heavy ion collisions; it was found that in commonly used simulations of the hydrodynamic regime, there were causality violations at early times.

\subsubsection{Resummed BRSSS}
The approach of Baier, Romatschke, Son, Starinets and Stephanov (BRSSS)~\cite{Baier:2007ix} is not, at its core, an attempt to derive MIS-type equations. Rather, it is instead a classification of all second-order terms in the derivative expansion that can be written down for a conformal hydrodynamic theory. For simplicity, in the following, we will restrict ourselves to an uncharged conformal fluid. Then $\nabla_\mu T^{\mu\nu} = 0$ is the only conservation equation, and $T^{\mu\nu}g_{\mu\nu} = 0$. We know that at first order in the derivative expansion, a conformal fluid has the following constitutive relations in Landau frame:
\begin{equation}
        T^{\mu\nu} = \epsilon u^\mu u^\nu + p \Delta^{\mu\nu} - \eta \sigma^{\mu\nu} + \oser{\de^2}\,,
\end{equation}
where $p =  \epsilon/d = a \,T^{d+1} f(\mu/T)$, with the final equality coming from the demand that $p$ scale homogeneously under a Weyl transformation. Let us now compare to the general form of $T^{\mu\nu}$ in Landau frame for a conformal fluid:
\begin{equation}
    T^{\mu\nu} = \epsilon u^\mu u^\nu + p \Delta^{\mu\nu} + \pi^{\mu\nu}\,.
\end{equation}
By direct comparison, we see that, 
\begin{equation}
    \pi^{\mu\nu} = -\eta \sigma^{\mu\nu} + \oser{\de^2}\,.
\end{equation}
Let us now consider the set of second-order transverse, traceless contributions one could make to $\pi^{\mu\nu}$. There are five independent tensors one can form which transform homogeneously under Weyl transformations:
\begin{subequations}
    \begin{align}
        {\cal O}^{\mu\nu}_1 &= \Delta^{\mu\nu\alpha\beta} u^\lambda \nabla_\lambda \sigma_{\alpha\beta} + \frac{1}{d} \sigma^{\mu\nu} \lr{\nabla_\lambda u^\lambda}\,,\\
        {\cal O}^{\mu\nu}_2 &= \Delta^{\mu\nu\alpha\beta} R_{\alpha\beta} - (d-1) \Delta^{\mu\nu\alpha\beta}u^\rho u^\sigma R_{\rho\alpha\beta\sigma}\,,\\
        {\cal O}_3^{\mu\nu} &= \Delta^{\mu\nu\alpha\beta} \sigma_{\alpha\lambda} \sigma_\beta^{\,\,\,\,\lambda}\,,\\
        {\cal O}_4^{\mu\nu} &= \Delta^{\mu\nu\alpha\beta} \sigma_{\alpha\lambda} \Omega_\beta^{\,\,\,\,\lambda}\,,\\
        {\cal O}_5^{\mu\nu} &= \Delta^{\mu\nu\alpha\beta} \Omega_{\alpha\lambda} \Omega_\beta^{\,\,\,\,\lambda}\,,
    \end{align}
\end{subequations}
where $\Omega^{\mu\nu} = \Delta^{\mu\alpha} \Delta^{\nu\beta} \nabla_{[\alpha} u_{\beta]}$ is the vorticity tensor, $R_{\mu\nu\alpha\beta}$ is the Riemann tensor, $R_{\mu\nu}$ is the Ricci tensor, and 
\begin{equation}
    \Delta^{\mu\nu\alpha\beta} = \frac{1}{2} \lr{\Delta^{\mu\alpha} \Delta^{\nu\beta} + \Delta^{\mu\beta} \Delta^{\nu\alpha} - \frac{2}{d} \Delta^{\mu\nu} \Delta^{\alpha\beta}}\,,
\end{equation}
is once again the transverse traceless projector. We therefore can write
\begin{equation}\label{app:hydro:BRSSS-nosum}
    \begin{split}
        \pi^{\mu\nu} &= -\eta \sigma^{\mu\nu}\\
        &+ \eta \tau_\pi {\cal O}^{\mu\nu}_1 + \kappa {\cal O}_2^{\mu\nu} + \lambda_1 {\cal O}_3^{\mu\nu} + \lambda_2 {\cal O}_4^{\mu\nu} + \lambda_3 {\cal O}_5^{\mu\nu}\,.
    \end{split}
\end{equation}
For future convenience, $\tau_\pi$ has been defined with a factor of $\eta$ pulled out. Note that the factors $\tau_\pi$, $\kappa$, $\lambda_{1,2,3}$ in equation~\eqref{app:hydro:BRSSS-nosum} are physical second-order transport coefficients, and can in principle be computed given a microscopic theory. 

Now, let us note the following. To a second-order error, $\sigma^{\mu\nu} = - \pi^{\mu\nu}/\eta$. Of course, this cannot be an exact statement; $\sigma^{\mu\nu}$ is first order in derivatives, while $\pi^{\mu\nu}$ is infinite order in derivatives. Nevertheless, we may use this relation to perform an infinite-order re-summation of~\eqref{app:hydro:BRSSS-nosum}. In practice, we replace $\sigma^{\mu\nu}$ everywhere it appears in the second-order part of~\eqref{app:hydro:BRSSS-nosum} with $-\pi^{\mu\nu}/\eta$. Following this prescription, we arrive (after also using the ideal-order equations of motion) at the following equation~\cite{Baier:2007ix}:
\begin{equation}\label{app:hydro:rBRSSS-eq}
    \begin{split}
        \pi^{\mu\nu} &= - \eta \sigma^{\mu\nu} - \tau_{\pi} \Delta^{\mu\nu\alpha\beta} \biggl[u^\lambda \nabla_\lambda \pi_{\alpha\beta} + \frac{d+1}{d} \pi_{\alpha\beta} \lr{\nabla_\lambda u^\lambda} \biggr] + \kappa {\cal O}_2^{\mu\nu}\\
        &+ \frac{\lambda_1}{\eta^2} \Delta^{\mu\nu\alpha\beta} \pi_{\alpha\lambda} \pi_{\beta}^{\,\,\,\,\lambda} - \frac{\lambda_2}{\eta} \Delta^{\mu\nu\alpha\beta} \pi_{\alpha\lambda} \Omega_\beta^{\,\,\,\,\lambda} + \lambda_3 {\cal O}_5^{\mu\nu\,,.}
    \end{split}
\end{equation}
These equations are the equations of the resummed BRSSS  (rBRSSS) theory~\cite{Romatschke:2017ejr}, though the resummed theory is often simply called the BRSSS theory. Pulling the time derivative of $\pi^{\mu\nu}$ over, we arrive at the standard form for MIS equations for a conformal fluid,
\begin{equation}
    \tau_\pi \dot{\pi}^{\mu\nu} + \pi^{\mu\nu} = - \eta \sigma^{\mu\nu }+\,...\,,
\end{equation}
where here $\dot{\pi}^{\mu\nu} = \Delta^{\mu\nu\alpha\beta}u^\lambda\nabla_\lambda \pi_{\alpha\beta}$, and the $...$ denote higher-order terms. Note that the higher-order terms included in~\eqref{app:hydro:rBRSSS-eq} are not quite the same as~\eqref{app:hydro:MIS1-EoM}.

\subsubsection{Generating Functional Approach}
The final approach we will consider in this section is a very recent one, taking advantage of the idea of ``non-equilibrium thermodynamics". Introduced in~\cite{Jain:2023obu}, the generating functional approach gives an alternative way of extracting phenomenological equations from the entropy current. Let us consider the generating functional of charged relativistic hydrodynamics. Up to first order in the derivative expansion, it is given by
\begin{equation}\label{app:hydro:gen_func_1}
     W[g,A] = \int d^{d+1}x \sqrt{-g} \lr{p[T,\mu] + ...}
\end{equation}
We would like to be able to describe the dissipative corrections to the stress-energy tensor $T^{\mu\nu}$ and the charge current $J^\mu$ using the language of extended thermodynamics. In that vein, we will modify~\eqref{app:hydro:gen_func_1} to include contributions from a symmetric 2-tensor $\kappa^{\mu\nu}$ and a vector $v^\mu$. In addition, let us impose that $\kappa^{\mu\nu}u_\nu = v^\mu u_\mu = 0$. With the restriction, $\kappa^{\mu\nu}$ in particular may be decomposed as:
\begin{equation}
    \kappa^{\mu\nu} = \frac{1}{d}\lr{\tr \kappa} \Delta^{\mu\nu} + \kappa^{\braket{\mu\nu}}
\end{equation}
where $\kappa^{\braket{\mu\nu}} = \Delta^{\mu\nu\alpha\beta}\kappa_{\alpha\beta}$. We see then that we have two independent parts of $\kappa$: the trace, and the traceless part. We would like to connect $\kappa_{\mu\nu}$ and $v_\mu$ to the viscous corrections to the stress-energy tensor and charge current. The constitutive relations in Landau frame for $T^{\mu\nu}$ and $J^\mu$ may be written as
\begin{equation}
    T^{\mu\nu} = \epsilon u^\mu u^\nu + p  \Delta^{\mu\nu} + \Pi^{\mu\nu}\,, \quad J^\mu = n u^\mu + n^\mu\,,
\end{equation}
where $\Pi^{\mu\nu} = \Pi \Delta^{\mu\nu} + \pi^{\mu\nu}$. We can write $\Pi^{\mu\nu}$ and $n^\mu$ in terms of $\kappa^{\mu\nu}$, $v^\mu$:
\begin{subequations}\label{app:MIS:Pin-original}
    \begin{align}
        \Pi^{\mu\nu} &= \alpha_\kappa^S \tr\kappa \Delta^{\mu\nu} + \alpha_\kappa^{T} \kappa^{\braket{\mu\nu}} + \Pi^{\mu\nu}_{\rm h.s.}\,,\\
        n^\mu &= \alpha_v v^\mu + n^\mu_{\rm h.s.}\,,
    \end{align}
\end{subequations}
where $\alpha_{\kappa}^{S,T}$ and $\alpha_v$ are arbitrary coefficients used to fix the relationship between $\Pi^{\mu\nu}, n^\mu$ and $\kappa^{\mu\nu},v^\mu$. The terms $\Pi^{\mu\nu}_{\rm h.s.}$ and $n^\mu_{\rm h.s.}$ denote hydrostatic contributions to $\Pi^{\mu\nu}$, $n^\mu$ from the generating functional due to the dependence of the isotropic pressure $p$ on $\kappa_{\mu\nu}$ and $v_\mu$. To determine the dependence of the pressure, let us begin by considering the independent scalars that can be formed from these quantities. The first obvious one is $v^2 = v_\mu v^\mu$. The second obvious one is $\tr\kappa$. Finally, we could use $\kappa^2 = \kappa_{\mu\nu} \kappa^{\mu\nu}$. However, let us note instead that
\begin{equation}
    \kappa_{\braket{\mu\nu}}\kappa^{\braket{\mu\nu}} = \lr{\kappa^{\mu\nu} - \frac{1}{d} \lr{\tr \kappa} \Delta^{\mu\nu}}\lr{\kappa_{\mu\nu} - \frac{1}{d} \lr{\tr \kappa} \Delta_{\mu\nu}} = \kappa^2 - \frac{1}{d} \lr{\tr \kappa}^2 
\end{equation}
Since $\tr \kappa$ and $\kappa_{\braket{\mu\nu}}$ are entirely independent, let us choose to use $\lr{\tr\kappa}^2$ and $\kappa_{\braket{\mu\nu}}\kappa^{\braket{\mu\nu}} = \kappa^2 - \frac{1}{d} \lr{\tr \kappa}^2$ as the two scalars from $\kappa^{\mu\nu}$. We may then introduce these degrees of freedom as extended thermodynamic variables. In other words, we write the first law of thermodynamics as
\begin{equation}\label{app:MIS:firstlaw}
    d\epsilon = T \,ds + \mu \,dn + \frac{1}{8} \chi^{S}_{\kappa} d\lr{\tr \kappa}^2 + \frac{1}{4} \chi^{T}_{\kappa} d\lr{\kappa^2 - \frac{1}{d} \lr{\tr \kappa}^2} + \frac{1}{2} \chi_v dv^2\,.
\end{equation}
where $\chi^{S,T}_{\kappa}$, $\chi_v$ are susceptibilities. Equation~\eqref{app:MIS:firstlaw} may be straightforwardly re-written as a Gibbs-Duhem relation using $\epsilon = -p + s T + n \mu$:
\begin{equation}
    dp = s dT + n d\mu - \frac{1}{8} \chi^{S}_{\kappa} d\lr{\tr \kappa}^2 - \frac{1}{4} \chi^{T}_{\kappa} d\lr{\kappa^2 - \frac{1}{d} \lr{\tr \kappa}^2} - \frac{1}{2} \chi_v dv^2\,.
\end{equation}
Let us note at this point that
\begin{subequations}
\begin{align}
    \delta (\tr{\kappa}) &= \lr{ \kappa^{\alpha\beta} }\delta g_{\alpha\beta}\,,\\ 
    \delta (\kappa^2 - \frac{1}{d} \lr{\tr \kappa}^2) &= \lr{\kappa^{\alpha}_{\,\,\,\,\lambda} \kappa^{\lambda\beta} - \frac{2}{d} \lr{\tr \kappa} \kappa^{\alpha\beta}} \delta g_{\alpha\beta}\,,\\
    \delta v^2 &= v^\alpha v^\beta \delta g_{\alpha\beta}
\end{align}
\end{subequations}
Let us then write the modified generating functional with $\kappa_{\mu\nu}$, $v_\mu$ dependence as
\begin{equation}
   W[A,g] = \int d^{d+1}x \sqrt{-g} \biggl[p\lr{T,\mu,v^2,\lr{\tr \kappa}^2,(\kappa^2 - \frac{1}{d} \lr{\tr \kappa}^2)}+...\biggr]
\end{equation}
It is convenient to split the pressure into an equilibrium part $P$, and a non-equilibrium part ${\cal P}$ which vanishes when $\kappa\to0$, $v\to 0$, i.e. $p = P + {\cal P}$. Varying, we subsequently find
\begin{subequations}
    \begin{align}
        T^{\mu\nu} &= \epsilon u^\mu u^\nu + P \Delta^{\mu\nu} \\
        &+ \biggl[{\cal P}\Delta^{\mu\nu} +\frac{\chi_\kappa^S}{2} \kappa^{\mu\nu} \tr\kappa + \chi_\kappa^T\lr{\kappa^{\mu}_{\,\,\,\,\lambda} \kappa^{\lambda \nu} - \frac{1}{d} \tr\kappa \kappa^{\mu\nu}} + \chi_v v^\mu v^\nu\biggr]\,,\nonumber\\
        J^\mu &= n u^\mu\,.
    \end{align}
\end{subequations}
We have now determined the ``hydrostatic" contribution to the constitutive relations, and equations~\eqref{app:MIS:Pin-original} become
\begin{subequations}
    \begin{align}
        \Pi^{\mu\nu} &= \alpha_\kappa^S \tr\kappa \Delta^{\mu\nu} + \alpha_\kappa^{T} \kappa^{\braket{\mu\nu}} \\
        &+ \biggl[{\cal P}\Delta^{\mu\nu} +\frac{\chi_\kappa^S}{2} \kappa^{\mu\nu} \tr\kappa + \chi_\kappa^T\lr{\kappa^{\mu}_{\,\,\,\,\lambda} \kappa^{\lambda \nu} - \frac{1}{d} \tr\kappa \kappa^{\mu\nu}} + \chi_v v^\mu v^\nu\biggr]\,,\nonumber\\
        n^\mu &= \alpha_v v^\mu\,,
    \end{align}
\end{subequations}
Now, let us turn to the equations of motion. The scalar conservation equations are given by
\begin{subequations}
    \begin{align}\label{app:hydro:scalar_genfunc}
        -u_\nu \nabla_\mu T^{\mu\nu} &= \nabla_\mu \lr{\epsilon u^\mu} + P \nabla_\mu u^\mu - u_\nu \nabla_\mu \Pi^{\mu\nu} =0\,,\\
        \nabla_\mu J^\mu &= \nabla_\mu \lr{n u^\mu} + n^\mu = 0\,.
    \end{align}
\end{subequations}
Let us now note that $\epsilon = - P - {\cal P} + s T + n \mu$, and $\nabla_\mu {\cal P} = \nabla_\mu p - \nabla_\mu P $. The two scalar equations~\eqref{app:hydro:scalar_genfunc} may then be used (with some work) to get a single equation for the entropy density $s$:
\begin{equation}\label{app:hydro:sumu-replace}
\begin{split}
     \nabla_\mu \lr{s u^\mu} =& - \lr{\frac{\chi_\kappa^S}{8} \Lied_\beta \tr\kappa^2 + \frac{\chi_\kappa^T}{4} \Lied_\beta\lr{\kappa^2 - \frac{1}{d} \tr\kappa^2} + \frac{\chi_v}{2} \Lied_\beta v^2} \\
     &+ {\cal P} \Delta^{\mu\nu}\nabla_\mu \beta_\nu + \frac{\mu}{T} \nabla_\mu n^\mu + \beta_\nu \nabla_\mu \Pi^{\mu\nu} \,,
\end{split}
\end{equation}
where $\Lied_\beta$ is the Lie derivative along $\beta^\mu$, and we have again used the fact that $(\nabla_\mu u^\mu)/T = \Delta^{\mu\nu}\nabla_\mu \beta_\nu$. Let us now also recall the definitions of $\Pi^{\mu\nu}$ and $n^\mu$. Taking the parameters $\chi_{\kappa,v}^{S,T}$ and $\alpha_{\kappa,v}^{S,T}$ to be constant for simplicity\footnote{This assumption is, in general, not true -- more general equations may be obtained by including derivatives of the $\chi$'s and $\alpha$'s.}, we can again find with some work that
\begin{subequations}
    \begin{align}
        \beta_\nu \nabla_\mu \Pi^{\mu\nu} &= - \frac{\alpha^S}{2} \tr\kappa \Delta^{\mu\nu} \nabla_\mu \beta_\nu - \alpha^T \kappa^{\braket{\mu\nu}} \nabla_\mu \beta_\nu - {\cal P} \Delta^{\mu\nu} \nabla_\mu \beta_\nu\nonumber\\
        &+\frac{\chi_\kappa^S}{2} \lr{- \frac{1}{2} \tr\kappa \Delta^{\mu\nu} \Lied_\beta \kappa_{\mu\nu} + \frac{1}{4} \Lied_{\beta} \lr{\tr \kappa}^2} \\
        &+ \chi_\kappa^T \lr{-\frac{1}{2} \kappa^{\braket{\mu\nu}}\Lied_{\beta}\kappa_{\mu\nu} + \frac{1}{4} \Lied_{\beta} \lr{\kappa^2 - \frac{1}{d} \lr{\tr \kappa}^2}}\,,\nonumber\\
        \frac{\mu}{T} \nabla_\mu n^\mu &= \frac{\mu}{T} \alpha_v \nabla_\mu v^\mu\,.
    \end{align}
\end{subequations}
Let us now return to the canonical form of the entropy-current~\eqref{ch2:hydro:canonical-entropy}:
\[
S^\mu = p \beta^\mu - T^{\mu\nu} \beta_\nu - \frac{\mu}{T} J^\mu
\]
Inserting the constitutive relations for $T^{\mu\nu}$ and $J^\mu$ leads to
\begin{equation}
    S^\mu = s u^\mu - \Pi^{\mu\nu} \beta_\nu - \frac{\mu}{T} n^\mu
\end{equation}
However, $\Pi^{\mu\nu}\beta_\nu = 0$, and so we are left with
\begin{equation}\label{app:hydro:entropy-temp-MIS-3}
    S^\mu = s u^\mu  - \frac{\mu}{T} n^\mu
\end{equation}
Recall that the divergence of the entropy current $\nabla_\mu S^\mu$ must be non-negative. Taking the divergence of~\eqref{app:hydro:entropy-temp-MIS-3} and replacing $\nabla_\mu \lr{s u^\mu}$ via equation~\eqref{app:hydro:sumu-replace}, we are left with
\begin{equation}
    \begin{split}
        \nabla_\mu S^\mu &= - \lr{\frac{\chi_\kappa^S}{8} \Lied_\beta \tr\kappa^2 + \frac{\chi_\kappa^T}{4} \Lied_\beta\lr{\kappa^2 - \frac{1}{d} \tr\kappa^2} + \frac{\chi_v}{2} \Lied_\beta v^2} \\
     &+ {\cal P} \Delta^{\mu\nu}\nabla_\mu \beta_\nu  + \beta_\nu \nabla_\mu \Pi^{\mu\nu} - n^\mu \nabla_\mu \lr{\frac{\mu}{T}} 
    \end{split}
\end{equation}
Inserting the definitions of $\Pi^{\mu\nu}$ and $n^\mu$ in terms of $\kappa^{\mu\nu}$ and $v^\mu$ yields (after some simplification)
\begin{equation}
    \begin{split}
        \nabla_\mu S^\mu =& - \frac{\alpha_\kappa^S}{4} \tr\kappa \Delta^{\mu\nu} \lr{2 \nabla_\mu \beta_\nu + \frac{\chi_\kappa^S}{\alpha_\kappa^S}  \Lied_\beta \kappa_{\mu\nu}} \\
        &- \frac{\alpha_\kappa^T}{2} \kappa_{\braket{\mu\nu}} \Delta^{\mu\nu\alpha\beta} \lr{ 2\nabla_\mu \beta_\nu + \frac{\chi_{\kappa}^T}{\alpha_\kappa^T}\Lied_{\beta}\kappa_{\mu\nu}}\\
        & - \alpha_v v^\mu \lr{ \nabla_\mu\lr{\frac{\mu}{T}} + \frac{\chi_v}{\alpha_v} \Lied_\beta v_\mu}
    \end{split}
\end{equation}
We need the divergence of the entropy current to be non-negative for all possible hydrodynamic field configurations. This demand is satisfied if we make the identification
\begin{subequations}\label{app:hydro:SKMIS-eqs}
    \begin{align}
        \alpha_\kappa^S \tr \kappa \Delta^{\mu\nu} &= - \zeta \Delta^{\mu\nu} \lr{ \nabla_\mu u_\nu + \frac{T\chi_\kappa^S}{2\alpha_\kappa^S}  \Lied_\beta \kappa_{\mu\nu}}\,,\\
        \alpha_\kappa^T \kappa^{\braket{\mu\nu}} &= - 2\eta \Delta^{\mu\nu\alpha\beta} \lr{ \nabla_\alpha u_\beta + \frac{T\chi_{\kappa}^T}{2\alpha_\kappa^T}\Lied_{\beta}\kappa_{\alpha\beta}}\,,\\
        \alpha_v v^\mu &= - \sigma T \Delta^{\mu\nu} \lr{ \nabla_\nu\lr{\frac{\mu}{T}} + \frac{\chi_v}{\alpha_v} \Lied_\beta v_\nu}\,,
    \end{align}
\end{subequations}
where we introduce parameters $\zeta$, $\sigma$, and $\eta$. The factors of $2$ and $d$ are purely for convention. We can identify $\zeta$ as the bulk viscosity, $\sigma$ with the charge conductivity, and $\eta$ with the shear viscosity. This is all well and good, but we want equations which describe the evolution of $\Pi^{\mu\nu}$ and $n^\mu$, not $\kappa^{\mu\nu}$ and $v^\mu$. However, because of the hydrostatic piece, the full relationship between $\{\Pi^{\mu\nu},n^\mu\}$ and $\{\chi^{\mu\nu}, v^\mu\}$ is non-linear. We can get a sense of their form by looking at the linearized equations (i.e. neglecting the hydrostatic contribution): the equations may then be written in terms of $\Pi^{\mu\nu}$ using the identification
\begin{equation}\label{app:hydro:SKMIS-linear-rel}
    \kappa^{\mu\nu} = \frac{1}{d} \tr\kappa \Delta^{\mu\nu} + \kappa^{\braket{\mu\nu}} = \frac{1}{d^2 \alpha_\kappa^S} \tr\Pi \Delta^{\mu\nu} + \frac{1}{\alpha_\kappa^T} \Pi^{\braket{\mu\nu}}
\end{equation}
Using equation~\eqref{app:hydro:SKMIS-linear-rel} on the equations~\eqref{app:hydro:SKMIS-eqs} yields the linear equations
\begin{subequations}
    \begin{align}
         \tau_{\Pi}^{S} T\Lied_{\beta} \lr{\frac{1}{d} \tr\Pi} + \frac{1}{d}\tr \Pi &= - \zeta \nabla_\mu u^\mu  + ...\,,\\
        \tau_{\Pi}^T T \Delta^{\mu\nu\alpha\beta}\Lied_{\beta} \Pi_{\braket{\alpha\beta}} + \Pi^{\braket{\mu\nu}} &= - \eta \sigma^{\mu\nu}+...\,,\\
        \tau_n T \Delta^{\mu\nu}\Lied_{\beta} n_\nu + n^\mu &= - \sigma T \Delta^{\mu\nu}\nabla_\nu \lr{\frac{\mu}{T}} +...\,,
    \end{align}
\end{subequations}
where $\tau_{\Pi}^S = \zeta \chi_{\kappa}^S/(\sqrt{2} \alpha_{\kappa}^{S})^2$, $\tau_{\Pi}^T = \eta \chi_{\kappa}^T/(\alpha_{\kappa}^T)^2$, and $\tau_{n} = \sigma \chi_v/\alpha_v^2$ denote relaxation times for the bulk viscosity, shear viscosity, and transverse charge current. The $...$ denote numerous non-linear terms which were not included. We also neglected certain derivatives of $u^\mu$, which would be included in the $...$ terms. For the full expressions in the case of a conformal fluid, the interested reader may refer to~\cite{Jain:2023obu}.

    \startappendix{Linear Response Theory}
\label{app:linear response}
Linear response theory (LRT) investigates how systems respond at a linear level after begin deformed by a source. This gives us a great deal of valuable information about the macroscopic properties of the system, from susceptibilities to dispersion relations to Kubo formulae and more. Here, we will attempt to convey the core idea behind the framework; a reader interested in more detail may refer to~\cite{Kovtun:2012rj,Forster}, upon which this appendix has been built. We will work with $d=3$, but results may be straightforwardly generalized to $d+1$ spacetime dimensions.

\section{Diffusion}

In what follows, we will consider the case of a system where we only care about perturbations in $U(1)$ charge. For example, we could be thinking about spin diffusion~\cite{Forster}. Regardless of the situation, we wish to consider a system with only one relevant conserved quantity: the one-point function of the $U(1)$ charge current operator $J^\mu$.

Let us begin by considering the diffusion equation, and a perturbation $J^\mu = J^\mu_0 + \delta J^\mu$. As we saw in Chapter~\ref{ch2:sec_diffusion}, this can be obtained from the one-point function by writing down the spatial component of the charge current in terms of the time component via Fick's law:
\begin{equation}
    \delta \braket{J^i} = - D \de^i \delta \braket{J^0}\,,
\end{equation}
where $D$ is the diffusion constant. Now, let us call $J^0$ the charge density $n$. Then, assuming $D$ is a constant yields the conservation equation
\begin{equation}\label{app:LRT:diffusion-equation}
\de_t \delta \braket{n(t,x^j)} - D \de_i \de^i \delta\braket{n(t,x^j)} = 0\,.
\end{equation}
For notational clarity, we hereafter drop the $\delta$. Let us now Fourier-transform this equation solely in the spatial components, according to
\begin{equation}
     \braket{n(t,x^j)} = \int \frac{d^3 x}{(2\pi)^3} e^{i k_j x^j} \braket{n(t,k^j)}\,,
\end{equation}
This yields the transformed equation
\begin{equation}
    \de_t \braket{n(t,k^j)} + D k^2 \braket{n(t,k^j)} = 0\,.
\end{equation}
This may be easily solved:
\begin{equation}~\label{app:LRT:fourier}
    \braket{n(t, k^j)} = e^{- D k^2 t} \braket{n(t=0,k^j)}\,.
\end{equation}
Let us now further define the Laplace transformation of $\braket{n(t,k^j)}$ as
\begin{equation}~\label{app:LRT:Laplace}
    \braket{n(z,k^j)} = \int_{0}^\infty dt\, e^{i z t} \braket{n(t,k^j)}\,.
\end{equation}
Inserting the solution~\eqref{app:LRT:fourier} into the Laplace transformation~\eqref{app:LRT:Laplace} yields
\begin{equation}\label{app:LRT:final-Laplace}
    \braket{n(z,k^j)} = \frac{\braket{n(t=0,k^j)}}{-i z+ D k^2}
\end{equation}
Note that $\braket{n(z,k^j)}$ has a pole in the plane of complex $z$ at $z = - i D k^2$. If $z$ were $\omega$, this would be the dispersion relation for the diffusion equation. However, referring to equation~\eqref{app:LRT:Laplace}, we can see that the integral is only convergent if $\Im(z) > 0$. We will later analytically continue to the lower-half complex plane; for now, we simply note the pole.

We would now like to introduce a source term to deform the theory, and see how the charge density responds. Recall that the charge density represents the expectation value of $J^0$. In general, if we introduce a perturbation to the Hamiltonian of a system, the expectation value of an operator ${\cal O}$ will change according to~\cite{Kovtun:2012rj}
\begin{equation}\label{app:LRT:master-eq}
    \delta \braket{{\cal O}(t,x^j)} = - i \int_{-\infty}^{t} dt'\, \braket{[{\cal O}(t,x^j),\delta H(t')]}\,.
\end{equation}
Let us introduce a source $\mu$ for the charge density -- this is the $U(1)$ chemical potential. Let us denote the Hamiltonian of the system without the external source by $H$; then the modified Hamiltonian ${\cal H} = H + \delta H$ is given by
\begin{equation}
    {\cal H} = H - \int d^dx\,\mu(t,x^j)\,n(t,x^j)\,.
\end{equation}
In what follows, we will be interested primarily in the one-point function and the retarded two-point function. 

\subsection{Static susceptibility}
Let us now consider computing the equilibrium expectation value $\braket{n(t,x)}_{\rm eq}$ in the presence of a constant $\mu(t,x^j) = \mu_0$ at $t=0$; we will denote this by $\braket{n(x^j)}_{\mu_0}$. Let us also denote the total charge operator by $N = \int d^dx \,n(x^j)$, and the spatially averaged charge density by $\bar{n} = N/V$. Taking the thermodynamic limit, a constant source term means we expect $n(x^j) \to \bar{n}$. Therefore, 
\begin{align*}
    \braket{n(x^j)}_{\mu_0} \to\braket{\bar{n}}_{\mu_0} &= \frac{ \tr\lr{ e^{-\beta \lr{H - \int d^d x \mu_0 \bar{n}}} \bar{n}}}{\tr\lr{ e^{-\beta \lr{H - \int d^d x \mu_0 \bar{n}}}}}= \frac{ \tr\lr{ e^{-\beta \lr{H - N \mu_0}} \bar{n}}}{\tr\lr{ e^{-\beta \lr{H - N\mu_0}}}}\,.
\end{align*}
We therefore can write
\begin{align*}
    \frac{1}{\beta}\pder{\braket{\bar{n}}_{\mu_0}}{\mu_0} &= \lr{\frac{ \tr\lr{ e^{-\beta \lr{H - N \mu_0}} \bar{n} N}}{\tr\lr{ e^{-\beta \lr{H - N \mu_0}}}}}  - \lr{\frac{\tr\lr{ e^{- \beta (H - N \mu_0)} \bar{n}}}{\tr\lr{ e^{-\beta \lr{H - N \mu_0}}}}}\lr{\frac{\tr\lr{ e^{- \beta (H - N \mu_0)} N}}{\tr\lr{ e^{-\beta \lr{H - N \mu_0}}}}}\\
    &= \frac{1}{V} \lr{\braket{N^2}_{\mu_0} - \braket{N}^2_{\mu_0}}\,.
\end{align*}
Taking the limit as $\mu_0 \to 0$, we find an expression for the static susceptibility $\chi_{nn}$:
\begin{equation}
    \pder{\braket{n}_{\mu_0}}{\mu_0} \biggl|_{\mu_0 = 0} = \frac{\beta}{V} \lr{\braket{N^2}_{\rm eq} - \braket{N}_{\rm eq}^2} \equiv \chi_{nn}\,.
\end{equation}
Since $\lr{\braket{N^2}_{\rm eq} - \braket{N}_{\rm eq}^2} \geq 0$ for all configurations, we must have $\chi_{nn} \geq 0$. We may then claim that, for a ``small enough" perturbation (i.e. for small $k$) $\mu(t=0,x^j)$, one can write the charge density as
\begin{equation}\label{app:LRT:static-susceptibility-diffusion}
    \braket{n(t=0,x^j)} = \chi_{nn} \mu(t=0,x^j).
\end{equation}
In the above, we have restored the spatial dependence of $n$ and $\mu$, with the understanding that $n$ and $\mu$ do not vary too quickly. 

\subsection{Two-point functions}
\begin{figure}[t]
    \centering
    \includegraphics[width=0.7\linewidth]{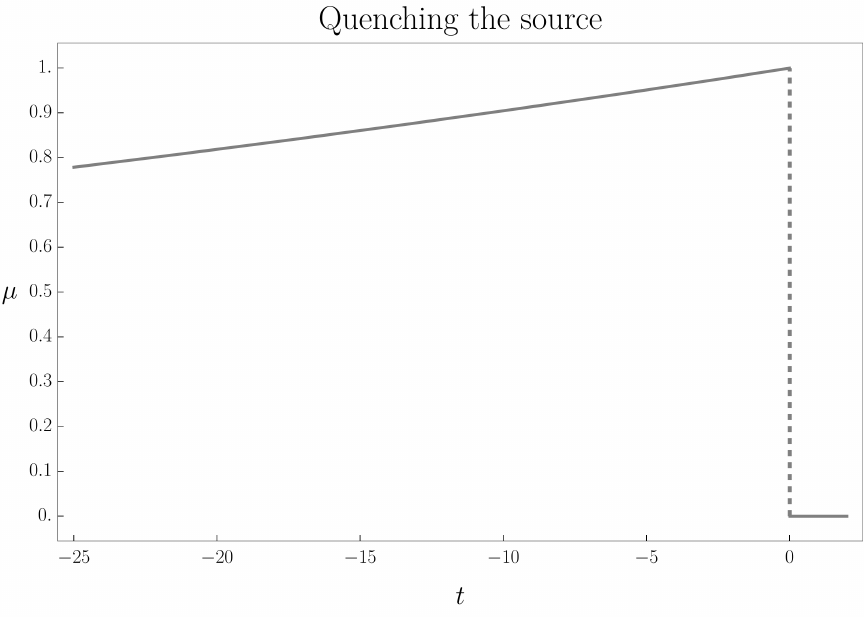}
    \caption{A plot of the behaviour of $\mu(t,x^j)$ with respect to time. The source is adiabatically turned on from time $t=-\infty$, and then is suddenly quenched at $t=0$. Linear response theory studies how the charge density reacts to the sudden change.}
    \label{fig:quench}
\end{figure}

Now, let us return to the equation~\eqref{app:LRT:master-eq}, and input $\delta H(t') = - \int d^{d}y \mu(t',y^j) n(t',y^j)$. We also make a particular assumption about the form of $\mu(t',y^j)$ in time -- that we adiabatically (i.e. slowly enough that the system never leaves equilibrium) turn on $\mu$ up until $t' = 0$, then immediately and sharply shut the source off, thrusting the system out of equilibrium. In other words, we define
\begin{equation}
    \mu(t', y'^j) = e^{\ce t'} \bar{\mu}(y^j) \theta(-t')\,,
\end{equation}
where we define $\bar{\mu}$ as the purely spatial part of $\mu$, and $\ce$ is an infinitesimally small positive constant. This type of sharp change in a parameter is sometimes called a quench; see Figure~\ref{fig:quench}.
We then have (slightly adjusting the bounds on the integral)
\begin{equation}
    \delta \braket{n(t,x^j)} =  i \int_{-\infty}^{0} dt' \int d^dy e^{\ce t'} \theta(t-t') \bar{\mu}(y^j)\braket{[n(t,x^j), n(t',y^j)]}
\end{equation}
Let us now define the following shorthand. We define the retarded two-point function $G_{nn}^R(t-t',x^j - y^j)$ by
\begin{equation}
    G_{nn}^{R}(t-t',x^j-y^j) \equiv - i \theta(t-t') \braket{[n(t,x^j),n(t',y^j)]}\,.
\end{equation}
We can then write
\begin{equation}\label{app:LRT:eq-1-diff}
    \delta\braket{n(t,x^j)} = -\int_{-\infty}^0 dt' \int d^dy e^{\ce t'} G_{nn}^{R}(t-t',x^j-y^j)\bar{\mu}(y^j) 
\end{equation}
Let us now Fourier-transform equation~\eqref{app:LRT:eq-1-diff} in space to find (dropping the $\delta$ as before for convenience)
\begin{equation}~\label{app:LRT:spatial_FT_diffusion}
    \braket{\tilde{n}(t,k^j)} = - \int_{-\infty}^0 dt' e^{\ce t'} \bar{\mu}(k^j) G^{R}_{nn}(t-t', k^j)\,.
\end{equation}
We then Fourier transform $G^{R}_{nn}(t-t',k^j)$ in time, giving
\begin{equation}
    G^{R}_{nn}(t-t,k^j) = \int_{-\infty}^{\infty} \frac{d\omega}{2\pi} G^R_{nn}(\omega,k^j) e^{-i \omega (t-t')}\,.
\end{equation}
Inserting this in to~\eqref{app:LRT:spatial_FT_diffusion} and evaluating the $t'$ integral yields
\begin{equation}
    \braket{\tilde{n}(t,k^j)} = -  \bar{\mu}(k^j)\int_{-\infty}^{\infty} \frac{d\omega}{2\pi} \frac{e^{-i \omega t} G^R_{nn}(\omega,k^j)}{i \omega +  \ce}
\end{equation}
Let us now Laplace transform both sides of this equation to get
\begin{equation}~\label{app:LRT:barn-no-eval}
    \braket{\bar{n}(z,k^j)} = - \bar{\mu}(k^j) \int_{-\infty}^\infty \frac{d\omega}{2\pi} \frac{G^R_{nn}(\omega,k^j)}{(i \omega + \ce)(i(\omega-z) + \ce)}\,,
\end{equation}
where a factor of $\ce>0$ has been added to the second pole for convergence. Now, we can note that $G^{R}_{nn}(\omega,k^j)$ is analytic in the upper-half plane of complex $\omega$, due to the fact that $G^{R}_{nn}(t-t',k^j)$ was only non-zero for $t-t'>0$. This implies that we can close the integral~\eqref{app:LRT:barn-no-eval} in the upper half-plane without worrying about branch cuts. Closing in the upper half plane, we pick up the poles at $\omega = i \ce$ and $\omega = z + i \ce$. By the residue theorem, we therefore have
\begin{equation}\label{app:LRT:solve-for-G-diffusion}
    \braket{\bar{n}(z,k^j)} = - \bar{\mu}(k^j) \lr{\frac{G^R_{nn}(z+i \ce,k^j) - G^R_{nn}(0 + i \ce,k^j)}{i z}}
\end{equation}
Let us now return to equation~\eqref{app:LRT:fourier}, and set $t=0$. Defining a new $t'' = - t'$, we find
\begin{equation}
    \braket{n(0,k^j)} = - \bar{\mu}(k^j) \int_0^\infty dt'' e^{-\ce t''} G_{nn}^R(t'',k^j)
\end{equation}
However, the integral on the right hand side is simply the Laplace transform with $z = i \ce$, and so we find
\begin{equation}
    \braket{n(0,k^j)} = - \bar{\mu}(k^j) G^R_{nn}(i \ce, k^j)\,.
\end{equation}
Comparing to the Fourier transform of equation~\eqref{app:LRT:static-susceptibility-diffusion}, we immediately see that we can identify
\begin{equation}
    G_{nn}^{R}(i\ce, k^j) = - \chi_{nn}\,.
\end{equation}
We can therefore solve for $G^R_{nn}(z+i \ce, k^j)$ in equation~\eqref{app:LRT:solve-for-G-diffusion}
\begin{equation}
    G^R_{nn}(z+i\ce, k^j) = -\frac{i z}{\bar{\mu}(k)} \braket{n(z,k^j)} - \chi_{nn}
\end{equation}
Finally, let us insert the expression for $\braket{n(z,k^j)}$ that we previously derived all the way back in equation~\eqref{app:LRT:final-Laplace}, adding in the fact that we know $n(0,k^j) = \chi_{nn} \bar{\mu}(k)$. We then arrive at the final expression for the retarded two-point function, letting $\ce \to 0$:
\begin{equation}
    G^R_{nn}(z,k^j) = \frac{\chi_{nn} D k^2}{i z - D k^2}\,.
\end{equation}
This expression is analytic in the upper-half complex plane in $z$, and undefined in the lower-half plane. Let us analytically continue to the lower half plane, writing instead $G^R_{nn}(\omega,k^j)$. Then $G^R_{nn}(\omega,k^j)$ has a pole in the lower-half complex plane at $\omega = - i D k^2$, which we may now take seriously as the diffusion pole. Finally, let us define $\sigma = \chi_{nn} D$. Then the retarded two-point function may be written
\begin{equation}
    G^R_{nn}(z,k^j) = \frac{\sigma k^2}{i z - (\sigma/\chi_{nn}) k^2}\,.
\end{equation}
The transport coefficient $\sigma$ is what we get if we re-write the diffusion equation~\eqref{app:LRT:diffusion-equation} in terms of the source $\mu$
\begin{equation}
    \de_t n(\mu) - \sigma \de_j \de^j \mu = 0\,.
\end{equation}
The transport coefficient $\sigma$ is not free to choose. Instead, it can be extracted directly from the retarded two-point function:
\begin{equation}\label{app:LRT:Kubo-formula}
    \sigma = - \lim_{\omega \to 0} \lim_{|k| \to 0} \frac{\omega}{k^2} \Im G_{nn}^{R}(\omega,k)\,.
\end{equation}
Expressions of the type~\eqref{app:LRT:Kubo-formula} are known as Kubo formulae. Since one can in principle compute $G_{nn}^{R}(\omega,k)$ from a microscopic theory, the Kubo formula~\eqref{app:LRT:Kubo-formula} fixes the value of $\sigma$ from the microscopic theory. This is why, in the main body of the dissertation, we did not have control over the transport coefficients.

\section{Hydrodynamics}
The generalization of the above to hydrodynamics is immediate. That said, we will not spend as much time on the subject here as it deserves, nor delve into as much detail. There are many more things to say; here, we hope to only convey the bare essentials. Let us consider a collection of degrees of freedom $\varphi_a$, which are coupled to classical sources $\lambda_a$ via
\begin{equation}
    \delta H = - \int d^3 y \lambda_a(t,y^j) \varphi_a(t,y^j)\,.
\end{equation}
One can then find
\begin{equation}
    \delta \braket{\varphi_a(t,x^j)}_{\lambda} = - \int_{-\infty}^{\infty} dt' \int d^3y G^R_{ab}(t-t', x^j-y^j) \lambda_b(t',y^j)\,,
\end{equation}
where we again define again
\begin{equation}
    G_{ab}^R(t-t',x^j-y^j) \equiv - i \theta(t-t') \braket{[\varphi_a(t,x^j), \varphi_b(t,y^j)]}\,.
\end{equation}
We note that, unlike in the previous case in which we only had $G^{R}_{nn}$, we now allow the possibility of cross correlation $G_{ab}^R$. As a simple example, we could consider the correlator $G^R_{\epsilon\,n}$, where $\epsilon$ is the energy density, and $n$ is the charge density as before. Let us again consider the fields at the time $t=0$ when we perform the quench, i.e. $\varphi_a(t=0,x^j)$. Assuming that the sources are small and do not vary too much, we can repeat the analysis of the previous section to find the fields in terms of the sources. In other words,
\begin{equation}
    \varphi_a(0,x^j) = \chi_{ab} \lambda_b(0,x^j)\,.
\end{equation}
Let us consider the specific case where $a=b$. Then
\begin{equation}
    \chi_{aa} = \pder{\varphi_a}{\lambda_a}\biggl|_{\lambda_a = 0} \geq 0\,,
\end{equation}
where there is no summation over repeated indices $a$. The inequality comes from the fact that, defining $\Phi_a = \int d^3x \varphi_a(0,x^j)$,
\begin{equation}
    \chi_{aa} \propto \braket{\Phi_a^2} - \braket{\Phi_a}^2 \geq 0\,.
\end{equation}
Looking back at the main body of the dissertation, this was the source of the demand that the diagonal terms in the static susceptibility~\eqref{ch2:hydro:static_susceptibility} were non-negative. 

Repeating the analysis of the previous section, we can also find that in the limit of small $k$ (i.e. assuming that spatial variations are small) that
\begin{equation}
    G_{ab}^R(z=i\ce, k^j) = - \chi_{ab}\,.
\end{equation}
With this in mind, let us consider the equations of motion which the fields $\varphi_a$ obey. We assume that they are first order in time, which is a crucial assumption -- in the body of the dissertation, we considered theories such as the BDNK theory of hydrodynamics which were second order in time. Such theories are not amenable to the description described hereafter. The fields obey
\begin{equation}
    \de_t \varphi_a +M_{ab}(\de_j) \varphi_b = 0\,.
\end{equation}
Fourier transforming in space yields
\begin{equation}
    \de_t \tilde{\varphi}_a(t,k^j) + M_{ab}(k^j) \tilde{\varphi}_b = 0\,.
\end{equation}
Now, let us apply a Laplace transform to the equation; then, integrating by parts and imposing that $\tilde{\varphi}_a(t=\infty,k^j) = 0$, we find
\begin{equation}
    \lr{ - i z \delta_{ab} + M_{ab}} \bar{\varphi}(z,k^j) = \tilde{\varphi}_a(t=0,k^j)\,.
\end{equation}
However, we know that $\tilde{\varphi}_a(t=0,k^j) = \chi_{ab} \lambda_b(t=0,k^j)$, and so we can write
\begin{equation}\label{app:LRT:varphi_final}
    \bar{\varphi}_a(z,k^j) = \lr{K^{-1}}_{ab} \chi_{bc} \lambda_c(t=0,k^j)\,,
\end{equation}
where we define $K^{-1}$ is the inverse of $K_{ab} = - i z \delta_{ab} + M_{ab}$. Now, following the same approach as in the previous section, we can also write the retarded two-point function in terms of $\bar{\varphi}_a(z,k^j)$ and $\chi_{nn}$. Doing so, and then comparing with~\eqref{app:LRT:varphi_final}, we can write down the final form of the retarded two-point function:
\begin{equation}\label{app:LRT:final-two-point}
    G^R_{ab}(z,k^j) = - \lr{\mathbb{1} + i z K^{-1}}_{ac} \chi_{cb}\,,
\end{equation}
There are two final points to be made here. First of all, one can use the retarded two-point function~\eqref{app:LRT:final-two-point} to extract the microscopic definitions of transport coefficients in the theory via Kubo formulae, in a directly analogous fashion to~\eqref{app:LRT:Kubo-formula}. Of course, knowledge of the equations under consideration are necessary to know exactly which limits to take.

The second point requires a little bit more care -- the application of Onsager relations.

\subsection{Onsager relations}
The microscopic theory which we wish to study is assumed to have time reversal symmetry. This demand has an immediate effect on the two-point functions. Let us begin by considering an arbitrary operator $A$ acted upon by the time-reversal operator $\Theta$. It is then the case that
\begin{equation}\label{app:LRT:A-identity}
    \braket{\beta|A|\alpha} = \braket{\tilde{\alpha}|\Theta A^\dagger \Theta^{-1}|\tilde{\beta}}\,,
\end{equation}
where $\ket{\tilde{\alpha}} = \Theta \ket{\alpha}$ is the time-reversed state. Let us now consider Hermitian operators $\varphi_a$ which transform in a well-prescribed way under the application of $\Theta$, namely
\begin{equation}
    \Theta \varphi_a(t,x^j) \Theta^{-1} = \eta_a \varphi_a(-t,x^j)
\end{equation}
where $\eta_{a} = \pm 1$ is the eigenvalue of $\varphi_a$ under $\Theta$. Let us suppose we are taking expectation values in the grand canonical ensemble. Since time reversal is assumed here to be a symmetry of the microscopic theory, we assume $[H,\Theta] =0$. We can also note that we are free to take the trace in whatever basis we want. We are free to use the energy eigenbasis $\ket{n}$, and we are also free to use their time-reversed counterparts $\ket{\tilde{n}}$. We may therefore note that
\begin{equation}
\begin{split}
    G_{ab}^R(t,x^j) &= - i \theta(t) \braket{[\varphi_a(t,x^j)\varphi_b(0,0)]}\\
    &= - i \theta(t) \sum_{n} \braket{n|e^{- \beta(H-\mu Q)}[\varphi_a(t,x^j),\varphi_b(0,0)]|n}\\
    &= - i \theta(t) \sum_{n} \braket{\tilde{n}|e^{-\beta(H-\mu Q)} \Theta [\varphi_a(t,x^j),\varphi_b(0,0)]^\dagger \Theta^{-1}|\tilde{n}}\\
    &= - i \theta(t) \eta_{a} \eta_b \sum_{n} \braket{\tilde{n}|e^{-\beta(H-\mu Q)}  [\varphi_b(0,0),\varphi_a(-t,x^j)] |\tilde{n}}\\
    &= - i \theta(t) \eta_{a} \eta_b \sum_{n} \braket{\tilde{n}|e^{-\beta(H-\mu Q)}  [\varphi_b(t,-x^j),\varphi_a(0,0)] |\tilde{n}}\\
     &= - i \eta_a \eta_b\theta(t) \braket{[\varphi_b(t,-x^j), \varphi_a(0,0)]} = \eta_a \eta_b G^R_{ba}(t,-x^j)\,,
\end{split}
\end{equation}
where in the third line we have used the identity~\eqref{app:LRT:A-identity}, in the fourth line we have used the fact that $\varphi_{a,b}$ are assumed Hermitian, and in the fifth line we have used spacetime translation invariance. Fourier transforming yields the relation~\cite{Kovtun:2012rj}
\begin{equation}
    G_{ab}^R(\omega,k^j) = \eta_a \eta_b G_{ba}^R(\omega,-k^j)\,.
\end{equation}
These are the Onsager relations. Let us now take the limit of $\omega \to 0$ and small $k$; then we know that $G_{ab}^R(0,k^j) = - \chi_{ab}$. We therefore have the condition on the static susceptibility that
\begin{equation}
    \chi_{ab} = \eta_a \eta_b \chi_{ba}\,.
\end{equation}
 Referring back to the main body of the dissertation, we see that this is the source for the demand that the static susceptibility matrix~\eqref{ch2:hydro:static_susceptibility} was symmetric, as both $\epsilon$ and $n$ are even under time reversal. If Kubo formulae for $G_{ab}^R$ and $G_{ba}^R$ yield differing transport coefficients, the Onsager relations connect them, as we saw indirectly in Chapter~\ref{chapter:extensions}.

    \startappendix{Routh-Hurwitz criteria}
\label{app:RH-criteria}

Here we present the RH criterion for polynomials of order two, three, and four. In the following, for definiteness, we will always assume that $a_n$ is positive. This fixes the criterion from all of the entries of ${\cal G}$ being the same sign to them all being positive.

\section{Order two}
Consider the polynomial
\begin{equation}
    P(z) = a_2 z^2 + a_1 z + a_0
\end{equation}
Then
\begin{subequations}
    \begin{align}
        P_0 &= a_2 z^2 + a_0\,,\\
        P_1 &= a_1 z\,,
    \end{align}
\end{subequations}
and
\begin{equation}
    P_2 = \text{Rem}(P_0,P_1)=a_0\,.
\end{equation}
The set ${\cal G}$ is then given by
\begin{equation}
    {\cal G} = \{a_2, a_1, a_0\}
\end{equation}
and the Routh-Hurwitz criterion is given by the demand that $a_0 >0$, $a_1>0$, and $a_2>0$.

\section{Order three}
Consider the polynomial
\begin{equation}
    P(z) = a_3 z^2 + a_2 z^2 + a_1 z + a_0\,.
\end{equation}
Then the subpolynomials are given by
\begin{subequations}
    \begin{align}
        P_0(z) &= a_3 z^3 + a_1 z\,,\\
        P_1(z) &= a_2 z^2 + a_0\,,\\
        P_2(z) &= \text{Rem}(P_0,P_1) = \lr{a_1 - \frac{a_2 a_3}{a_0}} z\,,\\
        P_3(z) &= \text{Rem}(P_1,P_2) = a_0\,.
    \end{align}
\end{subequations}
and so
\begin{equation}
    {\cal G} = \{ a_3, a_2, \lr{a_1 - \frac{a_0 a_3}{a_2}}, a_0\}
\end{equation}
leading to the conditions $a_3 >0$, $a_2>0$, $a_1 a_2 - a_0 a_3 > 0$, and $a_0 > 0$. We have made use of $a_2>0$ to multiply the third condition through by $a_0$.

\section{Order four}
Consider the polynomial
\begin{equation}
    P(z) = a_4 z^4 + a_3 z^3 + a_2 z^2 + a_1 z + a_0\,.
\end{equation}
Then the subpolynomials are given by
\begin{subequations}
    \begin{align}
        P_0 &= a_4 z^4 + a_2 z^2 + a_0\,,\\
        P_1 &= a_3 z^3 + a_1 z\,,\\
        P_2 &= \text{Rem}(P_0,P_1) = \lr{a_2 - \frac{a_1 a_4}{a_3}}z^2 + a_0\,,\\
        P_3 &= \text{Rem}(P_1,P_2) = \lr{a_1 - \frac{a_0 a_3}{a_2 - \frac{a_1 a_4}{a_3}}} z\,,\\
        P_4 &= \text{Rem}(P_2,P_3) = a_0\,.
    \end{align}
\end{subequations}
and so
\begin{equation}
    {\cal G} = \left\{ a_4, a_3,\lr{a_2 - \frac{a_1 a_4}{a_3}}, \lr{a_1 - \frac{a_0 a_3}{a_2 - \frac{a_1 a_4}{a_3}}}, a_0\right\} \,.
\end{equation}
Therefore, the RH conditions for an order-four polynomial are given by
\begin{equation}
    a_4 > 0, \quad a_3 >0,\quad a_2 a_3 - a_1 a_4 > 0, \quad a_1 a_2 a_3-a_1^2 a_4 - a_0 a_3^2  > 0, \quad a_0 > 0\,,
\end{equation}
where we have used $a_3>0$ to simplify the third inequality and $a_2 a_3 - a_1 a_4>0$ to then simplify the fourth inequality.



	\TOCadd{Bibliography}
	\bibliographystyle{JHEP}
	\bibliography{UvicThesis}

\providecommand{\href}[2]{#2}\begingroup\raggedright\begin{thebibliography}{100}

\bibitem{Hoult:2021gnb}
R.E.~Hoult and P.~Kovtun, \emph{{Causal first-order hydrodynamics from kinetic theory and holography}}, \href{https://doi.org/10.1103/PhysRevD.106.066023}{\emph{Phys. Rev. D} {\bfseries 106} (2022) 066023} [\href{https://arxiv.org/abs/2112.14042}{{\ttfamily 2112.14042}}].

\bibitem{Hoult:2023clg}
R.E.~Hoult and P.~Kovtun, \emph{{Causality and classical dispersion relations}}, \href{https://doi.org/10.1103/PhysRevD.109.046018}{\emph{Phys. Rev. D} {\bfseries 109} (2024) 046018} [\href{https://arxiv.org/abs/2309.11703}{{\ttfamily 2309.11703}}].

\bibitem{Hoult:2024cyx}
R.E.~Hoult and A.~Shukla, \emph{Causal and stable superfluid hydrodynamics}, \href{https://doi.org/10.1007/jhep04(2025)172}{\emph{Journal of High Energy Physics} {\bfseries 2025} (2025) } [\href{https://arxiv.org/abs/2410.22855}{{\ttfamily 2410.22855}}].

\bibitem{Hoult:2024qph}
R.E.~Hoult and P.~Kovtun, \emph{Causality in dissipative relativistic magnetohydrodynamics}, \href{https://doi.org/10.1007/jhep04(2025)009}{\emph{Journal of High Energy Physics} {\bfseries 2025} (2025) } [\href{https://arxiv.org/abs/2411.04966}{{\ttfamily 2411.04966}}].

\bibitem{Hoult:2020eho}
R.E.~Hoult and P.~Kovtun, \emph{{Stable and causal relativistic Navier-Stokes equations}}, \href{https://doi.org/10.1007/JHEP06(2020)067}{\emph{JHEP} {\bfseries 06} (2020) 067} [\href{https://arxiv.org/abs/2004.04102}{{\ttfamily 2004.04102}}].

\bibitem{Gavassino:2024EAS}
L.~Gavassino, ``\emph{Embedding Irreversability in General Relativity}.'' European Astronomical Society Plenary Session, July, 2024, \emph{\href{https://youtu.be/SQ81eBdndHU?si=CYPAcbdhLlrrj2pz\&t=1007}{Youtube}}.

\bibitem{Gavassino:2025hwz}
L.~Gavassino, \emph{{Non-covariant parabolic theories of relativistic diffusion}},  \href{https://arxiv.org/abs/2505.18815}{{\ttfamily 2505.18815}}.

\bibitem{Bistafa:NS2024}
S.R.~{Bistafa}, \emph{{200 Years of the Navier-Stokes Equation}}, \href{https://doi.org/10.48550/arXiv.2401.13669}{\emph{arXiv e-prints} (2023) arXiv:2401.13669} [\href{https://arxiv.org/abs/2401.13669}{{\ttfamily 2401.13669}}].

\bibitem{DARRIGOL2002}
O.~Darrigol, \emph{Between hydrodynamics and elasticity theory: The first five births of the navier-stokes equation}, \href{https://doi.org/10.1007/s004070200000}{\emph{Archive for History of Exact Sciences} {\bfseries 56} (2002) 95–150}.

\bibitem{Eckart-original}
C.~Eckart, \emph{The thermodynamics of irreversible processes. {III.} {R}elativistic theory of the simple fluid}, \href{https://doi.org/10.1103/PhysRev.58.919}{\emph{Phys. Rev.} {\bfseries 58} (1940) 919}.

\bibitem{LL6}
L.D.~Landau and E.M.~Lifshitz, \emph{Fluid Mechanics}, Pergamon (1987).

\bibitem{Romatschke:2017ejr}
P.~Romatschke and U.~Romatschke, \emph{{Relativistic Fluid Dynamics In and Out of Equilibrium}}, Cambridge University Press (2019), \href{https://doi.org/10.1017/9781108651998}{10.1017/9781108651998}, [\href{https://arxiv.org/abs/1712.05815}{{\ttfamily 1712.05815}}].

\bibitem{Rezzolla-Zanotti}
L.~Rezzolla and O.~Zanotti, \emph{Relativistic Hydrodynamics}, Oxford (2013).

\bibitem{Faber:2012rw}
J.A.~Faber and F.A.~Rasio, \emph{{Binary Neutron Star Mergers}}, \href{https://doi.org/10.12942/lrr-2012-8}{\emph{Living Rev. Rel.} {\bfseries 15} (2012) 8} [\href{https://arxiv.org/abs/1204.3858}{{\ttfamily 1204.3858}}].

\bibitem{Abramowicz:2011xu}
M.A.~Abramowicz and P.C.~Fragile, \emph{{Foundations of Black Hole Accretion Disk Theory}}, \href{https://doi.org/10.12942/lrr-2013-1}{\emph{Living Rev. Rel.} {\bfseries 16} (2013) 1} [\href{https://arxiv.org/abs/1104.5499}{{\ttfamily 1104.5499}}].

\bibitem{Policastro:2001yc}
G.~Policastro, D.T.~Son and A.O.~Starinets, \emph{{The Shear viscosity of strongly coupled N=4 supersymmetric Yang-Mills plasma}}, \href{https://doi.org/10.1103/PhysRevLett.87.081601}{\emph{Phys. Rev. Lett.} {\bfseries 87} (2001) 081601} [\href{https://arxiv.org/abs/hep-th/0104066}{{\ttfamily hep-th/0104066}}].

\bibitem{Policastro:2002se}
G.~Policastro, D.T.~Son and A.O.~Starinets, \emph{{From AdS / CFT correspondence to hydrodynamics}}, \href{https://doi.org/10.1088/1126-6708/2002/09/043}{\emph{JHEP} {\bfseries 09} (2002) 043} [\href{https://arxiv.org/abs/hep-th/0205052}{{\ttfamily hep-th/0205052}}].

\bibitem{Kovtun:2003wp}
P.~Kovtun, D.T.~Son and A.O.~Starinets, \emph{{Holography and hydrodynamics: Diffusion on stretched horizons}}, \href{https://doi.org/10.1088/1126-6708/2003/10/064}{\emph{JHEP} {\bfseries 10} (2003) 064} [\href{https://arxiv.org/abs/hep-th/0309213}{{\ttfamily hep-th/0309213}}].

\bibitem{Kovtun:2004de}
P.~Kovtun, D.T.~Son and A.O.~Starinets, \emph{{Viscosity in strongly interacting quantum field theories from black hole physics}}, \href{https://doi.org/10.1103/PhysRevLett.94.111601}{\emph{Phys. Rev. Lett.} {\bfseries 94} (2005) 111601} [\href{https://arxiv.org/abs/hep-th/0405231}{{\ttfamily hep-th/0405231}}].

\bibitem{Kovtun:2005ev}
P.K.~Kovtun and A.O.~Starinets, \emph{{Quasinormal modes and holography}}, \href{https://doi.org/10.1103/PhysRevD.72.086009}{\emph{Phys. Rev. D} {\bfseries 72} (2005) 086009} [\href{https://arxiv.org/abs/hep-th/0506184}{{\ttfamily hep-th/0506184}}].

\bibitem{Bhattacharyya:2008jc}
S.~Bhattacharyya, V.E.~Hubeny, S.~Minwalla and M.~Rangamani, \emph{{Nonlinear Fluid Dynamics from Gravity}}, \href{https://doi.org/10.1088/1126-6708/2008/02/045}{\emph{JHEP} {\bfseries 02} (2008) 045} [\href{https://arxiv.org/abs/0712.2456}{{\ttfamily 0712.2456}}].

\bibitem{Banerjee:2008th}
N.~Banerjee, J.~Bhattacharya, S.~Bhattacharyya, S.~Dutta, R.~Loganayagam and P.~Surowka, \emph{{Hydrodynamics from charged black branes}}, \href{https://doi.org/10.1007/JHEP01(2011)094}{\emph{JHEP} {\bfseries 01} (2011) 094} [\href{https://arxiv.org/abs/0809.2596}{{\ttfamily 0809.2596}}].

\bibitem{Erdmenger:2008rm}
J.~Erdmenger, M.~Haack, M.~Kaminski and A.~Yarom, \emph{{Fluid dynamics of R-charged black holes}}, \href{https://doi.org/10.1088/1126-6708/2009/01/055}{\emph{JHEP} {\bfseries 01} (2009) 055} [\href{https://arxiv.org/abs/0809.2488}{{\ttfamily 0809.2488}}].

\bibitem{Liu:2018kfw}
H.~Liu and P.~Glorioso, \emph{{Lectures on non-equilibrium effective field theories and fluctuating hydrodynamics}}, \href{https://doi.org/10.22323/1.305.0008}{\emph{PoS} {\bfseries TASI2017} (2018) 008} [\href{https://arxiv.org/abs/1805.09331}{{\ttfamily 1805.09331}}].

\bibitem{Hiscock:1985zz}
W.A.~Hiscock and L.~Lindblom, \emph{{Generic instabilities in first-order dissipative relativistic fluid theories}}, \href{https://doi.org/10.1103/PhysRevD.31.725}{\emph{Phys.Rev.} {\bfseries D31} (1985) 725}.

\bibitem{Hiscock:1987zz}
W.A.~Hiscock and L.~Lindblom, \emph{{Linear plane waves in dissipative relativistic fluids}}, \href{https://doi.org/10.1103/PhysRevD.35.3723}{\emph{Phys. Rev.} {\bfseries D35} (1987) 3723}.

\bibitem{Muller:1967zza}
I.~Muller, \emph{{Zum Paradoxon der Warmeleitungstheorie}}, \href{https://doi.org/10.1007/BF01326412}{\emph{Z. Phys.} {\bfseries 198} (1967) 329}.

\bibitem{Israel:1976tn}
W.~Israel, \emph{{Nonstationary irreversible thermodynamics: A Causal relativistic theory}}, \href{https://doi.org/10.1016/0003-4916(76)90064-6}{\emph{Annals Phys.} {\bfseries 100} (1976) 310}.

\bibitem{Israel-Stewart}
W.~Israel and J.M.~Stewart, \emph{Thermodynamics of nonstationary and transient effects in a relativistic gas}, {\emph{Phys. Lett. A} {\bfseries 58} (1976) 213}.

\bibitem{Bemfica:2017wps}
F.S.~Bemfica, M.M.~Disconzi and J.~Noronha, \emph{{Causality and existence of solutions of relativistic viscous fluid dynamics with gravity}}, \href{https://doi.org/10.1103/PhysRevD.98.104064}{\emph{Phys. Rev. D} {\bfseries 98} (2018) 104064} [\href{https://arxiv.org/abs/1708.06255}{{\ttfamily 1708.06255}}].

\bibitem{Kovtun:2019hdm}
P.~Kovtun, \emph{{First-order relativistic hydrodynamics is stable}}, \href{https://doi.org/10.1007/JHEP10(2019)034}{\emph{JHEP} {\bfseries 10} (2019) 034} [\href{https://arxiv.org/abs/1907.08191}{{\ttfamily 1907.08191}}].

\bibitem{Bemfica:2019knx}
F.S.~Bemfica, M.M.~Disconzi and J.~Noronha, \emph{{Nonlinear Causality of General First-Order Relativistic Viscous Hydrodynamics}}, \href{https://doi.org/10.1103/PhysRevD.100.104020}{\emph{Phys. Rev. D} {\bfseries 100} (2019) 104020} [\href{https://arxiv.org/abs/1907.12695}{{\ttfamily 1907.12695}}].

\bibitem{Bemfica:2020zjp}
F.S.~Bemfica, M.M.~Disconzi and J.~Noronha, \emph{{First-Order General-Relativistic Viscous Fluid Dynamics}}, \href{https://doi.org/10.1103/PhysRevX.12.021044}{\emph{Phys. Rev. X} {\bfseries 12} (2022) 021044} [\href{https://arxiv.org/abs/2009.11388}{{\ttfamily 2009.11388}}].

\bibitem{Grozdanov:2019uhi}
S.~Grozdanov, P.K.~Kovtun, A.O.~Starinets and P.~Tadi\'c, \emph{{The complex life of hydrodynamic modes}}, \href{https://doi.org/10.1007/JHEP11(2019)097}{\emph{JHEP} {\bfseries 11} (2019) 097} [\href{https://arxiv.org/abs/1904.12862}{{\ttfamily 1904.12862}}].

\bibitem{Gallegos:2022jow}
A.D.~Gallegos, U.~Gursoy and A.~Yarom, \emph{{Hydrodynamics, spin currents and torsion}}, \href{https://doi.org/10.1007/JHEP05(2023)139}{\emph{JHEP} {\bfseries 05} (2023) 139} [\href{https://arxiv.org/abs/2203.05044}{{\ttfamily 2203.05044}}].

\bibitem{Becattini:2023ouz}
F.~Becattini, A.~Daher and X.-L.~Sheng, \emph{{Entropy current and entropy production in relativistic spin hydrodynamics}}, \href{https://doi.org/10.1016/j.physletb.2024.138533}{\emph{Phys. Lett. B} {\bfseries 850} (2024) 138533} [\href{https://arxiv.org/abs/2309.05789}{{\ttfamily 2309.05789}}].

\bibitem{Belinfante:1940}
F.J.~{Belinfante}, \emph{{On the current and the density of the electric charge, the energy, the linear momentum and the angular momentum of arbitrary fields}}, \href{https://doi.org/10.1016/S0031-8914(40)90091-X}{\emph{Physica} {\bfseries 7} (1940) 449}.

\bibitem{Rosenfeld}
L.~Rosenfeld, \emph{{Sur le tenseur d’impulsion-energie}}, {\emph{Acad\'emie royale de Belgique} (1940) }.

\bibitem{DiFrancesco:1997nk}
P.~Di~Francesco, P.~Mathieu and D.~Senechal, \emph{{Conformal Field Theory}}, Graduate Texts in Contemporary Physics, Springer-Verlag, New York (1997), \href{https://doi.org/10.1007/978-1-4612-2256-9}{10.1007/978-1-4612-2256-9}.

\bibitem{LL2}
L.D.~Landau and E.M.~Lifschits, \emph{{The Classical Theory of Fields}}, Pergamon (1975).

\bibitem{Hernandez:2017mch}
J.~Hernandez and P.~Kovtun, \emph{{Relativistic magnetohydrodynamics}}, \href{https://doi.org/10.1007/JHEP05(2017)001}{\emph{JHEP} {\bfseries 05} (2017) 001} [\href{https://arxiv.org/abs/1703.08757}{{\ttfamily 1703.08757}}].

\bibitem{Jensen:2012jh}
K.~Jensen, M.~Kaminski, P.~Kovtun, R.~Meyer, A.~Ritz and A.~Yarom, \emph{{Towards hydrodynamics without an entropy current}}, \href{https://doi.org/10.1103/PhysRevLett.109.101601}{\emph{Phys. Rev. Lett.} {\bfseries 109} (2012) 101601} [\href{https://arxiv.org/abs/1203.3556}{{\ttfamily 1203.3556}}].

\bibitem{Banerjee:2012iz}
N.~Banerjee, J.~Bhattacharya, S.~Bhattacharyya, S.~Jain, S.~Minwalla and T.~Sharma, \emph{{Constraints on Fluid Dynamics from Equilibrium Partition Functions}}, \href{https://doi.org/10.1007/JHEP09(2012)046}{\emph{JHEP} {\bfseries 09} (2012) 046} [\href{https://arxiv.org/abs/1203.3544}{{\ttfamily 1203.3544}}].

\bibitem{Tolman:1930zza}
R.C.~Tolman, \emph{{On the Weight of Heat and Thermal Equilibrium in General Relativity}}, \href{https://doi.org/10.1103/PhysRev.35.904}{\emph{Phys. Rev.} {\bfseries 35} (1930) 904}.

\bibitem{Tolman}
R.C.~Tolman, \emph{Relativity, Thermodynamics, and Cosmology}, Oxford (1934).

\bibitem{Jensen:2013kka}
K.~Jensen, R.~Loganayagam and A.~Yarom, \emph{{Anomaly inflow and thermal equilibrium}}, \href{https://doi.org/10.1007/JHEP05(2014)134}{\emph{JHEP} {\bfseries 05} (2014) 134} [\href{https://arxiv.org/abs/1310.7024}{{\ttfamily 1310.7024}}].

\bibitem{Kovtun:2012rj}
P.~Kovtun, \emph{{Lectures on hydrodynamic fluctuations in relativistic theories}}, \href{https://doi.org/10.1088/1751-8113/45/47/473001}{\emph{J. Phys.} {\bfseries A45} (2012) 473001} [\href{https://arxiv.org/abs/1205.5040}{{\ttfamily 1205.5040}}].

\bibitem{Grozdanov:2019kge}
S.~Grozdanov, P.K.~Kovtun, A.O.~Starinets and P.~Tadi\'c, \emph{{Convergence of the Gradient Expansion in Hydrodynamics}}, \href{https://doi.org/10.1103/PhysRevLett.122.251601}{\emph{Phys. Rev. Lett.} {\bfseries 122} (2019) 251601} [\href{https://arxiv.org/abs/1904.01018}{{\ttfamily 1904.01018}}].

\bibitem{Heller:2021oxl}
M.P.~Heller, A.~Serantes, M.~Spali\'nski, V.~Svensson and B.~Withers, \emph{{Hydrodynamic Gradient Expansion Diverges beyond Bjorken Flow}}, \href{https://doi.org/10.1103/PhysRevLett.128.122302}{\emph{Phys. Rev. Lett.} {\bfseries 128} (2022) 122302} [\href{https://arxiv.org/abs/2110.07621}{{\ttfamily 2110.07621}}].

\bibitem{Heller:2021yjh}
M.P.~Heller, A.~Serantes, M.~Spali\'nski, V.~Svensson and B.~Withers, \emph{{Relativistic Hydrodynamics: A Singulant Perspective}}, \href{https://doi.org/10.1103/PhysRevX.12.041010}{\emph{Phys. Rev. X} {\bfseries 12} (2022) 041010} [\href{https://arxiv.org/abs/2112.12794}{{\ttfamily 2112.12794}}].

\bibitem{Taub:1948zz}
A.H.~Taub, \emph{{Relativistic Rankine-Hugoniot Equations}}, \href{https://doi.org/10.1103/PhysRev.74.328}{\emph{Phys. Rev.} {\bfseries 74} (1948) 328}.

\bibitem{IsraelAnileChoquetBruhat:1989}
A.M.~{Anile} and Y.~{Choquet-Bruhat}, \emph{{Relativistic Fluid Dynamics}}, vol.~1385 (1989), \href{https://doi.org/10.1007/BFb0084027}{10.1007/BFb0084027}.

\bibitem{Pandya:2021ief}
A.~Pandya and F.~Pretorius, \emph{{Numerical exploration of first-order relativistic hydrodynamics}}, \href{https://doi.org/10.1103/PhysRevD.104.023015}{\emph{Phys. Rev. D} {\bfseries 104} (2021) 023015} [\href{https://arxiv.org/abs/2104.00804}{{\ttfamily 2104.00804}}].

\bibitem{Pandya:2022pif}
A.~Pandya, E.R.~Most and F.~Pretorius, \emph{{Conservative finite volume scheme for first-order viscous relativistic hydrodynamics}}, \href{https://doi.org/10.1103/PhysRevD.105.123001}{\emph{Phys. Rev. D} {\bfseries 105} (2022) 123001} [\href{https://arxiv.org/abs/2201.12317}{{\ttfamily 2201.12317}}].

\bibitem{Pandya:2022sff}
A.~Pandya, E.R.~Most and F.~Pretorius, \emph{{Causal, stable first-order viscous relativistic hydrodynamics with ideal gas microphysics}}, \href{https://doi.org/10.1103/PhysRevD.106.123036}{\emph{Phys. Rev. D} {\bfseries 106} (2022) 123036} [\href{https://arxiv.org/abs/2209.09265}{{\ttfamily 2209.09265}}].

\bibitem{Freistuhler:2021lla}
H.~Freistuhler, \emph{{Nonexistence and existence of shock profiles in the Bemfica-Disconzi-Noronha model}}, \href{https://doi.org/10.1103/PhysRevD.103.124045}{\emph{Phys. Rev. D} {\bfseries 103} (2021) 124045} [\href{https://arxiv.org/abs/2103.16661}{{\ttfamily 2103.16661}}].

\bibitem{Kovtun:2022vas}
P.~Kovtun, \emph{{Temperature in relativistic fluids}}, \href{https://doi.org/10.1103/PhysRevD.107.086012}{\emph{Phys. Rev. D} {\bfseries 107} (2023) 086012} [\href{https://arxiv.org/abs/2210.15605}{{\ttfamily 2210.15605}}].

\bibitem{weinberg:1972}
{S.~Weinberg}, \emph{{Gravitation and Cosmology}}, {John Wiley \& Sons} ({1972}).

\bibitem{Courant-Hilbert}
R.~Courant and D.~Hilbert, \emph{Methods of Mathematical Physics II. Partial Differential Equations}, Wiley (1989).

\bibitem{Leray:1958}
J.~Leray, \emph{{Hyperbolic Differential Equations}}, Institute for Advanced Study (1953).

\bibitem{Lax:2006}
P.D.~Lax, \emph{{Hyperbolic Partial Differential Equations}}, American Mathematical Society (2006).

\bibitem{Disconzi:2023rtt}
M.M.~Disconzi, \emph{{Recent developments in mathematical aspects of relativistic fluids}}, \href{https://doi.org/10.1007/s41114-024-00052-x}{\emph{Living Rev. Rel.} {\bfseries 27} (2024) 6} [\href{https://arxiv.org/abs/2308.09844}{{\ttfamily 2308.09844}}].

\bibitem{BBM:1972}
T.B.~Benjamin, J.L.~Bona and J.J.~Mahoney, \emph{Model equations for long waves in nonlinear dispersive systems}, \href{https://doi.org/10.1098/rsta.1972.0032}{\emph{Philosophical Transactions of the Royal Society of London. Series A, Mathematical and Physical Sciences} {\bfseries 272} (1972) 47–78}.

\bibitem{Gavassino:2021owo}
L.~Gavassino, \emph{{Can We Make Sense of Dissipation without Causality?}}, \href{https://doi.org/10.1103/PhysRevX.12.041001}{\emph{Phys. Rev. X} {\bfseries 12} (2022) 041001} [\href{https://arxiv.org/abs/2111.05254}{{\ttfamily 2111.05254}}].

\bibitem{Hiscock:1983zz}
W.A.~Hiscock and L.~Lindblom, \emph{{Stability and causality in dissipative relativistic fluids}}, \href{https://doi.org/10.1016/0003-4916(83)90288-9}{\emph{Annals Phys.} {\bfseries 151} (1983) 466}.

\bibitem{Noronha:2021syv}
J.~Noronha, M.~Spali\'nski and E.~Speranza, \emph{{Transient Relativistic Fluid Dynamics in a General Hydrodynamic Frame}}, \href{https://doi.org/10.1103/PhysRevLett.128.252302}{\emph{Phys. Rev. Lett.} {\bfseries 128} (2022) 252302} [\href{https://arxiv.org/abs/2105.01034}{{\ttfamily 2105.01034}}].

\bibitem{Bemfica:2020xym}
F.S.~Bemfica, M.M.~Disconzi, V.~Hoang, J.~Noronha and M.~Radosz, \emph{{Nonlinear Constraints on Relativistic Fluids Far from Equilibrium}}, \href{https://doi.org/10.1103/PhysRevLett.126.222301}{\emph{Phys. Rev. Lett.} {\bfseries 126} (2021) 222301} [\href{https://arxiv.org/abs/2005.11632}{{\ttfamily 2005.11632}}].

\bibitem{Baier:2007ix}
R.~Baier, P.~Romatschke, D.T.~Son, A.O.~Starinets and M.A.~Stephanov, \emph{{Relativistic viscous hydrodynamics, conformal invariance, and holography}}, \href{https://doi.org/10.1088/1126-6708/2008/04/100}{\emph{JHEP} {\bfseries 04} (2008) 100} [\href{https://arxiv.org/abs/0712.2451}{{\ttfamily 0712.2451}}].

\bibitem{Jain:2023obu}
A.~Jain and P.~Kovtun, \emph{{Schwinger-Keldysh effective field theory for stable and causal relativistic hydrodynamics}}, \href{https://doi.org/10.1007/JHEP01(2024)162}{\emph{JHEP} {\bfseries 01} (2024) 162} [\href{https://arxiv.org/abs/2309.00511}{{\ttfamily 2309.00511}}].

\bibitem{Hoult2020:thesis}
R.E.~Hoult, \emph{Constraints effecting stability and causality of charged relativistic hydrodynamics},  Master's thesis, University of Victoria, July, 2020.

\bibitem{Abboud:2023hos}
N.~Abboud, E.~Speranza and J.~Noronha, \emph{{Causal and stable first-order chiral hydrodynamics}}, \href{https://doi.org/10.1103/PhysRevD.109.094007}{\emph{Phys. Rev. D} {\bfseries 109} (2024) 094007} [\href{https://arxiv.org/abs/2308.02928}{{\ttfamily 2308.02928}}].

\bibitem{routh1877treatise}
E.~Routh, \emph{A Treatise on the Stability of a Given State of Motion: Particularly Steady Motion. Being the Essay to which the Adams Prize was Adjudged in 1877, in the University of Cambridge}, Macmillan and Company (1877).

\bibitem{Hurwitz1895}
A.~Hurwitz, \emph{Ueber die bedingungen, unter welchen eine gleichung nur wurzeln mit negativen reellen theilen besitzt}, \href{https://doi.org/10.1007/bf01446812}{\emph{Mathematische Annalen} {\bfseries 46} (1895) 273–284}.

\bibitem{korn2013mathematical}
G.~Korn and T.~Korn, \emph{Mathematical Handbook for Scientists and Engineers: Definitions, Theorems, and Formulas for Reference and Review}, Dover Civil and Mechanical Engineering, Dover Publications (2013).

\bibitem{Hastir_2023}
A.~Hastir and R.~Muolo, \emph{A generalized routh–hurwitz criterion for the stability analysis of polynomials with complex coefficients: Application to the pi-control of vibrating structures}, \href{https://doi.org/10.1016/j.ifacsc.2023.100235}{\emph{IFAC Journal of Systems and Control} {\bfseries 26} (2023) 100235}.

\bibitem{Ogata:2009}
K.~Ogata, \emph{{Modern Control Engineering}}, Pearson (2009).

\bibitem{Gargantini-Schur}
I.~Gargantini, \emph{The numerical stability of the schur-cohn criterion}, \href{https://doi.org/10.1137/0708003}{\emph{SIAM Journal on Numerical Analysis} {\bfseries 8} (1971) 24} [\href{https://arxiv.org/abs/https://doi.org/10.1137/0708003}{{\ttfamily https://doi.org/10.1137/0708003}}].

\bibitem{Washington1982}
L.C.~Washington, \emph{Introduction to Cyclotomic Fields}, Springer US (1982), \href{https://doi.org/10.1007/978-1-4684-0133-2}{10.1007/978-1-4684-0133-2}.

\bibitem{Israel2009}
W.~Israel, \emph{Relativistic thermodynamics},  in \emph{E.C.G. Stueckelberg, An Unconventional Figure of Twentieth Century Physics}, p.~101–113, Birkh\"{a}user Basel (2009), \href{https://doi.org/10.1007/978-3-7643-8878-2_8}{DOI}.

\bibitem{Heller:2022ejw}
M.P.~Heller, A.~Serantes, M.~Spali\'nski and B.~Withers, \emph{{Rigorous Bounds on Transport from Causality}}, \href{https://doi.org/10.1103/PhysRevLett.130.261601}{\emph{Phys. Rev. Lett.} {\bfseries 130} (2023) 261601} [\href{https://arxiv.org/abs/2212.07434}{{\ttfamily 2212.07434}}].

\bibitem{Itzykson:1980rh}
C.~Itzykson and J.B.~Zuber, \emph{{Quantum Field Theory}}, International Series In Pure and Applied Physics, McGraw-Hill, New York (1980).

\bibitem{Streater-Wightman}
R.F.~Streater and A.S.~Wightman, \emph{{PCT, spin and statistics, and all that}}, Princeton (2000).

\bibitem{Gavassino:2023myj}
L.~Gavassino, \emph{{Bounds on transport from hydrodynamic stability}}, \href{https://doi.org/10.1016/j.physletb.2023.137854}{\emph{Phys. Lett. B} {\bfseries 840} (2023) 137854} [\href{https://arxiv.org/abs/2301.06651}{{\ttfamily 2301.06651}}].

\bibitem{Heller:2023jtd}
M.P.~Heller, A.~Serantes, M.~Spali\'nski and B.~Withers, \emph{{The Hydrohedron: Bootstrapping Relativistic Hydrodynamics}},  \href{https://arxiv.org/abs/2305.07703}{{\ttfamily 2305.07703}}.

\bibitem{Wang:2023csj}
D.-L.~Wang and S.~Pu, \emph{{Stability and causality criteria in linear mode analysis: stability means causality}},  \href{https://arxiv.org/abs/2309.11708}{{\ttfamily 2309.11708}}.

\bibitem{Cartwright:2024rus}
C.~Cartwright, \emph{{Example of the convergence of hydrodynamics in strong external fields}}, \href{https://doi.org/10.1103/PhysRevD.110.026021}{\emph{Phys. Rev. D} {\bfseries 110} (2024) 026021} [\href{https://arxiv.org/abs/2403.12638}{{\ttfamily 2403.12638}}].

\bibitem{WangPu}
D.-L.~Wang and S.~Pu, \emph{{Stability and causality criteria in linear mode analysis: stability means causality}}, .

\bibitem{Berti:2009kk}
E.~Berti, V.~Cardoso and A.O.~Starinets, \emph{{Quasinormal modes of black holes and black branes}}, \href{https://doi.org/10.1088/0264-9381/26/16/163001}{\emph{Class. Quant. Grav.} {\bfseries 26} (2009) 163001} [\href{https://arxiv.org/abs/0905.2975}{{\ttfamily 0905.2975}}].

\bibitem{Grozdanov:2023txs}
S.~Grozdanov and M.~Vrbica, \emph{{Pole-skipping of gravitational waves in the backgrounds of four-dimensional massive black holes}}, \href{https://doi.org/10.1140/epjc/s10052-023-12273-5}{\emph{Eur. Phys. J. C} {\bfseries 83} (2023) 1103} [\href{https://arxiv.org/abs/2303.15921}{{\ttfamily 2303.15921}}].

\bibitem{Wall-singular-points}
C.T.C.~Wall, \emph{{Singular Points of Plane Curves}}, Cambridge University Press (2004), \href{https://doi.org/10.1017/CBO9780511617560}{10.1017/CBO9780511617560}.

\bibitem{DeGroot:1980dk}
S.R.~De~Groot, \emph{{Relativistic Kinetic Theory. Principles and Applications}} (1980).

\bibitem{Cercignani1988}
C.~Cercignani, \emph{The Boltzmann Equation and Its Applications}, Springer New York (1988), \href{https://doi.org/10.1007/978-1-4612-1039-9}{10.1007/978-1-4612-1039-9}.

\bibitem{Cercignani2002}
C.~Cercignani and G.M.~Kremer, \emph{The Relativistic Boltzmann Equation: Theory and Applications}, Birkh\"{a}user Basel (2002), \href{https://doi.org/10.1007/978-3-0348-8165-4}{10.1007/978-3-0348-8165-4}.

\bibitem{Yvon:1935}
J.~Yvon, \emph{La th{\'e}orie statistique des fluides et l'{\'e}quation d'{\'e}tat}, Actualit{\'e}s scientifiques et industrielles : hydrodynamique, acoustique: Th{\'e}ories m{\'e}caniques, Hermann \& cie (1935).

\bibitem{Bogoliubov:1946}
N.N.~Bogoliubov, \emph{Kinetic equations}, {\emph{Journal of Physics USSR} {\bfseries 10} (1946) }.

\bibitem{Kirkwood1946}
J.G.~Kirkwood, \emph{The statistical mechanical theory of transport processes i. general theory}, \href{https://doi.org/10.1063/1.1724117}{\emph{The Journal of Chemical Physics} {\bfseries 14} (1946) 180–201}.

\bibitem{Born:1946}
M.~Born and H.S.~Green, \emph{A general kinetic theory of liquids. i. the molecular distribution functions}, {\emph{Proceedings of the Royal Society of London. Series A, Mathematical and Physical Sciences} {\bfseries 188} (1946) 10}.

\bibitem{Denicol:2012cn}
G.S.~Denicol, H.~Niemi, E.~Molnar and D.H.~Rischke, \emph{{Derivation of transient relativistic fluid dynamics from the Boltzmann equation}}, \href{https://doi.org/10.1103/PhysRevD.85.114047, 10.1103/PhysRevD.91.039902}{\emph{Phys. Rev.} {\bfseries D85} (2012) 114047} [\href{https://arxiv.org/abs/1202.4551}{{\ttfamily 1202.4551}}].

\bibitem{Grad1949}
H.~Grad, \emph{On the kinetic theory of rarefied gases}, \href{https://doi.org/10.1002/cpa.3160020403}{\emph{Communications on Pure and Applied Mathematics} {\bfseries 2} (1949) 331–407}.

\bibitem{Israel:1979wp}
W.~Israel and J.M.~Stewart, \emph{{Transient relativistic thermodynamics and kinetic theory}}, \href{https://doi.org/10.1016/0003-4916(79)90130-1}{\emph{Annals Phys.} {\bfseries 118} (1979) 341}.

\bibitem{Wagner:2022ayd}
D.~Wagner, A.~Palermo and V.E.~Ambru\c{s}, \emph{{Inverse-Reynolds-dominance approach to transient fluid dynamics}}, \href{https://doi.org/10.1103/PhysRevD.106.016013}{\emph{Phys. Rev. D} {\bfseries 106} (2022) 016013} [\href{https://arxiv.org/abs/2203.12608}{{\ttfamily 2203.12608}}].

\bibitem{Maldacena:1997re}
J.M.~Maldacena, \emph{{The Large N limit of superconformal field theories and supergravity}}, \href{https://doi.org/10.4310/ATMP.1998.v2.n2.a1}{\emph{Adv. Theor. Math. Phys.} {\bfseries 2} (1998) 231} [\href{https://arxiv.org/abs/hep-th/9711200}{{\ttfamily hep-th/9711200}}].

\bibitem{Natsuume2015}
M.~Natsuume, \emph{AdS/CFT Duality User Guide}, Springer Japan (2015), \href{https://doi.org/10.1007/978-4-431-55441-7}{10.1007/978-4-431-55441-7}.

\bibitem{Aharony:1999ti}
O.~Aharony, S.S.~Gubser, J.M.~Maldacena, H.~Ooguri and Y.~Oz, \emph{{Large N field theories, string theory and gravity}}, \href{https://doi.org/10.1016/S0370-1573(99)00083-6}{\emph{Phys. Rept.} {\bfseries 323} (2000) 183} [\href{https://arxiv.org/abs/hep-th/9905111}{{\ttfamily hep-th/9905111}}].

\bibitem{Deppe:2015qsa}
N.~Deppe and A.R.~Frey, \emph{{Classes of Stable Initial Data for Massless and Massive Scalars in Anti-de Sitter Spacetime}}, \href{https://doi.org/10.1007/JHEP12(2015)004}{\emph{JHEP} {\bfseries 12} (2015) 004} [\href{https://arxiv.org/abs/1508.02709}{{\ttfamily 1508.02709}}].

\bibitem{Cownden:2017bog}
B.~Cownden, N.~Deppe and A.R.~Frey, \emph{{Phase diagram of stability for massive scalars in anti\textendash{}de Sitter spacetime}}, \href{https://doi.org/10.1103/PhysRevD.102.026015}{\emph{Phys. Rev. D} {\bfseries 102} (2020) 026015} [\href{https://arxiv.org/abs/1711.00454}{{\ttfamily 1711.00454}}].

\bibitem{Cownden:2020oge}
B.~Cownden, \emph{{Gravitational collapse in anti-de Sitter spacetime}}, Ph.D. thesis, Manitoba U., 2020.

\bibitem{Bizon:2011gg}
P.~Bizon and A.~Rostworowski, \emph{{On weakly turbulent instability of anti-de Sitter space}}, \href{https://doi.org/10.1103/PhysRevLett.107.031102}{\emph{Phys. Rev. Lett.} {\bfseries 107} (2011) 031102} [\href{https://arxiv.org/abs/1104.3702}{{\ttfamily 1104.3702}}].

\bibitem{Carroll:2004st}
S.M.~Carroll, \emph{{Spacetime and Geometry}: {An Introduction to General Relativity}}, Cambridge University Press (7, 2019), \href{https://doi.org/10.1017/9781108770385}{10.1017/9781108770385}.

\bibitem{Iqbal:2008by}
N.~Iqbal and H.~Liu, \emph{{Universality of the hydrodynamic limit in AdS/CFT and the membrane paradigm}}, \href{https://doi.org/10.1103/PhysRevD.79.025023}{\emph{Phys. Rev. D} {\bfseries 79} (2009) 025023} [\href{https://arxiv.org/abs/0809.3808}{{\ttfamily 0809.3808}}].

\bibitem{Balasubramanian:1999re}
V.~Balasubramanian and P.~Kraus, \emph{{A Stress tensor for Anti-de Sitter gravity}}, \href{https://doi.org/10.1007/s002200050764}{\emph{Commun. Math. Phys.} {\bfseries 208} (1999) 413} [\href{https://arxiv.org/abs/hep-th/9902121}{{\ttfamily hep-th/9902121}}].

\bibitem{Bekenstein:1973ur}
J.D.~Bekenstein, \emph{{Black holes and entropy}}, \href{https://doi.org/10.1103/PhysRevD.7.2333}{\emph{Phys. Rev. D} {\bfseries 7} (1973) 2333}.

\bibitem{Hawking:1975vcx}
S.W.~Hawking, \emph{{Particle Creation by Black Holes}}, \href{https://doi.org/10.1007/BF02345020}{\emph{Commun. Math. Phys.} {\bfseries 43} (1975) 199}.

\bibitem{Thorne:1986iy}
K.S.~Thorne, R.H.~Price and D.A.~Macdonald, eds., \emph{{Black Holes: The Membrane Paradigm}} (1986).

\bibitem{Witten:2024upt}
E.~Witten, \emph{{Introduction to black hole thermodynamics}}, \href{https://doi.org/10.1140/epjp/s13360-025-06288-y}{\emph{Eur. Phys. J. Plus} {\bfseries 140} (2025) 430} [\href{https://arxiv.org/abs/2412.16795}{{\ttfamily 2412.16795}}].

\bibitem{Armas:2020mpr}
J.~Armas and A.~Jain, \emph{{Effective field theory for hydrodynamics without boosts}}, \href{https://doi.org/10.21468/SciPostPhys.11.3.054}{\emph{SciPost Phys.} {\bfseries 11} (2021) 054} [\href{https://arxiv.org/abs/2010.15782}{{\ttfamily 2010.15782}}].

\bibitem{Bhambure:2024axa}
J.~Bhambure, A.~Mazeliauskas, J.-F.~Paquet, R.~Singh, M.~Singh, D.~Teaney et~al., \emph{{Relativistic Viscous Hydrodynamics in the Density Frame: Numerical Tests and Comparisons}},  \href{https://arxiv.org/abs/2412.10303}{{\ttfamily 2412.10303}}.

\bibitem{Biswas:2020rps}
R.~Biswas, A.~Dash, N.~Haque, S.~Pu and V.~Roy, \emph{{Causality and stability in relativistic viscous non-resistive magneto-fluid dynamics}}, \href{https://doi.org/10.1007/JHEP10(2020)171}{\emph{JHEP} {\bfseries 10} (2020) 171} [\href{https://arxiv.org/abs/2007.05431}{{\ttfamily 2007.05431}}].

\bibitem{Gavassino:2021crz}
L.~Gavassino, M.~Antonelli and B.~Haskell, \emph{{Extending Israel and Stewart hydrodynamics to relativistic superfluids via Carter{\textquoteright}s multifluid approach}}, \href{https://doi.org/10.1103/PhysRevD.105.045011}{\emph{Phys. Rev. D} {\bfseries 105} (2022) 045011} [\href{https://arxiv.org/abs/2110.05546}{{\ttfamily 2110.05546}}].

\bibitem{Anile}
A.M.~Anile, \emph{Relativistic fluids and magneto-fluids}, Cambridge University Press (1989).

\bibitem{Elitzur1975}
S.~Elitzur, \emph{Impossibility of spontaneously breaking local symmetries}, \href{https://doi.org/10.1103/physrevd.12.3978}{\emph{Physical Review D} {\bfseries 12} (1975) 3978–3982}.

\bibitem{TongGT}
D.~Tong, ``Lectures on gauge theory.'' \url{https://www.damtp.cam.ac.uk/user/tong/gaugetheory.html}, 2018.

\bibitem{Schafer-Nameki:2023jdn}
S.~Schafer-Nameki, \emph{{ICTP lectures on (non-)invertible generalized symmetries}}, \href{https://doi.org/10.1016/j.physrep.2024.01.007}{\emph{Phys. Rept.} {\bfseries 1063} (2024) 1} [\href{https://arxiv.org/abs/2305.18296}{{\ttfamily 2305.18296}}].

\bibitem{Shao:2023gho}
S.-H.~Shao, \emph{{What's Done Cannot Be Undone: TASI Lectures on Non-Invertible Symmetries}},  \href{https://arxiv.org/abs/2308.00747}{{\ttfamily 2308.00747}}.

\bibitem{Schubring:2014iwa}
D.~Schubring, \emph{{Dissipative String Fluids}}, \href{https://doi.org/10.1103/PhysRevD.91.043518}{\emph{Phys. Rev.} {\bfseries D91} (2015) 043518} [\href{https://arxiv.org/abs/1412.3135}{{\ttfamily 1412.3135}}].

\bibitem{Grozdanov:2016tdf}
S.~Grozdanov, D.M.~Hofman and N.~Iqbal, \emph{{Generalized global symmetries and dissipative magnetohydrodynamics}}, \href{https://doi.org/10.1103/PhysRevD.95.096003}{\emph{Phys. Rev. D} {\bfseries 95} (2017) 096003} [\href{https://arxiv.org/abs/1610.07392}{{\ttfamily 1610.07392}}].

\bibitem{Armas:2018atq}
J.~Armas and A.~Jain, \emph{{Magnetohydrodynamics as superfluidity}}, \href{https://doi.org/10.1103/PhysRevLett.122.141603}{\emph{Phys. Rev. Lett.} {\bfseries 122} (2019) 141603} [\href{https://arxiv.org/abs/1808.01939}{{\ttfamily 1808.01939}}].

\bibitem{Armas:2018zbe}
J.~Armas and A.~Jain, \emph{{One-form superfluids \& magnetohydrodynamics}}, \href{https://doi.org/10.1007/JHEP01(2020)041}{\emph{JHEP} {\bfseries 01} (2020) 041} [\href{https://arxiv.org/abs/1811.04913}{{\ttfamily 1811.04913}}].

\bibitem{Gaiotto:2014kfa}
D.~Gaiotto, A.~Kapustin, N.~Seiberg and B.~Willett, \emph{{Generalized Global Symmetries}}, \href{https://doi.org/10.1007/JHEP02(2015)172}{\emph{JHEP} {\bfseries 02} (2015) 172} [\href{https://arxiv.org/abs/1412.5148}{{\ttfamily 1412.5148}}].

\bibitem{Lake:2018dqm}
E.~Lake, \emph{{Higher-form symmetries and spontaneous symmetry breaking}},  \href{https://arxiv.org/abs/1802.07747}{{\ttfamily 1802.07747}}.

\bibitem{Hofman:2018lfz}
D.M.~Hofman and N.~Iqbal, \emph{{Goldstone modes and photonization for higher form symmetries}}, \href{https://doi.org/10.21468/SciPostPhys.6.1.006}{\emph{SciPost Phys.} {\bfseries 6} (2019) 006} [\href{https://arxiv.org/abs/1802.09512}{{\ttfamily 1802.09512}}].

\bibitem{Armas:2022wvb}
J.~Armas and F.~Camilloni, \emph{{A stable and causal model of magnetohydrodynamics}}, \href{https://doi.org/10.1088/1475-7516/2022/10/039}{\emph{JCAP} {\bfseries 10} (2022) 039} [\href{https://arxiv.org/abs/2201.06847}{{\ttfamily 2201.06847}}].

\bibitem{ALFVN1942}
H.~ALFVÉN, \emph{Existence of electromagnetic-hydrodynamic waves}, \href{https://doi.org/10.1038/150405d0}{\emph{Nature} {\bfseries 150} (1942) 405–406}.

\bibitem{Fang:2024hxa}
Z.~Fang, K.~Hattori and J.~Hu, \emph{{First-order spin magnetohydrodynamics}},  \href{https://arxiv.org/abs/2410.00721}{{\ttfamily 2410.00721}}.

\bibitem{Fang:2024skm}
Z.~Fang, K.~Hattori and J.~Hu, \emph{{Analytic solutions for the linearized first-order magnetohydrodynamics and implications for causality and stability}}, \href{https://doi.org/10.1103/PhysRevD.110.056049}{\emph{Phys. Rev. D} {\bfseries 110} (2024) 056049} [\href{https://arxiv.org/abs/2402.18601}{{\ttfamily 2402.18601}}].

\bibitem{Porth:2016rfi}
O.~Porth, H.~Olivares, Y.~Mizuno, Z.~Younsi, L.~Rezzolla, M.~Moscibrodzka et~al., \emph{{The black hole accretion code}}, \href{https://doi.org/10.1186/s40668-017-0020-2}{\emph{Comput. Astrophys. Cosmol.} {\bfseries 4} (2017) 1} [\href{https://arxiv.org/abs/1611.09720}{{\ttfamily 1611.09720}}].

\bibitem{Kapitza1938}
P.~Kapitza, \emph{Viscosity of liquid helium below the $\lambda$-point}, \href{https://doi.org/10.1038/141074a0}{\emph{Nature} {\bfseries 141} (1938) 74–74}.

\bibitem{ALLEN1938}
J.F.~ALLEN and A.D.~MISENER, \emph{Flow phenomena in liquid helium ii}, \href{https://doi.org/10.1038/142643a0}{\emph{Nature} {\bfseries 142} (1938) 643–644}.

\bibitem{TISZA1}
L.~TISZA, \emph{Transport phenomena in helium ii}, \href{https://doi.org/10.1038/141913a0}{\emph{Nature} {\bfseries 141} (1938) 913–913}.

\bibitem{tisza:1938:cr1}
L.~Tisza, \emph{Sur la supraconductibilit e thermique de l'helium ii liquide et la statistique de bose-einstein}, {\emph{CR Acad. Sci} {\bfseries 207} (1938) 1035}.

\bibitem{tisza:1940:vol1}
L.~Tisza, \emph{Sur la th{\'e}orie des liquides quantiques. application a l'h{\'e}lium liquide}, {\emph{Journal de Physique et le Radium} {\bfseries 1} (1940) 164}.

\bibitem{tisza:1940:vol2}
L.~Tisza, \emph{Sur la th{\'e}orie des liquides quantiques. application {\`a} l'h{\'e}lium liquide. ii}, {\emph{Journal de Physique et le Radium} {\bfseries 1} (1940) 350}.

\bibitem{Landau:1941}
L.~Landau, \emph{Theory of the superfluidity of helium ii}, \href{https://doi.org/10.1103/PhysRev.60.356}{\emph{Phys. Rev.} {\bfseries 60} (1941) 356}.

\bibitem{Landau:1941ussr}
L.D.~Landau, \emph{{The theory of superfuidity of helium II}}, \href{https://doi.org/10.1016/B978-0-08-010586-4.50051-1}{\emph{J. Phys. (USSR)} {\bfseries 5} (1941) 71}.

\bibitem{MIGDAL:1959655}
A.~Migdal, \emph{Superfluidity and the moments of inertia of nuclei}, \href{https://doi.org/https://doi.org/10.1016/0029-5582(59)90264-0}{\emph{Nuclear Physics} {\bfseries 13} (1959) 655}.

\bibitem{BAYM1969}
G.~Baym, C.~Pethick and D.~Pines, \emph{Superfluidity in neutron stars}, \href{https://doi.org/10.1038/224673a0}{\emph{Nature} {\bfseries 224} (1969) 673}.

\bibitem{Haskell:2017lkl}
B.~Haskell and A.~Sedrakian, \emph{{Superfluidity and Superconductivity in Neutron Stars}}, \href{https://doi.org/10.1007/978-3-319-97616-7_8}{\emph{Astrophys. Space Sci. Libr.} {\bfseries 457} (2018) 401} [\href{https://arxiv.org/abs/1709.10340}{{\ttfamily 1709.10340}}].

\bibitem{Balibar:2017}
S.~Balibar, \emph{Laszlo {Tisza} and the two-fluid model of superfluidity}, \href{https://doi.org/10.1016/j.crhy.2017.10.016}{\emph{Comptes Rendus. Physique} {\bfseries 18} (2017) 586}.

\bibitem{ClarkThesis}
J.A.~Clark, Ph.D. thesis, MIT, Cambridge, USA, 1963.

\bibitem{putterman1974}
S.J.~Putterman, \emph{Superfluid Hydrodynamics}, vol.~3 of \emph{Low Temperature Physics Series}, North-Holland Publishing Company (1974).

\bibitem{Carter:1992}
B.~Carter and I.M.~Khalatnikov, \emph{Equivalence of convective and potential variational derivations of covariant superfluid dynamics}, \href{https://doi.org/10.1103/PhysRevD.45.4536}{\emph{Phys. Rev. D} {\bfseries 45} (1992) 4536}.

\bibitem{Herzog:2011ec}
C.P.~Herzog, N.~Lisker, P.~Surowka and A.~Yarom, \emph{{Transport in holographic superfluids}}, \href{https://doi.org/10.1007/JHEP08(2011)052}{\emph{JHEP} {\bfseries 08} (2011) 052} [\href{https://arxiv.org/abs/1101.3330}{{\ttfamily 1101.3330}}].

\bibitem{Bhattacharya:2011eea}
J.~Bhattacharya, S.~Bhattacharyya and S.~Minwalla, \emph{{Dissipative Superfluid dynamics from gravity}}, \href{https://doi.org/10.1007/JHEP04(2011)125}{\emph{JHEP} {\bfseries 04} (2011) 125} [\href{https://arxiv.org/abs/1101.3332}{{\ttfamily 1101.3332}}].

\bibitem{Bhattacharya:2011tra}
J.~Bhattacharya, S.~Bhattacharyya, S.~Minwalla and A.~Yarom, \emph{{A Theory of first order dissipative superfluid dynamics}}, \href{https://doi.org/10.1007/JHEP05(2014)147}{\emph{JHEP} {\bfseries 05} (2014) 147} [\href{https://arxiv.org/abs/1105.3733}{{\ttfamily 1105.3733}}].

\bibitem{Pekker-Varma}
D.~{Pekker} and C.M.~{Varma}, \emph{{Amplitude/Higgs Modes in Condensed Matter Physics}}, \href{https://doi.org/10.1146/annurev-conmatphys-031214-014350}{\emph{Annual Review of Condensed Matter Physics} {\bfseries 6} (2015) 269} [\href{https://arxiv.org/abs/1406.2968}{{\ttfamily 1406.2968}}].

\bibitem{Donos:2022xfd}
A.~Donos and C.~Pantelidou, \emph{{Higgs/amplitude mode dynamics from holography}}, \href{https://doi.org/10.1007/JHEP08(2022)246}{\emph{JHEP} {\bfseries 08} (2022) 246} [\href{https://arxiv.org/abs/2205.06294}{{\ttfamily 2205.06294}}].

\bibitem{Gouteraux:2022qix}
B.~Gout\'eraux, E.~Mefford and F.~Sottovia, \emph{{Critical superflows and thermodynamic instabilities in superfluids}}, \href{https://doi.org/10.1103/PhysRevD.108.L081903}{\emph{Phys. Rev. D} {\bfseries 108} (2023) L081903} [\href{https://arxiv.org/abs/2212.10410}{{\ttfamily 2212.10410}}].

\bibitem{Lier:2025wfw}
R.~Lier, A.~Jain, J.~Armas and O.~Porth, \emph{{Resistive relativistic magnetohydrodynamics without Amperes Law}},  \href{https://arxiv.org/abs/2501.04638}{{\ttfamily 2501.04638}}.

\bibitem{Baiotti:2016qnr}
L.~Baiotti and L.~Rezzolla, \emph{{Binary neutron star mergers: a review of Einstein{\textquoteright}s richest laboratory}}, \href{https://doi.org/10.1088/1361-6633/aa67bb}{\emph{Rept. Prog. Phys.} {\bfseries 80} (2017) 096901} [\href{https://arxiv.org/abs/1607.03540}{{\ttfamily 1607.03540}}].

\bibitem{Plumberg:2021bme}
C.~Plumberg, D.~Almaalol, T.~Dore, J.~Noronha and J.~Noronha-Hostler, \emph{{Causality violations in realistic simulations of heavy-ion collisions}}, \href{https://doi.org/10.1103/PhysRevC.105.L061901}{\emph{Phys. Rev. C} {\bfseries 105} (2022) L061901} [\href{https://arxiv.org/abs/2103.15889}{{\ttfamily 2103.15889}}].

\bibitem{Forster}
D.~Forster, \emph{Hydrodynamic Fluctuations, Broken Symmetry, And Correlation Functions}, Addison-Wesley (1990).

\end{thebibliography}\endgroup

\end{document}